%% file: main.tex
\documentclass[acmsmall,screen,nonacm]{acmart}

\pdfoutput=1

\usepackage{sosy-paper}
\usepackage{changepage}
\usepackage{syntax}
\usepackage{balance}
\usepackage{longtable}
\usepackage{rotating}
\usepackage{colortbl}
\usepackage{bm}
\usepackage[table]{xcolor}
\usepackage{tcolorbox}
\usetikzlibrary{calc}

\newcommand{\ttbox}[1]{\colorbox{black!8}{\footnotesize\texttt{#1}\tiny\strut}}
\newlength\multilen
\newcommand{\multi}[2][\multilen]{\parbox[c]{#1}{\baselineskip=0pt \vspace{2.5pt}\strut #2 \vspace{2.5pt}}}

\makeatletter
\def\myrulefill{\leavevmode\leaders\hrule height .7ex width 1ex depth -0.6ex\hfill\kern\z@}
\makeatother

\usepackage{xcolor}
\definecolor{commentgreen}{RGB}{255,0,0}

\renewcommand{\cmark}{{\color{green}\ding{51}}}
\renewcommand{\xmark}{{\color{red}\ding{55}}}

\def\orcidID#1{\href{https://orcid.org/#1}{\protect\raisebox{2.25pt}{\protect\includegraphics{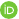}}}}

\title{A Modular Program-Transformation Framework for Reducing Specifications to Reachability}
\renewcommand{\shorttitle}{A Modular Framework for Transforming Specifications to Reachability}

\begin{document}

\author{Dirk Beyer}
\orcid{0000-0003-4832-7662}
\affiliation{%
  \department{Institute for Informatics}
  \institution{LMU Munich}
  \city{Munich}
  \country{Germany}
}
\email{dirk.beyer@sosy.ifi.lmu.de}

\author{Marek Jankola}
\orcid{0009-0008-7961-190X}
\affiliation{%
  \department{Institute for Informatics}
  \institution{LMU Munich}
  \city{Munich}
  \country{Germany}
}
\email{marek.jankola@sosy.ifi.lmu.de}

\author{Marian Lingsch-Rosenfeld}
\orcid{0000-0002-8172-3184}
\affiliation{%
  \department{Institute for Informatics}
  \institution{LMU Munich}
  \city{Munich}
  \country{Germany}
}
\email{marian.lingsch@sosy.ifi.lmu.de}

\author{Tian Xia}
\orcid{0009-0006-4767-8435}
\affiliation{%
  \department{Institute for Informatics}
  \institution{LMU Munich}
  \city{Munich}
  \country{Germany}
}
\email{xiatian1996cd@gmail.com}

\author{Xiyue Zheng}
\orcid{0009-0007-7298-6998}
\affiliation{%
  \department{Institute for Informatics}
  \institution{LMU Munich}
  \city{Munich}
  \country{Germany}
}
\email{xiyuezheng@outlook.com}

\renewcommand{\shortauthors}{D.~Beyer, M.~Jankola, M.~Lingsch-Rosenfeld, T.~Xia and X.~Zheng}

\newcommand{\mypaperkeywords}{
Formal Verification,
Model Checking,
Software Verification,
Program Analysis,
Specification,
Monitoring,
Specification Reduction,
Reachability
}

\input{commands.tex}

\begin{abstract}
    \input{sections/abstract.tex}
\end{abstract}

\ccsdesc[500]{Software and its engineering~Formal software verification}
\ccsdesc[300]{Software and its engineering~Formal methods}
\ccsdesc[300]{Theory of computation~Verification by model checking}
\ccsdesc[300]{Theory of computation~Program reasoning}

\keywords{\mypaperkeywords}

\maketitle

\keywords{\mypaperkeywords}

\input{sections/introduction.tex}
\input{sections/related-work.tex}
\input{sections/background.tex}

\input{sections/transformation.tex}
\input{sections/showcase.tex}
\input{sections/expressiveness.tex}
\input{sections/evaluation.tex}
\input{sections/threats-to-validity.tex}
\input{sections/conclusion.tex}

\inlineheadingbf{Data-Availability Statement}
TransVer is open-source and can be found under \url{https://gitlab.com/sosy-lab/software/transver}.

\inlineheadingbf{Funding Statement}
This project was funded in part by the Deutsche Forschungsgemeinschaft (DFG)
--- \href{http://gepris.dfg.de/gepris/projekt/378803395}{378803395} (ConVeY),

\bibliography{bib/dbeyer,bib/sw,bib/sw-tmp,bib/artifacts,bib/svcomp,bib/svcomp-artifacts,bib/testcomp,bib/testcomp-artifacts,bib/websites-tmp}

\vfill
\pagebreak

\end{document}

%% file: commands.tex
\newcommand{\rqeffectiveness}{RQ\,2}
\newcommand{\rqefficiency}{RQ\,3}
\newcommand{\rqapplicability}{RQ\,1}
\newcommand{\rqmodularity}{RQ\,4}

\newcommand{\llbox}[1]{
	\vspace*{2pt plus 3pt minus 1pt}
	\begin{tcolorbox}[width=\columnwidth, colframe=black, boxrule=0.25mm, top=1mm, left=1mm, right=1mm, bottom=1mm]
		#1
	\end{tcolorbox}
}

\newcommand{\reachs}{\color{red}R\color{black}}
\newcommand{\overf}{\color{brown}O\color{black}}
\newcommand{\mem}{\color{green}M\color{black}}
\newcommand{\term}{\color{blue}T\color{black}}

\newcommand\plotpath{figures/results/tex}
\input{\plotpath/plot-defs}
\input{\plotpath/data-commands}
\input{\plotpath/manual-commands}

%% file: figures/results/tex/plot-defs.tex
\pgfplotsset{quantile plot/.style={
    width=6cm,
    height=4cm,
    scale only axis,
    /pgfplots/table/x expr={\coordindex+1},
    /pgfplots/table/y index=3,
    /pgfplots/table/header=false,
    ylabel shift=-1em,
    ticklabel style={font={\smaller}},
    xmin=0,
    ymin=1,
    ymax=2000,
    legend cell align={left},
    legend style={at={(0,1)}, anchor=north west, outer xsep=5pt, outer ysep=5pt, fill=none, font={\smaller}},
    legend columns=3,
    cycle multiindex list={
      orange, red, blue, green, black\nextlist
      mark list*\nextlist
      solid, densely dashed, densely dashdotdotted, densely dotted},
  },
  every axis plot/.append style={thick}
}

\onlyifstandalone{
  \pgfplotsset{quantile plot/.append style={
      ylabel=CPU time (\second),
    },
  }
}

\newcommand\addgraph[2]{{
  \newcommand\csvfile{\plotpath/\detokenize{#2}}
  \IfFileExists\csvfile{
    \addplot+ table {\csvfile}; \addlegendentry{#1}
  }{
    \addplot coordinates {};
  }
}}

\newcommand\addgraphtable[4]{{
  \newcommand\csvfile{\plotpath/\detokenize{#4}}
  \IfFileExists\csvfile{
    \addplot+[#1] table[#2] {\csvfile}; \addlegendentry{#3}
  }{
    \addplot coordinates {};
  }
}}

%% file: figures/results/tex/data-commands.tex
\providecommand\StoreBenchExecResult[7]{\expandafter\newcommand\csname#1#2#3#4#5#6\endcsname{#7}}%
\StoreBenchExecResult{UltimateBenchmarkDefinition}{TRReachabilityReachsafetyLoops}{Total}{}{Count}{}{386}%
\StoreBenchExecResult{UltimateBenchmarkDefinition}{TRReachabilityReachsafetyLoops}{Total}{}{Cputime}{}{38565.473063695}%
\StoreBenchExecResult{UltimateBenchmarkDefinition}{TRReachabilityReachsafetyLoops}{Total}{}{Cputime}{Avg}{99.91055197848445595854922280}%
\StoreBenchExecResult{UltimateBenchmarkDefinition}{TRReachabilityReachsafetyLoops}{Total}{}{Cputime}{Median}{18.820008490}%
\StoreBenchExecResult{UltimateBenchmarkDefinition}{TRReachabilityReachsafetyLoops}{Total}{}{Cputime}{Min}{13.734675152}%
\StoreBenchExecResult{UltimateBenchmarkDefinition}{TRReachabilityReachsafetyLoops}{Total}{}{Cputime}{Max}{961.197408191}%
\StoreBenchExecResult{UltimateBenchmarkDefinition}{TRReachabilityReachsafetyLoops}{Total}{}{Cputime}{Stdev}{256.0420667079349644276637972}%
\StoreBenchExecResult{UltimateBenchmarkDefinition}{TRReachabilityReachsafetyLoops}{Total}{}{Walltime}{}{34085.443553328979670}%
\StoreBenchExecResult{UltimateBenchmarkDefinition}{TRReachabilityReachsafetyLoops}{Total}{}{Walltime}{Avg}{88.30425791017870380829015544}%
\StoreBenchExecResult{UltimateBenchmarkDefinition}{TRReachabilityReachsafetyLoops}{Total}{}{Walltime}{Median}{11.1392723130993545}%
\StoreBenchExecResult{UltimateBenchmarkDefinition}{TRReachabilityReachsafetyLoops}{Total}{}{Walltime}{Min}{8.417610278120264}%
\StoreBenchExecResult{UltimateBenchmarkDefinition}{TRReachabilityReachsafetyLoops}{Total}{}{Walltime}{Max}{943.4945834709797}%
\StoreBenchExecResult{UltimateBenchmarkDefinition}{TRReachabilityReachsafetyLoops}{Total}{}{Walltime}{Stdev}{247.1458913031718372382981215}%
\StoreBenchExecResult{UltimateBenchmarkDefinition}{TRReachabilityReachsafetyLoops}{Correct}{}{Count}{}{327}%
\StoreBenchExecResult{UltimateBenchmarkDefinition}{TRReachabilityReachsafetyLoops}{Correct}{}{Cputime}{}{9052.159638288}%
\StoreBenchExecResult{UltimateBenchmarkDefinition}{TRReachabilityReachsafetyLoops}{Correct}{}{Cputime}{Avg}{27.68244537702752293577981651}%
\StoreBenchExecResult{UltimateBenchmarkDefinition}{TRReachabilityReachsafetyLoops}{Correct}{}{Cputime}{Median}{17.656330436}%
\StoreBenchExecResult{UltimateBenchmarkDefinition}{TRReachabilityReachsafetyLoops}{Correct}{}{Cputime}{Min}{13.734675152}%
\StoreBenchExecResult{UltimateBenchmarkDefinition}{TRReachabilityReachsafetyLoops}{Correct}{}{Cputime}{Max}{827.376071622}%
\StoreBenchExecResult{UltimateBenchmarkDefinition}{TRReachabilityReachsafetyLoops}{Correct}{}{Cputime}{Stdev}{60.99183490174827765535156244}%
\StoreBenchExecResult{UltimateBenchmarkDefinition}{TRReachabilityReachsafetyLoops}{Correct}{}{Walltime}{}{6075.967262004967697}%
\StoreBenchExecResult{UltimateBenchmarkDefinition}{TRReachabilityReachsafetyLoops}{Correct}{}{Walltime}{Avg}{18.58093963915892262079510703}%
\StoreBenchExecResult{UltimateBenchmarkDefinition}{TRReachabilityReachsafetyLoops}{Correct}{}{Walltime}{Median}{10.50513687194325}%
\StoreBenchExecResult{UltimateBenchmarkDefinition}{TRReachabilityReachsafetyLoops}{Correct}{}{Walltime}{Min}{8.417610278120264}%
\StoreBenchExecResult{UltimateBenchmarkDefinition}{TRReachabilityReachsafetyLoops}{Correct}{}{Walltime}{Max}{801.8854906398337}%
\StoreBenchExecResult{UltimateBenchmarkDefinition}{TRReachabilityReachsafetyLoops}{Correct}{}{Walltime}{Stdev}{57.41646843109672043641238835}%
\StoreBenchExecResult{UltimateBenchmarkDefinition}{TRReachabilityReachsafetyLoops}{Correct}{False}{Count}{}{68}%
\StoreBenchExecResult{UltimateBenchmarkDefinition}{TRReachabilityReachsafetyLoops}{Correct}{False}{Cputime}{}{1852.458968640}%
\StoreBenchExecResult{UltimateBenchmarkDefinition}{TRReachabilityReachsafetyLoops}{Correct}{False}{Cputime}{Avg}{27.24204365647058823529411765}%
\StoreBenchExecResult{UltimateBenchmarkDefinition}{TRReachabilityReachsafetyLoops}{Correct}{False}{Cputime}{Median}{16.2637290885}%
\StoreBenchExecResult{UltimateBenchmarkDefinition}{TRReachabilityReachsafetyLoops}{Correct}{False}{Cputime}{Min}{14.931480819}%
\StoreBenchExecResult{UltimateBenchmarkDefinition}{TRReachabilityReachsafetyLoops}{Correct}{False}{Cputime}{Max}{151.89544018}%
\StoreBenchExecResult{UltimateBenchmarkDefinition}{TRReachabilityReachsafetyLoops}{Correct}{False}{Cputime}{Stdev}{27.17608800830456089837227694}%
\StoreBenchExecResult{UltimateBenchmarkDefinition}{TRReachabilityReachsafetyLoops}{Correct}{False}{Walltime}{}{1141.977043764898537}%
\StoreBenchExecResult{UltimateBenchmarkDefinition}{TRReachabilityReachsafetyLoops}{Correct}{False}{Walltime}{Avg}{16.79378005536615495588235294}%
\StoreBenchExecResult{UltimateBenchmarkDefinition}{TRReachabilityReachsafetyLoops}{Correct}{False}{Walltime}{Median}{9.7499743105145175}%
\StoreBenchExecResult{UltimateBenchmarkDefinition}{TRReachabilityReachsafetyLoops}{Correct}{False}{Walltime}{Min}{9.012153699994087}%
\StoreBenchExecResult{UltimateBenchmarkDefinition}{TRReachabilityReachsafetyLoops}{Correct}{False}{Walltime}{Max}{105.06277953390963}%
\StoreBenchExecResult{UltimateBenchmarkDefinition}{TRReachabilityReachsafetyLoops}{Correct}{False}{Walltime}{Stdev}{18.27000884432568588581951127}%

\StoreBenchExecResult{UltimateBenchmarkDefinition}{TRReachabilityReachsafetyLoops}{Correct}{True}{Count}{}{259}%
\StoreBenchExecResult{UltimateBenchmarkDefinition}{TRReachabilityReachsafetyLoops}{Correct}{True}{Cputime}{}{7199.700669648}%
\StoreBenchExecResult{UltimateBenchmarkDefinition}{TRReachabilityReachsafetyLoops}{Correct}{True}{Cputime}{Avg}{27.79807208358301158301158301}%
\StoreBenchExecResult{UltimateBenchmarkDefinition}{TRReachabilityReachsafetyLoops}{Correct}{True}{Cputime}{Median}{18.062888166}%
\StoreBenchExecResult{UltimateBenchmarkDefinition}{TRReachabilityReachsafetyLoops}{Correct}{True}{Cputime}{Min}{13.734675152}%
\StoreBenchExecResult{UltimateBenchmarkDefinition}{TRReachabilityReachsafetyLoops}{Correct}{True}{Cputime}{Max}{827.376071622}%
\StoreBenchExecResult{UltimateBenchmarkDefinition}{TRReachabilityReachsafetyLoops}{Correct}{True}{Cputime}{Stdev}{67.10229406944182886229379089}%
\StoreBenchExecResult{UltimateBenchmarkDefinition}{TRReachabilityReachsafetyLoops}{Correct}{True}{Walltime}{}{4933.990218240069160}%
\StoreBenchExecResult{UltimateBenchmarkDefinition}{TRReachabilityReachsafetyLoops}{Correct}{True}{Walltime}{Avg}{19.05015528278018980694980695}%
\StoreBenchExecResult{UltimateBenchmarkDefinition}{TRReachabilityReachsafetyLoops}{Correct}{True}{Walltime}{Median}{10.680176462046802}%
\StoreBenchExecResult{UltimateBenchmarkDefinition}{TRReachabilityReachsafetyLoops}{Correct}{True}{Walltime}{Min}{8.417610278120264}%
\StoreBenchExecResult{UltimateBenchmarkDefinition}{TRReachabilityReachsafetyLoops}{Correct}{True}{Walltime}{Max}{801.8854906398337}%
\StoreBenchExecResult{UltimateBenchmarkDefinition}{TRReachabilityReachsafetyLoops}{Correct}{True}{Walltime}{Stdev}{63.82386107001481941772162400}%

\StoreBenchExecResult{UltimateBenchmarkDefinition}{TRReachabilityReachsafetyLoops}{Error}{}{Count}{}{30}%
\StoreBenchExecResult{UltimateBenchmarkDefinition}{TRReachabilityReachsafetyLoops}{Error}{}{Cputime}{}{28813.743200669}%
\StoreBenchExecResult{UltimateBenchmarkDefinition}{TRReachabilityReachsafetyLoops}{Error}{}{Cputime}{Avg}{960.4581066889666666666666667}%
\StoreBenchExecResult{UltimateBenchmarkDefinition}{TRReachabilityReachsafetyLoops}{Error}{}{Cputime}{Median}{960.344853491}%
\StoreBenchExecResult{UltimateBenchmarkDefinition}{TRReachabilityReachsafetyLoops}{Error}{}{Cputime}{Min}{960.2404222}%
\StoreBenchExecResult{UltimateBenchmarkDefinition}{TRReachabilityReachsafetyLoops}{Error}{}{Cputime}{Max}{961.197408191}%
\StoreBenchExecResult{UltimateBenchmarkDefinition}{TRReachabilityReachsafetyLoops}{Error}{}{Cputime}{Stdev}{0.2552554284413256266531850470}%
\StoreBenchExecResult{UltimateBenchmarkDefinition}{TRReachabilityReachsafetyLoops}{Error}{}{Walltime}{}{27590.1726199153345}%
\StoreBenchExecResult{UltimateBenchmarkDefinition}{TRReachabilityReachsafetyLoops}{Error}{}{Walltime}{Avg}{919.6724206638444833333333333}%
\StoreBenchExecResult{UltimateBenchmarkDefinition}{TRReachabilityReachsafetyLoops}{Error}{}{Walltime}{Median}{928.9248429329600}%
\StoreBenchExecResult{UltimateBenchmarkDefinition}{TRReachabilityReachsafetyLoops}{Error}{}{Walltime}{Min}{840.5776327881031}%
\StoreBenchExecResult{UltimateBenchmarkDefinition}{TRReachabilityReachsafetyLoops}{Error}{}{Walltime}{Max}{943.4945834709797}%
\StoreBenchExecResult{UltimateBenchmarkDefinition}{TRReachabilityReachsafetyLoops}{Error}{}{Walltime}{Stdev}{23.31273652156701462997404850}%
\StoreBenchExecResult{UltimateBenchmarkDefinition}{TRReachabilityReachsafetyLoops}{Error}{Timeout}{Count}{}{30}%
\StoreBenchExecResult{UltimateBenchmarkDefinition}{TRReachabilityReachsafetyLoops}{Error}{Timeout}{Cputime}{}{28813.743200669}%
\StoreBenchExecResult{UltimateBenchmarkDefinition}{TRReachabilityReachsafetyLoops}{Error}{Timeout}{Cputime}{Avg}{960.4581066889666666666666667}%
\StoreBenchExecResult{UltimateBenchmarkDefinition}{TRReachabilityReachsafetyLoops}{Error}{Timeout}{Cputime}{Median}{960.344853491}%
\StoreBenchExecResult{UltimateBenchmarkDefinition}{TRReachabilityReachsafetyLoops}{Error}{Timeout}{Cputime}{Min}{960.2404222}%
\StoreBenchExecResult{UltimateBenchmarkDefinition}{TRReachabilityReachsafetyLoops}{Error}{Timeout}{Cputime}{Max}{961.197408191}%
\StoreBenchExecResult{UltimateBenchmarkDefinition}{TRReachabilityReachsafetyLoops}{Error}{Timeout}{Cputime}{Stdev}{0.2552554284413256266531850470}%
\StoreBenchExecResult{UltimateBenchmarkDefinition}{TRReachabilityReachsafetyLoops}{Error}{Timeout}{Walltime}{}{27590.1726199153345}%
\StoreBenchExecResult{UltimateBenchmarkDefinition}{TRReachabilityReachsafetyLoops}{Error}{Timeout}{Walltime}{Avg}{919.6724206638444833333333333}%
\StoreBenchExecResult{UltimateBenchmarkDefinition}{TRReachabilityReachsafetyLoops}{Error}{Timeout}{Walltime}{Median}{928.9248429329600}%
\StoreBenchExecResult{UltimateBenchmarkDefinition}{TRReachabilityReachsafetyLoops}{Error}{Timeout}{Walltime}{Min}{840.5776327881031}%
\StoreBenchExecResult{UltimateBenchmarkDefinition}{TRReachabilityReachsafetyLoops}{Error}{Timeout}{Walltime}{Max}{943.4945834709797}%
\StoreBenchExecResult{UltimateBenchmarkDefinition}{TRReachabilityReachsafetyLoops}{Error}{Timeout}{Walltime}{Stdev}{23.31273652156701462997404850}%
\StoreBenchExecResult{UltimateBenchmarkDefinition}{TRReachabilityReachsafetyLoops}{Unknown}{}{Count}{}{29}%
\StoreBenchExecResult{UltimateBenchmarkDefinition}{TRReachabilityReachsafetyLoops}{Unknown}{}{Cputime}{}{699.570224738}%
\StoreBenchExecResult{UltimateBenchmarkDefinition}{TRReachabilityReachsafetyLoops}{Unknown}{}{Cputime}{Avg}{24.12311119786206896551724138}%
\StoreBenchExecResult{UltimateBenchmarkDefinition}{TRReachabilityReachsafetyLoops}{Unknown}{}{Cputime}{Median}{24.117285407}%
\StoreBenchExecResult{UltimateBenchmarkDefinition}{TRReachabilityReachsafetyLoops}{Unknown}{}{Cputime}{Min}{23.265346192}%
\StoreBenchExecResult{UltimateBenchmarkDefinition}{TRReachabilityReachsafetyLoops}{Unknown}{}{Cputime}{Max}{25.563454557}%
\StoreBenchExecResult{UltimateBenchmarkDefinition}{TRReachabilityReachsafetyLoops}{Unknown}{}{Cputime}{Stdev}{0.5845002460532088899496100312}%
\StoreBenchExecResult{UltimateBenchmarkDefinition}{TRReachabilityReachsafetyLoops}{Unknown}{}{Walltime}{}{419.303671408677473}%
\StoreBenchExecResult{UltimateBenchmarkDefinition}{TRReachabilityReachsafetyLoops}{Unknown}{}{Walltime}{Avg}{14.45874728995439562068965517}%
\StoreBenchExecResult{UltimateBenchmarkDefinition}{TRReachabilityReachsafetyLoops}{Unknown}{}{Walltime}{Median}{14.447189366910607}%
\StoreBenchExecResult{UltimateBenchmarkDefinition}{TRReachabilityReachsafetyLoops}{Unknown}{}{Walltime}{Min}{14.085790056968108}%
\StoreBenchExecResult{UltimateBenchmarkDefinition}{TRReachabilityReachsafetyLoops}{Unknown}{}{Walltime}{Max}{14.825340517098084}%
\StoreBenchExecResult{UltimateBenchmarkDefinition}{TRReachabilityReachsafetyLoops}{Unknown}{}{Walltime}{Stdev}{0.2023410981668299124834615360}%
\StoreBenchExecResult{UltimateBenchmarkDefinition}{TRReachabilityReachsafetyLoops}{Unknown}{Unknown}{Count}{}{29}%
\StoreBenchExecResult{UltimateBenchmarkDefinition}{TRReachabilityReachsafetyLoops}{Unknown}{Unknown}{Cputime}{}{699.570224738}%
\StoreBenchExecResult{UltimateBenchmarkDefinition}{TRReachabilityReachsafetyLoops}{Unknown}{Unknown}{Cputime}{Avg}{24.12311119786206896551724138}%
\StoreBenchExecResult{UltimateBenchmarkDefinition}{TRReachabilityReachsafetyLoops}{Unknown}{Unknown}{Cputime}{Median}{24.117285407}%
\StoreBenchExecResult{UltimateBenchmarkDefinition}{TRReachabilityReachsafetyLoops}{Unknown}{Unknown}{Cputime}{Min}{23.265346192}%
\StoreBenchExecResult{UltimateBenchmarkDefinition}{TRReachabilityReachsafetyLoops}{Unknown}{Unknown}{Cputime}{Max}{25.563454557}%
\StoreBenchExecResult{UltimateBenchmarkDefinition}{TRReachabilityReachsafetyLoops}{Unknown}{Unknown}{Cputime}{Stdev}{0.5845002460532088899496100312}%
\StoreBenchExecResult{UltimateBenchmarkDefinition}{TRReachabilityReachsafetyLoops}{Unknown}{Unknown}{Walltime}{}{419.303671408677473}%
\StoreBenchExecResult{UltimateBenchmarkDefinition}{TRReachabilityReachsafetyLoops}{Unknown}{Unknown}{Walltime}{Avg}{14.45874728995439562068965517}%
\StoreBenchExecResult{UltimateBenchmarkDefinition}{TRReachabilityReachsafetyLoops}{Unknown}{Unknown}{Walltime}{Median}{14.447189366910607}%
\StoreBenchExecResult{UltimateBenchmarkDefinition}{TRReachabilityReachsafetyLoops}{Unknown}{Unknown}{Walltime}{Min}{14.085790056968108}%
\StoreBenchExecResult{UltimateBenchmarkDefinition}{TRReachabilityReachsafetyLoops}{Unknown}{Unknown}{Walltime}{Max}{14.825340517098084}%
\StoreBenchExecResult{UltimateBenchmarkDefinition}{TRReachabilityReachsafetyLoops}{Unknown}{Unknown}{Walltime}{Stdev}{0.2023410981668299124834615360}%
\ifdefined\UltimateUltimateTotalCount\else\edef\UltimateUltimateTotalCount{0}\fi
\ifdefined\UltimateUltimateCorrectCount\else\edef\UltimateUltimateCorrectCount{0}\fi
\ifdefined\UltimateUltimateCorrectTrueCount\else\edef\UltimateUltimateCorrectTrueCount{0}\fi
\ifdefined\UltimateUltimateCorrectFalseCount\else\edef\UltimateUltimateCorrectFalseCount{0}\fi
\ifdefined\UltimateUltimateWrongCount\else\edef\UltimateUltimateWrongCount{0}\fi
\ifdefined\UltimateUltimateWrongTrueCount\else\edef\UltimateUltimateWrongTrueCount{0}\fi
\ifdefined\UltimateUltimateWrongFalseCount\else\edef\UltimateUltimateWrongFalseCount{0}\fi
\ifdefined\UltimateUltimateErrorTimeoutCount\else\edef\UltimateUltimateErrorTimeoutCount{0}\fi
\ifdefined\UltimateUltimateErrorOutOfMemoryCount\else\edef\UltimateUltimateErrorOutOfMemoryCount{0}\fi
\ifdefined\UltimateUltimateCorrectCputime\else\edef\UltimateUltimateCorrectCputime{0}\fi
\ifdefined\UltimateUltimateCorrectCputimeAvg\else\edef\UltimateUltimateCorrectCputimeAvg{None}\fi
\ifdefined\UltimateUltimateCorrectWalltime\else\edef\UltimateUltimateCorrectWalltime{0}\fi
\ifdefined\UltimateUltimateCorrectWalltimeAvg\else\edef\UltimateUltimateCorrectWalltimeAvg{None}\fi
\edef\UltimateUltimateErrorOtherInconclusiveCount{\the\numexpr \UltimateUltimateTotalCount - \UltimateUltimateCorrectCount - \UltimateUltimateWrongTrueCount - \UltimateUltimateWrongFalseCount - \UltimateUltimateErrorTimeoutCount - \UltimateUltimateErrorOutOfMemoryCount \relax}
\providecommand\StoreBenchExecResult[7]{\expandafter\newcommand\csname#1#2#3#4#5#6\endcsname{#7}}%
\StoreBenchExecResult{CpacheckerBenchmarkDefinition}{TRReachabilityReachsafetyLoops}{Total}{}{Count}{}{386}%
\StoreBenchExecResult{CpacheckerBenchmarkDefinition}{TRReachabilityReachsafetyLoops}{Total}{}{Cputime}{}{224017.235774047}%
\StoreBenchExecResult{CpacheckerBenchmarkDefinition}{TRReachabilityReachsafetyLoops}{Total}{}{Cputime}{Avg}{580.3555330933860103626943005}%
\StoreBenchExecResult{CpacheckerBenchmarkDefinition}{TRReachabilityReachsafetyLoops}{Total}{}{Cputime}{Median}{903.0252000735}%
\StoreBenchExecResult{CpacheckerBenchmarkDefinition}{TRReachabilityReachsafetyLoops}{Total}{}{Cputime}{Min}{2.755320309}%
\StoreBenchExecResult{CpacheckerBenchmarkDefinition}{TRReachabilityReachsafetyLoops}{Total}{}{Cputime}{Max}{962.100257316}%
\StoreBenchExecResult{CpacheckerBenchmarkDefinition}{TRReachabilityReachsafetyLoops}{Total}{}{Cputime}{Stdev}{416.4759519033526649016480317}%
\StoreBenchExecResult{CpacheckerBenchmarkDefinition}{TRReachabilityReachsafetyLoops}{Total}{}{Walltime}{}{208475.9759639103425493}%
\StoreBenchExecResult{CpacheckerBenchmarkDefinition}{TRReachabilityReachsafetyLoops}{Total}{}{Walltime}{Avg}{540.0932019790423382106217617}%
\StoreBenchExecResult{CpacheckerBenchmarkDefinition}{TRReachabilityReachsafetyLoops}{Total}{}{Walltime}{Median}{846.45672906492835}%
\StoreBenchExecResult{CpacheckerBenchmarkDefinition}{TRReachabilityReachsafetyLoops}{Total}{}{Walltime}{Min}{1.479888177011162}%
\StoreBenchExecResult{CpacheckerBenchmarkDefinition}{TRReachabilityReachsafetyLoops}{Total}{}{Walltime}{Max}{928.1734276819043}%
\StoreBenchExecResult{CpacheckerBenchmarkDefinition}{TRReachabilityReachsafetyLoops}{Total}{}{Walltime}{Stdev}{392.7877215346956984394188484}%
\StoreBenchExecResult{CpacheckerBenchmarkDefinition}{TRReachabilityReachsafetyLoops}{Correct}{}{Count}{}{126}%
\StoreBenchExecResult{CpacheckerBenchmarkDefinition}{TRReachabilityReachsafetyLoops}{Correct}{}{Cputime}{}{14126.083647092}%
\StoreBenchExecResult{CpacheckerBenchmarkDefinition}{TRReachabilityReachsafetyLoops}{Correct}{}{Cputime}{Avg}{112.1117749769206349206349206}%
\StoreBenchExecResult{CpacheckerBenchmarkDefinition}{TRReachabilityReachsafetyLoops}{Correct}{}{Cputime}{Median}{9.5508585515}%
\StoreBenchExecResult{CpacheckerBenchmarkDefinition}{TRReachabilityReachsafetyLoops}{Correct}{}{Cputime}{Min}{6.542492465}%
\StoreBenchExecResult{CpacheckerBenchmarkDefinition}{TRReachabilityReachsafetyLoops}{Correct}{}{Cputime}{Max}{773.353600287}%
\StoreBenchExecResult{CpacheckerBenchmarkDefinition}{TRReachabilityReachsafetyLoops}{Correct}{}{Cputime}{Stdev}{178.7487796182673113337025440}%
\StoreBenchExecResult{CpacheckerBenchmarkDefinition}{TRReachabilityReachsafetyLoops}{Correct}{}{Walltime}{}{12492.6865892324131734}%
\StoreBenchExecResult{CpacheckerBenchmarkDefinition}{TRReachabilityReachsafetyLoops}{Correct}{}{Walltime}{Avg}{99.1483062637493109}%
\StoreBenchExecResult{CpacheckerBenchmarkDefinition}{TRReachabilityReachsafetyLoops}{Correct}{}{Walltime}{Median}{4.9448231015121565}%
\StoreBenchExecResult{CpacheckerBenchmarkDefinition}{TRReachabilityReachsafetyLoops}{Correct}{}{Walltime}{Min}{3.4230121129658073}%
\StoreBenchExecResult{CpacheckerBenchmarkDefinition}{TRReachabilityReachsafetyLoops}{Correct}{}{Walltime}{Max}{732.0504627909977}%
\StoreBenchExecResult{CpacheckerBenchmarkDefinition}{TRReachabilityReachsafetyLoops}{Correct}{}{Walltime}{Stdev}{166.5600880706341993892797389}%
\StoreBenchExecResult{CpacheckerBenchmarkDefinition}{TRReachabilityReachsafetyLoops}{Correct}{False}{Count}{}{66}%
\StoreBenchExecResult{CpacheckerBenchmarkDefinition}{TRReachabilityReachsafetyLoops}{Correct}{False}{Cputime}{}{1617.976647767}%
\StoreBenchExecResult{CpacheckerBenchmarkDefinition}{TRReachabilityReachsafetyLoops}{Correct}{False}{Cputime}{Avg}{24.51479769343939393939393939}%
\StoreBenchExecResult{CpacheckerBenchmarkDefinition}{TRReachabilityReachsafetyLoops}{Correct}{False}{Cputime}{Median}{8.590141111}%
\StoreBenchExecResult{CpacheckerBenchmarkDefinition}{TRReachabilityReachsafetyLoops}{Correct}{False}{Cputime}{Min}{7.326023136}%
\StoreBenchExecResult{CpacheckerBenchmarkDefinition}{TRReachabilityReachsafetyLoops}{Correct}{False}{Cputime}{Max}{238.76672744}%
\StoreBenchExecResult{CpacheckerBenchmarkDefinition}{TRReachabilityReachsafetyLoops}{Correct}{False}{Cputime}{Stdev}{49.06979773678834891165172236}%
\StoreBenchExecResult{CpacheckerBenchmarkDefinition}{TRReachabilityReachsafetyLoops}{Correct}{False}{Walltime}{}{1206.3822529888711984}%
\StoreBenchExecResult{CpacheckerBenchmarkDefinition}{TRReachabilityReachsafetyLoops}{Correct}{False}{Walltime}{Avg}{18.27851898467986664242424242}%
\StoreBenchExecResult{CpacheckerBenchmarkDefinition}{TRReachabilityReachsafetyLoops}{Correct}{False}{Walltime}{Median}{4.4667526100529355}%
\StoreBenchExecResult{CpacheckerBenchmarkDefinition}{TRReachabilityReachsafetyLoops}{Correct}{False}{Walltime}{Min}{3.8089780719019473}%
\StoreBenchExecResult{CpacheckerBenchmarkDefinition}{TRReachabilityReachsafetyLoops}{Correct}{False}{Walltime}{Max}{217.03101862594485}%
\StoreBenchExecResult{CpacheckerBenchmarkDefinition}{TRReachabilityReachsafetyLoops}{Correct}{False}{Walltime}{Stdev}{43.50735961566775023194202055}%

\StoreBenchExecResult{CpacheckerBenchmarkDefinition}{TRReachabilityReachsafetyLoops}{Correct}{True}{Count}{}{60}%
\StoreBenchExecResult{CpacheckerBenchmarkDefinition}{TRReachabilityReachsafetyLoops}{Correct}{True}{Cputime}{}{12508.106999325}%
\StoreBenchExecResult{CpacheckerBenchmarkDefinition}{TRReachabilityReachsafetyLoops}{Correct}{True}{Cputime}{Avg}{208.46844998875}%
\StoreBenchExecResult{CpacheckerBenchmarkDefinition}{TRReachabilityReachsafetyLoops}{Correct}{True}{Cputime}{Median}{146.279843298}%
\StoreBenchExecResult{CpacheckerBenchmarkDefinition}{TRReachabilityReachsafetyLoops}{Correct}{True}{Cputime}{Min}{6.542492465}%
\StoreBenchExecResult{CpacheckerBenchmarkDefinition}{TRReachabilityReachsafetyLoops}{Correct}{True}{Cputime}{Max}{773.353600287}%
\StoreBenchExecResult{CpacheckerBenchmarkDefinition}{TRReachabilityReachsafetyLoops}{Correct}{True}{Cputime}{Stdev}{216.1563628606875556096499547}%
\StoreBenchExecResult{CpacheckerBenchmarkDefinition}{TRReachabilityReachsafetyLoops}{Correct}{True}{Walltime}{}{11286.3043362435419750}%
\StoreBenchExecResult{CpacheckerBenchmarkDefinition}{TRReachabilityReachsafetyLoops}{Correct}{True}{Walltime}{Avg}{188.1050722707256995833333333}%
\StoreBenchExecResult{CpacheckerBenchmarkDefinition}{TRReachabilityReachsafetyLoops}{Correct}{True}{Walltime}{Median}{132.24475190648809}%
\StoreBenchExecResult{CpacheckerBenchmarkDefinition}{TRReachabilityReachsafetyLoops}{Correct}{True}{Walltime}{Min}{3.4230121129658073}%
\StoreBenchExecResult{CpacheckerBenchmarkDefinition}{TRReachabilityReachsafetyLoops}{Correct}{True}{Walltime}{Max}{732.0504627909977}%
\StoreBenchExecResult{CpacheckerBenchmarkDefinition}{TRReachabilityReachsafetyLoops}{Correct}{True}{Walltime}{Stdev}{202.6557471002654712166362064}%

\StoreBenchExecResult{CpacheckerBenchmarkDefinition}{TRReachabilityReachsafetyLoops}{Error}{}{Count}{}{260}%
\StoreBenchExecResult{CpacheckerBenchmarkDefinition}{TRReachabilityReachsafetyLoops}{Error}{}{Cputime}{}{209891.152126955}%
\StoreBenchExecResult{CpacheckerBenchmarkDefinition}{TRReachabilityReachsafetyLoops}{Error}{}{Cputime}{Avg}{807.27366202675}%
\StoreBenchExecResult{CpacheckerBenchmarkDefinition}{TRReachabilityReachsafetyLoops}{Error}{}{Cputime}{Median}{904.7269962755}%
\StoreBenchExecResult{CpacheckerBenchmarkDefinition}{TRReachabilityReachsafetyLoops}{Error}{}{Cputime}{Min}{2.755320309}%
\StoreBenchExecResult{CpacheckerBenchmarkDefinition}{TRReachabilityReachsafetyLoops}{Error}{}{Cputime}{Max}{962.100257316}%
\StoreBenchExecResult{CpacheckerBenchmarkDefinition}{TRReachabilityReachsafetyLoops}{Error}{}{Cputime}{Stdev}{290.3118841290209564131135251}%
\StoreBenchExecResult{CpacheckerBenchmarkDefinition}{TRReachabilityReachsafetyLoops}{Error}{}{Walltime}{}{195983.2893746779293759}%
\StoreBenchExecResult{CpacheckerBenchmarkDefinition}{TRReachabilityReachsafetyLoops}{Error}{}{Walltime}{Avg}{753.7818822102997283688461538}%
\StoreBenchExecResult{CpacheckerBenchmarkDefinition}{TRReachabilityReachsafetyLoops}{Error}{}{Walltime}{Median}{860.96013018954545}%
\StoreBenchExecResult{CpacheckerBenchmarkDefinition}{TRReachabilityReachsafetyLoops}{Error}{}{Walltime}{Min}{1.479888177011162}%
\StoreBenchExecResult{CpacheckerBenchmarkDefinition}{TRReachabilityReachsafetyLoops}{Error}{}{Walltime}{Max}{928.1734276819043}%
\StoreBenchExecResult{CpacheckerBenchmarkDefinition}{TRReachabilityReachsafetyLoops}{Error}{}{Walltime}{Stdev}{275.1683244150913433366933883}%
\StoreBenchExecResult{CpacheckerBenchmarkDefinition}{TRReachabilityReachsafetyLoops}{Error}{Assertion}{Count}{}{3}%
\StoreBenchExecResult{CpacheckerBenchmarkDefinition}{TRReachabilityReachsafetyLoops}{Error}{Assertion}{Cputime}{}{219.796042477}%
\StoreBenchExecResult{CpacheckerBenchmarkDefinition}{TRReachabilityReachsafetyLoops}{Error}{Assertion}{Cputime}{Avg}{73.26534749233333333333333333}%
\StoreBenchExecResult{CpacheckerBenchmarkDefinition}{TRReachabilityReachsafetyLoops}{Error}{Assertion}{Cputime}{Median}{7.731116676}%
\StoreBenchExecResult{CpacheckerBenchmarkDefinition}{TRReachabilityReachsafetyLoops}{Error}{Assertion}{Cputime}{Min}{6.189300814}%
\StoreBenchExecResult{CpacheckerBenchmarkDefinition}{TRReachabilityReachsafetyLoops}{Error}{Assertion}{Cputime}{Max}{205.875624987}%
\StoreBenchExecResult{CpacheckerBenchmarkDefinition}{TRReachabilityReachsafetyLoops}{Error}{Assertion}{Cputime}{Stdev}{93.77173906868192153717203226}%
\StoreBenchExecResult{CpacheckerBenchmarkDefinition}{TRReachabilityReachsafetyLoops}{Error}{Assertion}{Walltime}{}{192.9204487686511094}%
\StoreBenchExecResult{CpacheckerBenchmarkDefinition}{TRReachabilityReachsafetyLoops}{Error}{Assertion}{Walltime}{Avg}{64.30681625621703646666666667}%
\StoreBenchExecResult{CpacheckerBenchmarkDefinition}{TRReachabilityReachsafetyLoops}{Error}{Assertion}{Walltime}{Median}{3.998717252863571}%
\StoreBenchExecResult{CpacheckerBenchmarkDefinition}{TRReachabilityReachsafetyLoops}{Error}{Assertion}{Walltime}{Min}{3.2129207169637084}%
\StoreBenchExecResult{CpacheckerBenchmarkDefinition}{TRReachabilityReachsafetyLoops}{Error}{Assertion}{Walltime}{Max}{185.70881079882383}%
\StoreBenchExecResult{CpacheckerBenchmarkDefinition}{TRReachabilityReachsafetyLoops}{Error}{Assertion}{Walltime}{Stdev}{85.84477300430626807505690944}%
\StoreBenchExecResult{CpacheckerBenchmarkDefinition}{TRReachabilityReachsafetyLoops}{Error}{Error}{Count}{}{27}%
\StoreBenchExecResult{CpacheckerBenchmarkDefinition}{TRReachabilityReachsafetyLoops}{Error}{Error}{Cputime}{}{83.091277567}%
\StoreBenchExecResult{CpacheckerBenchmarkDefinition}{TRReachabilityReachsafetyLoops}{Error}{Error}{Cputime}{Avg}{3.077454724703703703703703704}%
\StoreBenchExecResult{CpacheckerBenchmarkDefinition}{TRReachabilityReachsafetyLoops}{Error}{Error}{Cputime}{Median}{3.08212593}%
\StoreBenchExecResult{CpacheckerBenchmarkDefinition}{TRReachabilityReachsafetyLoops}{Error}{Error}{Cputime}{Min}{2.755320309}%
\StoreBenchExecResult{CpacheckerBenchmarkDefinition}{TRReachabilityReachsafetyLoops}{Error}{Error}{Cputime}{Max}{3.539716082}%
\StoreBenchExecResult{CpacheckerBenchmarkDefinition}{TRReachabilityReachsafetyLoops}{Error}{Error}{Cputime}{Stdev}{0.2255248462087470588703426295}%
\StoreBenchExecResult{CpacheckerBenchmarkDefinition}{TRReachabilityReachsafetyLoops}{Error}{Error}{Walltime}{}{44.2915547247976065}%
\StoreBenchExecResult{CpacheckerBenchmarkDefinition}{TRReachabilityReachsafetyLoops}{Error}{Error}{Walltime}{Avg}{1.640427952770281722222222222}%
\StoreBenchExecResult{CpacheckerBenchmarkDefinition}{TRReachabilityReachsafetyLoops}{Error}{Error}{Walltime}{Median}{1.6366941430605948}%
\StoreBenchExecResult{CpacheckerBenchmarkDefinition}{TRReachabilityReachsafetyLoops}{Error}{Error}{Walltime}{Min}{1.479888177011162}%
\StoreBenchExecResult{CpacheckerBenchmarkDefinition}{TRReachabilityReachsafetyLoops}{Error}{Error}{Walltime}{Max}{1.8902588649652898}%
\StoreBenchExecResult{CpacheckerBenchmarkDefinition}{TRReachabilityReachsafetyLoops}{Error}{Error}{Walltime}{Stdev}{0.1167923290636382260651065179}%
\StoreBenchExecResult{CpacheckerBenchmarkDefinition}{TRReachabilityReachsafetyLoops}{Error}{Exception}{Count}{}{1}%
\StoreBenchExecResult{CpacheckerBenchmarkDefinition}{TRReachabilityReachsafetyLoops}{Error}{Exception}{Cputime}{}{408.590037547}%
\StoreBenchExecResult{CpacheckerBenchmarkDefinition}{TRReachabilityReachsafetyLoops}{Error}{Exception}{Cputime}{Avg}{408.590037547}%
\StoreBenchExecResult{CpacheckerBenchmarkDefinition}{TRReachabilityReachsafetyLoops}{Error}{Exception}{Cputime}{Median}{408.590037547}%
\StoreBenchExecResult{CpacheckerBenchmarkDefinition}{TRReachabilityReachsafetyLoops}{Error}{Exception}{Cputime}{Min}{408.590037547}%
\StoreBenchExecResult{CpacheckerBenchmarkDefinition}{TRReachabilityReachsafetyLoops}{Error}{Exception}{Cputime}{Max}{408.590037547}%
\StoreBenchExecResult{CpacheckerBenchmarkDefinition}{TRReachabilityReachsafetyLoops}{Error}{Exception}{Cputime}{Stdev}{0E-14}%
\StoreBenchExecResult{CpacheckerBenchmarkDefinition}{TRReachabilityReachsafetyLoops}{Error}{Exception}{Walltime}{}{347.07851111819036}%
\StoreBenchExecResult{CpacheckerBenchmarkDefinition}{TRReachabilityReachsafetyLoops}{Error}{Exception}{Walltime}{Avg}{347.07851111819036}%
\StoreBenchExecResult{CpacheckerBenchmarkDefinition}{TRReachabilityReachsafetyLoops}{Error}{Exception}{Walltime}{Median}{347.07851111819036}%
\StoreBenchExecResult{CpacheckerBenchmarkDefinition}{TRReachabilityReachsafetyLoops}{Error}{Exception}{Walltime}{Min}{347.07851111819036}%
\StoreBenchExecResult{CpacheckerBenchmarkDefinition}{TRReachabilityReachsafetyLoops}{Error}{Exception}{Walltime}{Max}{347.07851111819036}%
\StoreBenchExecResult{CpacheckerBenchmarkDefinition}{TRReachabilityReachsafetyLoops}{Error}{Exception}{Walltime}{Stdev}{0E-14}%
\StoreBenchExecResult{CpacheckerBenchmarkDefinition}{TRReachabilityReachsafetyLoops}{Error}{Timeout}{Count}{}{229}%
\StoreBenchExecResult{CpacheckerBenchmarkDefinition}{TRReachabilityReachsafetyLoops}{Error}{Timeout}{Cputime}{}{209179.674769364}%
\StoreBenchExecResult{CpacheckerBenchmarkDefinition}{TRReachabilityReachsafetyLoops}{Error}{Timeout}{Cputime}{Avg}{913.4483614382707423580786026}%
\StoreBenchExecResult{CpacheckerBenchmarkDefinition}{TRReachabilityReachsafetyLoops}{Error}{Timeout}{Cputime}{Median}{905.250327838}%
\StoreBenchExecResult{CpacheckerBenchmarkDefinition}{TRReachabilityReachsafetyLoops}{Error}{Timeout}{Cputime}{Min}{902.4049194}%
\StoreBenchExecResult{CpacheckerBenchmarkDefinition}{TRReachabilityReachsafetyLoops}{Error}{Timeout}{Cputime}{Max}{962.100257316}%
\StoreBenchExecResult{CpacheckerBenchmarkDefinition}{TRReachabilityReachsafetyLoops}{Error}{Timeout}{Cputime}{Stdev}{17.24970796734427521582389857}%
\StoreBenchExecResult{CpacheckerBenchmarkDefinition}{TRReachabilityReachsafetyLoops}{Error}{Timeout}{Walltime}{}{195398.9988600662903}%
\StoreBenchExecResult{CpacheckerBenchmarkDefinition}{TRReachabilityReachsafetyLoops}{Error}{Timeout}{Walltime}{Avg}{853.2707373802021410480349345}%
\StoreBenchExecResult{CpacheckerBenchmarkDefinition}{TRReachabilityReachsafetyLoops}{Error}{Timeout}{Walltime}{Median}{862.1913650578354}%
\StoreBenchExecResult{CpacheckerBenchmarkDefinition}{TRReachabilityReachsafetyLoops}{Error}{Timeout}{Walltime}{Min}{529.6526941661723}%
\StoreBenchExecResult{CpacheckerBenchmarkDefinition}{TRReachabilityReachsafetyLoops}{Error}{Timeout}{Walltime}{Max}{928.1734276819043}%
\StoreBenchExecResult{CpacheckerBenchmarkDefinition}{TRReachabilityReachsafetyLoops}{Error}{Timeout}{Walltime}{Stdev}{48.19592793915323260912252365}%
\ifdefined\CpacheckerCpacheckerTotalCount\else\edef\CpacheckerCpacheckerTotalCount{0}\fi
\ifdefined\CpacheckerCpacheckerCorrectCount\else\edef\CpacheckerCpacheckerCorrectCount{0}\fi
\ifdefined\CpacheckerCpacheckerCorrectTrueCount\else\edef\CpacheckerCpacheckerCorrectTrueCount{0}\fi
\ifdefined\CpacheckerCpacheckerCorrectFalseCount\else\edef\CpacheckerCpacheckerCorrectFalseCount{0}\fi
\ifdefined\CpacheckerCpacheckerWrongCount\else\edef\CpacheckerCpacheckerWrongCount{0}\fi
\ifdefined\CpacheckerCpacheckerWrongTrueCount\else\edef\CpacheckerCpacheckerWrongTrueCount{0}\fi
\ifdefined\CpacheckerCpacheckerWrongFalseCount\else\edef\CpacheckerCpacheckerWrongFalseCount{0}\fi
\ifdefined\CpacheckerCpacheckerErrorTimeoutCount\else\edef\CpacheckerCpacheckerErrorTimeoutCount{0}\fi
\ifdefined\CpacheckerCpacheckerErrorOutOfMemoryCount\else\edef\CpacheckerCpacheckerErrorOutOfMemoryCount{0}\fi
\ifdefined\CpacheckerCpacheckerCorrectCputime\else\edef\CpacheckerCpacheckerCorrectCputime{0}\fi
\ifdefined\CpacheckerCpacheckerCorrectCputimeAvg\else\edef\CpacheckerCpacheckerCorrectCputimeAvg{None}\fi
\ifdefined\CpacheckerCpacheckerCorrectWalltime\else\edef\CpacheckerCpacheckerCorrectWalltime{0}\fi
\ifdefined\CpacheckerCpacheckerCorrectWalltimeAvg\else\edef\CpacheckerCpacheckerCorrectWalltimeAvg{None}\fi
\edef\CpacheckerCpacheckerErrorOtherInconclusiveCount{\the\numexpr \CpacheckerCpacheckerTotalCount - \CpacheckerCpacheckerCorrectCount - \CpacheckerCpacheckerWrongTrueCount - \CpacheckerCpacheckerWrongFalseCount - \CpacheckerCpacheckerErrorTimeoutCount - \CpacheckerCpacheckerErrorOutOfMemoryCount \relax}
\providecommand\StoreBenchExecResult[7]{\expandafter\newcommand\csname#1#2#3#4#5#6\endcsname{#7}}%
\StoreBenchExecResult{UltimateBenchmarkDefinitiontermination}{TRTerminationReachsafetyLoops}{Total}{}{Count}{}{386}%
\StoreBenchExecResult{UltimateBenchmarkDefinitiontermination}{TRTerminationReachsafetyLoops}{Total}{}{Cputime}{}{18906.861776557}%
\StoreBenchExecResult{UltimateBenchmarkDefinitiontermination}{TRTerminationReachsafetyLoops}{Total}{}{Cputime}{Avg}{48.98150719315284974093264249}%
\StoreBenchExecResult{UltimateBenchmarkDefinitiontermination}{TRTerminationReachsafetyLoops}{Total}{}{Cputime}{Median}{18.852815901}%
\StoreBenchExecResult{UltimateBenchmarkDefinitiontermination}{TRTerminationReachsafetyLoops}{Total}{}{Cputime}{Min}{13.39842813}%
\StoreBenchExecResult{UltimateBenchmarkDefinitiontermination}{TRTerminationReachsafetyLoops}{Total}{}{Cputime}{Max}{960.753635499}%
\StoreBenchExecResult{UltimateBenchmarkDefinitiontermination}{TRTerminationReachsafetyLoops}{Total}{}{Cputime}{Stdev}{136.5475794690682906591122811}%
\StoreBenchExecResult{UltimateBenchmarkDefinitiontermination}{TRTerminationReachsafetyLoops}{Total}{}{Walltime}{}{14803.091328104259514}%
\StoreBenchExecResult{UltimateBenchmarkDefinitiontermination}{TRTerminationReachsafetyLoops}{Total}{}{Walltime}{Avg}{38.34997753394885884455958549}%
\StoreBenchExecResult{UltimateBenchmarkDefinitiontermination}{TRTerminationReachsafetyLoops}{Total}{}{Walltime}{Median}{11.311473489040509}%
\StoreBenchExecResult{UltimateBenchmarkDefinitiontermination}{TRTerminationReachsafetyLoops}{Total}{}{Walltime}{Min}{8.075321631040424}%
\StoreBenchExecResult{UltimateBenchmarkDefinitiontermination}{TRTerminationReachsafetyLoops}{Total}{}{Walltime}{Max}{952.9164266260341}%
\StoreBenchExecResult{UltimateBenchmarkDefinitiontermination}{TRTerminationReachsafetyLoops}{Total}{}{Walltime}{Stdev}{130.2443362494274767818076900}%
\StoreBenchExecResult{UltimateBenchmarkDefinitiontermination}{TRTerminationReachsafetyLoops}{Correct}{}{Count}{}{305}%
\StoreBenchExecResult{UltimateBenchmarkDefinitiontermination}{TRTerminationReachsafetyLoops}{Correct}{}{Cputime}{}{6853.656411350}%
\StoreBenchExecResult{UltimateBenchmarkDefinitiontermination}{TRTerminationReachsafetyLoops}{Correct}{}{Cputime}{Avg}{22.47100462737704918032786885}%
\StoreBenchExecResult{UltimateBenchmarkDefinitiontermination}{TRTerminationReachsafetyLoops}{Correct}{}{Cputime}{Median}{17.599762539}%
\StoreBenchExecResult{UltimateBenchmarkDefinitiontermination}{TRTerminationReachsafetyLoops}{Correct}{}{Cputime}{Min}{14.303195094}%
\StoreBenchExecResult{UltimateBenchmarkDefinitiontermination}{TRTerminationReachsafetyLoops}{Correct}{}{Cputime}{Max}{172.841004215}%
\StoreBenchExecResult{UltimateBenchmarkDefinitiontermination}{TRTerminationReachsafetyLoops}{Correct}{}{Cputime}{Stdev}{17.09107232680751993501092920}%
\StoreBenchExecResult{UltimateBenchmarkDefinitiontermination}{TRTerminationReachsafetyLoops}{Correct}{}{Walltime}{}{4224.330395949771626}%
\StoreBenchExecResult{UltimateBenchmarkDefinitiontermination}{TRTerminationReachsafetyLoops}{Correct}{}{Walltime}{Avg}{13.85026359327793975737704918}%
\StoreBenchExecResult{UltimateBenchmarkDefinitiontermination}{TRTerminationReachsafetyLoops}{Correct}{}{Walltime}{Median}{10.679181735031307}%
\StoreBenchExecResult{UltimateBenchmarkDefinitiontermination}{TRTerminationReachsafetyLoops}{Correct}{}{Walltime}{Min}{8.607518758857623}%
\StoreBenchExecResult{UltimateBenchmarkDefinitiontermination}{TRTerminationReachsafetyLoops}{Correct}{}{Walltime}{Max}{114.58341692388058}%
\StoreBenchExecResult{UltimateBenchmarkDefinitiontermination}{TRTerminationReachsafetyLoops}{Correct}{}{Walltime}{Stdev}{11.81271989069634150188986637}%
\StoreBenchExecResult{UltimateBenchmarkDefinitiontermination}{TRTerminationReachsafetyLoops}{Correct}{False}{Count}{}{59}%
\StoreBenchExecResult{UltimateBenchmarkDefinitiontermination}{TRTerminationReachsafetyLoops}{Correct}{False}{Cputime}{}{1308.868613631}%
\StoreBenchExecResult{UltimateBenchmarkDefinitiontermination}{TRTerminationReachsafetyLoops}{Correct}{False}{Cputime}{Avg}{22.18421379035593220338983051}%
\StoreBenchExecResult{UltimateBenchmarkDefinitiontermination}{TRTerminationReachsafetyLoops}{Correct}{False}{Cputime}{Median}{15.750553879}%
\StoreBenchExecResult{UltimateBenchmarkDefinitiontermination}{TRTerminationReachsafetyLoops}{Correct}{False}{Cputime}{Min}{14.303195094}%
\StoreBenchExecResult{UltimateBenchmarkDefinitiontermination}{TRTerminationReachsafetyLoops}{Correct}{False}{Cputime}{Max}{157.01784064}%
\StoreBenchExecResult{UltimateBenchmarkDefinitiontermination}{TRTerminationReachsafetyLoops}{Correct}{False}{Cputime}{Stdev}{20.02925359612633439358724442}%
\StoreBenchExecResult{UltimateBenchmarkDefinitiontermination}{TRTerminationReachsafetyLoops}{Correct}{False}{Walltime}{}{817.045512388693169}%
\StoreBenchExecResult{UltimateBenchmarkDefinitiontermination}{TRTerminationReachsafetyLoops}{Correct}{False}{Walltime}{Avg}{13.84822902353717235593220339}%
\StoreBenchExecResult{UltimateBenchmarkDefinitiontermination}{TRTerminationReachsafetyLoops}{Correct}{False}{Walltime}{Median}{9.377363633131608}%
\StoreBenchExecResult{UltimateBenchmarkDefinitiontermination}{TRTerminationReachsafetyLoops}{Correct}{False}{Walltime}{Min}{8.607518758857623}%
\StoreBenchExecResult{UltimateBenchmarkDefinitiontermination}{TRTerminationReachsafetyLoops}{Correct}{False}{Walltime}{Max}{111.77409276901744}%
\StoreBenchExecResult{UltimateBenchmarkDefinitiontermination}{TRTerminationReachsafetyLoops}{Correct}{False}{Walltime}{Stdev}{14.84672965935348839014890304}%

\StoreBenchExecResult{UltimateBenchmarkDefinitiontermination}{TRTerminationReachsafetyLoops}{Correct}{True}{Count}{}{246}%
\StoreBenchExecResult{UltimateBenchmarkDefinitiontermination}{TRTerminationReachsafetyLoops}{Correct}{True}{Cputime}{}{5544.787797719}%
\StoreBenchExecResult{UltimateBenchmarkDefinitiontermination}{TRTerminationReachsafetyLoops}{Correct}{True}{Cputime}{Avg}{22.53978779560569105691056911}%
\StoreBenchExecResult{UltimateBenchmarkDefinitiontermination}{TRTerminationReachsafetyLoops}{Correct}{True}{Cputime}{Median}{17.9698353235}%
\StoreBenchExecResult{UltimateBenchmarkDefinitiontermination}{TRTerminationReachsafetyLoops}{Correct}{True}{Cputime}{Min}{14.69651184}%
\StoreBenchExecResult{UltimateBenchmarkDefinitiontermination}{TRTerminationReachsafetyLoops}{Correct}{True}{Cputime}{Max}{172.841004215}%
\StoreBenchExecResult{UltimateBenchmarkDefinitiontermination}{TRTerminationReachsafetyLoops}{Correct}{True}{Cputime}{Stdev}{16.30711902824889749640137598}%
\StoreBenchExecResult{UltimateBenchmarkDefinitiontermination}{TRTerminationReachsafetyLoops}{Correct}{True}{Walltime}{}{3407.284883561078457}%
\StoreBenchExecResult{UltimateBenchmarkDefinitiontermination}{TRTerminationReachsafetyLoops}{Correct}{True}{Walltime}{Avg}{13.85075155919137584146341463}%
\StoreBenchExecResult{UltimateBenchmarkDefinitiontermination}{TRTerminationReachsafetyLoops}{Correct}{True}{Walltime}{Median}{10.839787341072224}%
\StoreBenchExecResult{UltimateBenchmarkDefinitiontermination}{TRTerminationReachsafetyLoops}{Correct}{True}{Walltime}{Min}{8.992324406979606}%
\StoreBenchExecResult{UltimateBenchmarkDefinitiontermination}{TRTerminationReachsafetyLoops}{Correct}{True}{Walltime}{Max}{114.58341692388058}%
\StoreBenchExecResult{UltimateBenchmarkDefinitiontermination}{TRTerminationReachsafetyLoops}{Correct}{True}{Walltime}{Stdev}{10.96088932230177806899570011}%

\StoreBenchExecResult{UltimateBenchmarkDefinitiontermination}{TRTerminationReachsafetyLoops}{Error}{}{Count}{}{81}%
\StoreBenchExecResult{UltimateBenchmarkDefinitiontermination}{TRTerminationReachsafetyLoops}{Error}{}{Cputime}{}{12053.205365207}%
\StoreBenchExecResult{UltimateBenchmarkDefinitiontermination}{TRTerminationReachsafetyLoops}{Error}{}{Cputime}{Avg}{148.8050045087283950617283951}%
\StoreBenchExecResult{UltimateBenchmarkDefinitiontermination}{TRTerminationReachsafetyLoops}{Error}{}{Cputime}{Median}{37.85653603}%
\StoreBenchExecResult{UltimateBenchmarkDefinitiontermination}{TRTerminationReachsafetyLoops}{Error}{}{Cputime}{Min}{13.39842813}%
\StoreBenchExecResult{UltimateBenchmarkDefinitiontermination}{TRTerminationReachsafetyLoops}{Error}{}{Cputime}{Max}{960.753635499}%
\StoreBenchExecResult{UltimateBenchmarkDefinitiontermination}{TRTerminationReachsafetyLoops}{Error}{}{Cputime}{Stdev}{274.1197362924924642078271590}%
\StoreBenchExecResult{UltimateBenchmarkDefinitiontermination}{TRTerminationReachsafetyLoops}{Error}{}{Walltime}{}{10578.760932154487888}%
\StoreBenchExecResult{UltimateBenchmarkDefinitiontermination}{TRTerminationReachsafetyLoops}{Error}{}{Walltime}{Avg}{130.6019868167220726913580247}%
\StoreBenchExecResult{UltimateBenchmarkDefinitiontermination}{TRTerminationReachsafetyLoops}{Error}{}{Walltime}{Median}{22.837063991930336}%
\StoreBenchExecResult{UltimateBenchmarkDefinitiontermination}{TRTerminationReachsafetyLoops}{Error}{}{Walltime}{Min}{8.075321631040424}%
\StoreBenchExecResult{UltimateBenchmarkDefinitiontermination}{TRTerminationReachsafetyLoops}{Error}{}{Walltime}{Max}{952.9164266260341}%
\StoreBenchExecResult{UltimateBenchmarkDefinitiontermination}{TRTerminationReachsafetyLoops}{Error}{}{Walltime}{Stdev}{263.7097112808042545041429273}%
\StoreBenchExecResult{UltimateBenchmarkDefinitiontermination}{TRTerminationReachsafetyLoops}{Error}{Error}{Count}{}{74}%
\StoreBenchExecResult{UltimateBenchmarkDefinitiontermination}{TRTerminationReachsafetyLoops}{Error}{Error}{Cputime}{}{5330.257568888}%
\StoreBenchExecResult{UltimateBenchmarkDefinitiontermination}{TRTerminationReachsafetyLoops}{Error}{Error}{Cputime}{Avg}{72.03050768767567567567567568}%
\StoreBenchExecResult{UltimateBenchmarkDefinitiontermination}{TRTerminationReachsafetyLoops}{Error}{Error}{Cputime}{Median}{35.610470582}%
\StoreBenchExecResult{UltimateBenchmarkDefinitiontermination}{TRTerminationReachsafetyLoops}{Error}{Error}{Cputime}{Min}{13.39842813}%
\StoreBenchExecResult{UltimateBenchmarkDefinitiontermination}{TRTerminationReachsafetyLoops}{Error}{Error}{Cputime}{Max}{601.694803674}%
\StoreBenchExecResult{UltimateBenchmarkDefinitiontermination}{TRTerminationReachsafetyLoops}{Error}{Error}{Cputime}{Stdev}{118.5068595131385349781516265}%
\StoreBenchExecResult{UltimateBenchmarkDefinitiontermination}{TRTerminationReachsafetyLoops}{Error}{Error}{Walltime}{}{4176.277034232160088}%
\StoreBenchExecResult{UltimateBenchmarkDefinitiontermination}{TRTerminationReachsafetyLoops}{Error}{Error}{Walltime}{Avg}{56.43617613827243362162162162}%
\StoreBenchExecResult{UltimateBenchmarkDefinitiontermination}{TRTerminationReachsafetyLoops}{Error}{Error}{Walltime}{Median}{20.9485189439728855}%
\StoreBenchExecResult{UltimateBenchmarkDefinitiontermination}{TRTerminationReachsafetyLoops}{Error}{Error}{Walltime}{Min}{8.075321631040424}%
\StoreBenchExecResult{UltimateBenchmarkDefinitiontermination}{TRTerminationReachsafetyLoops}{Error}{Error}{Walltime}{Max}{556.9116008030251}%
\StoreBenchExecResult{UltimateBenchmarkDefinitiontermination}{TRTerminationReachsafetyLoops}{Error}{Error}{Walltime}{Stdev}{110.5533069781202933294832193}%
\StoreBenchExecResult{UltimateBenchmarkDefinitiontermination}{TRTerminationReachsafetyLoops}{Error}{Timeout}{Count}{}{7}%
\StoreBenchExecResult{UltimateBenchmarkDefinitiontermination}{TRTerminationReachsafetyLoops}{Error}{Timeout}{Cputime}{}{6722.947796319}%
\StoreBenchExecResult{UltimateBenchmarkDefinitiontermination}{TRTerminationReachsafetyLoops}{Error}{Timeout}{Cputime}{Avg}{960.4211137598571428571428571}%
\StoreBenchExecResult{UltimateBenchmarkDefinitiontermination}{TRTerminationReachsafetyLoops}{Error}{Timeout}{Cputime}{Median}{960.34480893}%
\StoreBenchExecResult{UltimateBenchmarkDefinitiontermination}{TRTerminationReachsafetyLoops}{Error}{Timeout}{Cputime}{Min}{960.196866245}%
\StoreBenchExecResult{UltimateBenchmarkDefinitiontermination}{TRTerminationReachsafetyLoops}{Error}{Timeout}{Cputime}{Max}{960.753635499}%
\StoreBenchExecResult{UltimateBenchmarkDefinitiontermination}{TRTerminationReachsafetyLoops}{Error}{Timeout}{Cputime}{Stdev}{0.2137548936359289646183355173}%
\StoreBenchExecResult{UltimateBenchmarkDefinitiontermination}{TRTerminationReachsafetyLoops}{Error}{Timeout}{Walltime}{}{6402.4838979223278}%
\StoreBenchExecResult{UltimateBenchmarkDefinitiontermination}{TRTerminationReachsafetyLoops}{Error}{Timeout}{Walltime}{Avg}{914.6405568460468285714285714}%
\StoreBenchExecResult{UltimateBenchmarkDefinitiontermination}{TRTerminationReachsafetyLoops}{Error}{Timeout}{Walltime}{Median}{932.2524833751377}%
\StoreBenchExecResult{UltimateBenchmarkDefinitiontermination}{TRTerminationReachsafetyLoops}{Error}{Timeout}{Walltime}{Min}{795.0620779278688}%
\StoreBenchExecResult{UltimateBenchmarkDefinitiontermination}{TRTerminationReachsafetyLoops}{Error}{Timeout}{Walltime}{Max}{952.9164266260341}%
\StoreBenchExecResult{UltimateBenchmarkDefinitiontermination}{TRTerminationReachsafetyLoops}{Error}{Timeout}{Walltime}{Stdev}{51.38081496593776522537084662}%
\ifdefined\UltimateUltimateTotalCount\else\edef\UltimateUltimateTotalCount{0}\fi
\ifdefined\UltimateUltimateCorrectCount\else\edef\UltimateUltimateCorrectCount{0}\fi
\ifdefined\UltimateUltimateCorrectTrueCount\else\edef\UltimateUltimateCorrectTrueCount{0}\fi
\ifdefined\UltimateUltimateCorrectFalseCount\else\edef\UltimateUltimateCorrectFalseCount{0}\fi
\ifdefined\UltimateUltimateWrongCount\else\edef\UltimateUltimateWrongCount{0}\fi
\ifdefined\UltimateUltimateWrongTrueCount\else\edef\UltimateUltimateWrongTrueCount{0}\fi
\ifdefined\UltimateUltimateWrongFalseCount\else\edef\UltimateUltimateWrongFalseCount{0}\fi
\ifdefined\UltimateUltimateErrorTimeoutCount\else\edef\UltimateUltimateErrorTimeoutCount{0}\fi
\ifdefined\UltimateUltimateErrorOutOfMemoryCount\else\edef\UltimateUltimateErrorOutOfMemoryCount{0}\fi
\ifdefined\UltimateUltimateCorrectCputime\else\edef\UltimateUltimateCorrectCputime{0}\fi
\ifdefined\UltimateUltimateCorrectCputimeAvg\else\edef\UltimateUltimateCorrectCputimeAvg{None}\fi
\ifdefined\UltimateUltimateCorrectWalltime\else\edef\UltimateUltimateCorrectWalltime{0}\fi
\ifdefined\UltimateUltimateCorrectWalltimeAvg\else\edef\UltimateUltimateCorrectWalltimeAvg{None}\fi
\edef\UltimateUltimateErrorOtherInconclusiveCount{\the\numexpr \UltimateUltimateTotalCount - \UltimateUltimateCorrectCount - \UltimateUltimateWrongTrueCount - \UltimateUltimateWrongFalseCount - \UltimateUltimateErrorTimeoutCount - \UltimateUltimateErrorOutOfMemoryCount \relax}
\providecommand\StoreBenchExecResult[7]{\expandafter\newcommand\csname#1#2#3#4#5#6\endcsname{#7}}%
\StoreBenchExecResult{LsBenchmarkDefinitiontermination}{TRTerminationReachsafetyLoops}{Total}{}{Count}{}{386}%
\StoreBenchExecResult{LsBenchmarkDefinitiontermination}{TRTerminationReachsafetyLoops}{Total}{}{Cputime}{}{83608.093291485}%
\StoreBenchExecResult{LsBenchmarkDefinitiontermination}{TRTerminationReachsafetyLoops}{Total}{}{Cputime}{Avg}{216.6012779572150259067357513}%
\StoreBenchExecResult{LsBenchmarkDefinitiontermination}{TRTerminationReachsafetyLoops}{Total}{}{Cputime}{Median}{0.5553706225}%
\StoreBenchExecResult{LsBenchmarkDefinitiontermination}{TRTerminationReachsafetyLoops}{Total}{}{Cputime}{Min}{0.096467097}%
\StoreBenchExecResult{LsBenchmarkDefinitiontermination}{TRTerminationReachsafetyLoops}{Total}{}{Cputime}{Max}{962.269800888}%
\StoreBenchExecResult{LsBenchmarkDefinitiontermination}{TRTerminationReachsafetyLoops}{Total}{}{Cputime}{Stdev}{384.8288812181103681792148320}%
\StoreBenchExecResult{LsBenchmarkDefinitiontermination}{TRTerminationReachsafetyLoops}{Total}{}{Walltime}{}{76210.58442065864780069}%
\StoreBenchExecResult{LsBenchmarkDefinitiontermination}{TRTerminationReachsafetyLoops}{Total}{}{Walltime}{Avg}{197.4367472037788803126683938}%
\StoreBenchExecResult{LsBenchmarkDefinitiontermination}{TRTerminationReachsafetyLoops}{Total}{}{Walltime}{Median}{0.30059528548736125}%
\StoreBenchExecResult{LsBenchmarkDefinitiontermination}{TRTerminationReachsafetyLoops}{Total}{}{Walltime}{Min}{0.07004928402602673}%
\StoreBenchExecResult{LsBenchmarkDefinitiontermination}{TRTerminationReachsafetyLoops}{Total}{}{Walltime}{Max}{960.5419237220194}%
\StoreBenchExecResult{LsBenchmarkDefinitiontermination}{TRTerminationReachsafetyLoops}{Total}{}{Walltime}{Stdev}{364.6042492305068395069425819}%
\StoreBenchExecResult{LsBenchmarkDefinitiontermination}{TRTerminationReachsafetyLoops}{Correct}{}{Count}{}{262}%
\StoreBenchExecResult{LsBenchmarkDefinitiontermination}{TRTerminationReachsafetyLoops}{Correct}{}{Cputime}{}{205.283963594}%
\StoreBenchExecResult{LsBenchmarkDefinitiontermination}{TRTerminationReachsafetyLoops}{Correct}{}{Cputime}{Avg}{0.7835265786030534351145038168}%
\StoreBenchExecResult{LsBenchmarkDefinitiontermination}{TRTerminationReachsafetyLoops}{Correct}{}{Cputime}{Median}{0.381055074}%
\StoreBenchExecResult{LsBenchmarkDefinitiontermination}{TRTerminationReachsafetyLoops}{Correct}{}{Cputime}{Min}{0.135882784}%
\StoreBenchExecResult{LsBenchmarkDefinitiontermination}{TRTerminationReachsafetyLoops}{Correct}{}{Cputime}{Max}{67.134017394}%
\StoreBenchExecResult{LsBenchmarkDefinitiontermination}{TRTerminationReachsafetyLoops}{Correct}{}{Cputime}{Stdev}{4.135176609092797463396642771}%
\StoreBenchExecResult{LsBenchmarkDefinitiontermination}{TRTerminationReachsafetyLoops}{Correct}{}{Walltime}{}{141.56441645557060537}%
\StoreBenchExecResult{LsBenchmarkDefinitiontermination}{TRTerminationReachsafetyLoops}{Correct}{}{Walltime}{Avg}{0.5403222002121015472137404580}%
\StoreBenchExecResult{LsBenchmarkDefinitiontermination}{TRTerminationReachsafetyLoops}{Correct}{}{Walltime}{Median}{0.21030397352296859}%
\StoreBenchExecResult{LsBenchmarkDefinitiontermination}{TRTerminationReachsafetyLoops}{Correct}{}{Walltime}{Min}{0.09258801210671663}%
\StoreBenchExecResult{LsBenchmarkDefinitiontermination}{TRTerminationReachsafetyLoops}{Correct}{}{Walltime}{Max}{66.84917381382547}%
\StoreBenchExecResult{LsBenchmarkDefinitiontermination}{TRTerminationReachsafetyLoops}{Correct}{}{Walltime}{Stdev}{4.111449641254940541080416845}%
\StoreBenchExecResult{LsBenchmarkDefinitiontermination}{TRTerminationReachsafetyLoops}{Correct}{False}{Count}{}{70}%
\StoreBenchExecResult{LsBenchmarkDefinitiontermination}{TRTerminationReachsafetyLoops}{Correct}{False}{Cputime}{}{84.596549379}%
\StoreBenchExecResult{LsBenchmarkDefinitiontermination}{TRTerminationReachsafetyLoops}{Correct}{False}{Cputime}{Avg}{1.208522133985714285714285714}%
\StoreBenchExecResult{LsBenchmarkDefinitiontermination}{TRTerminationReachsafetyLoops}{Correct}{False}{Cputime}{Median}{0.1752962685}%
\StoreBenchExecResult{LsBenchmarkDefinitiontermination}{TRTerminationReachsafetyLoops}{Correct}{False}{Cputime}{Min}{0.135882784}%
\StoreBenchExecResult{LsBenchmarkDefinitiontermination}{TRTerminationReachsafetyLoops}{Correct}{False}{Cputime}{Max}{67.134017394}%
\StoreBenchExecResult{LsBenchmarkDefinitiontermination}{TRTerminationReachsafetyLoops}{Correct}{False}{Cputime}{Stdev}{7.939743474532389183867069390}%
\StoreBenchExecResult{LsBenchmarkDefinitiontermination}{TRTerminationReachsafetyLoops}{Correct}{False}{Walltime}{}{77.39417450851760482}%
\StoreBenchExecResult{LsBenchmarkDefinitiontermination}{TRTerminationReachsafetyLoops}{Correct}{False}{Walltime}{Avg}{1.105631064407394354571428571}%
\StoreBenchExecResult{LsBenchmarkDefinitiontermination}{TRTerminationReachsafetyLoops}{Correct}{False}{Walltime}{Median}{0.110646996879950165}%
\StoreBenchExecResult{LsBenchmarkDefinitiontermination}{TRTerminationReachsafetyLoops}{Correct}{False}{Walltime}{Min}{0.09318500105291605}%
\StoreBenchExecResult{LsBenchmarkDefinitiontermination}{TRTerminationReachsafetyLoops}{Correct}{False}{Walltime}{Max}{66.84917381382547}%
\StoreBenchExecResult{LsBenchmarkDefinitiontermination}{TRTerminationReachsafetyLoops}{Correct}{False}{Walltime}{Stdev}{7.915467722678127491558355139}%

\StoreBenchExecResult{LsBenchmarkDefinitiontermination}{TRTerminationReachsafetyLoops}{Correct}{True}{Count}{}{192}%
\StoreBenchExecResult{LsBenchmarkDefinitiontermination}{TRTerminationReachsafetyLoops}{Correct}{True}{Cputime}{}{120.687414215}%
\StoreBenchExecResult{LsBenchmarkDefinitiontermination}{TRTerminationReachsafetyLoops}{Correct}{True}{Cputime}{Avg}{0.6285802823697916666666666667}%
\StoreBenchExecResult{LsBenchmarkDefinitiontermination}{TRTerminationReachsafetyLoops}{Correct}{True}{Cputime}{Median}{0.465816798}%
\StoreBenchExecResult{LsBenchmarkDefinitiontermination}{TRTerminationReachsafetyLoops}{Correct}{True}{Cputime}{Min}{0.14593546}%
\StoreBenchExecResult{LsBenchmarkDefinitiontermination}{TRTerminationReachsafetyLoops}{Correct}{True}{Cputime}{Max}{3.192920308}%
\StoreBenchExecResult{LsBenchmarkDefinitiontermination}{TRTerminationReachsafetyLoops}{Correct}{True}{Cputime}{Stdev}{0.5108086094254811794023896557}%
\StoreBenchExecResult{LsBenchmarkDefinitiontermination}{TRTerminationReachsafetyLoops}{Correct}{True}{Walltime}{}{64.17024194705300055}%
\StoreBenchExecResult{LsBenchmarkDefinitiontermination}{TRTerminationReachsafetyLoops}{Correct}{True}{Walltime}{Avg}{0.33422001014090104453125}%
\StoreBenchExecResult{LsBenchmarkDefinitiontermination}{TRTerminationReachsafetyLoops}{Correct}{True}{Walltime}{Median}{0.252926057670265435}%
\StoreBenchExecResult{LsBenchmarkDefinitiontermination}{TRTerminationReachsafetyLoops}{Correct}{True}{Walltime}{Min}{0.09258801210671663}%
\StoreBenchExecResult{LsBenchmarkDefinitiontermination}{TRTerminationReachsafetyLoops}{Correct}{True}{Walltime}{Max}{1.616755205905065}%
\StoreBenchExecResult{LsBenchmarkDefinitiontermination}{TRTerminationReachsafetyLoops}{Correct}{True}{Walltime}{Stdev}{0.2551829580331481622958351262}%

\StoreBenchExecResult{LsBenchmarkDefinitiontermination}{TRTerminationReachsafetyLoops}{Error}{}{Count}{}{100}%
\StoreBenchExecResult{LsBenchmarkDefinitiontermination}{TRTerminationReachsafetyLoops}{Error}{}{Cputime}{}{83349.087808175}%
\StoreBenchExecResult{LsBenchmarkDefinitiontermination}{TRTerminationReachsafetyLoops}{Error}{}{Cputime}{Avg}{833.49087808175}%
\StoreBenchExecResult{LsBenchmarkDefinitiontermination}{TRTerminationReachsafetyLoops}{Error}{}{Cputime}{Median}{960.193727100}%
\StoreBenchExecResult{LsBenchmarkDefinitiontermination}{TRTerminationReachsafetyLoops}{Error}{}{Cputime}{Min}{178.113267749}%
\StoreBenchExecResult{LsBenchmarkDefinitiontermination}{TRTerminationReachsafetyLoops}{Error}{}{Cputime}{Max}{962.269800888}%
\StoreBenchExecResult{LsBenchmarkDefinitiontermination}{TRTerminationReachsafetyLoops}{Error}{}{Cputime}{Stdev}{240.7779356530029729318811654}%
\StoreBenchExecResult{LsBenchmarkDefinitiontermination}{TRTerminationReachsafetyLoops}{Error}{}{Walltime}{}{76040.35483828978592}%
\StoreBenchExecResult{LsBenchmarkDefinitiontermination}{TRTerminationReachsafetyLoops}{Error}{}{Walltime}{Avg}{760.4035483828978592}%
\StoreBenchExecResult{LsBenchmarkDefinitiontermination}{TRTerminationReachsafetyLoops}{Error}{}{Walltime}{Median}{956.6981456839712}%
\StoreBenchExecResult{LsBenchmarkDefinitiontermination}{TRTerminationReachsafetyLoops}{Error}{}{Walltime}{Min}{137.764233173104}%
\StoreBenchExecResult{LsBenchmarkDefinitiontermination}{TRTerminationReachsafetyLoops}{Error}{}{Walltime}{Max}{960.5419237220194}%
\StoreBenchExecResult{LsBenchmarkDefinitiontermination}{TRTerminationReachsafetyLoops}{Error}{}{Walltime}{Stdev}{292.1314756590975213676053174}%
\StoreBenchExecResult{LsBenchmarkDefinitiontermination}{TRTerminationReachsafetyLoops}{Error}{OutOfMemory}{Count}{}{26}%
\StoreBenchExecResult{LsBenchmarkDefinitiontermination}{TRTerminationReachsafetyLoops}{Error}{OutOfMemory}{Cputime}{}{12282.621402778}%
\StoreBenchExecResult{LsBenchmarkDefinitiontermination}{TRTerminationReachsafetyLoops}{Error}{OutOfMemory}{Cputime}{Avg}{472.4085154914615384615384615}%
\StoreBenchExecResult{LsBenchmarkDefinitiontermination}{TRTerminationReachsafetyLoops}{Error}{OutOfMemory}{Cputime}{Median}{422.6521282255}%
\StoreBenchExecResult{LsBenchmarkDefinitiontermination}{TRTerminationReachsafetyLoops}{Error}{OutOfMemory}{Cputime}{Min}{178.113267749}%
\StoreBenchExecResult{LsBenchmarkDefinitiontermination}{TRTerminationReachsafetyLoops}{Error}{OutOfMemory}{Cputime}{Max}{873.137655745}%
\StoreBenchExecResult{LsBenchmarkDefinitiontermination}{TRTerminationReachsafetyLoops}{Error}{OutOfMemory}{Cputime}{Stdev}{216.2947860805285950789948806}%
\StoreBenchExecResult{LsBenchmarkDefinitiontermination}{TRTerminationReachsafetyLoops}{Error}{OutOfMemory}{Walltime}{}{9780.88486710493445}%
\StoreBenchExecResult{LsBenchmarkDefinitiontermination}{TRTerminationReachsafetyLoops}{Error}{OutOfMemory}{Walltime}{Avg}{376.1878795040359403846153846}%
\StoreBenchExecResult{LsBenchmarkDefinitiontermination}{TRTerminationReachsafetyLoops}{Error}{OutOfMemory}{Walltime}{Median}{257.035139357089065}%
\StoreBenchExecResult{LsBenchmarkDefinitiontermination}{TRTerminationReachsafetyLoops}{Error}{OutOfMemory}{Walltime}{Min}{137.764233173104}%
\StoreBenchExecResult{LsBenchmarkDefinitiontermination}{TRTerminationReachsafetyLoops}{Error}{OutOfMemory}{Walltime}{Max}{867.2535497769713}%
\StoreBenchExecResult{LsBenchmarkDefinitiontermination}{TRTerminationReachsafetyLoops}{Error}{OutOfMemory}{Walltime}{Stdev}{243.3797470669717581395818156}%
\StoreBenchExecResult{LsBenchmarkDefinitiontermination}{TRTerminationReachsafetyLoops}{Error}{Timeout}{Count}{}{74}%
\StoreBenchExecResult{LsBenchmarkDefinitiontermination}{TRTerminationReachsafetyLoops}{Error}{Timeout}{Cputime}{}{71066.466405397}%
\StoreBenchExecResult{LsBenchmarkDefinitiontermination}{TRTerminationReachsafetyLoops}{Error}{Timeout}{Cputime}{Avg}{960.3576541269864864864864865}%
\StoreBenchExecResult{LsBenchmarkDefinitiontermination}{TRTerminationReachsafetyLoops}{Error}{Timeout}{Cputime}{Median}{960.216868523}%
\StoreBenchExecResult{LsBenchmarkDefinitiontermination}{TRTerminationReachsafetyLoops}{Error}{Timeout}{Cputime}{Min}{951.97599971}%
\StoreBenchExecResult{LsBenchmarkDefinitiontermination}{TRTerminationReachsafetyLoops}{Error}{Timeout}{Cputime}{Max}{962.269800888}%
\StoreBenchExecResult{LsBenchmarkDefinitiontermination}{TRTerminationReachsafetyLoops}{Error}{Timeout}{Cputime}{Stdev}{1.142603968309704185876271031}%
\StoreBenchExecResult{LsBenchmarkDefinitiontermination}{TRTerminationReachsafetyLoops}{Error}{Timeout}{Walltime}{}{66259.46997118485147}%
\StoreBenchExecResult{LsBenchmarkDefinitiontermination}{TRTerminationReachsafetyLoops}{Error}{Timeout}{Walltime}{Avg}{895.3982428538493441891891892}%
\StoreBenchExecResult{LsBenchmarkDefinitiontermination}{TRTerminationReachsafetyLoops}{Error}{Timeout}{Walltime}{Median}{959.09221429540775}%
\StoreBenchExecResult{LsBenchmarkDefinitiontermination}{TRTerminationReachsafetyLoops}{Error}{Timeout}{Walltime}{Min}{481.00158359413035}%
\StoreBenchExecResult{LsBenchmarkDefinitiontermination}{TRTerminationReachsafetyLoops}{Error}{Timeout}{Walltime}{Max}{960.5419237220194}%
\StoreBenchExecResult{LsBenchmarkDefinitiontermination}{TRTerminationReachsafetyLoops}{Error}{Timeout}{Walltime}{Stdev}{156.2783375647785257410214372}%
\StoreBenchExecResult{LsBenchmarkDefinitiontermination}{TRTerminationReachsafetyLoops}{Unknown}{}{Count}{}{24}%
\StoreBenchExecResult{LsBenchmarkDefinitiontermination}{TRTerminationReachsafetyLoops}{Unknown}{}{Cputime}{}{53.721519716}%
\StoreBenchExecResult{LsBenchmarkDefinitiontermination}{TRTerminationReachsafetyLoops}{Unknown}{}{Cputime}{Avg}{2.238396654833333333333333333}%
\StoreBenchExecResult{LsBenchmarkDefinitiontermination}{TRTerminationReachsafetyLoops}{Unknown}{}{Cputime}{Median}{0.268359160}%
\StoreBenchExecResult{LsBenchmarkDefinitiontermination}{TRTerminationReachsafetyLoops}{Unknown}{}{Cputime}{Min}{0.096467097}%
\StoreBenchExecResult{LsBenchmarkDefinitiontermination}{TRTerminationReachsafetyLoops}{Unknown}{}{Cputime}{Max}{28.408014754}%
\StoreBenchExecResult{LsBenchmarkDefinitiontermination}{TRTerminationReachsafetyLoops}{Unknown}{}{Cputime}{Stdev}{5.605820798896828603419850825}%
\StoreBenchExecResult{LsBenchmarkDefinitiontermination}{TRTerminationReachsafetyLoops}{Unknown}{}{Walltime}{}{28.66516591329127532}%
\StoreBenchExecResult{LsBenchmarkDefinitiontermination}{TRTerminationReachsafetyLoops}{Unknown}{}{Walltime}{Avg}{1.194381913053803138333333333}%
\StoreBenchExecResult{LsBenchmarkDefinitiontermination}{TRTerminationReachsafetyLoops}{Unknown}{}{Walltime}{Median}{0.155185430427081885}%
\StoreBenchExecResult{LsBenchmarkDefinitiontermination}{TRTerminationReachsafetyLoops}{Unknown}{}{Walltime}{Min}{0.07004928402602673}%
\StoreBenchExecResult{LsBenchmarkDefinitiontermination}{TRTerminationReachsafetyLoops}{Unknown}{}{Walltime}{Max}{14.203841261798516}%
\StoreBenchExecResult{LsBenchmarkDefinitiontermination}{TRTerminationReachsafetyLoops}{Unknown}{}{Walltime}{Stdev}{2.817850674888223221856643562}%
\StoreBenchExecResult{LsBenchmarkDefinitiontermination}{TRTerminationReachsafetyLoops}{Unknown}{Unknown}{Count}{}{24}%
\StoreBenchExecResult{LsBenchmarkDefinitiontermination}{TRTerminationReachsafetyLoops}{Unknown}{Unknown}{Cputime}{}{53.721519716}%
\StoreBenchExecResult{LsBenchmarkDefinitiontermination}{TRTerminationReachsafetyLoops}{Unknown}{Unknown}{Cputime}{Avg}{2.238396654833333333333333333}%
\StoreBenchExecResult{LsBenchmarkDefinitiontermination}{TRTerminationReachsafetyLoops}{Unknown}{Unknown}{Cputime}{Median}{0.268359160}%
\StoreBenchExecResult{LsBenchmarkDefinitiontermination}{TRTerminationReachsafetyLoops}{Unknown}{Unknown}{Cputime}{Min}{0.096467097}%
\StoreBenchExecResult{LsBenchmarkDefinitiontermination}{TRTerminationReachsafetyLoops}{Unknown}{Unknown}{Cputime}{Max}{28.408014754}%
\StoreBenchExecResult{LsBenchmarkDefinitiontermination}{TRTerminationReachsafetyLoops}{Unknown}{Unknown}{Cputime}{Stdev}{5.605820798896828603419850825}%
\StoreBenchExecResult{LsBenchmarkDefinitiontermination}{TRTerminationReachsafetyLoops}{Unknown}{Unknown}{Walltime}{}{28.66516591329127532}%
\StoreBenchExecResult{LsBenchmarkDefinitiontermination}{TRTerminationReachsafetyLoops}{Unknown}{Unknown}{Walltime}{Avg}{1.194381913053803138333333333}%
\StoreBenchExecResult{LsBenchmarkDefinitiontermination}{TRTerminationReachsafetyLoops}{Unknown}{Unknown}{Walltime}{Median}{0.155185430427081885}%
\StoreBenchExecResult{LsBenchmarkDefinitiontermination}{TRTerminationReachsafetyLoops}{Unknown}{Unknown}{Walltime}{Min}{0.07004928402602673}%
\StoreBenchExecResult{LsBenchmarkDefinitiontermination}{TRTerminationReachsafetyLoops}{Unknown}{Unknown}{Walltime}{Max}{14.203841261798516}%
\StoreBenchExecResult{LsBenchmarkDefinitiontermination}{TRTerminationReachsafetyLoops}{Unknown}{Unknown}{Walltime}{Stdev}{2.817850674888223221856643562}%
\ifdefined\LsLsTotalCount\else\edef\LsLsTotalCount{0}\fi
\ifdefined\LsLsCorrectCount\else\edef\LsLsCorrectCount{0}\fi
\ifdefined\LsLsCorrectTrueCount\else\edef\LsLsCorrectTrueCount{0}\fi
\ifdefined\LsLsCorrectFalseCount\else\edef\LsLsCorrectFalseCount{0}\fi
\ifdefined\LsLsWrongCount\else\edef\LsLsWrongCount{0}\fi
\ifdefined\LsLsWrongTrueCount\else\edef\LsLsWrongTrueCount{0}\fi
\ifdefined\LsLsWrongFalseCount\else\edef\LsLsWrongFalseCount{0}\fi
\ifdefined\LsLsErrorTimeoutCount\else\edef\LsLsErrorTimeoutCount{0}\fi
\ifdefined\LsLsErrorOutOfMemoryCount\else\edef\LsLsErrorOutOfMemoryCount{0}\fi
\ifdefined\LsLsCorrectCputime\else\edef\LsLsCorrectCputime{0}\fi
\ifdefined\LsLsCorrectCputimeAvg\else\edef\LsLsCorrectCputimeAvg{None}\fi
\ifdefined\LsLsCorrectWalltime\else\edef\LsLsCorrectWalltime{0}\fi
\ifdefined\LsLsCorrectWalltimeAvg\else\edef\LsLsCorrectWalltimeAvg{None}\fi
\edef\LsLsErrorOtherInconclusiveCount{\the\numexpr \LsLsTotalCount - \LsLsCorrectCount - \LsLsWrongTrueCount - \LsLsWrongFalseCount - \LsLsErrorTimeoutCount - \LsLsErrorOutOfMemoryCount \relax}
\providecommand\StoreBenchExecResult[7]{\expandafter\newcommand\csname#1#2#3#4#5#6\endcsname{#7}}%
\StoreBenchExecResult{CpacheckerBenchmarkDefinition}{ORReachabilityReachsafetyLoops}{Total}{}{Count}{}{890}%
\StoreBenchExecResult{CpacheckerBenchmarkDefinition}{ORReachabilityReachsafetyLoops}{Total}{}{Cputime}{}{125877.546357568}%
\StoreBenchExecResult{CpacheckerBenchmarkDefinition}{ORReachabilityReachsafetyLoops}{Total}{}{Cputime}{Avg}{141.4354453455820224719101124}%
\StoreBenchExecResult{CpacheckerBenchmarkDefinition}{ORReachabilityReachsafetyLoops}{Total}{}{Cputime}{Median}{9.4063032475}%
\StoreBenchExecResult{CpacheckerBenchmarkDefinition}{ORReachabilityReachsafetyLoops}{Total}{}{Cputime}{Min}{2.549944837}%
\StoreBenchExecResult{CpacheckerBenchmarkDefinition}{ORReachabilityReachsafetyLoops}{Total}{}{Cputime}{Max}{962.039333578}%
\StoreBenchExecResult{CpacheckerBenchmarkDefinition}{ORReachabilityReachsafetyLoops}{Total}{}{Cputime}{Stdev}{270.2880098191574687635306063}%
\StoreBenchExecResult{CpacheckerBenchmarkDefinition}{ORReachabilityReachsafetyLoops}{Total}{}{Walltime}{}{110940.6136451680216236}%
\StoreBenchExecResult{CpacheckerBenchmarkDefinition}{ORReachabilityReachsafetyLoops}{Total}{}{Walltime}{Avg}{124.6523748822112602512359551}%
\StoreBenchExecResult{CpacheckerBenchmarkDefinition}{ORReachabilityReachsafetyLoops}{Total}{}{Walltime}{Median}{4.972069392912090}%
\StoreBenchExecResult{CpacheckerBenchmarkDefinition}{ORReachabilityReachsafetyLoops}{Total}{}{Walltime}{Min}{1.3628894458524883}%
\StoreBenchExecResult{CpacheckerBenchmarkDefinition}{ORReachabilityReachsafetyLoops}{Total}{}{Walltime}{Max}{914.4986932629254}%
\StoreBenchExecResult{CpacheckerBenchmarkDefinition}{ORReachabilityReachsafetyLoops}{Total}{}{Walltime}{Stdev}{242.7024078288345883870716797}%
\StoreBenchExecResult{CpacheckerBenchmarkDefinition}{ORReachabilityReachsafetyLoops}{Correct}{}{Count}{}{674}%
\StoreBenchExecResult{CpacheckerBenchmarkDefinition}{ORReachabilityReachsafetyLoops}{Correct}{}{Cputime}{}{41558.487382337}%
\StoreBenchExecResult{CpacheckerBenchmarkDefinition}{ORReachabilityReachsafetyLoops}{Correct}{}{Cputime}{Avg}{61.65947682839317507418397626}%
\StoreBenchExecResult{CpacheckerBenchmarkDefinition}{ORReachabilityReachsafetyLoops}{Correct}{}{Cputime}{Median}{9.863775984}%
\StoreBenchExecResult{CpacheckerBenchmarkDefinition}{ORReachabilityReachsafetyLoops}{Correct}{}{Cputime}{Min}{6.176411091}%
\StoreBenchExecResult{CpacheckerBenchmarkDefinition}{ORReachabilityReachsafetyLoops}{Correct}{}{Cputime}{Max}{587.830496087}%
\StoreBenchExecResult{CpacheckerBenchmarkDefinition}{ORReachabilityReachsafetyLoops}{Correct}{}{Cputime}{Stdev}{78.45397729623349549786459289}%
\StoreBenchExecResult{CpacheckerBenchmarkDefinition}{ORReachabilityReachsafetyLoops}{Correct}{}{Walltime}{}{36006.6764766604174257}%
\StoreBenchExecResult{CpacheckerBenchmarkDefinition}{ORReachabilityReachsafetyLoops}{Correct}{}{Walltime}{Avg}{53.42236865973355701142433234}%
\StoreBenchExecResult{CpacheckerBenchmarkDefinition}{ORReachabilityReachsafetyLoops}{Correct}{}{Walltime}{Median}{5.158734572469257}%
\StoreBenchExecResult{CpacheckerBenchmarkDefinition}{ORReachabilityReachsafetyLoops}{Correct}{}{Walltime}{Min}{3.245244201971218}%
\StoreBenchExecResult{CpacheckerBenchmarkDefinition}{ORReachabilityReachsafetyLoops}{Correct}{}{Walltime}{Max}{487.88605564297177}%
\StoreBenchExecResult{CpacheckerBenchmarkDefinition}{ORReachabilityReachsafetyLoops}{Correct}{}{Walltime}{Stdev}{72.00040115055391058588482034}%
\StoreBenchExecResult{CpacheckerBenchmarkDefinition}{ORReachabilityReachsafetyLoops}{Correct}{False}{Count}{}{233}%
\StoreBenchExecResult{CpacheckerBenchmarkDefinition}{ORReachabilityReachsafetyLoops}{Correct}{False}{Cputime}{}{2784.413322046}%
\StoreBenchExecResult{CpacheckerBenchmarkDefinition}{ORReachabilityReachsafetyLoops}{Correct}{False}{Cputime}{Avg}{11.95027176843776824034334764}%
\StoreBenchExecResult{CpacheckerBenchmarkDefinition}{ORReachabilityReachsafetyLoops}{Correct}{False}{Cputime}{Median}{8.098398754}%
\StoreBenchExecResult{CpacheckerBenchmarkDefinition}{ORReachabilityReachsafetyLoops}{Correct}{False}{Cputime}{Min}{6.873407616}%
\StoreBenchExecResult{CpacheckerBenchmarkDefinition}{ORReachabilityReachsafetyLoops}{Correct}{False}{Cputime}{Max}{156.702922026}%
\StoreBenchExecResult{CpacheckerBenchmarkDefinition}{ORReachabilityReachsafetyLoops}{Correct}{False}{Cputime}{Stdev}{19.40882579843853148045329021}%
\StoreBenchExecResult{CpacheckerBenchmarkDefinition}{ORReachabilityReachsafetyLoops}{Correct}{False}{Walltime}{}{1666.7557192430831440}%
\StoreBenchExecResult{CpacheckerBenchmarkDefinition}{ORReachabilityReachsafetyLoops}{Correct}{False}{Walltime}{Avg}{7.153458022502502763948497854}%
\StoreBenchExecResult{CpacheckerBenchmarkDefinition}{ORReachabilityReachsafetyLoops}{Correct}{False}{Walltime}{Median}{4.223396683111787}%
\StoreBenchExecResult{CpacheckerBenchmarkDefinition}{ORReachabilityReachsafetyLoops}{Correct}{False}{Walltime}{Min}{3.576606296002865}%
\StoreBenchExecResult{CpacheckerBenchmarkDefinition}{ORReachabilityReachsafetyLoops}{Correct}{False}{Walltime}{Max}{143.2143928091973}%
\StoreBenchExecResult{CpacheckerBenchmarkDefinition}{ORReachabilityReachsafetyLoops}{Correct}{False}{Walltime}{Stdev}{16.19540574604004761693550978}%
\StoreBenchExecResult{CpacheckerBenchmarkDefinition}{ORReachabilityReachsafetyLoops}{Correct}{True}{Count}{}{441}%
\StoreBenchExecResult{CpacheckerBenchmarkDefinition}{ORReachabilityReachsafetyLoops}{Correct}{True}{Cputime}{}{38774.074060291}%
\StoreBenchExecResult{CpacheckerBenchmarkDefinition}{ORReachabilityReachsafetyLoops}{Correct}{True}{Cputime}{Avg}{87.92307043149886621315192744}%
\StoreBenchExecResult{CpacheckerBenchmarkDefinition}{ORReachabilityReachsafetyLoops}{Correct}{True}{Cputime}{Median}{67.195437602}%
\StoreBenchExecResult{CpacheckerBenchmarkDefinition}{ORReachabilityReachsafetyLoops}{Correct}{True}{Cputime}{Min}{6.176411091}%
\StoreBenchExecResult{CpacheckerBenchmarkDefinition}{ORReachabilityReachsafetyLoops}{Correct}{True}{Cputime}{Max}{587.830496087}%
\StoreBenchExecResult{CpacheckerBenchmarkDefinition}{ORReachabilityReachsafetyLoops}{Correct}{True}{Cputime}{Stdev}{84.92734869783933103948249487}%
\StoreBenchExecResult{CpacheckerBenchmarkDefinition}{ORReachabilityReachsafetyLoops}{Correct}{True}{Walltime}{}{34339.9207574173342817}%
\StoreBenchExecResult{CpacheckerBenchmarkDefinition}{ORReachabilityReachsafetyLoops}{Correct}{True}{Walltime}{Avg}{77.86830103722751537800453515}%
\StoreBenchExecResult{CpacheckerBenchmarkDefinition}{ORReachabilityReachsafetyLoops}{Correct}{True}{Walltime}{Median}{60.271086861146614}%
\StoreBenchExecResult{CpacheckerBenchmarkDefinition}{ORReachabilityReachsafetyLoops}{Correct}{True}{Walltime}{Min}{3.245244201971218}%
\StoreBenchExecResult{CpacheckerBenchmarkDefinition}{ORReachabilityReachsafetyLoops}{Correct}{True}{Walltime}{Max}{487.88605564297177}%
\StoreBenchExecResult{CpacheckerBenchmarkDefinition}{ORReachabilityReachsafetyLoops}{Correct}{True}{Walltime}{Stdev}{77.81874228402069641939221276}%
\StoreBenchExecResult{CpacheckerBenchmarkDefinition}{ORReachabilityReachsafetyLoops}{Error}{}{Count}{}{208}%
\StoreBenchExecResult{CpacheckerBenchmarkDefinition}{ORReachabilityReachsafetyLoops}{Error}{}{Cputime}{}{83975.397899253}%
\StoreBenchExecResult{CpacheckerBenchmarkDefinition}{ORReachabilityReachsafetyLoops}{Error}{}{Cputime}{Avg}{403.7278745156394230769230769}%
\StoreBenchExecResult{CpacheckerBenchmarkDefinition}{ORReachabilityReachsafetyLoops}{Error}{}{Cputime}{Median}{3.6019891975}%
\StoreBenchExecResult{CpacheckerBenchmarkDefinition}{ORReachabilityReachsafetyLoops}{Error}{}{Cputime}{Min}{2.549944837}%
\StoreBenchExecResult{CpacheckerBenchmarkDefinition}{ORReachabilityReachsafetyLoops}{Error}{}{Cputime}{Max}{962.039333578}%
\StoreBenchExecResult{CpacheckerBenchmarkDefinition}{ORReachabilityReachsafetyLoops}{Error}{}{Cputime}{Stdev}{450.2406442613082657718569363}%
\StoreBenchExecResult{CpacheckerBenchmarkDefinition}{ORReachabilityReachsafetyLoops}{Error}{}{Walltime}{}{74635.4294286808001759}%
\StoreBenchExecResult{CpacheckerBenchmarkDefinition}{ORReachabilityReachsafetyLoops}{Error}{}{Walltime}{Avg}{358.8241799455807700764423077}%
\StoreBenchExecResult{CpacheckerBenchmarkDefinition}{ORReachabilityReachsafetyLoops}{Error}{}{Walltime}{Median}{1.9061726131476462}%
\StoreBenchExecResult{CpacheckerBenchmarkDefinition}{ORReachabilityReachsafetyLoops}{Error}{}{Walltime}{Min}{1.3628894458524883}%
\StoreBenchExecResult{CpacheckerBenchmarkDefinition}{ORReachabilityReachsafetyLoops}{Error}{}{Walltime}{Max}{914.4986932629254}%
\StoreBenchExecResult{CpacheckerBenchmarkDefinition}{ORReachabilityReachsafetyLoops}{Error}{}{Walltime}{Stdev}{404.4106088983408229158856267}%
\StoreBenchExecResult{CpacheckerBenchmarkDefinition}{ORReachabilityReachsafetyLoops}{Error}{Error}{Count}{}{114}%
\StoreBenchExecResult{CpacheckerBenchmarkDefinition}{ORReachabilityReachsafetyLoops}{Error}{Error}{Cputime}{}{356.077493223}%
\StoreBenchExecResult{CpacheckerBenchmarkDefinition}{ORReachabilityReachsafetyLoops}{Error}{Error}{Cputime}{Avg}{3.123486782657894736842105263}%
\StoreBenchExecResult{CpacheckerBenchmarkDefinition}{ORReachabilityReachsafetyLoops}{Error}{Error}{Cputime}{Median}{3.0752970005}%
\StoreBenchExecResult{CpacheckerBenchmarkDefinition}{ORReachabilityReachsafetyLoops}{Error}{Error}{Cputime}{Min}{2.549944837}%
\StoreBenchExecResult{CpacheckerBenchmarkDefinition}{ORReachabilityReachsafetyLoops}{Error}{Error}{Cputime}{Max}{4.31357683}%
\StoreBenchExecResult{CpacheckerBenchmarkDefinition}{ORReachabilityReachsafetyLoops}{Error}{Error}{Cputime}{Stdev}{0.3080219300000200888757791103}%
\StoreBenchExecResult{CpacheckerBenchmarkDefinition}{ORReachabilityReachsafetyLoops}{Error}{Error}{Walltime}{}{189.5836389900650829}%
\StoreBenchExecResult{CpacheckerBenchmarkDefinition}{ORReachabilityReachsafetyLoops}{Error}{Error}{Walltime}{Avg}{1.663014377105834060526315789}%
\StoreBenchExecResult{CpacheckerBenchmarkDefinition}{ORReachabilityReachsafetyLoops}{Error}{Error}{Walltime}{Median}{1.63807277602609245}%
\StoreBenchExecResult{CpacheckerBenchmarkDefinition}{ORReachabilityReachsafetyLoops}{Error}{Error}{Walltime}{Min}{1.3628894458524883}%
\StoreBenchExecResult{CpacheckerBenchmarkDefinition}{ORReachabilityReachsafetyLoops}{Error}{Error}{Walltime}{Max}{2.2841865280643106}%
\StoreBenchExecResult{CpacheckerBenchmarkDefinition}{ORReachabilityReachsafetyLoops}{Error}{Error}{Walltime}{Stdev}{0.1579405673037066631901569391}%
\StoreBenchExecResult{CpacheckerBenchmarkDefinition}{ORReachabilityReachsafetyLoops}{Error}{Exception}{Count}{}{1}%
\StoreBenchExecResult{CpacheckerBenchmarkDefinition}{ORReachabilityReachsafetyLoops}{Error}{Exception}{Cputime}{}{15.907113442}%
\StoreBenchExecResult{CpacheckerBenchmarkDefinition}{ORReachabilityReachsafetyLoops}{Error}{Exception}{Cputime}{Avg}{15.907113442}%
\StoreBenchExecResult{CpacheckerBenchmarkDefinition}{ORReachabilityReachsafetyLoops}{Error}{Exception}{Cputime}{Median}{15.907113442}%
\StoreBenchExecResult{CpacheckerBenchmarkDefinition}{ORReachabilityReachsafetyLoops}{Error}{Exception}{Cputime}{Min}{15.907113442}%
\StoreBenchExecResult{CpacheckerBenchmarkDefinition}{ORReachabilityReachsafetyLoops}{Error}{Exception}{Cputime}{Max}{15.907113442}%
\StoreBenchExecResult{CpacheckerBenchmarkDefinition}{ORReachabilityReachsafetyLoops}{Error}{Exception}{Cputime}{Stdev}{0E-14}%
\StoreBenchExecResult{CpacheckerBenchmarkDefinition}{ORReachabilityReachsafetyLoops}{Error}{Exception}{Walltime}{}{8.148837212007493}%
\StoreBenchExecResult{CpacheckerBenchmarkDefinition}{ORReachabilityReachsafetyLoops}{Error}{Exception}{Walltime}{Avg}{8.148837212007493}%
\StoreBenchExecResult{CpacheckerBenchmarkDefinition}{ORReachabilityReachsafetyLoops}{Error}{Exception}{Walltime}{Median}{8.148837212007493}%
\StoreBenchExecResult{CpacheckerBenchmarkDefinition}{ORReachabilityReachsafetyLoops}{Error}{Exception}{Walltime}{Min}{8.148837212007493}%
\StoreBenchExecResult{CpacheckerBenchmarkDefinition}{ORReachabilityReachsafetyLoops}{Error}{Exception}{Walltime}{Max}{8.148837212007493}%
\StoreBenchExecResult{CpacheckerBenchmarkDefinition}{ORReachabilityReachsafetyLoops}{Error}{Exception}{Walltime}{Stdev}{0E-15}%
\StoreBenchExecResult{CpacheckerBenchmarkDefinition}{ORReachabilityReachsafetyLoops}{Error}{OutOfJavaMemory}{Count}{}{2}%
\StoreBenchExecResult{CpacheckerBenchmarkDefinition}{ORReachabilityReachsafetyLoops}{Error}{OutOfJavaMemory}{Cputime}{}{503.611646428}%
\StoreBenchExecResult{CpacheckerBenchmarkDefinition}{ORReachabilityReachsafetyLoops}{Error}{OutOfJavaMemory}{Cputime}{Avg}{251.805823214}%
\StoreBenchExecResult{CpacheckerBenchmarkDefinition}{ORReachabilityReachsafetyLoops}{Error}{OutOfJavaMemory}{Cputime}{Median}{251.805823214}%
\StoreBenchExecResult{CpacheckerBenchmarkDefinition}{ORReachabilityReachsafetyLoops}{Error}{OutOfJavaMemory}{Cputime}{Min}{209.725237719}%
\StoreBenchExecResult{CpacheckerBenchmarkDefinition}{ORReachabilityReachsafetyLoops}{Error}{OutOfJavaMemory}{Cputime}{Max}{293.886408709}%
\StoreBenchExecResult{CpacheckerBenchmarkDefinition}{ORReachabilityReachsafetyLoops}{Error}{OutOfJavaMemory}{Cputime}{Stdev}{42.080585495000}%
\StoreBenchExecResult{CpacheckerBenchmarkDefinition}{ORReachabilityReachsafetyLoops}{Error}{OutOfJavaMemory}{Walltime}{}{439.0378281108569}%
\StoreBenchExecResult{CpacheckerBenchmarkDefinition}{ORReachabilityReachsafetyLoops}{Error}{OutOfJavaMemory}{Walltime}{Avg}{219.51891405542845}%
\StoreBenchExecResult{CpacheckerBenchmarkDefinition}{ORReachabilityReachsafetyLoops}{Error}{OutOfJavaMemory}{Walltime}{Median}{219.51891405542845}%
\StoreBenchExecResult{CpacheckerBenchmarkDefinition}{ORReachabilityReachsafetyLoops}{Error}{OutOfJavaMemory}{Walltime}{Min}{173.6730599249713}%
\StoreBenchExecResult{CpacheckerBenchmarkDefinition}{ORReachabilityReachsafetyLoops}{Error}{OutOfJavaMemory}{Walltime}{Max}{265.3647681858856}%
\StoreBenchExecResult{CpacheckerBenchmarkDefinition}{ORReachabilityReachsafetyLoops}{Error}{OutOfJavaMemory}{Walltime}{Stdev}{45.84585413045715000000000000}%
\StoreBenchExecResult{CpacheckerBenchmarkDefinition}{ORReachabilityReachsafetyLoops}{Error}{OutOfMemory}{Count}{}{1}%
\StoreBenchExecResult{CpacheckerBenchmarkDefinition}{ORReachabilityReachsafetyLoops}{Error}{OutOfMemory}{Cputime}{}{772.714053719}%
\StoreBenchExecResult{CpacheckerBenchmarkDefinition}{ORReachabilityReachsafetyLoops}{Error}{OutOfMemory}{Cputime}{Avg}{772.714053719}%
\StoreBenchExecResult{CpacheckerBenchmarkDefinition}{ORReachabilityReachsafetyLoops}{Error}{OutOfMemory}{Cputime}{Median}{772.714053719}%
\StoreBenchExecResult{CpacheckerBenchmarkDefinition}{ORReachabilityReachsafetyLoops}{Error}{OutOfMemory}{Cputime}{Min}{772.714053719}%
\StoreBenchExecResult{CpacheckerBenchmarkDefinition}{ORReachabilityReachsafetyLoops}{Error}{OutOfMemory}{Cputime}{Max}{772.714053719}%
\StoreBenchExecResult{CpacheckerBenchmarkDefinition}{ORReachabilityReachsafetyLoops}{Error}{OutOfMemory}{Cputime}{Stdev}{0E-14}%
\StoreBenchExecResult{CpacheckerBenchmarkDefinition}{ORReachabilityReachsafetyLoops}{Error}{OutOfMemory}{Walltime}{}{739.4678694810718}%
\StoreBenchExecResult{CpacheckerBenchmarkDefinition}{ORReachabilityReachsafetyLoops}{Error}{OutOfMemory}{Walltime}{Avg}{739.4678694810718}%
\StoreBenchExecResult{CpacheckerBenchmarkDefinition}{ORReachabilityReachsafetyLoops}{Error}{OutOfMemory}{Walltime}{Median}{739.4678694810718}%
\StoreBenchExecResult{CpacheckerBenchmarkDefinition}{ORReachabilityReachsafetyLoops}{Error}{OutOfMemory}{Walltime}{Min}{739.4678694810718}%
\StoreBenchExecResult{CpacheckerBenchmarkDefinition}{ORReachabilityReachsafetyLoops}{Error}{OutOfMemory}{Walltime}{Max}{739.4678694810718}%
\StoreBenchExecResult{CpacheckerBenchmarkDefinition}{ORReachabilityReachsafetyLoops}{Error}{OutOfMemory}{Walltime}{Stdev}{0E-14}%
\StoreBenchExecResult{CpacheckerBenchmarkDefinition}{ORReachabilityReachsafetyLoops}{Error}{Timeout}{Count}{}{90}%
\StoreBenchExecResult{CpacheckerBenchmarkDefinition}{ORReachabilityReachsafetyLoops}{Error}{Timeout}{Cputime}{}{82327.087592441}%
\StoreBenchExecResult{CpacheckerBenchmarkDefinition}{ORReachabilityReachsafetyLoops}{Error}{Timeout}{Cputime}{Avg}{914.7454176937888888888888889}%
\StoreBenchExecResult{CpacheckerBenchmarkDefinition}{ORReachabilityReachsafetyLoops}{Error}{Timeout}{Cputime}{Median}{906.299254846}%
\StoreBenchExecResult{CpacheckerBenchmarkDefinition}{ORReachabilityReachsafetyLoops}{Error}{Timeout}{Cputime}{Min}{902.766697679}%
\StoreBenchExecResult{CpacheckerBenchmarkDefinition}{ORReachabilityReachsafetyLoops}{Error}{Timeout}{Cputime}{Max}{962.039333578}%
\StoreBenchExecResult{CpacheckerBenchmarkDefinition}{ORReachabilityReachsafetyLoops}{Error}{Timeout}{Cputime}{Stdev}{18.59703325306007341289181897}%
\StoreBenchExecResult{CpacheckerBenchmarkDefinition}{ORReachabilityReachsafetyLoops}{Error}{Timeout}{Walltime}{}{73259.1912548867989}%
\StoreBenchExecResult{CpacheckerBenchmarkDefinition}{ORReachabilityReachsafetyLoops}{Error}{Timeout}{Walltime}{Avg}{813.9910139431866544444444444}%
\StoreBenchExecResult{CpacheckerBenchmarkDefinition}{ORReachabilityReachsafetyLoops}{Error}{Timeout}{Walltime}{Median}{844.71557440655305}%
\StoreBenchExecResult{CpacheckerBenchmarkDefinition}{ORReachabilityReachsafetyLoops}{Error}{Timeout}{Walltime}{Min}{617.7711704291869}%
\StoreBenchExecResult{CpacheckerBenchmarkDefinition}{ORReachabilityReachsafetyLoops}{Error}{Timeout}{Walltime}{Max}{914.4986932629254}%
\StoreBenchExecResult{CpacheckerBenchmarkDefinition}{ORReachabilityReachsafetyLoops}{Error}{Timeout}{Walltime}{Stdev}{75.92883967370354554792740442}%
\StoreBenchExecResult{CpacheckerBenchmarkDefinition}{ORReachabilityReachsafetyLoops}{Wrong}{}{Count}{}{8}%
\StoreBenchExecResult{CpacheckerBenchmarkDefinition}{ORReachabilityReachsafetyLoops}{Wrong}{}{Cputime}{}{343.661075978}%
\StoreBenchExecResult{CpacheckerBenchmarkDefinition}{ORReachabilityReachsafetyLoops}{Wrong}{}{Cputime}{Avg}{42.95763449725}%
\StoreBenchExecResult{CpacheckerBenchmarkDefinition}{ORReachabilityReachsafetyLoops}{Wrong}{}{Cputime}{Median}{8.3339697215}%
\StoreBenchExecResult{CpacheckerBenchmarkDefinition}{ORReachabilityReachsafetyLoops}{Wrong}{}{Cputime}{Min}{7.72398783}%
\StoreBenchExecResult{CpacheckerBenchmarkDefinition}{ORReachabilityReachsafetyLoops}{Wrong}{}{Cputime}{Max}{147.702717706}%
\StoreBenchExecResult{CpacheckerBenchmarkDefinition}{ORReachabilityReachsafetyLoops}{Wrong}{}{Cputime}{Stdev}{60.25286526772583418176310132}%
\StoreBenchExecResult{CpacheckerBenchmarkDefinition}{ORReachabilityReachsafetyLoops}{Wrong}{}{Walltime}{}{298.507739826804022}%
\StoreBenchExecResult{CpacheckerBenchmarkDefinition}{ORReachabilityReachsafetyLoops}{Wrong}{}{Walltime}{Avg}{37.31346747835050275}%
\StoreBenchExecResult{CpacheckerBenchmarkDefinition}{ORReachabilityReachsafetyLoops}{Wrong}{}{Walltime}{Median}{4.3209694939432665}%
\StoreBenchExecResult{CpacheckerBenchmarkDefinition}{ORReachabilityReachsafetyLoops}{Wrong}{}{Walltime}{Min}{4.03397434996441}%
\StoreBenchExecResult{CpacheckerBenchmarkDefinition}{ORReachabilityReachsafetyLoops}{Wrong}{}{Walltime}{Max}{136.70706312707625}%
\StoreBenchExecResult{CpacheckerBenchmarkDefinition}{ORReachabilityReachsafetyLoops}{Wrong}{}{Walltime}{Stdev}{57.28833127548977837709656810}%
\StoreBenchExecResult{CpacheckerBenchmarkDefinition}{ORReachabilityReachsafetyLoops}{Wrong}{False}{Count}{}{4}%
\StoreBenchExecResult{CpacheckerBenchmarkDefinition}{ORReachabilityReachsafetyLoops}{Wrong}{False}{Cputime}{}{32.358488903}%
\StoreBenchExecResult{CpacheckerBenchmarkDefinition}{ORReachabilityReachsafetyLoops}{Wrong}{False}{Cputime}{Avg}{8.08962222575}%
\StoreBenchExecResult{CpacheckerBenchmarkDefinition}{ORReachabilityReachsafetyLoops}{Wrong}{False}{Cputime}{Median}{8.034803418}%
\StoreBenchExecResult{CpacheckerBenchmarkDefinition}{ORReachabilityReachsafetyLoops}{Wrong}{False}{Cputime}{Min}{7.72398783}%
\StoreBenchExecResult{CpacheckerBenchmarkDefinition}{ORReachabilityReachsafetyLoops}{Wrong}{False}{Cputime}{Max}{8.564894237}%
\StoreBenchExecResult{CpacheckerBenchmarkDefinition}{ORReachabilityReachsafetyLoops}{Wrong}{False}{Cputime}{Stdev}{0.3025150872385507166963106795}%
\StoreBenchExecResult{CpacheckerBenchmarkDefinition}{ORReachabilityReachsafetyLoops}{Wrong}{False}{Walltime}{}{16.786926537053659}%
\StoreBenchExecResult{CpacheckerBenchmarkDefinition}{ORReachabilityReachsafetyLoops}{Wrong}{False}{Walltime}{Avg}{4.19673163426341475}%
\StoreBenchExecResult{CpacheckerBenchmarkDefinition}{ORReachabilityReachsafetyLoops}{Wrong}{False}{Walltime}{Median}{4.149594503454864}%
\StoreBenchExecResult{CpacheckerBenchmarkDefinition}{ORReachabilityReachsafetyLoops}{Wrong}{False}{Walltime}{Min}{4.03397434996441}%
\StoreBenchExecResult{CpacheckerBenchmarkDefinition}{ORReachabilityReachsafetyLoops}{Wrong}{False}{Walltime}{Max}{4.453763180179521}%
\StoreBenchExecResult{CpacheckerBenchmarkDefinition}{ORReachabilityReachsafetyLoops}{Wrong}{False}{Walltime}{Stdev}{0.1562837076949490988869281382}%
\StoreBenchExecResult{CpacheckerBenchmarkDefinition}{ORReachabilityReachsafetyLoops}{Wrong}{True}{Count}{}{4}%
\StoreBenchExecResult{CpacheckerBenchmarkDefinition}{ORReachabilityReachsafetyLoops}{Wrong}{True}{Cputime}{}{311.302587075}%
\StoreBenchExecResult{CpacheckerBenchmarkDefinition}{ORReachabilityReachsafetyLoops}{Wrong}{True}{Cputime}{Avg}{77.82564676875}%
\StoreBenchExecResult{CpacheckerBenchmarkDefinition}{ORReachabilityReachsafetyLoops}{Wrong}{True}{Cputime}{Median}{77.6366893405}%
\StoreBenchExecResult{CpacheckerBenchmarkDefinition}{ORReachabilityReachsafetyLoops}{Wrong}{True}{Cputime}{Min}{8.326490688}%
\StoreBenchExecResult{CpacheckerBenchmarkDefinition}{ORReachabilityReachsafetyLoops}{Wrong}{True}{Cputime}{Max}{147.702717706}%
\StoreBenchExecResult{CpacheckerBenchmarkDefinition}{ORReachabilityReachsafetyLoops}{Wrong}{True}{Cputime}{Stdev}{69.49221158541788841003766184}%
\StoreBenchExecResult{CpacheckerBenchmarkDefinition}{ORReachabilityReachsafetyLoops}{Wrong}{True}{Walltime}{}{281.720813289750363}%
\StoreBenchExecResult{CpacheckerBenchmarkDefinition}{ORReachabilityReachsafetyLoops}{Wrong}{True}{Walltime}{Avg}{70.43020332243759075}%
\StoreBenchExecResult{CpacheckerBenchmarkDefinition}{ORReachabilityReachsafetyLoops}{Wrong}{True}{Walltime}{Median}{70.354861009283922}%
\StoreBenchExecResult{CpacheckerBenchmarkDefinition}{ORReachabilityReachsafetyLoops}{Wrong}{True}{Walltime}{Min}{4.304028144106269}%
\StoreBenchExecResult{CpacheckerBenchmarkDefinition}{ORReachabilityReachsafetyLoops}{Wrong}{True}{Walltime}{Max}{136.70706312707625}%
\StoreBenchExecResult{CpacheckerBenchmarkDefinition}{ORReachabilityReachsafetyLoops}{Wrong}{True}{Walltime}{Stdev}{66.10934117149515693542250896}%
\ifdefined\CpacheckerCpacheckerTotalCount\else\edef\CpacheckerCpacheckerTotalCount{0}\fi
\ifdefined\CpacheckerCpacheckerCorrectCount\else\edef\CpacheckerCpacheckerCorrectCount{0}\fi
\ifdefined\CpacheckerCpacheckerCorrectTrueCount\else\edef\CpacheckerCpacheckerCorrectTrueCount{0}\fi
\ifdefined\CpacheckerCpacheckerCorrectFalseCount\else\edef\CpacheckerCpacheckerCorrectFalseCount{0}\fi
\ifdefined\CpacheckerCpacheckerWrongCount\else\edef\CpacheckerCpacheckerWrongCount{0}\fi
\ifdefined\CpacheckerCpacheckerWrongTrueCount\else\edef\CpacheckerCpacheckerWrongTrueCount{0}\fi
\ifdefined\CpacheckerCpacheckerWrongFalseCount\else\edef\CpacheckerCpacheckerWrongFalseCount{0}\fi
\ifdefined\CpacheckerCpacheckerErrorTimeoutCount\else\edef\CpacheckerCpacheckerErrorTimeoutCount{0}\fi
\ifdefined\CpacheckerCpacheckerErrorOutOfMemoryCount\else\edef\CpacheckerCpacheckerErrorOutOfMemoryCount{0}\fi
\ifdefined\CpacheckerCpacheckerCorrectCputime\else\edef\CpacheckerCpacheckerCorrectCputime{0}\fi
\ifdefined\CpacheckerCpacheckerCorrectCputimeAvg\else\edef\CpacheckerCpacheckerCorrectCputimeAvg{None}\fi
\ifdefined\CpacheckerCpacheckerCorrectWalltime\else\edef\CpacheckerCpacheckerCorrectWalltime{0}\fi
\ifdefined\CpacheckerCpacheckerCorrectWalltimeAvg\else\edef\CpacheckerCpacheckerCorrectWalltimeAvg{None}\fi
\edef\CpacheckerCpacheckerErrorOtherInconclusiveCount{\the\numexpr \CpacheckerCpacheckerTotalCount - \CpacheckerCpacheckerCorrectCount - \CpacheckerCpacheckerWrongTrueCount - \CpacheckerCpacheckerWrongFalseCount - \CpacheckerCpacheckerErrorTimeoutCount - \CpacheckerCpacheckerErrorOutOfMemoryCount \relax}
\providecommand\StoreBenchExecResult[7]{\expandafter\newcommand\csname#1#2#3#4#5#6\endcsname{#7}}%
\StoreBenchExecResult{UltimateBenchmarkDefinition}{ORReachabilityReachsafetyLoops}{Total}{}{Count}{}{890}%
\StoreBenchExecResult{UltimateBenchmarkDefinition}{ORReachabilityReachsafetyLoops}{Total}{}{Cputime}{}{124938.006819651}%
\StoreBenchExecResult{UltimateBenchmarkDefinition}{ORReachabilityReachsafetyLoops}{Total}{}{Cputime}{Avg}{140.3797829434280898876404494}%
\StoreBenchExecResult{UltimateBenchmarkDefinition}{ORReachabilityReachsafetyLoops}{Total}{}{Cputime}{Median}{17.049305691}%
\StoreBenchExecResult{UltimateBenchmarkDefinition}{ORReachabilityReachsafetyLoops}{Total}{}{Cputime}{Min}{13.433207601}%
\StoreBenchExecResult{UltimateBenchmarkDefinition}{ORReachabilityReachsafetyLoops}{Total}{}{Cputime}{Max}{962.737602838}%
\StoreBenchExecResult{UltimateBenchmarkDefinition}{ORReachabilityReachsafetyLoops}{Total}{}{Cputime}{Stdev}{308.0474223006306444504981752}%
\StoreBenchExecResult{UltimateBenchmarkDefinition}{ORReachabilityReachsafetyLoops}{Total}{}{Walltime}{}{104505.346997242886263}%
\StoreBenchExecResult{UltimateBenchmarkDefinition}{ORReachabilityReachsafetyLoops}{Total}{}{Walltime}{Avg}{117.4217381991493104078651685}%
\StoreBenchExecResult{UltimateBenchmarkDefinition}{ORReachabilityReachsafetyLoops}{Total}{}{Walltime}{Median}{10.084573500440456}%
\StoreBenchExecResult{UltimateBenchmarkDefinition}{ORReachabilityReachsafetyLoops}{Total}{}{Walltime}{Min}{8.147516114870086}%
\StoreBenchExecResult{UltimateBenchmarkDefinition}{ORReachabilityReachsafetyLoops}{Total}{}{Walltime}{Max}{941.6511245388538}%
\StoreBenchExecResult{UltimateBenchmarkDefinition}{ORReachabilityReachsafetyLoops}{Total}{}{Walltime}{Stdev}{273.0611823554759595783468567}%
\StoreBenchExecResult{UltimateBenchmarkDefinition}{ORReachabilityReachsafetyLoops}{Correct}{}{Count}{}{654}%
\StoreBenchExecResult{UltimateBenchmarkDefinition}{ORReachabilityReachsafetyLoops}{Correct}{}{Cputime}{}{16470.533230671}%
\StoreBenchExecResult{UltimateBenchmarkDefinition}{ORReachabilityReachsafetyLoops}{Correct}{}{Cputime}{Avg}{25.18430157594954128440366972}%
\StoreBenchExecResult{UltimateBenchmarkDefinition}{ORReachabilityReachsafetyLoops}{Correct}{}{Cputime}{Median}{15.7467592695}%
\StoreBenchExecResult{UltimateBenchmarkDefinition}{ORReachabilityReachsafetyLoops}{Correct}{}{Cputime}{Min}{13.433207601}%
\StoreBenchExecResult{UltimateBenchmarkDefinition}{ORReachabilityReachsafetyLoops}{Correct}{}{Cputime}{Max}{802.196462836}%
\StoreBenchExecResult{UltimateBenchmarkDefinition}{ORReachabilityReachsafetyLoops}{Correct}{}{Cputime}{Stdev}{53.52782789368719503712998922}%
\StoreBenchExecResult{UltimateBenchmarkDefinition}{ORReachabilityReachsafetyLoops}{Correct}{}{Walltime}{}{10968.362362579442539}%
\StoreBenchExecResult{UltimateBenchmarkDefinition}{ORReachabilityReachsafetyLoops}{Correct}{}{Walltime}{Avg}{16.77119627305725158868501529}%
\StoreBenchExecResult{UltimateBenchmarkDefinition}{ORReachabilityReachsafetyLoops}{Correct}{}{Walltime}{Median}{9.380672903382219}%
\StoreBenchExecResult{UltimateBenchmarkDefinition}{ORReachabilityReachsafetyLoops}{Correct}{}{Walltime}{Min}{8.147516114870086}%
\StoreBenchExecResult{UltimateBenchmarkDefinition}{ORReachabilityReachsafetyLoops}{Correct}{}{Walltime}{Max}{743.3737132109236}%
\StoreBenchExecResult{UltimateBenchmarkDefinition}{ORReachabilityReachsafetyLoops}{Correct}{}{Walltime}{Stdev}{47.97039871841958269837165559}%
\StoreBenchExecResult{UltimateBenchmarkDefinition}{ORReachabilityReachsafetyLoops}{Correct}{False}{Count}{}{216}%
\StoreBenchExecResult{UltimateBenchmarkDefinition}{ORReachabilityReachsafetyLoops}{Correct}{False}{Cputime}{}{4906.151124028}%
\StoreBenchExecResult{UltimateBenchmarkDefinition}{ORReachabilityReachsafetyLoops}{Correct}{False}{Cputime}{Avg}{22.71366261124074074074074074}%
\StoreBenchExecResult{UltimateBenchmarkDefinition}{ORReachabilityReachsafetyLoops}{Correct}{False}{Cputime}{Median}{15.4855550645}%
\StoreBenchExecResult{UltimateBenchmarkDefinition}{ORReachabilityReachsafetyLoops}{Correct}{False}{Cputime}{Min}{14.351386654}%
\StoreBenchExecResult{UltimateBenchmarkDefinition}{ORReachabilityReachsafetyLoops}{Correct}{False}{Cputime}{Max}{670.9407678}%
\StoreBenchExecResult{UltimateBenchmarkDefinition}{ORReachabilityReachsafetyLoops}{Correct}{False}{Cputime}{Stdev}{52.24811908412534142249091279}%
\StoreBenchExecResult{UltimateBenchmarkDefinition}{ORReachabilityReachsafetyLoops}{Correct}{False}{Walltime}{}{3293.328181278659020}%
\StoreBenchExecResult{UltimateBenchmarkDefinition}{ORReachabilityReachsafetyLoops}{Correct}{False}{Walltime}{Avg}{15.24688972814193990740740741}%
\StoreBenchExecResult{UltimateBenchmarkDefinition}{ORReachabilityReachsafetyLoops}{Correct}{False}{Walltime}{Median}{9.2989315540762615}%
\StoreBenchExecResult{UltimateBenchmarkDefinition}{ORReachabilityReachsafetyLoops}{Correct}{False}{Walltime}{Min}{8.678607093868777}%
\StoreBenchExecResult{UltimateBenchmarkDefinition}{ORReachabilityReachsafetyLoops}{Correct}{False}{Walltime}{Max}{620.6902852510102}%
\StoreBenchExecResult{UltimateBenchmarkDefinition}{ORReachabilityReachsafetyLoops}{Correct}{False}{Walltime}{Stdev}{48.47476356381182245395233414}%

\StoreBenchExecResult{UltimateBenchmarkDefinition}{ORReachabilityReachsafetyLoops}{Correct}{True}{Count}{}{438}%
\StoreBenchExecResult{UltimateBenchmarkDefinition}{ORReachabilityReachsafetyLoops}{Correct}{True}{Cputime}{}{11564.382106643}%
\StoreBenchExecResult{UltimateBenchmarkDefinition}{ORReachabilityReachsafetyLoops}{Correct}{True}{Cputime}{Avg}{26.40269887361415525114155251}%
\StoreBenchExecResult{UltimateBenchmarkDefinition}{ORReachabilityReachsafetyLoops}{Correct}{True}{Cputime}{Median}{16.0013486585}%
\StoreBenchExecResult{UltimateBenchmarkDefinition}{ORReachabilityReachsafetyLoops}{Correct}{True}{Cputime}{Min}{13.433207601}%
\StoreBenchExecResult{UltimateBenchmarkDefinition}{ORReachabilityReachsafetyLoops}{Correct}{True}{Cputime}{Max}{802.196462836}%
\StoreBenchExecResult{UltimateBenchmarkDefinition}{ORReachabilityReachsafetyLoops}{Correct}{True}{Cputime}{Stdev}{54.10626315865945187937770206}%
\StoreBenchExecResult{UltimateBenchmarkDefinition}{ORReachabilityReachsafetyLoops}{Correct}{True}{Walltime}{}{7675.034181300783519}%
\StoreBenchExecResult{UltimateBenchmarkDefinition}{ORReachabilityReachsafetyLoops}{Correct}{True}{Walltime}{Avg}{17.52290908972781625342465753}%
\StoreBenchExecResult{UltimateBenchmarkDefinition}{ORReachabilityReachsafetyLoops}{Correct}{True}{Walltime}{Median}{9.5254587384406475}%
\StoreBenchExecResult{UltimateBenchmarkDefinition}{ORReachabilityReachsafetyLoops}{Correct}{True}{Walltime}{Min}{8.147516114870086}%
\StoreBenchExecResult{UltimateBenchmarkDefinition}{ORReachabilityReachsafetyLoops}{Correct}{True}{Walltime}{Max}{743.3737132109236}%
\StoreBenchExecResult{UltimateBenchmarkDefinition}{ORReachabilityReachsafetyLoops}{Correct}{True}{Walltime}{Stdev}{47.70177817013787395610730557}%
\StoreBenchExecResult{UltimateBenchmarkDefinition}{ORReachabilityReachsafetyLoops}{Error}{}{Count}{}{115}%
\StoreBenchExecResult{UltimateBenchmarkDefinition}{ORReachabilityReachsafetyLoops}{Error}{}{Cputime}{}{105561.020716608}%
\StoreBenchExecResult{UltimateBenchmarkDefinition}{ORReachabilityReachsafetyLoops}{Error}{}{Cputime}{Avg}{917.9219192748521739130434783}%
\StoreBenchExecResult{UltimateBenchmarkDefinition}{ORReachabilityReachsafetyLoops}{Error}{}{Cputime}{Median}{960.693157992}%
\StoreBenchExecResult{UltimateBenchmarkDefinition}{ORReachabilityReachsafetyLoops}{Error}{}{Cputime}{Min}{23.966567822}%
\StoreBenchExecResult{UltimateBenchmarkDefinition}{ORReachabilityReachsafetyLoops}{Error}{}{Cputime}{Max}{962.737602838}%
\StoreBenchExecResult{UltimateBenchmarkDefinition}{ORReachabilityReachsafetyLoops}{Error}{}{Cputime}{Stdev}{154.3075853396442772203515089}%
\StoreBenchExecResult{UltimateBenchmarkDefinition}{ORReachabilityReachsafetyLoops}{Error}{}{Walltime}{}{91800.11946802237042}%
\StoreBenchExecResult{UltimateBenchmarkDefinition}{ORReachabilityReachsafetyLoops}{Error}{}{Walltime}{Avg}{798.2619084175858297391304348}%
\StoreBenchExecResult{UltimateBenchmarkDefinition}{ORReachabilityReachsafetyLoops}{Error}{}{Walltime}{Median}{836.2333922979888}%
\StoreBenchExecResult{UltimateBenchmarkDefinition}{ORReachabilityReachsafetyLoops}{Error}{}{Walltime}{Min}{14.51106445491314}%
\StoreBenchExecResult{UltimateBenchmarkDefinition}{ORReachabilityReachsafetyLoops}{Error}{}{Walltime}{Max}{941.6511245388538}%
\StoreBenchExecResult{UltimateBenchmarkDefinition}{ORReachabilityReachsafetyLoops}{Error}{}{Walltime}{Stdev}{177.8444568393819729113746469}%
\StoreBenchExecResult{UltimateBenchmarkDefinition}{ORReachabilityReachsafetyLoops}{Error}{Error}{Count}{}{2}%
\StoreBenchExecResult{UltimateBenchmarkDefinition}{ORReachabilityReachsafetyLoops}{Error}{Error}{Cputime}{}{47.976026310}%
\StoreBenchExecResult{UltimateBenchmarkDefinition}{ORReachabilityReachsafetyLoops}{Error}{Error}{Cputime}{Avg}{23.988013155}%
\StoreBenchExecResult{UltimateBenchmarkDefinition}{ORReachabilityReachsafetyLoops}{Error}{Error}{Cputime}{Median}{23.988013155}%
\StoreBenchExecResult{UltimateBenchmarkDefinition}{ORReachabilityReachsafetyLoops}{Error}{Error}{Cputime}{Min}{23.966567822}%
\StoreBenchExecResult{UltimateBenchmarkDefinition}{ORReachabilityReachsafetyLoops}{Error}{Error}{Cputime}{Max}{24.009458488}%
\StoreBenchExecResult{UltimateBenchmarkDefinition}{ORReachabilityReachsafetyLoops}{Error}{Error}{Cputime}{Stdev}{0.02144533300000}%
\StoreBenchExecResult{UltimateBenchmarkDefinition}{ORReachabilityReachsafetyLoops}{Error}{Error}{Walltime}{}{29.39328122092411}%
\StoreBenchExecResult{UltimateBenchmarkDefinition}{ORReachabilityReachsafetyLoops}{Error}{Error}{Walltime}{Avg}{14.696640610462055}%
\StoreBenchExecResult{UltimateBenchmarkDefinition}{ORReachabilityReachsafetyLoops}{Error}{Error}{Walltime}{Median}{14.696640610462055}%
\StoreBenchExecResult{UltimateBenchmarkDefinition}{ORReachabilityReachsafetyLoops}{Error}{Error}{Walltime}{Min}{14.51106445491314}%
\StoreBenchExecResult{UltimateBenchmarkDefinition}{ORReachabilityReachsafetyLoops}{Error}{Error}{Walltime}{Max}{14.88221676601097}%
\StoreBenchExecResult{UltimateBenchmarkDefinition}{ORReachabilityReachsafetyLoops}{Error}{Error}{Walltime}{Stdev}{0.1855761555489150000000000000}%
\StoreBenchExecResult{UltimateBenchmarkDefinition}{ORReachabilityReachsafetyLoops}{Error}{OutOfMemory}{Count}{}{9}%
\StoreBenchExecResult{UltimateBenchmarkDefinition}{ORReachabilityReachsafetyLoops}{Error}{OutOfMemory}{Cputime}{}{5578.371108317}%
\StoreBenchExecResult{UltimateBenchmarkDefinition}{ORReachabilityReachsafetyLoops}{Error}{OutOfMemory}{Cputime}{Avg}{619.8190120352222222222222222}%
\StoreBenchExecResult{UltimateBenchmarkDefinition}{ORReachabilityReachsafetyLoops}{Error}{OutOfMemory}{Cputime}{Median}{579.855185067}%
\StoreBenchExecResult{UltimateBenchmarkDefinition}{ORReachabilityReachsafetyLoops}{Error}{OutOfMemory}{Cputime}{Min}{495.017609723}%
\StoreBenchExecResult{UltimateBenchmarkDefinition}{ORReachabilityReachsafetyLoops}{Error}{OutOfMemory}{Cputime}{Max}{850.304615235}%
\StoreBenchExecResult{UltimateBenchmarkDefinition}{ORReachabilityReachsafetyLoops}{Error}{OutOfMemory}{Cputime}{Stdev}{128.2306446699470034962169279}%
\StoreBenchExecResult{UltimateBenchmarkDefinition}{ORReachabilityReachsafetyLoops}{Error}{OutOfMemory}{Walltime}{}{3897.02624404290688}%
\StoreBenchExecResult{UltimateBenchmarkDefinition}{ORReachabilityReachsafetyLoops}{Error}{OutOfMemory}{Walltime}{Avg}{433.0029160047674311111111111}%
\StoreBenchExecResult{UltimateBenchmarkDefinition}{ORReachabilityReachsafetyLoops}{Error}{OutOfMemory}{Walltime}{Median}{377.4341328900773}%
\StoreBenchExecResult{UltimateBenchmarkDefinition}{ORReachabilityReachsafetyLoops}{Error}{OutOfMemory}{Walltime}{Min}{308.93616327713244}%
\StoreBenchExecResult{UltimateBenchmarkDefinition}{ORReachabilityReachsafetyLoops}{Error}{OutOfMemory}{Walltime}{Max}{721.3108708199579}%
\StoreBenchExecResult{UltimateBenchmarkDefinition}{ORReachabilityReachsafetyLoops}{Error}{OutOfMemory}{Walltime}{Stdev}{137.8500526897356281232257902}%
\StoreBenchExecResult{UltimateBenchmarkDefinition}{ORReachabilityReachsafetyLoops}{Error}{Timeout}{Count}{}{104}%
\StoreBenchExecResult{UltimateBenchmarkDefinition}{ORReachabilityReachsafetyLoops}{Error}{Timeout}{Cputime}{}{99934.673581981}%
\StoreBenchExecResult{UltimateBenchmarkDefinition}{ORReachabilityReachsafetyLoops}{Error}{Timeout}{Cputime}{Avg}{960.9103229036634615384615385}%
\StoreBenchExecResult{UltimateBenchmarkDefinition}{ORReachabilityReachsafetyLoops}{Error}{Timeout}{Cputime}{Median}{960.7397572275}%
\StoreBenchExecResult{UltimateBenchmarkDefinition}{ORReachabilityReachsafetyLoops}{Error}{Timeout}{Cputime}{Min}{960.128670392}%
\StoreBenchExecResult{UltimateBenchmarkDefinition}{ORReachabilityReachsafetyLoops}{Error}{Timeout}{Cputime}{Max}{962.737602838}%
\StoreBenchExecResult{UltimateBenchmarkDefinition}{ORReachabilityReachsafetyLoops}{Error}{Timeout}{Cputime}{Stdev}{0.6326449207312872485619269842}%
\StoreBenchExecResult{UltimateBenchmarkDefinition}{ORReachabilityReachsafetyLoops}{Error}{Timeout}{Walltime}{}{87873.69994275853943}%
\StoreBenchExecResult{UltimateBenchmarkDefinition}{ORReachabilityReachsafetyLoops}{Error}{Timeout}{Walltime}{Avg}{844.9394225265244175961538462}%
\StoreBenchExecResult{UltimateBenchmarkDefinition}{ORReachabilityReachsafetyLoops}{Error}{Timeout}{Walltime}{Median}{847.5248842699220}%
\StoreBenchExecResult{UltimateBenchmarkDefinition}{ORReachabilityReachsafetyLoops}{Error}{Timeout}{Walltime}{Min}{487.0194578268565}%
\StoreBenchExecResult{UltimateBenchmarkDefinition}{ORReachabilityReachsafetyLoops}{Error}{Timeout}{Walltime}{Max}{941.6511245388538}%
\StoreBenchExecResult{UltimateBenchmarkDefinition}{ORReachabilityReachsafetyLoops}{Error}{Timeout}{Walltime}{Stdev}{88.30675380647535660261829995}%
\StoreBenchExecResult{UltimateBenchmarkDefinition}{ORReachabilityReachsafetyLoops}{Unknown}{}{Count}{}{119}%
\StoreBenchExecResult{UltimateBenchmarkDefinition}{ORReachabilityReachsafetyLoops}{Unknown}{}{Cputime}{}{2876.562596026}%
\StoreBenchExecResult{UltimateBenchmarkDefinition}{ORReachabilityReachsafetyLoops}{Unknown}{}{Cputime}{Avg}{24.17279492458823529411764706}%
\StoreBenchExecResult{UltimateBenchmarkDefinition}{ORReachabilityReachsafetyLoops}{Unknown}{}{Cputime}{Median}{24.026532216}%
\StoreBenchExecResult{UltimateBenchmarkDefinition}{ORReachabilityReachsafetyLoops}{Unknown}{}{Cputime}{Min}{22.702831951}%
\StoreBenchExecResult{UltimateBenchmarkDefinition}{ORReachabilityReachsafetyLoops}{Unknown}{}{Cputime}{Max}{28.052482841}%
\StoreBenchExecResult{UltimateBenchmarkDefinition}{ORReachabilityReachsafetyLoops}{Unknown}{}{Cputime}{Stdev}{0.7739770656061442716695593738}%
\StoreBenchExecResult{UltimateBenchmarkDefinition}{ORReachabilityReachsafetyLoops}{Unknown}{}{Walltime}{}{1718.584469052031639}%
\StoreBenchExecResult{UltimateBenchmarkDefinition}{ORReachabilityReachsafetyLoops}{Unknown}{}{Walltime}{Avg}{14.44188629455488772268907563}%
\StoreBenchExecResult{UltimateBenchmarkDefinition}{ORReachabilityReachsafetyLoops}{Unknown}{}{Walltime}{Median}{14.40673146606423}%
\StoreBenchExecResult{UltimateBenchmarkDefinition}{ORReachabilityReachsafetyLoops}{Unknown}{}{Walltime}{Min}{13.757377540925518}%
\StoreBenchExecResult{UltimateBenchmarkDefinition}{ORReachabilityReachsafetyLoops}{Unknown}{}{Walltime}{Max}{16.818794796941802}%
\StoreBenchExecResult{UltimateBenchmarkDefinition}{ORReachabilityReachsafetyLoops}{Unknown}{}{Walltime}{Stdev}{0.3300783259197136443099386219}%
\StoreBenchExecResult{UltimateBenchmarkDefinition}{ORReachabilityReachsafetyLoops}{Unknown}{Unknown}{Count}{}{119}%
\StoreBenchExecResult{UltimateBenchmarkDefinition}{ORReachabilityReachsafetyLoops}{Unknown}{Unknown}{Cputime}{}{2876.562596026}%
\StoreBenchExecResult{UltimateBenchmarkDefinition}{ORReachabilityReachsafetyLoops}{Unknown}{Unknown}{Cputime}{Avg}{24.17279492458823529411764706}%
\StoreBenchExecResult{UltimateBenchmarkDefinition}{ORReachabilityReachsafetyLoops}{Unknown}{Unknown}{Cputime}{Median}{24.026532216}%
\StoreBenchExecResult{UltimateBenchmarkDefinition}{ORReachabilityReachsafetyLoops}{Unknown}{Unknown}{Cputime}{Min}{22.702831951}%
\StoreBenchExecResult{UltimateBenchmarkDefinition}{ORReachabilityReachsafetyLoops}{Unknown}{Unknown}{Cputime}{Max}{28.052482841}%
\StoreBenchExecResult{UltimateBenchmarkDefinition}{ORReachabilityReachsafetyLoops}{Unknown}{Unknown}{Cputime}{Stdev}{0.7739770656061442716695593738}%
\StoreBenchExecResult{UltimateBenchmarkDefinition}{ORReachabilityReachsafetyLoops}{Unknown}{Unknown}{Walltime}{}{1718.584469052031639}%
\StoreBenchExecResult{UltimateBenchmarkDefinition}{ORReachabilityReachsafetyLoops}{Unknown}{Unknown}{Walltime}{Avg}{14.44188629455488772268907563}%
\StoreBenchExecResult{UltimateBenchmarkDefinition}{ORReachabilityReachsafetyLoops}{Unknown}{Unknown}{Walltime}{Median}{14.40673146606423}%
\StoreBenchExecResult{UltimateBenchmarkDefinition}{ORReachabilityReachsafetyLoops}{Unknown}{Unknown}{Walltime}{Min}{13.757377540925518}%
\StoreBenchExecResult{UltimateBenchmarkDefinition}{ORReachabilityReachsafetyLoops}{Unknown}{Unknown}{Walltime}{Max}{16.818794796941802}%
\StoreBenchExecResult{UltimateBenchmarkDefinition}{ORReachabilityReachsafetyLoops}{Unknown}{Unknown}{Walltime}{Stdev}{0.3300783259197136443099386219}%
\StoreBenchExecResult{UltimateBenchmarkDefinition}{ORReachabilityReachsafetyLoops}{Wrong}{}{Count}{}{2}%
\StoreBenchExecResult{UltimateBenchmarkDefinition}{ORReachabilityReachsafetyLoops}{Wrong}{}{Cputime}{}{29.890276346}%
\StoreBenchExecResult{UltimateBenchmarkDefinition}{ORReachabilityReachsafetyLoops}{Wrong}{}{Cputime}{Avg}{14.945138173}%
\StoreBenchExecResult{UltimateBenchmarkDefinition}{ORReachabilityReachsafetyLoops}{Wrong}{}{Cputime}{Median}{14.945138173}%
\StoreBenchExecResult{UltimateBenchmarkDefinition}{ORReachabilityReachsafetyLoops}{Wrong}{}{Cputime}{Min}{14.867223728}%
\StoreBenchExecResult{UltimateBenchmarkDefinition}{ORReachabilityReachsafetyLoops}{Wrong}{}{Cputime}{Max}{15.023052618}%
\StoreBenchExecResult{UltimateBenchmarkDefinition}{ORReachabilityReachsafetyLoops}{Wrong}{}{Cputime}{Stdev}{0.07791444500000}%
\StoreBenchExecResult{UltimateBenchmarkDefinition}{ORReachabilityReachsafetyLoops}{Wrong}{}{Walltime}{}{18.280697589041665}%
\StoreBenchExecResult{UltimateBenchmarkDefinition}{ORReachabilityReachsafetyLoops}{Wrong}{}{Walltime}{Avg}{9.1403487945208325}%
\StoreBenchExecResult{UltimateBenchmarkDefinition}{ORReachabilityReachsafetyLoops}{Wrong}{}{Walltime}{Median}{9.1403487945208325}%
\StoreBenchExecResult{UltimateBenchmarkDefinition}{ORReachabilityReachsafetyLoops}{Wrong}{}{Walltime}{Min}{8.949022069107741}%
\StoreBenchExecResult{UltimateBenchmarkDefinition}{ORReachabilityReachsafetyLoops}{Wrong}{}{Walltime}{Max}{9.331675519933924}%
\StoreBenchExecResult{UltimateBenchmarkDefinition}{ORReachabilityReachsafetyLoops}{Wrong}{}{Walltime}{Stdev}{0.1913267254130915000000000000}%
\StoreBenchExecResult{UltimateBenchmarkDefinition}{ORReachabilityReachsafetyLoops}{Wrong}{True}{Count}{}{2}%
\StoreBenchExecResult{UltimateBenchmarkDefinition}{ORReachabilityReachsafetyLoops}{Wrong}{True}{Cputime}{}{29.890276346}%
\StoreBenchExecResult{UltimateBenchmarkDefinition}{ORReachabilityReachsafetyLoops}{Wrong}{True}{Cputime}{Avg}{14.945138173}%
\StoreBenchExecResult{UltimateBenchmarkDefinition}{ORReachabilityReachsafetyLoops}{Wrong}{True}{Cputime}{Median}{14.945138173}%
\StoreBenchExecResult{UltimateBenchmarkDefinition}{ORReachabilityReachsafetyLoops}{Wrong}{True}{Cputime}{Min}{14.867223728}%
\StoreBenchExecResult{UltimateBenchmarkDefinition}{ORReachabilityReachsafetyLoops}{Wrong}{True}{Cputime}{Max}{15.023052618}%
\StoreBenchExecResult{UltimateBenchmarkDefinition}{ORReachabilityReachsafetyLoops}{Wrong}{True}{Cputime}{Stdev}{0.07791444500000}%
\StoreBenchExecResult{UltimateBenchmarkDefinition}{ORReachabilityReachsafetyLoops}{Wrong}{True}{Walltime}{}{18.280697589041665}%
\StoreBenchExecResult{UltimateBenchmarkDefinition}{ORReachabilityReachsafetyLoops}{Wrong}{True}{Walltime}{Avg}{9.1403487945208325}%
\StoreBenchExecResult{UltimateBenchmarkDefinition}{ORReachabilityReachsafetyLoops}{Wrong}{True}{Walltime}{Median}{9.1403487945208325}%
\StoreBenchExecResult{UltimateBenchmarkDefinition}{ORReachabilityReachsafetyLoops}{Wrong}{True}{Walltime}{Min}{8.949022069107741}%
\StoreBenchExecResult{UltimateBenchmarkDefinition}{ORReachabilityReachsafetyLoops}{Wrong}{True}{Walltime}{Max}{9.331675519933924}%
\StoreBenchExecResult{UltimateBenchmarkDefinition}{ORReachabilityReachsafetyLoops}{Wrong}{True}{Walltime}{Stdev}{0.1913267254130915000000000000}%
\ifdefined\UltimateUltimateTotalCount\else\edef\UltimateUltimateTotalCount{0}\fi
\ifdefined\UltimateUltimateCorrectCount\else\edef\UltimateUltimateCorrectCount{0}\fi
\ifdefined\UltimateUltimateCorrectTrueCount\else\edef\UltimateUltimateCorrectTrueCount{0}\fi
\ifdefined\UltimateUltimateCorrectFalseCount\else\edef\UltimateUltimateCorrectFalseCount{0}\fi
\ifdefined\UltimateUltimateWrongCount\else\edef\UltimateUltimateWrongCount{0}\fi
\ifdefined\UltimateUltimateWrongTrueCount\else\edef\UltimateUltimateWrongTrueCount{0}\fi
\ifdefined\UltimateUltimateWrongFalseCount\else\edef\UltimateUltimateWrongFalseCount{0}\fi
\ifdefined\UltimateUltimateErrorTimeoutCount\else\edef\UltimateUltimateErrorTimeoutCount{0}\fi
\ifdefined\UltimateUltimateErrorOutOfMemoryCount\else\edef\UltimateUltimateErrorOutOfMemoryCount{0}\fi
\ifdefined\UltimateUltimateCorrectCputime\else\edef\UltimateUltimateCorrectCputime{0}\fi
\ifdefined\UltimateUltimateCorrectCputimeAvg\else\edef\UltimateUltimateCorrectCputimeAvg{None}\fi
\ifdefined\UltimateUltimateCorrectWalltime\else\edef\UltimateUltimateCorrectWalltime{0}\fi
\ifdefined\UltimateUltimateCorrectWalltimeAvg\else\edef\UltimateUltimateCorrectWalltimeAvg{None}\fi
\edef\UltimateUltimateErrorOtherInconclusiveCount{\the\numexpr \UltimateUltimateTotalCount - \UltimateUltimateCorrectCount - \UltimateUltimateWrongTrueCount - \UltimateUltimateWrongFalseCount - \UltimateUltimateErrorTimeoutCount - \UltimateUltimateErrorOutOfMemoryCount \relax}
\providecommand\StoreBenchExecResult[7]{\expandafter\newcommand\csname#1#2#3#4#5#6\endcsname{#7}}%
\StoreBenchExecResult{UltimateBenchmarkDefinition}{ORNooverflowReachsafetyLoops}{Total}{}{Count}{}{890}%
\StoreBenchExecResult{UltimateBenchmarkDefinition}{ORNooverflowReachsafetyLoops}{Total}{}{Cputime}{}{211016.825569021}%
\StoreBenchExecResult{UltimateBenchmarkDefinition}{ORNooverflowReachsafetyLoops}{Total}{}{Cputime}{Avg}{237.0975568191247191011235955}%
\StoreBenchExecResult{UltimateBenchmarkDefinition}{ORNooverflowReachsafetyLoops}{Total}{}{Cputime}{Median}{17.6822655995}%
\StoreBenchExecResult{UltimateBenchmarkDefinition}{ORNooverflowReachsafetyLoops}{Total}{}{Cputime}{Min}{13.735585593}%
\StoreBenchExecResult{UltimateBenchmarkDefinition}{ORNooverflowReachsafetyLoops}{Total}{}{Cputime}{Max}{961.611707382}%
\StoreBenchExecResult{UltimateBenchmarkDefinition}{ORNooverflowReachsafetyLoops}{Total}{}{Cputime}{Stdev}{386.4229674698807343610275664}%
\StoreBenchExecResult{UltimateBenchmarkDefinition}{ORNooverflowReachsafetyLoops}{Total}{}{Walltime}{}{189896.284401565557989}%
\StoreBenchExecResult{UltimateBenchmarkDefinition}{ORNooverflowReachsafetyLoops}{Total}{}{Walltime}{Avg}{213.3666116871523123471910112}%
\StoreBenchExecResult{UltimateBenchmarkDefinition}{ORNooverflowReachsafetyLoops}{Total}{}{Walltime}{Median}{10.641837867558934}%
\StoreBenchExecResult{UltimateBenchmarkDefinition}{ORNooverflowReachsafetyLoops}{Total}{}{Walltime}{Min}{8.285295977024361}%
\StoreBenchExecResult{UltimateBenchmarkDefinition}{ORNooverflowReachsafetyLoops}{Total}{}{Walltime}{Max}{941.8804043841083}%
\StoreBenchExecResult{UltimateBenchmarkDefinition}{ORNooverflowReachsafetyLoops}{Total}{}{Walltime}{Stdev}{359.8801264085130354826983956}%
\StoreBenchExecResult{UltimateBenchmarkDefinition}{ORNooverflowReachsafetyLoops}{Correct}{}{Count}{}{676}%
\StoreBenchExecResult{UltimateBenchmarkDefinition}{ORNooverflowReachsafetyLoops}{Correct}{}{Cputime}{}{24670.399368634}%
\StoreBenchExecResult{UltimateBenchmarkDefinition}{ORNooverflowReachsafetyLoops}{Correct}{}{Cputime}{Avg}{36.49467362223964497041420118}%
\StoreBenchExecResult{UltimateBenchmarkDefinition}{ORNooverflowReachsafetyLoops}{Correct}{}{Cputime}{Median}{16.113299339}%
\StoreBenchExecResult{UltimateBenchmarkDefinition}{ORNooverflowReachsafetyLoops}{Correct}{}{Cputime}{Min}{13.735585593}%
\StoreBenchExecResult{UltimateBenchmarkDefinition}{ORNooverflowReachsafetyLoops}{Correct}{}{Cputime}{Max}{854.382334895}%
\StoreBenchExecResult{UltimateBenchmarkDefinition}{ORNooverflowReachsafetyLoops}{Correct}{}{Cputime}{Stdev}{80.82612111829864357148828269}%
\StoreBenchExecResult{UltimateBenchmarkDefinition}{ORNooverflowReachsafetyLoops}{Correct}{}{Walltime}{}{18219.335484369425180}%
\StoreBenchExecResult{UltimateBenchmarkDefinition}{ORNooverflowReachsafetyLoops}{Correct}{}{Walltime}{Avg}{26.95167971060565855029585799}%
\StoreBenchExecResult{UltimateBenchmarkDefinition}{ORNooverflowReachsafetyLoops}{Correct}{}{Walltime}{Median}{9.5734958979301155}%
\StoreBenchExecResult{UltimateBenchmarkDefinition}{ORNooverflowReachsafetyLoops}{Correct}{}{Walltime}{Min}{8.285295977024361}%
\StoreBenchExecResult{UltimateBenchmarkDefinition}{ORNooverflowReachsafetyLoops}{Correct}{}{Walltime}{Max}{804.2648605310824}%
\StoreBenchExecResult{UltimateBenchmarkDefinition}{ORNooverflowReachsafetyLoops}{Correct}{}{Walltime}{Stdev}{74.41353884533195022224533951}%
\StoreBenchExecResult{UltimateBenchmarkDefinition}{ORNooverflowReachsafetyLoops}{Correct}{False}{Count}{}{250}%
\StoreBenchExecResult{UltimateBenchmarkDefinition}{ORNooverflowReachsafetyLoops}{Correct}{False}{Cputime}{}{5307.202599077}%
\StoreBenchExecResult{UltimateBenchmarkDefinition}{ORNooverflowReachsafetyLoops}{Correct}{False}{Cputime}{Avg}{21.228810396308}%
\StoreBenchExecResult{UltimateBenchmarkDefinition}{ORNooverflowReachsafetyLoops}{Correct}{False}{Cputime}{Median}{15.498091702}%
\StoreBenchExecResult{UltimateBenchmarkDefinition}{ORNooverflowReachsafetyLoops}{Correct}{False}{Cputime}{Min}{14.01944414}%
\StoreBenchExecResult{UltimateBenchmarkDefinition}{ORNooverflowReachsafetyLoops}{Correct}{False}{Cputime}{Max}{437.89470318}%
\StoreBenchExecResult{UltimateBenchmarkDefinition}{ORNooverflowReachsafetyLoops}{Correct}{False}{Cputime}{Stdev}{32.76575736035480163896726195}%
\StoreBenchExecResult{UltimateBenchmarkDefinition}{ORNooverflowReachsafetyLoops}{Correct}{False}{Walltime}{}{3462.076013662386715}%
\StoreBenchExecResult{UltimateBenchmarkDefinition}{ORNooverflowReachsafetyLoops}{Correct}{False}{Walltime}{Avg}{13.84830405464954686}%
\StoreBenchExecResult{UltimateBenchmarkDefinition}{ORNooverflowReachsafetyLoops}{Correct}{False}{Walltime}{Median}{9.268391452496871}%
\StoreBenchExecResult{UltimateBenchmarkDefinition}{ORNooverflowReachsafetyLoops}{Correct}{False}{Walltime}{Min}{8.425731634022668}%
\StoreBenchExecResult{UltimateBenchmarkDefinition}{ORNooverflowReachsafetyLoops}{Correct}{False}{Walltime}{Max}{400.6642468061764}%
\StoreBenchExecResult{UltimateBenchmarkDefinition}{ORNooverflowReachsafetyLoops}{Correct}{False}{Walltime}{Stdev}{29.23756519973027464912777179}%

\StoreBenchExecResult{UltimateBenchmarkDefinition}{ORNooverflowReachsafetyLoops}{Correct}{True}{Count}{}{426}%
\StoreBenchExecResult{UltimateBenchmarkDefinition}{ORNooverflowReachsafetyLoops}{Correct}{True}{Cputime}{}{19363.196769557}%
\StoreBenchExecResult{UltimateBenchmarkDefinition}{ORNooverflowReachsafetyLoops}{Correct}{True}{Cputime}{Avg}{45.45351354356103286384976526}%
\StoreBenchExecResult{UltimateBenchmarkDefinition}{ORNooverflowReachsafetyLoops}{Correct}{True}{Cputime}{Median}{17.0310193425}%
\StoreBenchExecResult{UltimateBenchmarkDefinition}{ORNooverflowReachsafetyLoops}{Correct}{True}{Cputime}{Min}{13.735585593}%
\StoreBenchExecResult{UltimateBenchmarkDefinition}{ORNooverflowReachsafetyLoops}{Correct}{True}{Cputime}{Max}{854.382334895}%
\StoreBenchExecResult{UltimateBenchmarkDefinition}{ORNooverflowReachsafetyLoops}{Correct}{True}{Cputime}{Stdev}{97.56860050738513885221144377}%
\StoreBenchExecResult{UltimateBenchmarkDefinition}{ORNooverflowReachsafetyLoops}{Correct}{True}{Walltime}{}{14757.259470707038465}%
\StoreBenchExecResult{UltimateBenchmarkDefinition}{ORNooverflowReachsafetyLoops}{Correct}{True}{Walltime}{Avg}{34.64145415658929217136150235}%
\StoreBenchExecResult{UltimateBenchmarkDefinition}{ORNooverflowReachsafetyLoops}{Correct}{True}{Walltime}{Median}{10.2089670425048095}%
\StoreBenchExecResult{UltimateBenchmarkDefinition}{ORNooverflowReachsafetyLoops}{Correct}{True}{Walltime}{Min}{8.285295977024361}%
\StoreBenchExecResult{UltimateBenchmarkDefinition}{ORNooverflowReachsafetyLoops}{Correct}{True}{Walltime}{Max}{804.2648605310824}%
\StoreBenchExecResult{UltimateBenchmarkDefinition}{ORNooverflowReachsafetyLoops}{Correct}{True}{Walltime}{Stdev}{90.14127437273236362852323403}%

\StoreBenchExecResult{UltimateBenchmarkDefinition}{ORNooverflowReachsafetyLoops}{Error}{}{Count}{}{204}%
\StoreBenchExecResult{UltimateBenchmarkDefinition}{ORNooverflowReachsafetyLoops}{Error}{}{Cputime}{}{186103.208695353}%
\StoreBenchExecResult{UltimateBenchmarkDefinition}{ORNooverflowReachsafetyLoops}{Error}{}{Cputime}{Avg}{912.2706308595735294117647059}%
\StoreBenchExecResult{UltimateBenchmarkDefinition}{ORNooverflowReachsafetyLoops}{Error}{}{Cputime}{Median}{960.544188656}%
\StoreBenchExecResult{UltimateBenchmarkDefinition}{ORNooverflowReachsafetyLoops}{Error}{}{Cputime}{Min}{37.304787906}%
\StoreBenchExecResult{UltimateBenchmarkDefinition}{ORNooverflowReachsafetyLoops}{Error}{}{Cputime}{Max}{961.611707382}%
\StoreBenchExecResult{UltimateBenchmarkDefinition}{ORNooverflowReachsafetyLoops}{Error}{}{Cputime}{Stdev}{195.9118198442314967787052738}%
\StoreBenchExecResult{UltimateBenchmarkDefinition}{ORNooverflowReachsafetyLoops}{Error}{}{Walltime}{}{171531.136260121130019}%
\StoreBenchExecResult{UltimateBenchmarkDefinition}{ORNooverflowReachsafetyLoops}{Error}{}{Walltime}{Avg}{840.83890323588789225}%
\StoreBenchExecResult{UltimateBenchmarkDefinition}{ORNooverflowReachsafetyLoops}{Error}{}{Walltime}{Median}{900.1083296923898}%
\StoreBenchExecResult{UltimateBenchmarkDefinition}{ORNooverflowReachsafetyLoops}{Error}{}{Walltime}{Min}{24.822406917111948}%
\StoreBenchExecResult{UltimateBenchmarkDefinition}{ORNooverflowReachsafetyLoops}{Error}{}{Walltime}{Max}{941.8804043841083}%
\StoreBenchExecResult{UltimateBenchmarkDefinition}{ORNooverflowReachsafetyLoops}{Error}{}{Walltime}{Stdev}{189.4035423896173454368147519}%
\StoreBenchExecResult{UltimateBenchmarkDefinition}{ORNooverflowReachsafetyLoops}{Error}{Error}{Count}{}{9}%
\StoreBenchExecResult{UltimateBenchmarkDefinition}{ORNooverflowReachsafetyLoops}{Error}{Error}{Cputime}{}{391.400677252}%
\StoreBenchExecResult{UltimateBenchmarkDefinition}{ORNooverflowReachsafetyLoops}{Error}{Error}{Cputime}{Avg}{43.48896413911111111111111111}%
\StoreBenchExecResult{UltimateBenchmarkDefinition}{ORNooverflowReachsafetyLoops}{Error}{Error}{Cputime}{Median}{41.495998801}%
\StoreBenchExecResult{UltimateBenchmarkDefinition}{ORNooverflowReachsafetyLoops}{Error}{Error}{Cputime}{Min}{37.304787906}%
\StoreBenchExecResult{UltimateBenchmarkDefinition}{ORNooverflowReachsafetyLoops}{Error}{Error}{Cputime}{Max}{54.975351766}%
\StoreBenchExecResult{UltimateBenchmarkDefinition}{ORNooverflowReachsafetyLoops}{Error}{Error}{Cputime}{Stdev}{5.964672582922403295736169680}%
\StoreBenchExecResult{UltimateBenchmarkDefinition}{ORNooverflowReachsafetyLoops}{Error}{Error}{Walltime}{}{281.958126340294259}%
\StoreBenchExecResult{UltimateBenchmarkDefinition}{ORNooverflowReachsafetyLoops}{Error}{Error}{Walltime}{Avg}{31.32868070447713988888888889}%
\StoreBenchExecResult{UltimateBenchmarkDefinition}{ORNooverflowReachsafetyLoops}{Error}{Error}{Walltime}{Median}{28.57151674805209}%
\StoreBenchExecResult{UltimateBenchmarkDefinition}{ORNooverflowReachsafetyLoops}{Error}{Error}{Walltime}{Min}{24.822406917111948}%
\StoreBenchExecResult{UltimateBenchmarkDefinition}{ORNooverflowReachsafetyLoops}{Error}{Error}{Walltime}{Max}{42.59183071809821}%
\StoreBenchExecResult{UltimateBenchmarkDefinition}{ORNooverflowReachsafetyLoops}{Error}{Error}{Walltime}{Stdev}{6.177157512199104966725790748}%
\StoreBenchExecResult{UltimateBenchmarkDefinition}{ORNooverflowReachsafetyLoops}{Error}{OutOfMemory}{Count}{}{5}%
\StoreBenchExecResult{UltimateBenchmarkDefinition}{ORNooverflowReachsafetyLoops}{Error}{OutOfMemory}{Cputime}{}{3185.074000133}%
\StoreBenchExecResult{UltimateBenchmarkDefinition}{ORNooverflowReachsafetyLoops}{Error}{OutOfMemory}{Cputime}{Avg}{637.0148000266}%
\StoreBenchExecResult{UltimateBenchmarkDefinition}{ORNooverflowReachsafetyLoops}{Error}{OutOfMemory}{Cputime}{Median}{737.77993453}%
\StoreBenchExecResult{UltimateBenchmarkDefinition}{ORNooverflowReachsafetyLoops}{Error}{OutOfMemory}{Cputime}{Min}{320.11481317}%
\StoreBenchExecResult{UltimateBenchmarkDefinition}{ORNooverflowReachsafetyLoops}{Error}{OutOfMemory}{Cputime}{Max}{842.322503076}%
\StoreBenchExecResult{UltimateBenchmarkDefinition}{ORNooverflowReachsafetyLoops}{Error}{OutOfMemory}{Cputime}{Stdev}{206.2010456165194639973303494}%
\StoreBenchExecResult{UltimateBenchmarkDefinition}{ORNooverflowReachsafetyLoops}{Error}{OutOfMemory}{Walltime}{}{2524.79555687378156}%
\StoreBenchExecResult{UltimateBenchmarkDefinition}{ORNooverflowReachsafetyLoops}{Error}{OutOfMemory}{Walltime}{Avg}{504.959111374756312}%
\StoreBenchExecResult{UltimateBenchmarkDefinition}{ORNooverflowReachsafetyLoops}{Error}{OutOfMemory}{Walltime}{Median}{603.0371559797786}%
\StoreBenchExecResult{UltimateBenchmarkDefinition}{ORNooverflowReachsafetyLoops}{Error}{OutOfMemory}{Walltime}{Min}{230.56642307899892}%
\StoreBenchExecResult{UltimateBenchmarkDefinition}{ORNooverflowReachsafetyLoops}{Error}{OutOfMemory}{Walltime}{Max}{676.2588077599648}%
\StoreBenchExecResult{UltimateBenchmarkDefinition}{ORNooverflowReachsafetyLoops}{Error}{OutOfMemory}{Walltime}{Stdev}{170.3167412792629317386712623}%
\StoreBenchExecResult{UltimateBenchmarkDefinition}{ORNooverflowReachsafetyLoops}{Error}{Timeout}{Count}{}{190}%
\StoreBenchExecResult{UltimateBenchmarkDefinition}{ORNooverflowReachsafetyLoops}{Error}{Timeout}{Cputime}{}{182526.734017968}%
\StoreBenchExecResult{UltimateBenchmarkDefinition}{ORNooverflowReachsafetyLoops}{Error}{Timeout}{Cputime}{Avg}{960.6670211472}%
\StoreBenchExecResult{UltimateBenchmarkDefinition}{ORNooverflowReachsafetyLoops}{Error}{Timeout}{Cputime}{Median}{960.5828243675}%
\StoreBenchExecResult{UltimateBenchmarkDefinition}{ORNooverflowReachsafetyLoops}{Error}{Timeout}{Cputime}{Min}{960.172427346}%
\StoreBenchExecResult{UltimateBenchmarkDefinition}{ORNooverflowReachsafetyLoops}{Error}{Timeout}{Cputime}{Max}{961.611707382}%
\StoreBenchExecResult{UltimateBenchmarkDefinition}{ORNooverflowReachsafetyLoops}{Error}{Timeout}{Cputime}{Stdev}{0.2929798517435610505996359381}%
\StoreBenchExecResult{UltimateBenchmarkDefinition}{ORNooverflowReachsafetyLoops}{Error}{Timeout}{Walltime}{}{168724.3825769070542}%
\StoreBenchExecResult{UltimateBenchmarkDefinition}{ORNooverflowReachsafetyLoops}{Error}{Timeout}{Walltime}{Avg}{888.0230661942476536842105263}%
\StoreBenchExecResult{UltimateBenchmarkDefinition}{ORNooverflowReachsafetyLoops}{Error}{Timeout}{Walltime}{Median}{902.49502817401665}%
\StoreBenchExecResult{UltimateBenchmarkDefinition}{ORNooverflowReachsafetyLoops}{Error}{Timeout}{Walltime}{Min}{756.7623061570339}%
\StoreBenchExecResult{UltimateBenchmarkDefinition}{ORNooverflowReachsafetyLoops}{Error}{Timeout}{Walltime}{Max}{941.8804043841083}%
\StoreBenchExecResult{UltimateBenchmarkDefinition}{ORNooverflowReachsafetyLoops}{Error}{Timeout}{Walltime}{Stdev}{38.93380572251165763498667101}%
\StoreBenchExecResult{UltimateBenchmarkDefinition}{ORNooverflowReachsafetyLoops}{Unknown}{}{Count}{}{10}%
\StoreBenchExecResult{UltimateBenchmarkDefinition}{ORNooverflowReachsafetyLoops}{Unknown}{}{Cputime}{}{243.217505034}%
\StoreBenchExecResult{UltimateBenchmarkDefinition}{ORNooverflowReachsafetyLoops}{Unknown}{}{Cputime}{Avg}{24.3217505034}%
\StoreBenchExecResult{UltimateBenchmarkDefinition}{ORNooverflowReachsafetyLoops}{Unknown}{}{Cputime}{Median}{24.1839676945}%
\StoreBenchExecResult{UltimateBenchmarkDefinition}{ORNooverflowReachsafetyLoops}{Unknown}{}{Cputime}{Min}{22.937637558}%
\StoreBenchExecResult{UltimateBenchmarkDefinition}{ORNooverflowReachsafetyLoops}{Unknown}{}{Cputime}{Max}{27.02069417}%
\StoreBenchExecResult{UltimateBenchmarkDefinition}{ORNooverflowReachsafetyLoops}{Unknown}{}{Cputime}{Stdev}{1.078983149280839453295822140}%
\StoreBenchExecResult{UltimateBenchmarkDefinition}{ORNooverflowReachsafetyLoops}{Unknown}{}{Walltime}{}{145.812657075002790}%
\StoreBenchExecResult{UltimateBenchmarkDefinition}{ORNooverflowReachsafetyLoops}{Unknown}{}{Walltime}{Avg}{14.581265707500279}%
\StoreBenchExecResult{UltimateBenchmarkDefinition}{ORNooverflowReachsafetyLoops}{Unknown}{}{Walltime}{Median}{14.488763862405904}%
\StoreBenchExecResult{UltimateBenchmarkDefinition}{ORNooverflowReachsafetyLoops}{Unknown}{}{Walltime}{Min}{14.003448761068285}%
\StoreBenchExecResult{UltimateBenchmarkDefinition}{ORNooverflowReachsafetyLoops}{Unknown}{}{Walltime}{Max}{16.287989598000422}%
\StoreBenchExecResult{UltimateBenchmarkDefinition}{ORNooverflowReachsafetyLoops}{Unknown}{}{Walltime}{Stdev}{0.6110164456336264692499352322}%
\StoreBenchExecResult{UltimateBenchmarkDefinition}{ORNooverflowReachsafetyLoops}{Unknown}{Unknown}{Count}{}{10}%
\StoreBenchExecResult{UltimateBenchmarkDefinition}{ORNooverflowReachsafetyLoops}{Unknown}{Unknown}{Cputime}{}{243.217505034}%
\StoreBenchExecResult{UltimateBenchmarkDefinition}{ORNooverflowReachsafetyLoops}{Unknown}{Unknown}{Cputime}{Avg}{24.3217505034}%
\StoreBenchExecResult{UltimateBenchmarkDefinition}{ORNooverflowReachsafetyLoops}{Unknown}{Unknown}{Cputime}{Median}{24.1839676945}%
\StoreBenchExecResult{UltimateBenchmarkDefinition}{ORNooverflowReachsafetyLoops}{Unknown}{Unknown}{Cputime}{Min}{22.937637558}%
\StoreBenchExecResult{UltimateBenchmarkDefinition}{ORNooverflowReachsafetyLoops}{Unknown}{Unknown}{Cputime}{Max}{27.02069417}%
\StoreBenchExecResult{UltimateBenchmarkDefinition}{ORNooverflowReachsafetyLoops}{Unknown}{Unknown}{Cputime}{Stdev}{1.078983149280839453295822140}%
\StoreBenchExecResult{UltimateBenchmarkDefinition}{ORNooverflowReachsafetyLoops}{Unknown}{Unknown}{Walltime}{}{145.812657075002790}%
\StoreBenchExecResult{UltimateBenchmarkDefinition}{ORNooverflowReachsafetyLoops}{Unknown}{Unknown}{Walltime}{Avg}{14.581265707500279}%
\StoreBenchExecResult{UltimateBenchmarkDefinition}{ORNooverflowReachsafetyLoops}{Unknown}{Unknown}{Walltime}{Median}{14.488763862405904}%
\StoreBenchExecResult{UltimateBenchmarkDefinition}{ORNooverflowReachsafetyLoops}{Unknown}{Unknown}{Walltime}{Min}{14.003448761068285}%
\StoreBenchExecResult{UltimateBenchmarkDefinition}{ORNooverflowReachsafetyLoops}{Unknown}{Unknown}{Walltime}{Max}{16.287989598000422}%
\StoreBenchExecResult{UltimateBenchmarkDefinition}{ORNooverflowReachsafetyLoops}{Unknown}{Unknown}{Walltime}{Stdev}{0.6110164456336264692499352322}%
\ifdefined\UltimateUltimateTotalCount\else\edef\UltimateUltimateTotalCount{0}\fi
\ifdefined\UltimateUltimateCorrectCount\else\edef\UltimateUltimateCorrectCount{0}\fi
\ifdefined\UltimateUltimateCorrectTrueCount\else\edef\UltimateUltimateCorrectTrueCount{0}\fi
\ifdefined\UltimateUltimateCorrectFalseCount\else\edef\UltimateUltimateCorrectFalseCount{0}\fi
\ifdefined\UltimateUltimateWrongCount\else\edef\UltimateUltimateWrongCount{0}\fi
\ifdefined\UltimateUltimateWrongTrueCount\else\edef\UltimateUltimateWrongTrueCount{0}\fi
\ifdefined\UltimateUltimateWrongFalseCount\else\edef\UltimateUltimateWrongFalseCount{0}\fi
\ifdefined\UltimateUltimateErrorTimeoutCount\else\edef\UltimateUltimateErrorTimeoutCount{0}\fi
\ifdefined\UltimateUltimateErrorOutOfMemoryCount\else\edef\UltimateUltimateErrorOutOfMemoryCount{0}\fi
\ifdefined\UltimateUltimateCorrectCputime\else\edef\UltimateUltimateCorrectCputime{0}\fi
\ifdefined\UltimateUltimateCorrectCputimeAvg\else\edef\UltimateUltimateCorrectCputimeAvg{None}\fi
\ifdefined\UltimateUltimateCorrectWalltime\else\edef\UltimateUltimateCorrectWalltime{0}\fi
\ifdefined\UltimateUltimateCorrectWalltimeAvg\else\edef\UltimateUltimateCorrectWalltimeAvg{None}\fi
\edef\UltimateUltimateErrorOtherInconclusiveCount{\the\numexpr \UltimateUltimateTotalCount - \UltimateUltimateCorrectCount - \UltimateUltimateWrongTrueCount - \UltimateUltimateWrongFalseCount - \UltimateUltimateErrorTimeoutCount - \UltimateUltimateErrorOutOfMemoryCount \relax}
\providecommand\StoreBenchExecResult[7]{\expandafter\newcommand\csname#1#2#3#4#5#6\endcsname{#7}}%
\StoreBenchExecResult{TaipanBenchmarkDefinition}{ORNooverflowReachsafetyLoops}{Total}{}{Count}{}{890}%
\StoreBenchExecResult{TaipanBenchmarkDefinition}{ORNooverflowReachsafetyLoops}{Total}{}{Cputime}{}{200725.796478679}%
\StoreBenchExecResult{TaipanBenchmarkDefinition}{ORNooverflowReachsafetyLoops}{Total}{}{Cputime}{Avg}{225.5346027850325842696629213}%
\StoreBenchExecResult{TaipanBenchmarkDefinition}{ORNooverflowReachsafetyLoops}{Total}{}{Cputime}{Median}{19.1866033065}%
\StoreBenchExecResult{TaipanBenchmarkDefinition}{ORNooverflowReachsafetyLoops}{Total}{}{Cputime}{Min}{13.540077929}%
\StoreBenchExecResult{TaipanBenchmarkDefinition}{ORNooverflowReachsafetyLoops}{Total}{}{Cputime}{Max}{961.696309251}%
\StoreBenchExecResult{TaipanBenchmarkDefinition}{ORNooverflowReachsafetyLoops}{Total}{}{Cputime}{Stdev}{372.7734446574072533913039461}%
\StoreBenchExecResult{TaipanBenchmarkDefinition}{ORNooverflowReachsafetyLoops}{Total}{}{Walltime}{}{178916.267898258054521}%
\StoreBenchExecResult{TaipanBenchmarkDefinition}{ORNooverflowReachsafetyLoops}{Total}{}{Walltime}{Avg}{201.0295144924247803606741573}%
\StoreBenchExecResult{TaipanBenchmarkDefinition}{ORNooverflowReachsafetyLoops}{Total}{}{Walltime}{Median}{11.3039628855185585}%
\StoreBenchExecResult{TaipanBenchmarkDefinition}{ORNooverflowReachsafetyLoops}{Total}{}{Walltime}{Min}{8.241916667902842}%
\StoreBenchExecResult{TaipanBenchmarkDefinition}{ORNooverflowReachsafetyLoops}{Total}{}{Walltime}{Max}{938.9779192779679}%
\StoreBenchExecResult{TaipanBenchmarkDefinition}{ORNooverflowReachsafetyLoops}{Total}{}{Walltime}{Stdev}{345.3789540260309323304551730}%
\StoreBenchExecResult{TaipanBenchmarkDefinition}{ORNooverflowReachsafetyLoops}{Correct}{}{Count}{}{692}%
\StoreBenchExecResult{TaipanBenchmarkDefinition}{ORNooverflowReachsafetyLoops}{Correct}{}{Cputime}{}{30682.870942194}%
\StoreBenchExecResult{TaipanBenchmarkDefinition}{ORNooverflowReachsafetyLoops}{Correct}{}{Cputime}{Avg}{44.33940887600289017341040462}%
\StoreBenchExecResult{TaipanBenchmarkDefinition}{ORNooverflowReachsafetyLoops}{Correct}{}{Cputime}{Median}{17.6367573875}%
\StoreBenchExecResult{TaipanBenchmarkDefinition}{ORNooverflowReachsafetyLoops}{Correct}{}{Cputime}{Min}{13.540077929}%
\StoreBenchExecResult{TaipanBenchmarkDefinition}{ORNooverflowReachsafetyLoops}{Correct}{}{Cputime}{Max}{812.395207492}%
\StoreBenchExecResult{TaipanBenchmarkDefinition}{ORNooverflowReachsafetyLoops}{Correct}{}{Cputime}{Stdev}{95.56750037610972014453725741}%
\StoreBenchExecResult{TaipanBenchmarkDefinition}{ORNooverflowReachsafetyLoops}{Correct}{}{Walltime}{}{23085.971454446669635}%
\StoreBenchExecResult{TaipanBenchmarkDefinition}{ORNooverflowReachsafetyLoops}{Correct}{}{Walltime}{Avg}{33.36123042550096768063583815}%
\StoreBenchExecResult{TaipanBenchmarkDefinition}{ORNooverflowReachsafetyLoops}{Correct}{}{Walltime}{Median}{10.4619471456389875}%
\StoreBenchExecResult{TaipanBenchmarkDefinition}{ORNooverflowReachsafetyLoops}{Correct}{}{Walltime}{Min}{8.241916667902842}%
\StoreBenchExecResult{TaipanBenchmarkDefinition}{ORNooverflowReachsafetyLoops}{Correct}{}{Walltime}{Max}{702.3550239149481}%
\StoreBenchExecResult{TaipanBenchmarkDefinition}{ORNooverflowReachsafetyLoops}{Correct}{}{Walltime}{Stdev}{86.78468360367990486945328847}%
\StoreBenchExecResult{TaipanBenchmarkDefinition}{ORNooverflowReachsafetyLoops}{Correct}{False}{Count}{}{251}%
\StoreBenchExecResult{TaipanBenchmarkDefinition}{ORNooverflowReachsafetyLoops}{Correct}{False}{Cputime}{}{5711.612803663}%
\StoreBenchExecResult{TaipanBenchmarkDefinition}{ORNooverflowReachsafetyLoops}{Correct}{False}{Cputime}{Avg}{22.75542949666533864541832669}%
\StoreBenchExecResult{TaipanBenchmarkDefinition}{ORNooverflowReachsafetyLoops}{Correct}{False}{Cputime}{Median}{16.270092909}%
\StoreBenchExecResult{TaipanBenchmarkDefinition}{ORNooverflowReachsafetyLoops}{Correct}{False}{Cputime}{Min}{14.249440434}%
\StoreBenchExecResult{TaipanBenchmarkDefinition}{ORNooverflowReachsafetyLoops}{Correct}{False}{Cputime}{Max}{553.356650036}%
\StoreBenchExecResult{TaipanBenchmarkDefinition}{ORNooverflowReachsafetyLoops}{Correct}{False}{Cputime}{Stdev}{40.54257135043545981018273575}%
\StoreBenchExecResult{TaipanBenchmarkDefinition}{ORNooverflowReachsafetyLoops}{Correct}{False}{Walltime}{}{3773.748614377575002}%
\StoreBenchExecResult{TaipanBenchmarkDefinition}{ORNooverflowReachsafetyLoops}{Correct}{False}{Walltime}{Avg}{15.03485503736085658167330677}%
\StoreBenchExecResult{TaipanBenchmarkDefinition}{ORNooverflowReachsafetyLoops}{Correct}{False}{Walltime}{Median}{9.705018050037324}%
\StoreBenchExecResult{TaipanBenchmarkDefinition}{ORNooverflowReachsafetyLoops}{Correct}{False}{Walltime}{Min}{8.600078058894724}%
\StoreBenchExecResult{TaipanBenchmarkDefinition}{ORNooverflowReachsafetyLoops}{Correct}{False}{Walltime}{Max}{516.1671544779092}%
\StoreBenchExecResult{TaipanBenchmarkDefinition}{ORNooverflowReachsafetyLoops}{Correct}{False}{Walltime}{Stdev}{36.91175705521255350889128945}%

\StoreBenchExecResult{TaipanBenchmarkDefinition}{ORNooverflowReachsafetyLoops}{Correct}{True}{Count}{}{441}%
\StoreBenchExecResult{TaipanBenchmarkDefinition}{ORNooverflowReachsafetyLoops}{Correct}{True}{Cputime}{}{24971.258138531}%
\StoreBenchExecResult{TaipanBenchmarkDefinition}{ORNooverflowReachsafetyLoops}{Correct}{True}{Cputime}{Avg}{56.62416811458276643990929705}%
\StoreBenchExecResult{TaipanBenchmarkDefinition}{ORNooverflowReachsafetyLoops}{Correct}{True}{Cputime}{Median}{18.613091843}%
\StoreBenchExecResult{TaipanBenchmarkDefinition}{ORNooverflowReachsafetyLoops}{Correct}{True}{Cputime}{Min}{13.540077929}%
\StoreBenchExecResult{TaipanBenchmarkDefinition}{ORNooverflowReachsafetyLoops}{Correct}{True}{Cputime}{Max}{812.395207492}%
\StoreBenchExecResult{TaipanBenchmarkDefinition}{ORNooverflowReachsafetyLoops}{Correct}{True}{Cputime}{Stdev}{113.9288329450554614516364000}%
\StoreBenchExecResult{TaipanBenchmarkDefinition}{ORNooverflowReachsafetyLoops}{Correct}{True}{Walltime}{}{19312.222840069094633}%
\StoreBenchExecResult{TaipanBenchmarkDefinition}{ORNooverflowReachsafetyLoops}{Correct}{True}{Walltime}{Avg}{43.79188852623377467800453515}%
\StoreBenchExecResult{TaipanBenchmarkDefinition}{ORNooverflowReachsafetyLoops}{Correct}{True}{Walltime}{Median}{11.034483257913962}%
\StoreBenchExecResult{TaipanBenchmarkDefinition}{ORNooverflowReachsafetyLoops}{Correct}{True}{Walltime}{Min}{8.241916667902842}%
\StoreBenchExecResult{TaipanBenchmarkDefinition}{ORNooverflowReachsafetyLoops}{Correct}{True}{Walltime}{Max}{702.3550239149481}%
\StoreBenchExecResult{TaipanBenchmarkDefinition}{ORNooverflowReachsafetyLoops}{Correct}{True}{Walltime}{Stdev}{103.6476706281733767068362301}%

\StoreBenchExecResult{TaipanBenchmarkDefinition}{ORNooverflowReachsafetyLoops}{Error}{}{Count}{}{188}%
\StoreBenchExecResult{TaipanBenchmarkDefinition}{ORNooverflowReachsafetyLoops}{Error}{}{Cputime}{}{169792.794477683}%
\StoreBenchExecResult{TaipanBenchmarkDefinition}{ORNooverflowReachsafetyLoops}{Error}{}{Cputime}{Avg}{903.1531621153351063829787234}%
\StoreBenchExecResult{TaipanBenchmarkDefinition}{ORNooverflowReachsafetyLoops}{Error}{}{Cputime}{Median}{960.6455619935}%
\StoreBenchExecResult{TaipanBenchmarkDefinition}{ORNooverflowReachsafetyLoops}{Error}{}{Cputime}{Min}{37.372769141}%
\StoreBenchExecResult{TaipanBenchmarkDefinition}{ORNooverflowReachsafetyLoops}{Error}{}{Cputime}{Max}{961.696309251}%
\StoreBenchExecResult{TaipanBenchmarkDefinition}{ORNooverflowReachsafetyLoops}{Error}{}{Cputime}{Stdev}{205.1109522757561917729907594}%
\StoreBenchExecResult{TaipanBenchmarkDefinition}{ORNooverflowReachsafetyLoops}{Error}{}{Walltime}{}{155679.543583321617840}%
\StoreBenchExecResult{TaipanBenchmarkDefinition}{ORNooverflowReachsafetyLoops}{Error}{}{Walltime}{Avg}{828.0826786346894565957446809}%
\StoreBenchExecResult{TaipanBenchmarkDefinition}{ORNooverflowReachsafetyLoops}{Error}{}{Walltime}{Median}{894.81955080549235}%
\StoreBenchExecResult{TaipanBenchmarkDefinition}{ORNooverflowReachsafetyLoops}{Error}{}{Walltime}{Min}{24.61909121903591}%
\StoreBenchExecResult{TaipanBenchmarkDefinition}{ORNooverflowReachsafetyLoops}{Error}{}{Walltime}{Max}{938.9779192779679}%
\StoreBenchExecResult{TaipanBenchmarkDefinition}{ORNooverflowReachsafetyLoops}{Error}{}{Walltime}{Stdev}{196.1422576037100759607386733}%
\StoreBenchExecResult{TaipanBenchmarkDefinition}{ORNooverflowReachsafetyLoops}{Error}{Error}{Count}{}{12}%
\StoreBenchExecResult{TaipanBenchmarkDefinition}{ORNooverflowReachsafetyLoops}{Error}{Error}{Cputime}{}{2191.539788554}%
\StoreBenchExecResult{TaipanBenchmarkDefinition}{ORNooverflowReachsafetyLoops}{Error}{Error}{Cputime}{Avg}{182.6283157128333333333333333}%
\StoreBenchExecResult{TaipanBenchmarkDefinition}{ORNooverflowReachsafetyLoops}{Error}{Error}{Cputime}{Median}{45.452424824}%
\StoreBenchExecResult{TaipanBenchmarkDefinition}{ORNooverflowReachsafetyLoops}{Error}{Error}{Cputime}{Min}{37.372769141}%
\StoreBenchExecResult{TaipanBenchmarkDefinition}{ORNooverflowReachsafetyLoops}{Error}{Error}{Cputime}{Max}{725.316518743}%
\StoreBenchExecResult{TaipanBenchmarkDefinition}{ORNooverflowReachsafetyLoops}{Error}{Error}{Cputime}{Stdev}{247.3760490306011582927295466}%
\StoreBenchExecResult{TaipanBenchmarkDefinition}{ORNooverflowReachsafetyLoops}{Error}{Error}{Walltime}{}{1925.102358491858410}%
\StoreBenchExecResult{TaipanBenchmarkDefinition}{ORNooverflowReachsafetyLoops}{Error}{Error}{Walltime}{Avg}{160.4251965409882008333333333}%
\StoreBenchExecResult{TaipanBenchmarkDefinition}{ORNooverflowReachsafetyLoops}{Error}{Error}{Walltime}{Median}{33.2461165615823115}%
\StoreBenchExecResult{TaipanBenchmarkDefinition}{ORNooverflowReachsafetyLoops}{Error}{Error}{Walltime}{Min}{24.61909121903591}%
\StoreBenchExecResult{TaipanBenchmarkDefinition}{ORNooverflowReachsafetyLoops}{Error}{Error}{Walltime}{Max}{665.7968550140504}%
\StoreBenchExecResult{TaipanBenchmarkDefinition}{ORNooverflowReachsafetyLoops}{Error}{Error}{Walltime}{Stdev}{229.5338508958565201889199897}%
\StoreBenchExecResult{TaipanBenchmarkDefinition}{ORNooverflowReachsafetyLoops}{Error}{OutOfMemory}{Count}{}{5}%
\StoreBenchExecResult{TaipanBenchmarkDefinition}{ORNooverflowReachsafetyLoops}{Error}{OutOfMemory}{Cputime}{}{3395.997225259}%
\StoreBenchExecResult{TaipanBenchmarkDefinition}{ORNooverflowReachsafetyLoops}{Error}{OutOfMemory}{Cputime}{Avg}{679.1994450518}%
\StoreBenchExecResult{TaipanBenchmarkDefinition}{ORNooverflowReachsafetyLoops}{Error}{OutOfMemory}{Cputime}{Median}{619.614669022}%
\StoreBenchExecResult{TaipanBenchmarkDefinition}{ORNooverflowReachsafetyLoops}{Error}{OutOfMemory}{Cputime}{Min}{478.399994579}%
\StoreBenchExecResult{TaipanBenchmarkDefinition}{ORNooverflowReachsafetyLoops}{Error}{OutOfMemory}{Cputime}{Max}{872.142777246}%
\StoreBenchExecResult{TaipanBenchmarkDefinition}{ORNooverflowReachsafetyLoops}{Error}{OutOfMemory}{Cputime}{Stdev}{163.0768114153398207009842538}%
\StoreBenchExecResult{TaipanBenchmarkDefinition}{ORNooverflowReachsafetyLoops}{Error}{OutOfMemory}{Walltime}{}{2737.87388375587763}%
\StoreBenchExecResult{TaipanBenchmarkDefinition}{ORNooverflowReachsafetyLoops}{Error}{OutOfMemory}{Walltime}{Avg}{547.574776751175526}%
\StoreBenchExecResult{TaipanBenchmarkDefinition}{ORNooverflowReachsafetyLoops}{Error}{OutOfMemory}{Walltime}{Median}{512.8831050740555}%
\StoreBenchExecResult{TaipanBenchmarkDefinition}{ORNooverflowReachsafetyLoops}{Error}{OutOfMemory}{Walltime}{Min}{389.7914792308584}%
\StoreBenchExecResult{TaipanBenchmarkDefinition}{ORNooverflowReachsafetyLoops}{Error}{OutOfMemory}{Walltime}{Max}{732.3480666070245}%
\StoreBenchExecResult{TaipanBenchmarkDefinition}{ORNooverflowReachsafetyLoops}{Error}{OutOfMemory}{Walltime}{Stdev}{140.6012927137331271662806799}%
\StoreBenchExecResult{TaipanBenchmarkDefinition}{ORNooverflowReachsafetyLoops}{Error}{Timeout}{Count}{}{171}%
\StoreBenchExecResult{TaipanBenchmarkDefinition}{ORNooverflowReachsafetyLoops}{Error}{Timeout}{Cputime}{}{164205.257463870}%
\StoreBenchExecResult{TaipanBenchmarkDefinition}{ORNooverflowReachsafetyLoops}{Error}{Timeout}{Cputime}{Avg}{960.2646635314035087719298246}%
\StoreBenchExecResult{TaipanBenchmarkDefinition}{ORNooverflowReachsafetyLoops}{Error}{Timeout}{Cputime}{Median}{960.672990637}%
\StoreBenchExecResult{TaipanBenchmarkDefinition}{ORNooverflowReachsafetyLoops}{Error}{Timeout}{Cputime}{Min}{906.385333954}%
\StoreBenchExecResult{TaipanBenchmarkDefinition}{ORNooverflowReachsafetyLoops}{Error}{Timeout}{Cputime}{Max}{961.696309251}%
\StoreBenchExecResult{TaipanBenchmarkDefinition}{ORNooverflowReachsafetyLoops}{Error}{Timeout}{Cputime}{Stdev}{4.550975186542939910433855228}%
\StoreBenchExecResult{TaipanBenchmarkDefinition}{ORNooverflowReachsafetyLoops}{Error}{Timeout}{Walltime}{}{151016.5673410738818}%
\StoreBenchExecResult{TaipanBenchmarkDefinition}{ORNooverflowReachsafetyLoops}{Error}{Timeout}{Walltime}{Avg}{883.1378207080343964912280702}%
\StoreBenchExecResult{TaipanBenchmarkDefinition}{ORNooverflowReachsafetyLoops}{Error}{Timeout}{Walltime}{Median}{896.4684017270338}%
\StoreBenchExecResult{TaipanBenchmarkDefinition}{ORNooverflowReachsafetyLoops}{Error}{Timeout}{Walltime}{Min}{781.8883396100719}%
\StoreBenchExecResult{TaipanBenchmarkDefinition}{ORNooverflowReachsafetyLoops}{Error}{Timeout}{Walltime}{Max}{938.9779192779679}%
\StoreBenchExecResult{TaipanBenchmarkDefinition}{ORNooverflowReachsafetyLoops}{Error}{Timeout}{Walltime}{Stdev}{37.51710882374794383544921112}%
\StoreBenchExecResult{TaipanBenchmarkDefinition}{ORNooverflowReachsafetyLoops}{Unknown}{}{Count}{}{10}%
\StoreBenchExecResult{TaipanBenchmarkDefinition}{ORNooverflowReachsafetyLoops}{Unknown}{}{Cputime}{}{250.131058802}%
\StoreBenchExecResult{TaipanBenchmarkDefinition}{ORNooverflowReachsafetyLoops}{Unknown}{}{Cputime}{Avg}{25.0131058802}%
\StoreBenchExecResult{TaipanBenchmarkDefinition}{ORNooverflowReachsafetyLoops}{Unknown}{}{Cputime}{Median}{24.2099122395}%
\StoreBenchExecResult{TaipanBenchmarkDefinition}{ORNooverflowReachsafetyLoops}{Unknown}{}{Cputime}{Min}{23.076864015}%
\StoreBenchExecResult{TaipanBenchmarkDefinition}{ORNooverflowReachsafetyLoops}{Unknown}{}{Cputime}{Max}{29.123024211}%
\StoreBenchExecResult{TaipanBenchmarkDefinition}{ORNooverflowReachsafetyLoops}{Unknown}{}{Cputime}{Stdev}{2.020506612363509498287120885}%
\StoreBenchExecResult{TaipanBenchmarkDefinition}{ORNooverflowReachsafetyLoops}{Unknown}{}{Walltime}{}{150.752860489767046}%
\StoreBenchExecResult{TaipanBenchmarkDefinition}{ORNooverflowReachsafetyLoops}{Unknown}{}{Walltime}{Avg}{15.0752860489767046}%
\StoreBenchExecResult{TaipanBenchmarkDefinition}{ORNooverflowReachsafetyLoops}{Unknown}{}{Walltime}{Median}{14.855535398470238}%
\StoreBenchExecResult{TaipanBenchmarkDefinition}{ORNooverflowReachsafetyLoops}{Unknown}{}{Walltime}{Min}{14.076829242985696}%
\StoreBenchExecResult{TaipanBenchmarkDefinition}{ORNooverflowReachsafetyLoops}{Unknown}{}{Walltime}{Max}{17.510611704085022}%
\StoreBenchExecResult{TaipanBenchmarkDefinition}{ORNooverflowReachsafetyLoops}{Unknown}{}{Walltime}{Stdev}{1.074529229575867695976012653}%
\StoreBenchExecResult{TaipanBenchmarkDefinition}{ORNooverflowReachsafetyLoops}{Unknown}{Unknown}{Count}{}{10}%
\StoreBenchExecResult{TaipanBenchmarkDefinition}{ORNooverflowReachsafetyLoops}{Unknown}{Unknown}{Cputime}{}{250.131058802}%
\StoreBenchExecResult{TaipanBenchmarkDefinition}{ORNooverflowReachsafetyLoops}{Unknown}{Unknown}{Cputime}{Avg}{25.0131058802}%
\StoreBenchExecResult{TaipanBenchmarkDefinition}{ORNooverflowReachsafetyLoops}{Unknown}{Unknown}{Cputime}{Median}{24.2099122395}%
\StoreBenchExecResult{TaipanBenchmarkDefinition}{ORNooverflowReachsafetyLoops}{Unknown}{Unknown}{Cputime}{Min}{23.076864015}%
\StoreBenchExecResult{TaipanBenchmarkDefinition}{ORNooverflowReachsafetyLoops}{Unknown}{Unknown}{Cputime}{Max}{29.123024211}%
\StoreBenchExecResult{TaipanBenchmarkDefinition}{ORNooverflowReachsafetyLoops}{Unknown}{Unknown}{Cputime}{Stdev}{2.020506612363509498287120885}%
\StoreBenchExecResult{TaipanBenchmarkDefinition}{ORNooverflowReachsafetyLoops}{Unknown}{Unknown}{Walltime}{}{150.752860489767046}%
\StoreBenchExecResult{TaipanBenchmarkDefinition}{ORNooverflowReachsafetyLoops}{Unknown}{Unknown}{Walltime}{Avg}{15.0752860489767046}%
\StoreBenchExecResult{TaipanBenchmarkDefinition}{ORNooverflowReachsafetyLoops}{Unknown}{Unknown}{Walltime}{Median}{14.855535398470238}%
\StoreBenchExecResult{TaipanBenchmarkDefinition}{ORNooverflowReachsafetyLoops}{Unknown}{Unknown}{Walltime}{Min}{14.076829242985696}%
\StoreBenchExecResult{TaipanBenchmarkDefinition}{ORNooverflowReachsafetyLoops}{Unknown}{Unknown}{Walltime}{Max}{17.510611704085022}%
\StoreBenchExecResult{TaipanBenchmarkDefinition}{ORNooverflowReachsafetyLoops}{Unknown}{Unknown}{Walltime}{Stdev}{1.074529229575867695976012653}%
\ifdefined\TaipanTaipanTotalCount\else\edef\TaipanTaipanTotalCount{0}\fi
\ifdefined\TaipanTaipanCorrectCount\else\edef\TaipanTaipanCorrectCount{0}\fi
\ifdefined\TaipanTaipanCorrectTrueCount\else\edef\TaipanTaipanCorrectTrueCount{0}\fi
\ifdefined\TaipanTaipanCorrectFalseCount\else\edef\TaipanTaipanCorrectFalseCount{0}\fi
\ifdefined\TaipanTaipanWrongCount\else\edef\TaipanTaipanWrongCount{0}\fi
\ifdefined\TaipanTaipanWrongTrueCount\else\edef\TaipanTaipanWrongTrueCount{0}\fi
\ifdefined\TaipanTaipanWrongFalseCount\else\edef\TaipanTaipanWrongFalseCount{0}\fi
\ifdefined\TaipanTaipanErrorTimeoutCount\else\edef\TaipanTaipanErrorTimeoutCount{0}\fi
\ifdefined\TaipanTaipanErrorOutOfMemoryCount\else\edef\TaipanTaipanErrorOutOfMemoryCount{0}\fi
\ifdefined\TaipanTaipanCorrectCputime\else\edef\TaipanTaipanCorrectCputime{0}\fi
\ifdefined\TaipanTaipanCorrectCputimeAvg\else\edef\TaipanTaipanCorrectCputimeAvg{None}\fi
\ifdefined\TaipanTaipanCorrectWalltime\else\edef\TaipanTaipanCorrectWalltime{0}\fi
\ifdefined\TaipanTaipanCorrectWalltimeAvg\else\edef\TaipanTaipanCorrectWalltimeAvg{None}\fi
\edef\TaipanTaipanErrorOtherInconclusiveCount{\the\numexpr \TaipanTaipanTotalCount - \TaipanTaipanCorrectCount - \TaipanTaipanWrongTrueCount - \TaipanTaipanWrongFalseCount - \TaipanTaipanErrorTimeoutCount - \TaipanTaipanErrorOutOfMemoryCount \relax}
\providecommand\StoreBenchExecResult[7]{\expandafter\newcommand\csname#1#2#3#4#5#6\endcsname{#7}}%
\StoreBenchExecResult{CpacheckerBenchmarkDefinition}{ORNooverflowReachsafetyLoops}{Total}{}{Count}{}{890}%
\StoreBenchExecResult{CpacheckerBenchmarkDefinition}{ORNooverflowReachsafetyLoops}{Total}{}{Cputime}{}{241395.825857223}%
\StoreBenchExecResult{CpacheckerBenchmarkDefinition}{ORNooverflowReachsafetyLoops}{Total}{}{Cputime}{Avg}{271.2312650081157303370786517}%
\StoreBenchExecResult{CpacheckerBenchmarkDefinition}{ORNooverflowReachsafetyLoops}{Total}{}{Cputime}{Median}{8.1236162345}%
\StoreBenchExecResult{CpacheckerBenchmarkDefinition}{ORNooverflowReachsafetyLoops}{Total}{}{Cputime}{Min}{2.671261697}%
\StoreBenchExecResult{CpacheckerBenchmarkDefinition}{ORNooverflowReachsafetyLoops}{Total}{}{Cputime}{Max}{960.732177482}%
\StoreBenchExecResult{CpacheckerBenchmarkDefinition}{ORNooverflowReachsafetyLoops}{Total}{}{Cputime}{Stdev}{405.6158114883045853890050250}%
\StoreBenchExecResult{CpacheckerBenchmarkDefinition}{ORNooverflowReachsafetyLoops}{Total}{}{Walltime}{}{234252.2377908274060952}%
\StoreBenchExecResult{CpacheckerBenchmarkDefinition}{ORNooverflowReachsafetyLoops}{Total}{}{Walltime}{Avg}{263.2047615627274225788764045}%
\StoreBenchExecResult{CpacheckerBenchmarkDefinition}{ORNooverflowReachsafetyLoops}{Total}{}{Walltime}{Median}{4.2263415574561805}%
\StoreBenchExecResult{CpacheckerBenchmarkDefinition}{ORNooverflowReachsafetyLoops}{Total}{}{Walltime}{Min}{1.4463007519952953}%
\StoreBenchExecResult{CpacheckerBenchmarkDefinition}{ORNooverflowReachsafetyLoops}{Total}{}{Walltime}{Max}{930.0198502680287}%
\StoreBenchExecResult{CpacheckerBenchmarkDefinition}{ORNooverflowReachsafetyLoops}{Total}{}{Walltime}{Stdev}{399.2300702090535534467303079}%
\StoreBenchExecResult{CpacheckerBenchmarkDefinition}{ORNooverflowReachsafetyLoops}{Correct}{}{Count}{}{571}%
\StoreBenchExecResult{CpacheckerBenchmarkDefinition}{ORNooverflowReachsafetyLoops}{Correct}{}{Cputime}{}{8346.399339663}%
\StoreBenchExecResult{CpacheckerBenchmarkDefinition}{ORNooverflowReachsafetyLoops}{Correct}{}{Cputime}{Avg}{14.61716171569702276707530648}%
\StoreBenchExecResult{CpacheckerBenchmarkDefinition}{ORNooverflowReachsafetyLoops}{Correct}{}{Cputime}{Median}{7.259721504}%
\StoreBenchExecResult{CpacheckerBenchmarkDefinition}{ORNooverflowReachsafetyLoops}{Correct}{}{Cputime}{Min}{5.973973463}%
\StoreBenchExecResult{CpacheckerBenchmarkDefinition}{ORNooverflowReachsafetyLoops}{Correct}{}{Cputime}{Max}{571.083265282}%
\StoreBenchExecResult{CpacheckerBenchmarkDefinition}{ORNooverflowReachsafetyLoops}{Correct}{}{Cputime}{Stdev}{50.38688195896618600837221200}%
\StoreBenchExecResult{CpacheckerBenchmarkDefinition}{ORNooverflowReachsafetyLoops}{Correct}{}{Walltime}{}{6035.7552683698012688}%
\StoreBenchExecResult{CpacheckerBenchmarkDefinition}{ORNooverflowReachsafetyLoops}{Correct}{}{Walltime}{Avg}{10.57049959434290940245183888}%
\StoreBenchExecResult{CpacheckerBenchmarkDefinition}{ORNooverflowReachsafetyLoops}{Correct}{}{Walltime}{Median}{3.789366374956444}%
\StoreBenchExecResult{CpacheckerBenchmarkDefinition}{ORNooverflowReachsafetyLoops}{Correct}{}{Walltime}{Min}{3.1397447609342635}%
\StoreBenchExecResult{CpacheckerBenchmarkDefinition}{ORNooverflowReachsafetyLoops}{Correct}{}{Walltime}{Max}{546.2322233151644}%
\StoreBenchExecResult{CpacheckerBenchmarkDefinition}{ORNooverflowReachsafetyLoops}{Correct}{}{Walltime}{Stdev}{48.42270773749125892368396345}%
\StoreBenchExecResult{CpacheckerBenchmarkDefinition}{ORNooverflowReachsafetyLoops}{Correct}{False}{Count}{}{218}%
\StoreBenchExecResult{CpacheckerBenchmarkDefinition}{ORNooverflowReachsafetyLoops}{Correct}{False}{Cputime}{}{2301.782810700}%
\StoreBenchExecResult{CpacheckerBenchmarkDefinition}{ORNooverflowReachsafetyLoops}{Correct}{False}{Cputime}{Avg}{10.55863674633027522935779817}%
\StoreBenchExecResult{CpacheckerBenchmarkDefinition}{ORNooverflowReachsafetyLoops}{Correct}{False}{Cputime}{Median}{7.170801360}%
\StoreBenchExecResult{CpacheckerBenchmarkDefinition}{ORNooverflowReachsafetyLoops}{Correct}{False}{Cputime}{Min}{5.973973463}%
\StoreBenchExecResult{CpacheckerBenchmarkDefinition}{ORNooverflowReachsafetyLoops}{Correct}{False}{Cputime}{Max}{413.696860296}%
\StoreBenchExecResult{CpacheckerBenchmarkDefinition}{ORNooverflowReachsafetyLoops}{Correct}{False}{Cputime}{Stdev}{28.93157189857737425601613955}%
\StoreBenchExecResult{CpacheckerBenchmarkDefinition}{ORNooverflowReachsafetyLoops}{Correct}{False}{Walltime}{}{1477.1513596088625544}%
\StoreBenchExecResult{CpacheckerBenchmarkDefinition}{ORNooverflowReachsafetyLoops}{Correct}{False}{Walltime}{Avg}{6.775923667930562176146788991}%
\StoreBenchExecResult{CpacheckerBenchmarkDefinition}{ORNooverflowReachsafetyLoops}{Correct}{False}{Walltime}{Median}{3.737814674503170}%
\StoreBenchExecResult{CpacheckerBenchmarkDefinition}{ORNooverflowReachsafetyLoops}{Correct}{False}{Walltime}{Min}{3.1397447609342635}%
\StoreBenchExecResult{CpacheckerBenchmarkDefinition}{ORNooverflowReachsafetyLoops}{Correct}{False}{Walltime}{Max}{393.60304761398584}%
\StoreBenchExecResult{CpacheckerBenchmarkDefinition}{ORNooverflowReachsafetyLoops}{Correct}{False}{Walltime}{Stdev}{27.69518277344154938229786317}%

\StoreBenchExecResult{CpacheckerBenchmarkDefinition}{ORNooverflowReachsafetyLoops}{Correct}{True}{Count}{}{353}%
\StoreBenchExecResult{CpacheckerBenchmarkDefinition}{ORNooverflowReachsafetyLoops}{Correct}{True}{Cputime}{}{6044.616528963}%
\StoreBenchExecResult{CpacheckerBenchmarkDefinition}{ORNooverflowReachsafetyLoops}{Correct}{True}{Cputime}{Avg}{17.12355957213314447592067989}%
\StoreBenchExecResult{CpacheckerBenchmarkDefinition}{ORNooverflowReachsafetyLoops}{Correct}{True}{Cputime}{Median}{7.378520693}%
\StoreBenchExecResult{CpacheckerBenchmarkDefinition}{ORNooverflowReachsafetyLoops}{Correct}{True}{Cputime}{Min}{6.073677736}%
\StoreBenchExecResult{CpacheckerBenchmarkDefinition}{ORNooverflowReachsafetyLoops}{Correct}{True}{Cputime}{Max}{571.083265282}%
\StoreBenchExecResult{CpacheckerBenchmarkDefinition}{ORNooverflowReachsafetyLoops}{Correct}{True}{Cputime}{Stdev}{59.77754415189040222060853090}%
\StoreBenchExecResult{CpacheckerBenchmarkDefinition}{ORNooverflowReachsafetyLoops}{Correct}{True}{Walltime}{}{4558.6039087609387144}%
\StoreBenchExecResult{CpacheckerBenchmarkDefinition}{ORNooverflowReachsafetyLoops}{Correct}{True}{Walltime}{Avg}{12.91389209280719182549575071}%
\StoreBenchExecResult{CpacheckerBenchmarkDefinition}{ORNooverflowReachsafetyLoops}{Correct}{True}{Walltime}{Median}{3.825262292055413}%
\StoreBenchExecResult{CpacheckerBenchmarkDefinition}{ORNooverflowReachsafetyLoops}{Correct}{True}{Walltime}{Min}{3.1729543760884553}%
\StoreBenchExecResult{CpacheckerBenchmarkDefinition}{ORNooverflowReachsafetyLoops}{Correct}{True}{Walltime}{Max}{546.2322233151644}%
\StoreBenchExecResult{CpacheckerBenchmarkDefinition}{ORNooverflowReachsafetyLoops}{Correct}{True}{Walltime}{Stdev}{57.48675652387725610968801676}%

\StoreBenchExecResult{CpacheckerBenchmarkDefinition}{ORNooverflowReachsafetyLoops}{Error}{}{Count}{}{306}%
\StoreBenchExecResult{CpacheckerBenchmarkDefinition}{ORNooverflowReachsafetyLoops}{Error}{}{Cputime}{}{232957.025736514}%
\StoreBenchExecResult{CpacheckerBenchmarkDefinition}{ORNooverflowReachsafetyLoops}{Error}{}{Cputime}{Avg}{761.2974697271699346405228758}%
\StoreBenchExecResult{CpacheckerBenchmarkDefinition}{ORNooverflowReachsafetyLoops}{Error}{}{Cputime}{Median}{903.0191329685}%
\StoreBenchExecResult{CpacheckerBenchmarkDefinition}{ORNooverflowReachsafetyLoops}{Error}{}{Cputime}{Min}{2.671261697}%
\StoreBenchExecResult{CpacheckerBenchmarkDefinition}{ORNooverflowReachsafetyLoops}{Error}{}{Cputime}{Max}{960.732177482}%
\StoreBenchExecResult{CpacheckerBenchmarkDefinition}{ORNooverflowReachsafetyLoops}{Error}{}{Cputime}{Stdev}{328.2887498028202923114885140}%
\StoreBenchExecResult{CpacheckerBenchmarkDefinition}{ORNooverflowReachsafetyLoops}{Error}{}{Walltime}{}{228168.4434135521292401}%
\StoreBenchExecResult{CpacheckerBenchmarkDefinition}{ORNooverflowReachsafetyLoops}{Error}{}{Walltime}{Avg}{745.6485078874252589545751634}%
\StoreBenchExecResult{CpacheckerBenchmarkDefinition}{ORNooverflowReachsafetyLoops}{Error}{}{Walltime}{Median}{888.44623028102795}%
\StoreBenchExecResult{CpacheckerBenchmarkDefinition}{ORNooverflowReachsafetyLoops}{Error}{}{Walltime}{Min}{1.4463007519952953}%
\StoreBenchExecResult{CpacheckerBenchmarkDefinition}{ORNooverflowReachsafetyLoops}{Error}{}{Walltime}{Max}{930.0198502680287}%
\StoreBenchExecResult{CpacheckerBenchmarkDefinition}{ORNooverflowReachsafetyLoops}{Error}{}{Walltime}{Stdev}{323.2410755931563083156043509}%
\StoreBenchExecResult{CpacheckerBenchmarkDefinition}{ORNooverflowReachsafetyLoops}{Error}{Assertion}{Count}{}{1}%
\StoreBenchExecResult{CpacheckerBenchmarkDefinition}{ORNooverflowReachsafetyLoops}{Error}{Assertion}{Cputime}{}{10.199141244}%
\StoreBenchExecResult{CpacheckerBenchmarkDefinition}{ORNooverflowReachsafetyLoops}{Error}{Assertion}{Cputime}{Avg}{10.199141244}%
\StoreBenchExecResult{CpacheckerBenchmarkDefinition}{ORNooverflowReachsafetyLoops}{Error}{Assertion}{Cputime}{Median}{10.199141244}%
\StoreBenchExecResult{CpacheckerBenchmarkDefinition}{ORNooverflowReachsafetyLoops}{Error}{Assertion}{Cputime}{Min}{10.199141244}%
\StoreBenchExecResult{CpacheckerBenchmarkDefinition}{ORNooverflowReachsafetyLoops}{Error}{Assertion}{Cputime}{Max}{10.199141244}%
\StoreBenchExecResult{CpacheckerBenchmarkDefinition}{ORNooverflowReachsafetyLoops}{Error}{Assertion}{Cputime}{Stdev}{0E-14}%
\StoreBenchExecResult{CpacheckerBenchmarkDefinition}{ORNooverflowReachsafetyLoops}{Error}{Assertion}{Walltime}{}{5.4721552659757435}%
\StoreBenchExecResult{CpacheckerBenchmarkDefinition}{ORNooverflowReachsafetyLoops}{Error}{Assertion}{Walltime}{Avg}{5.4721552659757435}%
\StoreBenchExecResult{CpacheckerBenchmarkDefinition}{ORNooverflowReachsafetyLoops}{Error}{Assertion}{Walltime}{Median}{5.4721552659757435}%
\StoreBenchExecResult{CpacheckerBenchmarkDefinition}{ORNooverflowReachsafetyLoops}{Error}{Assertion}{Walltime}{Min}{5.4721552659757435}%
\StoreBenchExecResult{CpacheckerBenchmarkDefinition}{ORNooverflowReachsafetyLoops}{Error}{Assertion}{Walltime}{Max}{5.4721552659757435}%
\StoreBenchExecResult{CpacheckerBenchmarkDefinition}{ORNooverflowReachsafetyLoops}{Error}{Assertion}{Walltime}{Stdev}{0E-16}%
\StoreBenchExecResult{CpacheckerBenchmarkDefinition}{ORNooverflowReachsafetyLoops}{Error}{Error}{Count}{}{48}%
\StoreBenchExecResult{CpacheckerBenchmarkDefinition}{ORNooverflowReachsafetyLoops}{Error}{Error}{Cputime}{}{456.926184951}%
\StoreBenchExecResult{CpacheckerBenchmarkDefinition}{ORNooverflowReachsafetyLoops}{Error}{Error}{Cputime}{Avg}{9.5192955198125}%
\StoreBenchExecResult{CpacheckerBenchmarkDefinition}{ORNooverflowReachsafetyLoops}{Error}{Error}{Cputime}{Median}{8.9883054805}%
\StoreBenchExecResult{CpacheckerBenchmarkDefinition}{ORNooverflowReachsafetyLoops}{Error}{Error}{Cputime}{Min}{2.671261697}%
\StoreBenchExecResult{CpacheckerBenchmarkDefinition}{ORNooverflowReachsafetyLoops}{Error}{Error}{Cputime}{Max}{21.378778012}%
\StoreBenchExecResult{CpacheckerBenchmarkDefinition}{ORNooverflowReachsafetyLoops}{Error}{Error}{Cputime}{Stdev}{5.219940406310111934918819299}%
\StoreBenchExecResult{CpacheckerBenchmarkDefinition}{ORNooverflowReachsafetyLoops}{Error}{Error}{Walltime}{}{280.5179739932063966}%
\StoreBenchExecResult{CpacheckerBenchmarkDefinition}{ORNooverflowReachsafetyLoops}{Error}{Error}{Walltime}{Avg}{5.844124458191799929166666667}%
\StoreBenchExecResult{CpacheckerBenchmarkDefinition}{ORNooverflowReachsafetyLoops}{Error}{Error}{Walltime}{Median}{4.8189312529284505}%
\StoreBenchExecResult{CpacheckerBenchmarkDefinition}{ORNooverflowReachsafetyLoops}{Error}{Error}{Walltime}{Min}{1.4463007519952953}%
\StoreBenchExecResult{CpacheckerBenchmarkDefinition}{ORNooverflowReachsafetyLoops}{Error}{Error}{Walltime}{Max}{16.01454222202301}%
\StoreBenchExecResult{CpacheckerBenchmarkDefinition}{ORNooverflowReachsafetyLoops}{Error}{Error}{Walltime}{Stdev}{4.119734646406430744540952086}%
\StoreBenchExecResult{CpacheckerBenchmarkDefinition}{ORNooverflowReachsafetyLoops}{Error}{Timeout}{Count}{}{257}%
\StoreBenchExecResult{CpacheckerBenchmarkDefinition}{ORNooverflowReachsafetyLoops}{Error}{Timeout}{Cputime}{}{232489.900410319}%
\StoreBenchExecResult{CpacheckerBenchmarkDefinition}{ORNooverflowReachsafetyLoops}{Error}{Timeout}{Cputime}{Avg}{904.6299626860661478599221790}%
\StoreBenchExecResult{CpacheckerBenchmarkDefinition}{ORNooverflowReachsafetyLoops}{Error}{Timeout}{Cputime}{Median}{903.165936832}%
\StoreBenchExecResult{CpacheckerBenchmarkDefinition}{ORNooverflowReachsafetyLoops}{Error}{Timeout}{Cputime}{Min}{902.035793151}%
\StoreBenchExecResult{CpacheckerBenchmarkDefinition}{ORNooverflowReachsafetyLoops}{Error}{Timeout}{Cputime}{Max}{960.732177482}%
\StoreBenchExecResult{CpacheckerBenchmarkDefinition}{ORNooverflowReachsafetyLoops}{Error}{Timeout}{Cputime}{Stdev}{4.495902894665123028080593346}%
\StoreBenchExecResult{CpacheckerBenchmarkDefinition}{ORNooverflowReachsafetyLoops}{Error}{Timeout}{Walltime}{}{227882.4532842929471}%
\StoreBenchExecResult{CpacheckerBenchmarkDefinition}{ORNooverflowReachsafetyLoops}{Error}{Timeout}{Walltime}{Avg}{886.7021528571710003891050584}%
\StoreBenchExecResult{CpacheckerBenchmarkDefinition}{ORNooverflowReachsafetyLoops}{Error}{Timeout}{Walltime}{Median}{890.7651590749156}%
\StoreBenchExecResult{CpacheckerBenchmarkDefinition}{ORNooverflowReachsafetyLoops}{Error}{Timeout}{Walltime}{Min}{829.9525031491648}%
\StoreBenchExecResult{CpacheckerBenchmarkDefinition}{ORNooverflowReachsafetyLoops}{Error}{Timeout}{Walltime}{Max}{930.0198502680287}%
\StoreBenchExecResult{CpacheckerBenchmarkDefinition}{ORNooverflowReachsafetyLoops}{Error}{Timeout}{Walltime}{Stdev}{12.39117656430149928310999321}%
\StoreBenchExecResult{CpacheckerBenchmarkDefinition}{ORNooverflowReachsafetyLoops}{Unknown}{}{Count}{}{13}%
\StoreBenchExecResult{CpacheckerBenchmarkDefinition}{ORNooverflowReachsafetyLoops}{Unknown}{}{Cputime}{}{92.400781046}%
\StoreBenchExecResult{CpacheckerBenchmarkDefinition}{ORNooverflowReachsafetyLoops}{Unknown}{}{Cputime}{Avg}{7.107752388153846153846153846}%
\StoreBenchExecResult{CpacheckerBenchmarkDefinition}{ORNooverflowReachsafetyLoops}{Unknown}{}{Cputime}{Median}{7.151514658}%
\StoreBenchExecResult{CpacheckerBenchmarkDefinition}{ORNooverflowReachsafetyLoops}{Unknown}{}{Cputime}{Min}{6.505028884}%
\StoreBenchExecResult{CpacheckerBenchmarkDefinition}{ORNooverflowReachsafetyLoops}{Unknown}{}{Cputime}{Max}{7.56679779}%
\StoreBenchExecResult{CpacheckerBenchmarkDefinition}{ORNooverflowReachsafetyLoops}{Unknown}{}{Cputime}{Stdev}{0.2944980046283523551695683300}%
\StoreBenchExecResult{CpacheckerBenchmarkDefinition}{ORNooverflowReachsafetyLoops}{Unknown}{}{Walltime}{}{48.0391089054755863}%
\StoreBenchExecResult{CpacheckerBenchmarkDefinition}{ORNooverflowReachsafetyLoops}{Unknown}{}{Walltime}{Avg}{3.695316069651968176923076923}%
\StoreBenchExecResult{CpacheckerBenchmarkDefinition}{ORNooverflowReachsafetyLoops}{Unknown}{}{Walltime}{Median}{3.689197395928204}%
\StoreBenchExecResult{CpacheckerBenchmarkDefinition}{ORNooverflowReachsafetyLoops}{Unknown}{}{Walltime}{Min}{3.386993857799098}%
\StoreBenchExecResult{CpacheckerBenchmarkDefinition}{ORNooverflowReachsafetyLoops}{Unknown}{}{Walltime}{Max}{3.928338489960879}%
\StoreBenchExecResult{CpacheckerBenchmarkDefinition}{ORNooverflowReachsafetyLoops}{Unknown}{}{Walltime}{Stdev}{0.1525979112916455328895739168}%
\StoreBenchExecResult{CpacheckerBenchmarkDefinition}{ORNooverflowReachsafetyLoops}{Unknown}{Unknown}{Count}{}{13}%
\StoreBenchExecResult{CpacheckerBenchmarkDefinition}{ORNooverflowReachsafetyLoops}{Unknown}{Unknown}{Cputime}{}{92.400781046}%
\StoreBenchExecResult{CpacheckerBenchmarkDefinition}{ORNooverflowReachsafetyLoops}{Unknown}{Unknown}{Cputime}{Avg}{7.107752388153846153846153846}%
\StoreBenchExecResult{CpacheckerBenchmarkDefinition}{ORNooverflowReachsafetyLoops}{Unknown}{Unknown}{Cputime}{Median}{7.151514658}%
\StoreBenchExecResult{CpacheckerBenchmarkDefinition}{ORNooverflowReachsafetyLoops}{Unknown}{Unknown}{Cputime}{Min}{6.505028884}%
\StoreBenchExecResult{CpacheckerBenchmarkDefinition}{ORNooverflowReachsafetyLoops}{Unknown}{Unknown}{Cputime}{Max}{7.56679779}%
\StoreBenchExecResult{CpacheckerBenchmarkDefinition}{ORNooverflowReachsafetyLoops}{Unknown}{Unknown}{Cputime}{Stdev}{0.2944980046283523551695683300}%
\StoreBenchExecResult{CpacheckerBenchmarkDefinition}{ORNooverflowReachsafetyLoops}{Unknown}{Unknown}{Walltime}{}{48.0391089054755863}%
\StoreBenchExecResult{CpacheckerBenchmarkDefinition}{ORNooverflowReachsafetyLoops}{Unknown}{Unknown}{Walltime}{Avg}{3.695316069651968176923076923}%
\StoreBenchExecResult{CpacheckerBenchmarkDefinition}{ORNooverflowReachsafetyLoops}{Unknown}{Unknown}{Walltime}{Median}{3.689197395928204}%
\StoreBenchExecResult{CpacheckerBenchmarkDefinition}{ORNooverflowReachsafetyLoops}{Unknown}{Unknown}{Walltime}{Min}{3.386993857799098}%
\StoreBenchExecResult{CpacheckerBenchmarkDefinition}{ORNooverflowReachsafetyLoops}{Unknown}{Unknown}{Walltime}{Max}{3.928338489960879}%
\StoreBenchExecResult{CpacheckerBenchmarkDefinition}{ORNooverflowReachsafetyLoops}{Unknown}{Unknown}{Walltime}{Stdev}{0.1525979112916455328895739168}%
\ifdefined\CpacheckerCpacheckerTotalCount\else\edef\CpacheckerCpacheckerTotalCount{0}\fi
\ifdefined\CpacheckerCpacheckerCorrectCount\else\edef\CpacheckerCpacheckerCorrectCount{0}\fi
\ifdefined\CpacheckerCpacheckerCorrectTrueCount\else\edef\CpacheckerCpacheckerCorrectTrueCount{0}\fi
\ifdefined\CpacheckerCpacheckerCorrectFalseCount\else\edef\CpacheckerCpacheckerCorrectFalseCount{0}\fi
\ifdefined\CpacheckerCpacheckerWrongCount\else\edef\CpacheckerCpacheckerWrongCount{0}\fi
\ifdefined\CpacheckerCpacheckerWrongTrueCount\else\edef\CpacheckerCpacheckerWrongTrueCount{0}\fi
\ifdefined\CpacheckerCpacheckerWrongFalseCount\else\edef\CpacheckerCpacheckerWrongFalseCount{0}\fi
\ifdefined\CpacheckerCpacheckerErrorTimeoutCount\else\edef\CpacheckerCpacheckerErrorTimeoutCount{0}\fi
\ifdefined\CpacheckerCpacheckerErrorOutOfMemoryCount\else\edef\CpacheckerCpacheckerErrorOutOfMemoryCount{0}\fi
\ifdefined\CpacheckerCpacheckerCorrectCputime\else\edef\CpacheckerCpacheckerCorrectCputime{0}\fi
\ifdefined\CpacheckerCpacheckerCorrectCputimeAvg\else\edef\CpacheckerCpacheckerCorrectCputimeAvg{None}\fi
\ifdefined\CpacheckerCpacheckerCorrectWalltime\else\edef\CpacheckerCpacheckerCorrectWalltime{0}\fi
\ifdefined\CpacheckerCpacheckerCorrectWalltimeAvg\else\edef\CpacheckerCpacheckerCorrectWalltimeAvg{None}\fi
\edef\CpacheckerCpacheckerErrorOtherInconclusiveCount{\the\numexpr \CpacheckerCpacheckerTotalCount - \CpacheckerCpacheckerCorrectCount - \CpacheckerCpacheckerWrongTrueCount - \CpacheckerCpacheckerWrongFalseCount - \CpacheckerCpacheckerErrorTimeoutCount - \CpacheckerCpacheckerErrorOutOfMemoryCount \relax}
\providecommand\StoreBenchExecResult[7]{\expandafter\newcommand\csname#1#2#3#4#5#6\endcsname{#7}}%
\StoreBenchExecResult{ExperimentCpacheckerVerificationMemcleanupTransformed}{MemoryCleanupTransformedMemsafetyMemcleanup}{Total}{}{Count}{}{41}%
\StoreBenchExecResult{ExperimentCpacheckerVerificationMemcleanupTransformed}{MemoryCleanupTransformedMemsafetyMemcleanup}{Total}{}{Cputime}{}{5514.946101445}%
\StoreBenchExecResult{ExperimentCpacheckerVerificationMemcleanupTransformed}{MemoryCleanupTransformedMemsafetyMemcleanup}{Total}{}{Cputime}{Avg}{134.5108805230487804878048780}%
\StoreBenchExecResult{ExperimentCpacheckerVerificationMemcleanupTransformed}{MemoryCleanupTransformedMemsafetyMemcleanup}{Total}{}{Cputime}{Median}{14.536297125}%
\StoreBenchExecResult{ExperimentCpacheckerVerificationMemcleanupTransformed}{MemoryCleanupTransformedMemsafetyMemcleanup}{Total}{}{Cputime}{Min}{4.100548963}%
\StoreBenchExecResult{ExperimentCpacheckerVerificationMemcleanupTransformed}{MemoryCleanupTransformedMemsafetyMemcleanup}{Total}{}{Cputime}{Max}{916.061040746}%
\StoreBenchExecResult{ExperimentCpacheckerVerificationMemcleanupTransformed}{MemoryCleanupTransformedMemsafetyMemcleanup}{Total}{}{Cputime}{Stdev}{288.8921836457287915122754081}%
\StoreBenchExecResult{ExperimentCpacheckerVerificationMemcleanupTransformed}{MemoryCleanupTransformedMemsafetyMemcleanup}{Total}{}{Walltime}{}{4815.9154051060394716}%
\StoreBenchExecResult{ExperimentCpacheckerVerificationMemcleanupTransformed}{MemoryCleanupTransformedMemsafetyMemcleanup}{Total}{}{Walltime}{Avg}{117.4613513440497432097560976}%
\StoreBenchExecResult{ExperimentCpacheckerVerificationMemcleanupTransformed}{MemoryCleanupTransformedMemsafetyMemcleanup}{Total}{}{Walltime}{Median}{7.493426157998329}%
\StoreBenchExecResult{ExperimentCpacheckerVerificationMemcleanupTransformed}{MemoryCleanupTransformedMemsafetyMemcleanup}{Total}{}{Walltime}{Min}{2.175350575998891}%
\StoreBenchExecResult{ExperimentCpacheckerVerificationMemcleanupTransformed}{MemoryCleanupTransformedMemsafetyMemcleanup}{Total}{}{Walltime}{Max}{881.4796751100002}%
\StoreBenchExecResult{ExperimentCpacheckerVerificationMemcleanupTransformed}{MemoryCleanupTransformedMemsafetyMemcleanup}{Total}{}{Walltime}{Stdev}{272.2500174549438200604757121}%
\StoreBenchExecResult{ExperimentCpacheckerVerificationMemcleanupTransformed}{MemoryCleanupTransformedMemsafetyMemcleanup}{Correct}{}{Count}{}{28}%
\StoreBenchExecResult{ExperimentCpacheckerVerificationMemcleanupTransformed}{MemoryCleanupTransformedMemsafetyMemcleanup}{Correct}{}{Cputime}{}{763.571112319}%
\StoreBenchExecResult{ExperimentCpacheckerVerificationMemcleanupTransformed}{MemoryCleanupTransformedMemsafetyMemcleanup}{Correct}{}{Cputime}{Avg}{27.27039686853571428571428571}%
\StoreBenchExecResult{ExperimentCpacheckerVerificationMemcleanupTransformed}{MemoryCleanupTransformedMemsafetyMemcleanup}{Correct}{}{Cputime}{Median}{13.139767219}%
\StoreBenchExecResult{ExperimentCpacheckerVerificationMemcleanupTransformed}{MemoryCleanupTransformedMemsafetyMemcleanup}{Correct}{}{Cputime}{Min}{7.864610432}%
\StoreBenchExecResult{ExperimentCpacheckerVerificationMemcleanupTransformed}{MemoryCleanupTransformedMemsafetyMemcleanup}{Correct}{}{Cputime}{Max}{127.56748287}%
\StoreBenchExecResult{ExperimentCpacheckerVerificationMemcleanupTransformed}{MemoryCleanupTransformedMemsafetyMemcleanup}{Correct}{}{Cputime}{Stdev}{28.42265856632827637843729336}%
\StoreBenchExecResult{ExperimentCpacheckerVerificationMemcleanupTransformed}{MemoryCleanupTransformedMemsafetyMemcleanup}{Correct}{}{Walltime}{}{446.5449453770270266}%
\StoreBenchExecResult{ExperimentCpacheckerVerificationMemcleanupTransformed}{MemoryCleanupTransformedMemsafetyMemcleanup}{Correct}{}{Walltime}{Avg}{15.94803376346525095}%
\StoreBenchExecResult{ExperimentCpacheckerVerificationMemcleanupTransformed}{MemoryCleanupTransformedMemsafetyMemcleanup}{Correct}{}{Walltime}{Median}{6.8013920775010775}%
\StoreBenchExecResult{ExperimentCpacheckerVerificationMemcleanupTransformed}{MemoryCleanupTransformedMemsafetyMemcleanup}{Correct}{}{Walltime}{Min}{4.08793805300229}%
\StoreBenchExecResult{ExperimentCpacheckerVerificationMemcleanupTransformed}{MemoryCleanupTransformedMemsafetyMemcleanup}{Correct}{}{Walltime}{Max}{91.72551677600131}%
\StoreBenchExecResult{ExperimentCpacheckerVerificationMemcleanupTransformed}{MemoryCleanupTransformedMemsafetyMemcleanup}{Correct}{}{Walltime}{Stdev}{19.38761329977964321825298572}%
\StoreBenchExecResult{ExperimentCpacheckerVerificationMemcleanupTransformed}{MemoryCleanupTransformedMemsafetyMemcleanup}{Correct}{False}{Count}{}{28}%
\StoreBenchExecResult{ExperimentCpacheckerVerificationMemcleanupTransformed}{MemoryCleanupTransformedMemsafetyMemcleanup}{Correct}{False}{Cputime}{}{763.571112319}%
\StoreBenchExecResult{ExperimentCpacheckerVerificationMemcleanupTransformed}{MemoryCleanupTransformedMemsafetyMemcleanup}{Correct}{False}{Cputime}{Avg}{27.27039686853571428571428571}%
\StoreBenchExecResult{ExperimentCpacheckerVerificationMemcleanupTransformed}{MemoryCleanupTransformedMemsafetyMemcleanup}{Correct}{False}{Cputime}{Median}{13.139767219}%
\StoreBenchExecResult{ExperimentCpacheckerVerificationMemcleanupTransformed}{MemoryCleanupTransformedMemsafetyMemcleanup}{Correct}{False}{Cputime}{Min}{7.864610432}%
\StoreBenchExecResult{ExperimentCpacheckerVerificationMemcleanupTransformed}{MemoryCleanupTransformedMemsafetyMemcleanup}{Correct}{False}{Cputime}{Max}{127.56748287}%
\StoreBenchExecResult{ExperimentCpacheckerVerificationMemcleanupTransformed}{MemoryCleanupTransformedMemsafetyMemcleanup}{Correct}{False}{Cputime}{Stdev}{28.42265856632827637843729336}%
\StoreBenchExecResult{ExperimentCpacheckerVerificationMemcleanupTransformed}{MemoryCleanupTransformedMemsafetyMemcleanup}{Correct}{False}{Walltime}{}{446.5449453770270266}%
\StoreBenchExecResult{ExperimentCpacheckerVerificationMemcleanupTransformed}{MemoryCleanupTransformedMemsafetyMemcleanup}{Correct}{False}{Walltime}{Avg}{15.94803376346525095}%
\StoreBenchExecResult{ExperimentCpacheckerVerificationMemcleanupTransformed}{MemoryCleanupTransformedMemsafetyMemcleanup}{Correct}{False}{Walltime}{Median}{6.8013920775010775}%
\StoreBenchExecResult{ExperimentCpacheckerVerificationMemcleanupTransformed}{MemoryCleanupTransformedMemsafetyMemcleanup}{Correct}{False}{Walltime}{Min}{4.08793805300229}%
\StoreBenchExecResult{ExperimentCpacheckerVerificationMemcleanupTransformed}{MemoryCleanupTransformedMemsafetyMemcleanup}{Correct}{False}{Walltime}{Max}{91.72551677600131}%
\StoreBenchExecResult{ExperimentCpacheckerVerificationMemcleanupTransformed}{MemoryCleanupTransformedMemsafetyMemcleanup}{Correct}{False}{Walltime}{Stdev}{19.38761329977964321825298572}%
\StoreBenchExecResult{ExperimentCpacheckerVerificationMemcleanupTransformed}{MemoryCleanupTransformedMemsafetyMemcleanup}{Error}{}{Count}{}{11}%
\StoreBenchExecResult{ExperimentCpacheckerVerificationMemcleanupTransformed}{MemoryCleanupTransformedMemsafetyMemcleanup}{Error}{}{Cputime}{}{4600.268322622}%
\StoreBenchExecResult{ExperimentCpacheckerVerificationMemcleanupTransformed}{MemoryCleanupTransformedMemsafetyMemcleanup}{Error}{}{Cputime}{Avg}{418.2062111474545454545454545}%
\StoreBenchExecResult{ExperimentCpacheckerVerificationMemcleanupTransformed}{MemoryCleanupTransformedMemsafetyMemcleanup}{Error}{}{Cputime}{Median}{20.499199474}%
\StoreBenchExecResult{ExperimentCpacheckerVerificationMemcleanupTransformed}{MemoryCleanupTransformedMemsafetyMemcleanup}{Error}{}{Cputime}{Min}{4.100548963}%
\StoreBenchExecResult{ExperimentCpacheckerVerificationMemcleanupTransformed}{MemoryCleanupTransformedMemsafetyMemcleanup}{Error}{}{Cputime}{Max}{916.061040746}%
\StoreBenchExecResult{ExperimentCpacheckerVerificationMemcleanupTransformed}{MemoryCleanupTransformedMemsafetyMemcleanup}{Error}{}{Cputime}{Stdev}{445.5841390149893356683336078}%
\StoreBenchExecResult{ExperimentCpacheckerVerificationMemcleanupTransformed}{MemoryCleanupTransformedMemsafetyMemcleanup}{Error}{}{Walltime}{}{4262.214253785008555}%
\StoreBenchExecResult{ExperimentCpacheckerVerificationMemcleanupTransformed}{MemoryCleanupTransformedMemsafetyMemcleanup}{Error}{}{Walltime}{Avg}{387.4740230713644140909090909}%
\StoreBenchExecResult{ExperimentCpacheckerVerificationMemcleanupTransformed}{MemoryCleanupTransformedMemsafetyMemcleanup}{Error}{}{Walltime}{Median}{15.154532435000874}%
\StoreBenchExecResult{ExperimentCpacheckerVerificationMemcleanupTransformed}{MemoryCleanupTransformedMemsafetyMemcleanup}{Error}{}{Walltime}{Min}{2.175350575998891}%
\StoreBenchExecResult{ExperimentCpacheckerVerificationMemcleanupTransformed}{MemoryCleanupTransformedMemsafetyMemcleanup}{Error}{}{Walltime}{Max}{881.4796751100002}%
\StoreBenchExecResult{ExperimentCpacheckerVerificationMemcleanupTransformed}{MemoryCleanupTransformedMemsafetyMemcleanup}{Error}{}{Walltime}{Stdev}{418.7769944552310034218685430}%
\StoreBenchExecResult{ExperimentCpacheckerVerificationMemcleanupTransformed}{MemoryCleanupTransformedMemsafetyMemcleanup}{Error}{Error}{Count}{}{5}%
\StoreBenchExecResult{ExperimentCpacheckerVerificationMemcleanupTransformed}{MemoryCleanupTransformedMemsafetyMemcleanup}{Error}{Error}{Cputime}{}{49.808180409}%
\StoreBenchExecResult{ExperimentCpacheckerVerificationMemcleanupTransformed}{MemoryCleanupTransformedMemsafetyMemcleanup}{Error}{Error}{Cputime}{Avg}{9.9616360818}%
\StoreBenchExecResult{ExperimentCpacheckerVerificationMemcleanupTransformed}{MemoryCleanupTransformedMemsafetyMemcleanup}{Error}{Error}{Cputime}{Median}{7.889358343}%
\StoreBenchExecResult{ExperimentCpacheckerVerificationMemcleanupTransformed}{MemoryCleanupTransformedMemsafetyMemcleanup}{Error}{Error}{Cputime}{Min}{4.100548963}%
\StoreBenchExecResult{ExperimentCpacheckerVerificationMemcleanupTransformed}{MemoryCleanupTransformedMemsafetyMemcleanup}{Error}{Error}{Cputime}{Max}{20.499199474}%
\StoreBenchExecResult{ExperimentCpacheckerVerificationMemcleanupTransformed}{MemoryCleanupTransformedMemsafetyMemcleanup}{Error}{Error}{Cputime}{Stdev}{5.573225527199761102500966089}%
\StoreBenchExecResult{ExperimentCpacheckerVerificationMemcleanupTransformed}{MemoryCleanupTransformedMemsafetyMemcleanup}{Error}{Error}{Walltime}{}{30.455846966004174}%
\StoreBenchExecResult{ExperimentCpacheckerVerificationMemcleanupTransformed}{MemoryCleanupTransformedMemsafetyMemcleanup}{Error}{Error}{Walltime}{Avg}{6.0911693932008348}%
\StoreBenchExecResult{ExperimentCpacheckerVerificationMemcleanupTransformed}{MemoryCleanupTransformedMemsafetyMemcleanup}{Error}{Error}{Walltime}{Median}{4.160050869002589}%
\StoreBenchExecResult{ExperimentCpacheckerVerificationMemcleanupTransformed}{MemoryCleanupTransformedMemsafetyMemcleanup}{Error}{Error}{Walltime}{Min}{2.175350575998891}%
\StoreBenchExecResult{ExperimentCpacheckerVerificationMemcleanupTransformed}{MemoryCleanupTransformedMemsafetyMemcleanup}{Error}{Error}{Walltime}{Max}{15.154532435000874}%
\StoreBenchExecResult{ExperimentCpacheckerVerificationMemcleanupTransformed}{MemoryCleanupTransformedMemsafetyMemcleanup}{Error}{Error}{Walltime}{Stdev}{4.626109738370463539606332059}%
\StoreBenchExecResult{ExperimentCpacheckerVerificationMemcleanupTransformed}{MemoryCleanupTransformedMemsafetyMemcleanup}{Error}{Exception}{Count}{}{1}%
\StoreBenchExecResult{ExperimentCpacheckerVerificationMemcleanupTransformed}{MemoryCleanupTransformedMemsafetyMemcleanup}{Error}{Exception}{Cputime}{}{19.062168093}%
\StoreBenchExecResult{ExperimentCpacheckerVerificationMemcleanupTransformed}{MemoryCleanupTransformedMemsafetyMemcleanup}{Error}{Exception}{Cputime}{Avg}{19.062168093}%
\StoreBenchExecResult{ExperimentCpacheckerVerificationMemcleanupTransformed}{MemoryCleanupTransformedMemsafetyMemcleanup}{Error}{Exception}{Cputime}{Median}{19.062168093}%
\StoreBenchExecResult{ExperimentCpacheckerVerificationMemcleanupTransformed}{MemoryCleanupTransformedMemsafetyMemcleanup}{Error}{Exception}{Cputime}{Min}{19.062168093}%
\StoreBenchExecResult{ExperimentCpacheckerVerificationMemcleanupTransformed}{MemoryCleanupTransformedMemsafetyMemcleanup}{Error}{Exception}{Cputime}{Max}{19.062168093}%
\StoreBenchExecResult{ExperimentCpacheckerVerificationMemcleanupTransformed}{MemoryCleanupTransformedMemsafetyMemcleanup}{Error}{Exception}{Cputime}{Stdev}{0E-14}%
\StoreBenchExecResult{ExperimentCpacheckerVerificationMemcleanupTransformed}{MemoryCleanupTransformedMemsafetyMemcleanup}{Error}{Exception}{Walltime}{}{9.737350502000481}%
\StoreBenchExecResult{ExperimentCpacheckerVerificationMemcleanupTransformed}{MemoryCleanupTransformedMemsafetyMemcleanup}{Error}{Exception}{Walltime}{Avg}{9.737350502000481}%
\StoreBenchExecResult{ExperimentCpacheckerVerificationMemcleanupTransformed}{MemoryCleanupTransformedMemsafetyMemcleanup}{Error}{Exception}{Walltime}{Median}{9.737350502000481}%
\StoreBenchExecResult{ExperimentCpacheckerVerificationMemcleanupTransformed}{MemoryCleanupTransformedMemsafetyMemcleanup}{Error}{Exception}{Walltime}{Min}{9.737350502000481}%
\StoreBenchExecResult{ExperimentCpacheckerVerificationMemcleanupTransformed}{MemoryCleanupTransformedMemsafetyMemcleanup}{Error}{Exception}{Walltime}{Max}{9.737350502000481}%
\StoreBenchExecResult{ExperimentCpacheckerVerificationMemcleanupTransformed}{MemoryCleanupTransformedMemsafetyMemcleanup}{Error}{Exception}{Walltime}{Stdev}{0E-15}%
\StoreBenchExecResult{ExperimentCpacheckerVerificationMemcleanupTransformed}{MemoryCleanupTransformedMemsafetyMemcleanup}{Error}{Timeout}{Count}{}{5}%
\StoreBenchExecResult{ExperimentCpacheckerVerificationMemcleanupTransformed}{MemoryCleanupTransformedMemsafetyMemcleanup}{Error}{Timeout}{Cputime}{}{4531.397974120}%
\StoreBenchExecResult{ExperimentCpacheckerVerificationMemcleanupTransformed}{MemoryCleanupTransformedMemsafetyMemcleanup}{Error}{Timeout}{Cputime}{Avg}{906.279594824}%
\StoreBenchExecResult{ExperimentCpacheckerVerificationMemcleanupTransformed}{MemoryCleanupTransformedMemsafetyMemcleanup}{Error}{Timeout}{Cputime}{Median}{903.768606327}%
\StoreBenchExecResult{ExperimentCpacheckerVerificationMemcleanupTransformed}{MemoryCleanupTransformedMemsafetyMemcleanup}{Error}{Timeout}{Cputime}{Min}{902.237290507}%
\StoreBenchExecResult{ExperimentCpacheckerVerificationMemcleanupTransformed}{MemoryCleanupTransformedMemsafetyMemcleanup}{Error}{Timeout}{Cputime}{Max}{916.061040746}%
\StoreBenchExecResult{ExperimentCpacheckerVerificationMemcleanupTransformed}{MemoryCleanupTransformedMemsafetyMemcleanup}{Error}{Timeout}{Cputime}{Stdev}{5.097026961579798673137117971}%
\StoreBenchExecResult{ExperimentCpacheckerVerificationMemcleanupTransformed}{MemoryCleanupTransformedMemsafetyMemcleanup}{Error}{Timeout}{Walltime}{}{4222.0210563170039}%
\StoreBenchExecResult{ExperimentCpacheckerVerificationMemcleanupTransformed}{MemoryCleanupTransformedMemsafetyMemcleanup}{Error}{Timeout}{Walltime}{Avg}{844.40421126340078}%
\StoreBenchExecResult{ExperimentCpacheckerVerificationMemcleanupTransformed}{MemoryCleanupTransformedMemsafetyMemcleanup}{Error}{Timeout}{Walltime}{Median}{867.4125139170064}%
\StoreBenchExecResult{ExperimentCpacheckerVerificationMemcleanupTransformed}{MemoryCleanupTransformedMemsafetyMemcleanup}{Error}{Timeout}{Walltime}{Min}{735.7481325249973}%
\StoreBenchExecResult{ExperimentCpacheckerVerificationMemcleanupTransformed}{MemoryCleanupTransformedMemsafetyMemcleanup}{Error}{Timeout}{Walltime}{Max}{881.4796751100002}%
\StoreBenchExecResult{ExperimentCpacheckerVerificationMemcleanupTransformed}{MemoryCleanupTransformedMemsafetyMemcleanup}{Error}{Timeout}{Walltime}{Stdev}{55.01544123179195853526536254}%
\StoreBenchExecResult{ExperimentCpacheckerVerificationMemcleanupTransformed}{MemoryCleanupTransformedMemsafetyMemcleanup}{Wrong}{}{Count}{}{2}%
\StoreBenchExecResult{ExperimentCpacheckerVerificationMemcleanupTransformed}{MemoryCleanupTransformedMemsafetyMemcleanup}{Wrong}{}{Cputime}{}{151.106666504}%
\StoreBenchExecResult{ExperimentCpacheckerVerificationMemcleanupTransformed}{MemoryCleanupTransformedMemsafetyMemcleanup}{Wrong}{}{Cputime}{Avg}{75.553333252}%
\StoreBenchExecResult{ExperimentCpacheckerVerificationMemcleanupTransformed}{MemoryCleanupTransformedMemsafetyMemcleanup}{Wrong}{}{Cputime}{Median}{75.553333252}%
\StoreBenchExecResult{ExperimentCpacheckerVerificationMemcleanupTransformed}{MemoryCleanupTransformedMemsafetyMemcleanup}{Wrong}{}{Cputime}{Min}{54.2318559}%
\StoreBenchExecResult{ExperimentCpacheckerVerificationMemcleanupTransformed}{MemoryCleanupTransformedMemsafetyMemcleanup}{Wrong}{}{Cputime}{Max}{96.874810604}%
\StoreBenchExecResult{ExperimentCpacheckerVerificationMemcleanupTransformed}{MemoryCleanupTransformedMemsafetyMemcleanup}{Wrong}{}{Cputime}{Stdev}{21.3214773520000}%
\StoreBenchExecResult{ExperimentCpacheckerVerificationMemcleanupTransformed}{MemoryCleanupTransformedMemsafetyMemcleanup}{Wrong}{}{Walltime}{}{107.15620594400389}%
\StoreBenchExecResult{ExperimentCpacheckerVerificationMemcleanupTransformed}{MemoryCleanupTransformedMemsafetyMemcleanup}{Wrong}{}{Walltime}{Avg}{53.578102972001945}%
\StoreBenchExecResult{ExperimentCpacheckerVerificationMemcleanupTransformed}{MemoryCleanupTransformedMemsafetyMemcleanup}{Wrong}{}{Walltime}{Median}{53.578102972001945}%
\StoreBenchExecResult{ExperimentCpacheckerVerificationMemcleanupTransformed}{MemoryCleanupTransformedMemsafetyMemcleanup}{Wrong}{}{Walltime}{Min}{36.16646767099883}%
\StoreBenchExecResult{ExperimentCpacheckerVerificationMemcleanupTransformed}{MemoryCleanupTransformedMemsafetyMemcleanup}{Wrong}{}{Walltime}{Max}{70.98973827300506}%
\StoreBenchExecResult{ExperimentCpacheckerVerificationMemcleanupTransformed}{MemoryCleanupTransformedMemsafetyMemcleanup}{Wrong}{}{Walltime}{Stdev}{17.41163530100311500000000000}%
\StoreBenchExecResult{ExperimentCpacheckerVerificationMemcleanupTransformed}{MemoryCleanupTransformedMemsafetyMemcleanup}{Wrong}{False}{Count}{}{1}%
\StoreBenchExecResult{ExperimentCpacheckerVerificationMemcleanupTransformed}{MemoryCleanupTransformedMemsafetyMemcleanup}{Wrong}{False}{Cputime}{}{54.2318559}%
\StoreBenchExecResult{ExperimentCpacheckerVerificationMemcleanupTransformed}{MemoryCleanupTransformedMemsafetyMemcleanup}{Wrong}{False}{Cputime}{Avg}{54.2318559}%
\StoreBenchExecResult{ExperimentCpacheckerVerificationMemcleanupTransformed}{MemoryCleanupTransformedMemsafetyMemcleanup}{Wrong}{False}{Cputime}{Median}{54.2318559}%
\StoreBenchExecResult{ExperimentCpacheckerVerificationMemcleanupTransformed}{MemoryCleanupTransformedMemsafetyMemcleanup}{Wrong}{False}{Cputime}{Min}{54.2318559}%
\StoreBenchExecResult{ExperimentCpacheckerVerificationMemcleanupTransformed}{MemoryCleanupTransformedMemsafetyMemcleanup}{Wrong}{False}{Cputime}{Max}{54.2318559}%
\StoreBenchExecResult{ExperimentCpacheckerVerificationMemcleanupTransformed}{MemoryCleanupTransformedMemsafetyMemcleanup}{Wrong}{False}{Cputime}{Stdev}{0E-14}%
\StoreBenchExecResult{ExperimentCpacheckerVerificationMemcleanupTransformed}{MemoryCleanupTransformedMemsafetyMemcleanup}{Wrong}{False}{Walltime}{}{36.16646767099883}%
\StoreBenchExecResult{ExperimentCpacheckerVerificationMemcleanupTransformed}{MemoryCleanupTransformedMemsafetyMemcleanup}{Wrong}{False}{Walltime}{Avg}{36.16646767099883}%
\StoreBenchExecResult{ExperimentCpacheckerVerificationMemcleanupTransformed}{MemoryCleanupTransformedMemsafetyMemcleanup}{Wrong}{False}{Walltime}{Median}{36.16646767099883}%
\StoreBenchExecResult{ExperimentCpacheckerVerificationMemcleanupTransformed}{MemoryCleanupTransformedMemsafetyMemcleanup}{Wrong}{False}{Walltime}{Min}{36.16646767099883}%
\StoreBenchExecResult{ExperimentCpacheckerVerificationMemcleanupTransformed}{MemoryCleanupTransformedMemsafetyMemcleanup}{Wrong}{False}{Walltime}{Max}{36.16646767099883}%
\StoreBenchExecResult{ExperimentCpacheckerVerificationMemcleanupTransformed}{MemoryCleanupTransformedMemsafetyMemcleanup}{Wrong}{False}{Walltime}{Stdev}{0E-14}%
\StoreBenchExecResult{ExperimentCpacheckerVerificationMemcleanupTransformed}{MemoryCleanupTransformedMemsafetyMemcleanup}{Wrong}{True}{Count}{}{1}%
\StoreBenchExecResult{ExperimentCpacheckerVerificationMemcleanupTransformed}{MemoryCleanupTransformedMemsafetyMemcleanup}{Wrong}{True}{Cputime}{}{96.874810604}%
\StoreBenchExecResult{ExperimentCpacheckerVerificationMemcleanupTransformed}{MemoryCleanupTransformedMemsafetyMemcleanup}{Wrong}{True}{Cputime}{Avg}{96.874810604}%
\StoreBenchExecResult{ExperimentCpacheckerVerificationMemcleanupTransformed}{MemoryCleanupTransformedMemsafetyMemcleanup}{Wrong}{True}{Cputime}{Median}{96.874810604}%
\StoreBenchExecResult{ExperimentCpacheckerVerificationMemcleanupTransformed}{MemoryCleanupTransformedMemsafetyMemcleanup}{Wrong}{True}{Cputime}{Min}{96.874810604}%
\StoreBenchExecResult{ExperimentCpacheckerVerificationMemcleanupTransformed}{MemoryCleanupTransformedMemsafetyMemcleanup}{Wrong}{True}{Cputime}{Max}{96.874810604}%
\StoreBenchExecResult{ExperimentCpacheckerVerificationMemcleanupTransformed}{MemoryCleanupTransformedMemsafetyMemcleanup}{Wrong}{True}{Cputime}{Stdev}{0E-14}%
\StoreBenchExecResult{ExperimentCpacheckerVerificationMemcleanupTransformed}{MemoryCleanupTransformedMemsafetyMemcleanup}{Wrong}{True}{Walltime}{}{70.98973827300506}%
\StoreBenchExecResult{ExperimentCpacheckerVerificationMemcleanupTransformed}{MemoryCleanupTransformedMemsafetyMemcleanup}{Wrong}{True}{Walltime}{Avg}{70.98973827300506}%
\StoreBenchExecResult{ExperimentCpacheckerVerificationMemcleanupTransformed}{MemoryCleanupTransformedMemsafetyMemcleanup}{Wrong}{True}{Walltime}{Median}{70.98973827300506}%
\StoreBenchExecResult{ExperimentCpacheckerVerificationMemcleanupTransformed}{MemoryCleanupTransformedMemsafetyMemcleanup}{Wrong}{True}{Walltime}{Min}{70.98973827300506}%
\StoreBenchExecResult{ExperimentCpacheckerVerificationMemcleanupTransformed}{MemoryCleanupTransformedMemsafetyMemcleanup}{Wrong}{True}{Walltime}{Max}{70.98973827300506}%
\StoreBenchExecResult{ExperimentCpacheckerVerificationMemcleanupTransformed}{MemoryCleanupTransformedMemsafetyMemcleanup}{Wrong}{True}{Walltime}{Stdev}{0E-14}%

\ifdefined\CpacheckerCpacheckerTotalCount\else\edef\CpacheckerCpacheckerTotalCount{0}\fi
\ifdefined\CpacheckerCpacheckerCorrectCount\else\edef\CpacheckerCpacheckerCorrectCount{0}\fi
\ifdefined\CpacheckerCpacheckerCorrectTrueCount\else\edef\CpacheckerCpacheckerCorrectTrueCount{0}\fi
\ifdefined\CpacheckerCpacheckerCorrectFalseCount\else\edef\CpacheckerCpacheckerCorrectFalseCount{0}\fi
\ifdefined\CpacheckerCpacheckerWrongCount\else\edef\CpacheckerCpacheckerWrongCount{0}\fi
\ifdefined\CpacheckerCpacheckerWrongTrueCount\else\edef\CpacheckerCpacheckerWrongTrueCount{0}\fi
\ifdefined\CpacheckerCpacheckerWrongFalseCount\else\edef\CpacheckerCpacheckerWrongFalseCount{0}\fi
\ifdefined\CpacheckerCpacheckerErrorTimeoutCount\else\edef\CpacheckerCpacheckerErrorTimeoutCount{0}\fi
\ifdefined\CpacheckerCpacheckerErrorOutOfMemoryCount\else\edef\CpacheckerCpacheckerErrorOutOfMemoryCount{0}\fi
\ifdefined\CpacheckerCpacheckerCorrectCputime\else\edef\CpacheckerCpacheckerCorrectCputime{0}\fi
\ifdefined\CpacheckerCpacheckerCorrectCputimeAvg\else\edef\CpacheckerCpacheckerCorrectCputimeAvg{None}\fi
\ifdefined\CpacheckerCpacheckerCorrectWalltime\else\edef\CpacheckerCpacheckerCorrectWalltime{0}\fi
\ifdefined\CpacheckerCpacheckerCorrectWalltimeAvg\else\edef\CpacheckerCpacheckerCorrectWalltimeAvg{None}\fi
\edef\CpacheckerCpacheckerErrorOtherInconclusiveCount{\the\numexpr \CpacheckerCpacheckerTotalCount - \CpacheckerCpacheckerCorrectCount - \CpacheckerCpacheckerWrongTrueCount - \CpacheckerCpacheckerWrongFalseCount - \CpacheckerCpacheckerErrorTimeoutCount - \CpacheckerCpacheckerErrorOutOfMemoryCount \relax}
\providecommand\StoreBenchExecResult[7]{\expandafter\newcommand\csname#1#2#3#4#5#6\endcsname{#7}}%
\StoreBenchExecResult{ExperimentUautomizerVerificationMemcleanupTransformed}{MemoryCleanupTransformedMemsafetyMemcleanup}{Total}{}{Count}{}{41}%
\StoreBenchExecResult{ExperimentUautomizerVerificationMemcleanupTransformed}{MemoryCleanupTransformedMemsafetyMemcleanup}{Total}{}{Cputime}{}{11993.984423313}%
\StoreBenchExecResult{ExperimentUautomizerVerificationMemcleanupTransformed}{MemoryCleanupTransformedMemsafetyMemcleanup}{Total}{}{Cputime}{Avg}{292.5362054466585365853658537}%
\StoreBenchExecResult{ExperimentUautomizerVerificationMemcleanupTransformed}{MemoryCleanupTransformedMemsafetyMemcleanup}{Total}{}{Cputime}{Median}{34.249648478}%
\StoreBenchExecResult{ExperimentUautomizerVerificationMemcleanupTransformed}{MemoryCleanupTransformedMemsafetyMemcleanup}{Total}{}{Cputime}{Min}{20.756152144}%
\StoreBenchExecResult{ExperimentUautomizerVerificationMemcleanupTransformed}{MemoryCleanupTransformedMemsafetyMemcleanup}{Total}{}{Cputime}{Max}{961.204997659}%
\StoreBenchExecResult{ExperimentUautomizerVerificationMemcleanupTransformed}{MemoryCleanupTransformedMemsafetyMemcleanup}{Total}{}{Cputime}{Stdev}{406.2269227678286414810694995}%
\StoreBenchExecResult{ExperimentUautomizerVerificationMemcleanupTransformed}{MemoryCleanupTransformedMemsafetyMemcleanup}{Total}{}{Walltime}{}{10310.497651460966357}%
\StoreBenchExecResult{ExperimentUautomizerVerificationMemcleanupTransformed}{MemoryCleanupTransformedMemsafetyMemcleanup}{Total}{}{Walltime}{Avg}{251.4755524746577160243902439}%
\StoreBenchExecResult{ExperimentUautomizerVerificationMemcleanupTransformed}{MemoryCleanupTransformedMemsafetyMemcleanup}{Total}{}{Walltime}{Median}{19.82690336000087}%
\StoreBenchExecResult{ExperimentUautomizerVerificationMemcleanupTransformed}{MemoryCleanupTransformedMemsafetyMemcleanup}{Total}{}{Walltime}{Min}{12.158048999997845}%
\StoreBenchExecResult{ExperimentUautomizerVerificationMemcleanupTransformed}{MemoryCleanupTransformedMemsafetyMemcleanup}{Total}{}{Walltime}{Max}{893.3044641550005}%
\StoreBenchExecResult{ExperimentUautomizerVerificationMemcleanupTransformed}{MemoryCleanupTransformedMemsafetyMemcleanup}{Total}{}{Walltime}{Stdev}{365.9040276732057359425747445}%
\StoreBenchExecResult{ExperimentUautomizerVerificationMemcleanupTransformed}{MemoryCleanupTransformedMemsafetyMemcleanup}{Correct}{}{Count}{}{21}%
\StoreBenchExecResult{ExperimentUautomizerVerificationMemcleanupTransformed}{MemoryCleanupTransformedMemsafetyMemcleanup}{Correct}{}{Cputime}{}{961.309961282}%
\StoreBenchExecResult{ExperimentUautomizerVerificationMemcleanupTransformed}{MemoryCleanupTransformedMemsafetyMemcleanup}{Correct}{}{Cputime}{Avg}{45.77666482295238095238095238}%
\StoreBenchExecResult{ExperimentUautomizerVerificationMemcleanupTransformed}{MemoryCleanupTransformedMemsafetyMemcleanup}{Correct}{}{Cputime}{Median}{24.106825283}%
\StoreBenchExecResult{ExperimentUautomizerVerificationMemcleanupTransformed}{MemoryCleanupTransformedMemsafetyMemcleanup}{Correct}{}{Cputime}{Min}{20.756152144}%
\StoreBenchExecResult{ExperimentUautomizerVerificationMemcleanupTransformed}{MemoryCleanupTransformedMemsafetyMemcleanup}{Correct}{}{Cputime}{Max}{212.786169979}%
\StoreBenchExecResult{ExperimentUautomizerVerificationMemcleanupTransformed}{MemoryCleanupTransformedMemsafetyMemcleanup}{Correct}{}{Cputime}{Stdev}{43.69918574900474103226236398}%
\StoreBenchExecResult{ExperimentUautomizerVerificationMemcleanupTransformed}{MemoryCleanupTransformedMemsafetyMemcleanup}{Correct}{}{Walltime}{}{641.954908076979343}%
\StoreBenchExecResult{ExperimentUautomizerVerificationMemcleanupTransformed}{MemoryCleanupTransformedMemsafetyMemcleanup}{Correct}{}{Walltime}{Avg}{30.56928133699901633333333333}%
\StoreBenchExecResult{ExperimentUautomizerVerificationMemcleanupTransformed}{MemoryCleanupTransformedMemsafetyMemcleanup}{Correct}{}{Walltime}{Median}{14.20083991499996}%
\StoreBenchExecResult{ExperimentUautomizerVerificationMemcleanupTransformed}{MemoryCleanupTransformedMemsafetyMemcleanup}{Correct}{}{Walltime}{Min}{12.158048999997845}%
\StoreBenchExecResult{ExperimentUautomizerVerificationMemcleanupTransformed}{MemoryCleanupTransformedMemsafetyMemcleanup}{Correct}{}{Walltime}{Max}{191.24272116399516}%
\StoreBenchExecResult{ExperimentUautomizerVerificationMemcleanupTransformed}{MemoryCleanupTransformedMemsafetyMemcleanup}{Correct}{}{Walltime}{Stdev}{39.28742642692172245439227231}%
\StoreBenchExecResult{ExperimentUautomizerVerificationMemcleanupTransformed}{MemoryCleanupTransformedMemsafetyMemcleanup}{Correct}{False}{Count}{}{21}%
\StoreBenchExecResult{ExperimentUautomizerVerificationMemcleanupTransformed}{MemoryCleanupTransformedMemsafetyMemcleanup}{Correct}{False}{Cputime}{}{961.309961282}%
\StoreBenchExecResult{ExperimentUautomizerVerificationMemcleanupTransformed}{MemoryCleanupTransformedMemsafetyMemcleanup}{Correct}{False}{Cputime}{Avg}{45.77666482295238095238095238}%
\StoreBenchExecResult{ExperimentUautomizerVerificationMemcleanupTransformed}{MemoryCleanupTransformedMemsafetyMemcleanup}{Correct}{False}{Cputime}{Median}{24.106825283}%
\StoreBenchExecResult{ExperimentUautomizerVerificationMemcleanupTransformed}{MemoryCleanupTransformedMemsafetyMemcleanup}{Correct}{False}{Cputime}{Min}{20.756152144}%
\StoreBenchExecResult{ExperimentUautomizerVerificationMemcleanupTransformed}{MemoryCleanupTransformedMemsafetyMemcleanup}{Correct}{False}{Cputime}{Max}{212.786169979}%
\StoreBenchExecResult{ExperimentUautomizerVerificationMemcleanupTransformed}{MemoryCleanupTransformedMemsafetyMemcleanup}{Correct}{False}{Cputime}{Stdev}{43.69918574900474103226236398}%
\StoreBenchExecResult{ExperimentUautomizerVerificationMemcleanupTransformed}{MemoryCleanupTransformedMemsafetyMemcleanup}{Correct}{False}{Walltime}{}{641.954908076979343}%
\StoreBenchExecResult{ExperimentUautomizerVerificationMemcleanupTransformed}{MemoryCleanupTransformedMemsafetyMemcleanup}{Correct}{False}{Walltime}{Avg}{30.56928133699901633333333333}%
\StoreBenchExecResult{ExperimentUautomizerVerificationMemcleanupTransformed}{MemoryCleanupTransformedMemsafetyMemcleanup}{Correct}{False}{Walltime}{Median}{14.20083991499996}%
\StoreBenchExecResult{ExperimentUautomizerVerificationMemcleanupTransformed}{MemoryCleanupTransformedMemsafetyMemcleanup}{Correct}{False}{Walltime}{Min}{12.158048999997845}%
\StoreBenchExecResult{ExperimentUautomizerVerificationMemcleanupTransformed}{MemoryCleanupTransformedMemsafetyMemcleanup}{Correct}{False}{Walltime}{Max}{191.24272116399516}%
\StoreBenchExecResult{ExperimentUautomizerVerificationMemcleanupTransformed}{MemoryCleanupTransformedMemsafetyMemcleanup}{Correct}{False}{Walltime}{Stdev}{39.28742642692172245439227231}%

\StoreBenchExecResult{ExperimentUautomizerVerificationMemcleanupTransformed}{MemoryCleanupTransformedMemsafetyMemcleanup}{Error}{}{Count}{}{11}%
\StoreBenchExecResult{ExperimentUautomizerVerificationMemcleanupTransformed}{MemoryCleanupTransformedMemsafetyMemcleanup}{Error}{}{Cputime}{}{10567.515816837}%
\StoreBenchExecResult{ExperimentUautomizerVerificationMemcleanupTransformed}{MemoryCleanupTransformedMemsafetyMemcleanup}{Error}{}{Cputime}{Avg}{960.6832560760909090909090909}%
\StoreBenchExecResult{ExperimentUautomizerVerificationMemcleanupTransformed}{MemoryCleanupTransformedMemsafetyMemcleanup}{Error}{}{Cputime}{Median}{960.628873787}%
\StoreBenchExecResult{ExperimentUautomizerVerificationMemcleanupTransformed}{MemoryCleanupTransformedMemsafetyMemcleanup}{Error}{}{Cputime}{Min}{960.226767862}%
\StoreBenchExecResult{ExperimentUautomizerVerificationMemcleanupTransformed}{MemoryCleanupTransformedMemsafetyMemcleanup}{Error}{}{Cputime}{Max}{961.204997659}%
\StoreBenchExecResult{ExperimentUautomizerVerificationMemcleanupTransformed}{MemoryCleanupTransformedMemsafetyMemcleanup}{Error}{}{Cputime}{Stdev}{0.2959875190547794185071214373}%
\StoreBenchExecResult{ExperimentUautomizerVerificationMemcleanupTransformed}{MemoryCleanupTransformedMemsafetyMemcleanup}{Error}{}{Walltime}{}{9386.0918756399844}%
\StoreBenchExecResult{ExperimentUautomizerVerificationMemcleanupTransformed}{MemoryCleanupTransformedMemsafetyMemcleanup}{Error}{}{Walltime}{Avg}{853.2810796036349454545454545}%
\StoreBenchExecResult{ExperimentUautomizerVerificationMemcleanupTransformed}{MemoryCleanupTransformedMemsafetyMemcleanup}{Error}{}{Walltime}{Median}{843.7804206229994}%
\StoreBenchExecResult{ExperimentUautomizerVerificationMemcleanupTransformed}{MemoryCleanupTransformedMemsafetyMemcleanup}{Error}{}{Walltime}{Min}{827.6369727940037}%
\StoreBenchExecResult{ExperimentUautomizerVerificationMemcleanupTransformed}{MemoryCleanupTransformedMemsafetyMemcleanup}{Error}{}{Walltime}{Max}{893.3044641550005}%
\StoreBenchExecResult{ExperimentUautomizerVerificationMemcleanupTransformed}{MemoryCleanupTransformedMemsafetyMemcleanup}{Error}{}{Walltime}{Stdev}{24.13827267221679019625167519}%
\StoreBenchExecResult{ExperimentUautomizerVerificationMemcleanupTransformed}{MemoryCleanupTransformedMemsafetyMemcleanup}{Error}{Timeout}{Count}{}{11}%
\StoreBenchExecResult{ExperimentUautomizerVerificationMemcleanupTransformed}{MemoryCleanupTransformedMemsafetyMemcleanup}{Error}{Timeout}{Cputime}{}{10567.515816837}%
\StoreBenchExecResult{ExperimentUautomizerVerificationMemcleanupTransformed}{MemoryCleanupTransformedMemsafetyMemcleanup}{Error}{Timeout}{Cputime}{Avg}{960.6832560760909090909090909}%
\StoreBenchExecResult{ExperimentUautomizerVerificationMemcleanupTransformed}{MemoryCleanupTransformedMemsafetyMemcleanup}{Error}{Timeout}{Cputime}{Median}{960.628873787}%
\StoreBenchExecResult{ExperimentUautomizerVerificationMemcleanupTransformed}{MemoryCleanupTransformedMemsafetyMemcleanup}{Error}{Timeout}{Cputime}{Min}{960.226767862}%
\StoreBenchExecResult{ExperimentUautomizerVerificationMemcleanupTransformed}{MemoryCleanupTransformedMemsafetyMemcleanup}{Error}{Timeout}{Cputime}{Max}{961.204997659}%
\StoreBenchExecResult{ExperimentUautomizerVerificationMemcleanupTransformed}{MemoryCleanupTransformedMemsafetyMemcleanup}{Error}{Timeout}{Cputime}{Stdev}{0.2959875190547794185071214373}%
\StoreBenchExecResult{ExperimentUautomizerVerificationMemcleanupTransformed}{MemoryCleanupTransformedMemsafetyMemcleanup}{Error}{Timeout}{Walltime}{}{9386.0918756399844}%
\StoreBenchExecResult{ExperimentUautomizerVerificationMemcleanupTransformed}{MemoryCleanupTransformedMemsafetyMemcleanup}{Error}{Timeout}{Walltime}{Avg}{853.2810796036349454545454545}%
\StoreBenchExecResult{ExperimentUautomizerVerificationMemcleanupTransformed}{MemoryCleanupTransformedMemsafetyMemcleanup}{Error}{Timeout}{Walltime}{Median}{843.7804206229994}%
\StoreBenchExecResult{ExperimentUautomizerVerificationMemcleanupTransformed}{MemoryCleanupTransformedMemsafetyMemcleanup}{Error}{Timeout}{Walltime}{Min}{827.6369727940037}%
\StoreBenchExecResult{ExperimentUautomizerVerificationMemcleanupTransformed}{MemoryCleanupTransformedMemsafetyMemcleanup}{Error}{Timeout}{Walltime}{Max}{893.3044641550005}%
\StoreBenchExecResult{ExperimentUautomizerVerificationMemcleanupTransformed}{MemoryCleanupTransformedMemsafetyMemcleanup}{Error}{Timeout}{Walltime}{Stdev}{24.13827267221679019625167519}%
\StoreBenchExecResult{ExperimentUautomizerVerificationMemcleanupTransformed}{MemoryCleanupTransformedMemsafetyMemcleanup}{Unknown}{}{Count}{}{7}%
\StoreBenchExecResult{ExperimentUautomizerVerificationMemcleanupTransformed}{MemoryCleanupTransformedMemsafetyMemcleanup}{Unknown}{}{Cputime}{}{407.771811847}%
\StoreBenchExecResult{ExperimentUautomizerVerificationMemcleanupTransformed}{MemoryCleanupTransformedMemsafetyMemcleanup}{Unknown}{}{Cputime}{Avg}{58.25311597814285714285714286}%
\StoreBenchExecResult{ExperimentUautomizerVerificationMemcleanupTransformed}{MemoryCleanupTransformedMemsafetyMemcleanup}{Unknown}{}{Cputime}{Median}{33.573158688}%
\StoreBenchExecResult{ExperimentUautomizerVerificationMemcleanupTransformed}{MemoryCleanupTransformedMemsafetyMemcleanup}{Unknown}{}{Cputime}{Min}{26.150028761}%
\StoreBenchExecResult{ExperimentUautomizerVerificationMemcleanupTransformed}{MemoryCleanupTransformedMemsafetyMemcleanup}{Unknown}{}{Cputime}{Max}{130.575928168}%
\StoreBenchExecResult{ExperimentUautomizerVerificationMemcleanupTransformed}{MemoryCleanupTransformedMemsafetyMemcleanup}{Unknown}{}{Cputime}{Stdev}{43.03115410928964140595053712}%
\StoreBenchExecResult{ExperimentUautomizerVerificationMemcleanupTransformed}{MemoryCleanupTransformedMemsafetyMemcleanup}{Unknown}{}{Walltime}{}{249.451663079002181}%
\StoreBenchExecResult{ExperimentUautomizerVerificationMemcleanupTransformed}{MemoryCleanupTransformedMemsafetyMemcleanup}{Unknown}{}{Walltime}{Avg}{35.635951868428883}%
\StoreBenchExecResult{ExperimentUautomizerVerificationMemcleanupTransformed}{MemoryCleanupTransformedMemsafetyMemcleanup}{Unknown}{}{Walltime}{Median}{19.624667734002287}%
\StoreBenchExecResult{ExperimentUautomizerVerificationMemcleanupTransformed}{MemoryCleanupTransformedMemsafetyMemcleanup}{Unknown}{}{Walltime}{Min}{15.504505842996878}%
\StoreBenchExecResult{ExperimentUautomizerVerificationMemcleanupTransformed}{MemoryCleanupTransformedMemsafetyMemcleanup}{Unknown}{}{Walltime}{Max}{79.0845748530046}%
\StoreBenchExecResult{ExperimentUautomizerVerificationMemcleanupTransformed}{MemoryCleanupTransformedMemsafetyMemcleanup}{Unknown}{}{Walltime}{Stdev}{27.51220702377354684547211352}%
\StoreBenchExecResult{ExperimentUautomizerVerificationMemcleanupTransformed}{MemoryCleanupTransformedMemsafetyMemcleanup}{Unknown}{Unknown}{Count}{}{7}%
\StoreBenchExecResult{ExperimentUautomizerVerificationMemcleanupTransformed}{MemoryCleanupTransformedMemsafetyMemcleanup}{Unknown}{Unknown}{Cputime}{}{407.771811847}%
\StoreBenchExecResult{ExperimentUautomizerVerificationMemcleanupTransformed}{MemoryCleanupTransformedMemsafetyMemcleanup}{Unknown}{Unknown}{Cputime}{Avg}{58.25311597814285714285714286}%
\StoreBenchExecResult{ExperimentUautomizerVerificationMemcleanupTransformed}{MemoryCleanupTransformedMemsafetyMemcleanup}{Unknown}{Unknown}{Cputime}{Median}{33.573158688}%
\StoreBenchExecResult{ExperimentUautomizerVerificationMemcleanupTransformed}{MemoryCleanupTransformedMemsafetyMemcleanup}{Unknown}{Unknown}{Cputime}{Min}{26.150028761}%
\StoreBenchExecResult{ExperimentUautomizerVerificationMemcleanupTransformed}{MemoryCleanupTransformedMemsafetyMemcleanup}{Unknown}{Unknown}{Cputime}{Max}{130.575928168}%
\StoreBenchExecResult{ExperimentUautomizerVerificationMemcleanupTransformed}{MemoryCleanupTransformedMemsafetyMemcleanup}{Unknown}{Unknown}{Cputime}{Stdev}{43.03115410928964140595053712}%
\StoreBenchExecResult{ExperimentUautomizerVerificationMemcleanupTransformed}{MemoryCleanupTransformedMemsafetyMemcleanup}{Unknown}{Unknown}{Walltime}{}{249.451663079002181}%
\StoreBenchExecResult{ExperimentUautomizerVerificationMemcleanupTransformed}{MemoryCleanupTransformedMemsafetyMemcleanup}{Unknown}{Unknown}{Walltime}{Avg}{35.635951868428883}%
\StoreBenchExecResult{ExperimentUautomizerVerificationMemcleanupTransformed}{MemoryCleanupTransformedMemsafetyMemcleanup}{Unknown}{Unknown}{Walltime}{Median}{19.624667734002287}%
\StoreBenchExecResult{ExperimentUautomizerVerificationMemcleanupTransformed}{MemoryCleanupTransformedMemsafetyMemcleanup}{Unknown}{Unknown}{Walltime}{Min}{15.504505842996878}%
\StoreBenchExecResult{ExperimentUautomizerVerificationMemcleanupTransformed}{MemoryCleanupTransformedMemsafetyMemcleanup}{Unknown}{Unknown}{Walltime}{Max}{79.0845748530046}%
\StoreBenchExecResult{ExperimentUautomizerVerificationMemcleanupTransformed}{MemoryCleanupTransformedMemsafetyMemcleanup}{Unknown}{Unknown}{Walltime}{Stdev}{27.51220702377354684547211352}%
\StoreBenchExecResult{ExperimentUautomizerVerificationMemcleanupTransformed}{MemoryCleanupTransformedMemsafetyMemcleanup}{Wrong}{}{Count}{}{2}%
\StoreBenchExecResult{ExperimentUautomizerVerificationMemcleanupTransformed}{MemoryCleanupTransformedMemsafetyMemcleanup}{Wrong}{}{Cputime}{}{57.386833347}%
\StoreBenchExecResult{ExperimentUautomizerVerificationMemcleanupTransformed}{MemoryCleanupTransformedMemsafetyMemcleanup}{Wrong}{}{Cputime}{Avg}{28.6934166735}%
\StoreBenchExecResult{ExperimentUautomizerVerificationMemcleanupTransformed}{MemoryCleanupTransformedMemsafetyMemcleanup}{Wrong}{}{Cputime}{Median}{28.6934166735}%
\StoreBenchExecResult{ExperimentUautomizerVerificationMemcleanupTransformed}{MemoryCleanupTransformedMemsafetyMemcleanup}{Wrong}{}{Cputime}{Min}{27.943680645}%
\StoreBenchExecResult{ExperimentUautomizerVerificationMemcleanupTransformed}{MemoryCleanupTransformedMemsafetyMemcleanup}{Wrong}{}{Cputime}{Max}{29.443152702}%
\StoreBenchExecResult{ExperimentUautomizerVerificationMemcleanupTransformed}{MemoryCleanupTransformedMemsafetyMemcleanup}{Wrong}{}{Cputime}{Stdev}{0.74973602850000}%
\StoreBenchExecResult{ExperimentUautomizerVerificationMemcleanupTransformed}{MemoryCleanupTransformedMemsafetyMemcleanup}{Wrong}{}{Walltime}{}{32.999204665000433}%
\StoreBenchExecResult{ExperimentUautomizerVerificationMemcleanupTransformed}{MemoryCleanupTransformedMemsafetyMemcleanup}{Wrong}{}{Walltime}{Avg}{16.4996023325002165}%
\StoreBenchExecResult{ExperimentUautomizerVerificationMemcleanupTransformed}{MemoryCleanupTransformedMemsafetyMemcleanup}{Wrong}{}{Walltime}{Median}{16.4996023325002165}%
\StoreBenchExecResult{ExperimentUautomizerVerificationMemcleanupTransformed}{MemoryCleanupTransformedMemsafetyMemcleanup}{Wrong}{}{Walltime}{Min}{16.105498122997233}%
\StoreBenchExecResult{ExperimentUautomizerVerificationMemcleanupTransformed}{MemoryCleanupTransformedMemsafetyMemcleanup}{Wrong}{}{Walltime}{Max}{16.8937065420032}%
\StoreBenchExecResult{ExperimentUautomizerVerificationMemcleanupTransformed}{MemoryCleanupTransformedMemsafetyMemcleanup}{Wrong}{}{Walltime}{Stdev}{0.3941042095029835000000000000}%
\StoreBenchExecResult{ExperimentUautomizerVerificationMemcleanupTransformed}{MemoryCleanupTransformedMemsafetyMemcleanup}{Wrong}{True}{Count}{}{2}%
\StoreBenchExecResult{ExperimentUautomizerVerificationMemcleanupTransformed}{MemoryCleanupTransformedMemsafetyMemcleanup}{Wrong}{True}{Cputime}{}{57.386833347}%
\StoreBenchExecResult{ExperimentUautomizerVerificationMemcleanupTransformed}{MemoryCleanupTransformedMemsafetyMemcleanup}{Wrong}{True}{Cputime}{Avg}{28.6934166735}%
\StoreBenchExecResult{ExperimentUautomizerVerificationMemcleanupTransformed}{MemoryCleanupTransformedMemsafetyMemcleanup}{Wrong}{True}{Cputime}{Median}{28.6934166735}%
\StoreBenchExecResult{ExperimentUautomizerVerificationMemcleanupTransformed}{MemoryCleanupTransformedMemsafetyMemcleanup}{Wrong}{True}{Cputime}{Min}{27.943680645}%
\StoreBenchExecResult{ExperimentUautomizerVerificationMemcleanupTransformed}{MemoryCleanupTransformedMemsafetyMemcleanup}{Wrong}{True}{Cputime}{Max}{29.443152702}%
\StoreBenchExecResult{ExperimentUautomizerVerificationMemcleanupTransformed}{MemoryCleanupTransformedMemsafetyMemcleanup}{Wrong}{True}{Cputime}{Stdev}{0.74973602850000}%
\StoreBenchExecResult{ExperimentUautomizerVerificationMemcleanupTransformed}{MemoryCleanupTransformedMemsafetyMemcleanup}{Wrong}{True}{Walltime}{}{32.999204665000433}%
\StoreBenchExecResult{ExperimentUautomizerVerificationMemcleanupTransformed}{MemoryCleanupTransformedMemsafetyMemcleanup}{Wrong}{True}{Walltime}{Avg}{16.4996023325002165}%
\StoreBenchExecResult{ExperimentUautomizerVerificationMemcleanupTransformed}{MemoryCleanupTransformedMemsafetyMemcleanup}{Wrong}{True}{Walltime}{Median}{16.4996023325002165}%
\StoreBenchExecResult{ExperimentUautomizerVerificationMemcleanupTransformed}{MemoryCleanupTransformedMemsafetyMemcleanup}{Wrong}{True}{Walltime}{Min}{16.105498122997233}%
\StoreBenchExecResult{ExperimentUautomizerVerificationMemcleanupTransformed}{MemoryCleanupTransformedMemsafetyMemcleanup}{Wrong}{True}{Walltime}{Max}{16.8937065420032}%
\StoreBenchExecResult{ExperimentUautomizerVerificationMemcleanupTransformed}{MemoryCleanupTransformedMemsafetyMemcleanup}{Wrong}{True}{Walltime}{Stdev}{0.3941042095029835000000000000}%

\ifdefined\UltimateUltimateTotalCount\else\edef\UltimateUltimateTotalCount{0}\fi
\ifdefined\UltimateUltimateCorrectCount\else\edef\UltimateUltimateCorrectCount{0}\fi
\ifdefined\UltimateUltimateCorrectTrueCount\else\edef\UltimateUltimateCorrectTrueCount{0}\fi
\ifdefined\UltimateUltimateCorrectFalseCount\else\edef\UltimateUltimateCorrectFalseCount{0}\fi
\ifdefined\UltimateUltimateWrongCount\else\edef\UltimateUltimateWrongCount{0}\fi
\ifdefined\UltimateUltimateWrongTrueCount\else\edef\UltimateUltimateWrongTrueCount{0}\fi
\ifdefined\UltimateUltimateWrongFalseCount\else\edef\UltimateUltimateWrongFalseCount{0}\fi
\ifdefined\UltimateUltimateErrorTimeoutCount\else\edef\UltimateUltimateErrorTimeoutCount{0}\fi
\ifdefined\UltimateUltimateErrorOutOfMemoryCount\else\edef\UltimateUltimateErrorOutOfMemoryCount{0}\fi
\ifdefined\UltimateUltimateCorrectCputime\else\edef\UltimateUltimateCorrectCputime{0}\fi
\ifdefined\UltimateUltimateCorrectCputimeAvg\else\edef\UltimateUltimateCorrectCputimeAvg{None}\fi
\ifdefined\UltimateUltimateCorrectWalltime\else\edef\UltimateUltimateCorrectWalltime{0}\fi
\ifdefined\UltimateUltimateCorrectWalltimeAvg\else\edef\UltimateUltimateCorrectWalltimeAvg{None}\fi
\edef\UltimateUltimateErrorOtherInconclusiveCount{\the\numexpr \UltimateUltimateTotalCount - \UltimateUltimateCorrectCount - \UltimateUltimateWrongTrueCount - \UltimateUltimateWrongFalseCount - \UltimateUltimateErrorTimeoutCount - \UltimateUltimateErrorOutOfMemoryCount \relax}
\providecommand\StoreBenchExecResult[7]{\expandafter\newcommand\csname#1#2#3#4#5#6\endcsname{#7}}%
\StoreBenchExecResult{ExperimentCbmcVerificationMemcleanupTransformed}{MemoryCleanupTransformedMemsafetyMemcleanup}{Total}{}{Count}{}{41}%
\StoreBenchExecResult{ExperimentCbmcVerificationMemcleanupTransformed}{MemoryCleanupTransformedMemsafetyMemcleanup}{Total}{}{Cputime}{}{2815.119269269}%
\StoreBenchExecResult{ExperimentCbmcVerificationMemcleanupTransformed}{MemoryCleanupTransformedMemsafetyMemcleanup}{Total}{}{Cputime}{Avg}{68.66144559192682926829268293}%
\StoreBenchExecResult{ExperimentCbmcVerificationMemcleanupTransformed}{MemoryCleanupTransformedMemsafetyMemcleanup}{Total}{}{Cputime}{Median}{1.483218297}%
\StoreBenchExecResult{ExperimentCbmcVerificationMemcleanupTransformed}{MemoryCleanupTransformedMemsafetyMemcleanup}{Total}{}{Cputime}{Min}{0.099712325}%
\StoreBenchExecResult{ExperimentCbmcVerificationMemcleanupTransformed}{MemoryCleanupTransformedMemsafetyMemcleanup}{Total}{}{Cputime}{Max}{874.661650733}%
\StoreBenchExecResult{ExperimentCbmcVerificationMemcleanupTransformed}{MemoryCleanupTransformedMemsafetyMemcleanup}{Total}{}{Cputime}{Stdev}{185.4284960994333307548928228}%
\StoreBenchExecResult{ExperimentCbmcVerificationMemcleanupTransformed}{MemoryCleanupTransformedMemsafetyMemcleanup}{Total}{}{Walltime}{}{2814.73385073903050444}%
\StoreBenchExecResult{ExperimentCbmcVerificationMemcleanupTransformed}{MemoryCleanupTransformedMemsafetyMemcleanup}{Total}{}{Walltime}{Avg}{68.65204513997635376682926829}%
\StoreBenchExecResult{ExperimentCbmcVerificationMemcleanupTransformed}{MemoryCleanupTransformedMemsafetyMemcleanup}{Total}{}{Walltime}{Median}{1.4607806950007216}%
\StoreBenchExecResult{ExperimentCbmcVerificationMemcleanupTransformed}{MemoryCleanupTransformedMemsafetyMemcleanup}{Total}{}{Walltime}{Min}{0.10684325100010028}%
\StoreBenchExecResult{ExperimentCbmcVerificationMemcleanupTransformed}{MemoryCleanupTransformedMemsafetyMemcleanup}{Total}{}{Walltime}{Max}{875.0819925460019}%
\StoreBenchExecResult{ExperimentCbmcVerificationMemcleanupTransformed}{MemoryCleanupTransformedMemsafetyMemcleanup}{Total}{}{Walltime}{Stdev}{185.4381035186657919075218882}%
\StoreBenchExecResult{ExperimentCbmcVerificationMemcleanupTransformed}{MemoryCleanupTransformedMemsafetyMemcleanup}{Correct}{}{Count}{}{36}%
\StoreBenchExecResult{ExperimentCbmcVerificationMemcleanupTransformed}{MemoryCleanupTransformedMemsafetyMemcleanup}{Correct}{}{Cputime}{}{1266.778373661}%
\StoreBenchExecResult{ExperimentCbmcVerificationMemcleanupTransformed}{MemoryCleanupTransformedMemsafetyMemcleanup}{Correct}{}{Cputime}{Avg}{35.18828815725}%
\StoreBenchExecResult{ExperimentCbmcVerificationMemcleanupTransformed}{MemoryCleanupTransformedMemsafetyMemcleanup}{Correct}{}{Cputime}{Median}{1.5820880065}%
\StoreBenchExecResult{ExperimentCbmcVerificationMemcleanupTransformed}{MemoryCleanupTransformedMemsafetyMemcleanup}{Correct}{}{Cputime}{Min}{0.338664537}%
\StoreBenchExecResult{ExperimentCbmcVerificationMemcleanupTransformed}{MemoryCleanupTransformedMemsafetyMemcleanup}{Correct}{}{Cputime}{Max}{360.331329191}%
\StoreBenchExecResult{ExperimentCbmcVerificationMemcleanupTransformed}{MemoryCleanupTransformedMemsafetyMemcleanup}{Correct}{}{Cputime}{Stdev}{97.43851086656440292763804904}%
\StoreBenchExecResult{ExperimentCbmcVerificationMemcleanupTransformed}{MemoryCleanupTransformedMemsafetyMemcleanup}{Correct}{}{Walltime}{}{1266.71281217002251456}%
\StoreBenchExecResult{ExperimentCbmcVerificationMemcleanupTransformed}{MemoryCleanupTransformedMemsafetyMemcleanup}{Correct}{}{Walltime}{Avg}{35.18646700472284762666666667}%
\StoreBenchExecResult{ExperimentCbmcVerificationMemcleanupTransformed}{MemoryCleanupTransformedMemsafetyMemcleanup}{Correct}{}{Walltime}{Median}{1.56600754050305115}%
\StoreBenchExecResult{ExperimentCbmcVerificationMemcleanupTransformed}{MemoryCleanupTransformedMemsafetyMemcleanup}{Correct}{}{Walltime}{Min}{0.319393484001921}%
\StoreBenchExecResult{ExperimentCbmcVerificationMemcleanupTransformed}{MemoryCleanupTransformedMemsafetyMemcleanup}{Correct}{}{Walltime}{Max}{360.32384715900116}%
\StoreBenchExecResult{ExperimentCbmcVerificationMemcleanupTransformed}{MemoryCleanupTransformedMemsafetyMemcleanup}{Correct}{}{Walltime}{Stdev}{97.48498751390924711497528166}%
\StoreBenchExecResult{ExperimentCbmcVerificationMemcleanupTransformed}{MemoryCleanupTransformedMemsafetyMemcleanup}{Correct}{False}{Count}{}{35}%
\StoreBenchExecResult{ExperimentCbmcVerificationMemcleanupTransformed}{MemoryCleanupTransformedMemsafetyMemcleanup}{Correct}{False}{Cputime}{}{1266.272495489}%
\StoreBenchExecResult{ExperimentCbmcVerificationMemcleanupTransformed}{MemoryCleanupTransformedMemsafetyMemcleanup}{Correct}{False}{Cputime}{Avg}{36.17921415682857142857142857}%
\StoreBenchExecResult{ExperimentCbmcVerificationMemcleanupTransformed}{MemoryCleanupTransformedMemsafetyMemcleanup}{Correct}{False}{Cputime}{Median}{1.680957716}%
\StoreBenchExecResult{ExperimentCbmcVerificationMemcleanupTransformed}{MemoryCleanupTransformedMemsafetyMemcleanup}{Correct}{False}{Cputime}{Min}{0.338664537}%
\StoreBenchExecResult{ExperimentCbmcVerificationMemcleanupTransformed}{MemoryCleanupTransformedMemsafetyMemcleanup}{Correct}{False}{Cputime}{Max}{360.331329191}%
\StoreBenchExecResult{ExperimentCbmcVerificationMemcleanupTransformed}{MemoryCleanupTransformedMemsafetyMemcleanup}{Correct}{False}{Cputime}{Stdev}{98.64166681652793837972447175}%
\StoreBenchExecResult{ExperimentCbmcVerificationMemcleanupTransformed}{MemoryCleanupTransformedMemsafetyMemcleanup}{Correct}{False}{Walltime}{}{1266.21907144002761730}%
\StoreBenchExecResult{ExperimentCbmcVerificationMemcleanupTransformed}{MemoryCleanupTransformedMemsafetyMemcleanup}{Correct}{False}{Walltime}{Avg}{36.17768775542936049428571429}%
\StoreBenchExecResult{ExperimentCbmcVerificationMemcleanupTransformed}{MemoryCleanupTransformedMemsafetyMemcleanup}{Correct}{False}{Walltime}{Median}{1.6712343860053807}%
\StoreBenchExecResult{ExperimentCbmcVerificationMemcleanupTransformed}{MemoryCleanupTransformedMemsafetyMemcleanup}{Correct}{False}{Walltime}{Min}{0.319393484001921}%
\StoreBenchExecResult{ExperimentCbmcVerificationMemcleanupTransformed}{MemoryCleanupTransformedMemsafetyMemcleanup}{Correct}{False}{Walltime}{Max}{360.32384715900116}%
\StoreBenchExecResult{ExperimentCbmcVerificationMemcleanupTransformed}{MemoryCleanupTransformedMemsafetyMemcleanup}{Correct}{False}{Walltime}{Stdev}{98.68878168328356730364533982}%
\StoreBenchExecResult{ExperimentCbmcVerificationMemcleanupTransformed}{MemoryCleanupTransformedMemsafetyMemcleanup}{Correct}{True}{Count}{}{1}%
\StoreBenchExecResult{ExperimentCbmcVerificationMemcleanupTransformed}{MemoryCleanupTransformedMemsafetyMemcleanup}{Correct}{True}{Cputime}{}{0.505878172}%
\StoreBenchExecResult{ExperimentCbmcVerificationMemcleanupTransformed}{MemoryCleanupTransformedMemsafetyMemcleanup}{Correct}{True}{Cputime}{Avg}{0.505878172}%
\StoreBenchExecResult{ExperimentCbmcVerificationMemcleanupTransformed}{MemoryCleanupTransformedMemsafetyMemcleanup}{Correct}{True}{Cputime}{Median}{0.505878172}%
\StoreBenchExecResult{ExperimentCbmcVerificationMemcleanupTransformed}{MemoryCleanupTransformedMemsafetyMemcleanup}{Correct}{True}{Cputime}{Min}{0.505878172}%
\StoreBenchExecResult{ExperimentCbmcVerificationMemcleanupTransformed}{MemoryCleanupTransformedMemsafetyMemcleanup}{Correct}{True}{Cputime}{Max}{0.505878172}%
\StoreBenchExecResult{ExperimentCbmcVerificationMemcleanupTransformed}{MemoryCleanupTransformedMemsafetyMemcleanup}{Correct}{True}{Cputime}{Stdev}{0E-14}%
\StoreBenchExecResult{ExperimentCbmcVerificationMemcleanupTransformed}{MemoryCleanupTransformedMemsafetyMemcleanup}{Correct}{True}{Walltime}{}{0.49374072999489726}%
\StoreBenchExecResult{ExperimentCbmcVerificationMemcleanupTransformed}{MemoryCleanupTransformedMemsafetyMemcleanup}{Correct}{True}{Walltime}{Avg}{0.49374072999489726}%
\StoreBenchExecResult{ExperimentCbmcVerificationMemcleanupTransformed}{MemoryCleanupTransformedMemsafetyMemcleanup}{Correct}{True}{Walltime}{Median}{0.49374072999489726}%
\StoreBenchExecResult{ExperimentCbmcVerificationMemcleanupTransformed}{MemoryCleanupTransformedMemsafetyMemcleanup}{Correct}{True}{Walltime}{Min}{0.49374072999489726}%
\StoreBenchExecResult{ExperimentCbmcVerificationMemcleanupTransformed}{MemoryCleanupTransformedMemsafetyMemcleanup}{Correct}{True}{Walltime}{Max}{0.49374072999489726}%
\StoreBenchExecResult{ExperimentCbmcVerificationMemcleanupTransformed}{MemoryCleanupTransformedMemsafetyMemcleanup}{Correct}{True}{Walltime}{Stdev}{0E-17}%
\StoreBenchExecResult{ExperimentCbmcVerificationMemcleanupTransformed}{MemoryCleanupTransformedMemsafetyMemcleanup}{Error}{}{Count}{}{3}%
\StoreBenchExecResult{ExperimentCbmcVerificationMemcleanupTransformed}{MemoryCleanupTransformedMemsafetyMemcleanup}{Error}{}{Cputime}{}{1546.787498296}%
\StoreBenchExecResult{ExperimentCbmcVerificationMemcleanupTransformed}{MemoryCleanupTransformedMemsafetyMemcleanup}{Error}{}{Cputime}{Avg}{515.5958327653333333333333333}%
\StoreBenchExecResult{ExperimentCbmcVerificationMemcleanupTransformed}{MemoryCleanupTransformedMemsafetyMemcleanup}{Error}{}{Cputime}{Median}{672.026135238}%
\StoreBenchExecResult{ExperimentCbmcVerificationMemcleanupTransformed}{MemoryCleanupTransformedMemsafetyMemcleanup}{Error}{}{Cputime}{Min}{0.099712325}%
\StoreBenchExecResult{ExperimentCbmcVerificationMemcleanupTransformed}{MemoryCleanupTransformedMemsafetyMemcleanup}{Error}{}{Cputime}{Max}{874.661650733}%
\StoreBenchExecResult{ExperimentCbmcVerificationMemcleanupTransformed}{MemoryCleanupTransformedMemsafetyMemcleanup}{Error}{}{Cputime}{Stdev}{373.7802167734019990524327761}%
\StoreBenchExecResult{ExperimentCbmcVerificationMemcleanupTransformed}{MemoryCleanupTransformedMemsafetyMemcleanup}{Error}{}{Walltime}{}{1546.49749091700510028}%
\StoreBenchExecResult{ExperimentCbmcVerificationMemcleanupTransformed}{MemoryCleanupTransformedMemsafetyMemcleanup}{Error}{}{Walltime}{Avg}{515.4991636390017000933333333}%
\StoreBenchExecResult{ExperimentCbmcVerificationMemcleanupTransformed}{MemoryCleanupTransformedMemsafetyMemcleanup}{Error}{}{Walltime}{Median}{671.3086551200031}%
\StoreBenchExecResult{ExperimentCbmcVerificationMemcleanupTransformed}{MemoryCleanupTransformedMemsafetyMemcleanup}{Error}{}{Walltime}{Min}{0.10684325100010028}%
\StoreBenchExecResult{ExperimentCbmcVerificationMemcleanupTransformed}{MemoryCleanupTransformedMemsafetyMemcleanup}{Error}{}{Walltime}{Max}{875.0819925460019}%
\StoreBenchExecResult{ExperimentCbmcVerificationMemcleanupTransformed}{MemoryCleanupTransformedMemsafetyMemcleanup}{Error}{}{Walltime}{Stdev}{373.8117406731440602363543007}%
\StoreBenchExecResult{ExperimentCbmcVerificationMemcleanupTransformed}{MemoryCleanupTransformedMemsafetyMemcleanup}{Error}{Error}{Count}{}{2}%
\StoreBenchExecResult{ExperimentCbmcVerificationMemcleanupTransformed}{MemoryCleanupTransformedMemsafetyMemcleanup}{Error}{Error}{Cputime}{}{874.761363058}%
\StoreBenchExecResult{ExperimentCbmcVerificationMemcleanupTransformed}{MemoryCleanupTransformedMemsafetyMemcleanup}{Error}{Error}{Cputime}{Avg}{437.380681529}%
\StoreBenchExecResult{ExperimentCbmcVerificationMemcleanupTransformed}{MemoryCleanupTransformedMemsafetyMemcleanup}{Error}{Error}{Cputime}{Median}{437.380681529}%
\StoreBenchExecResult{ExperimentCbmcVerificationMemcleanupTransformed}{MemoryCleanupTransformedMemsafetyMemcleanup}{Error}{Error}{Cputime}{Min}{0.099712325}%
\StoreBenchExecResult{ExperimentCbmcVerificationMemcleanupTransformed}{MemoryCleanupTransformedMemsafetyMemcleanup}{Error}{Error}{Cputime}{Max}{874.661650733}%
\StoreBenchExecResult{ExperimentCbmcVerificationMemcleanupTransformed}{MemoryCleanupTransformedMemsafetyMemcleanup}{Error}{Error}{Cputime}{Stdev}{437.28096920400}%
\StoreBenchExecResult{ExperimentCbmcVerificationMemcleanupTransformed}{MemoryCleanupTransformedMemsafetyMemcleanup}{Error}{Error}{Walltime}{}{875.18883579700200028}%
\StoreBenchExecResult{ExperimentCbmcVerificationMemcleanupTransformed}{MemoryCleanupTransformedMemsafetyMemcleanup}{Error}{Error}{Walltime}{Avg}{437.59441789850100014}%
\StoreBenchExecResult{ExperimentCbmcVerificationMemcleanupTransformed}{MemoryCleanupTransformedMemsafetyMemcleanup}{Error}{Error}{Walltime}{Median}{437.59441789850100014}%
\StoreBenchExecResult{ExperimentCbmcVerificationMemcleanupTransformed}{MemoryCleanupTransformedMemsafetyMemcleanup}{Error}{Error}{Walltime}{Min}{0.10684325100010028}%
\StoreBenchExecResult{ExperimentCbmcVerificationMemcleanupTransformed}{MemoryCleanupTransformedMemsafetyMemcleanup}{Error}{Error}{Walltime}{Max}{875.0819925460019}%
\StoreBenchExecResult{ExperimentCbmcVerificationMemcleanupTransformed}{MemoryCleanupTransformedMemsafetyMemcleanup}{Error}{Error}{Walltime}{Stdev}{437.4875746475008998600000000}%
\StoreBenchExecResult{ExperimentCbmcVerificationMemcleanupTransformed}{MemoryCleanupTransformedMemsafetyMemcleanup}{Error}{OutOfMemory}{Count}{}{1}%
\StoreBenchExecResult{ExperimentCbmcVerificationMemcleanupTransformed}{MemoryCleanupTransformedMemsafetyMemcleanup}{Error}{OutOfMemory}{Cputime}{}{672.026135238}%
\StoreBenchExecResult{ExperimentCbmcVerificationMemcleanupTransformed}{MemoryCleanupTransformedMemsafetyMemcleanup}{Error}{OutOfMemory}{Cputime}{Avg}{672.026135238}%
\StoreBenchExecResult{ExperimentCbmcVerificationMemcleanupTransformed}{MemoryCleanupTransformedMemsafetyMemcleanup}{Error}{OutOfMemory}{Cputime}{Median}{672.026135238}%
\StoreBenchExecResult{ExperimentCbmcVerificationMemcleanupTransformed}{MemoryCleanupTransformedMemsafetyMemcleanup}{Error}{OutOfMemory}{Cputime}{Min}{672.026135238}%
\StoreBenchExecResult{ExperimentCbmcVerificationMemcleanupTransformed}{MemoryCleanupTransformedMemsafetyMemcleanup}{Error}{OutOfMemory}{Cputime}{Max}{672.026135238}%
\StoreBenchExecResult{ExperimentCbmcVerificationMemcleanupTransformed}{MemoryCleanupTransformedMemsafetyMemcleanup}{Error}{OutOfMemory}{Cputime}{Stdev}{0E-14}%
\StoreBenchExecResult{ExperimentCbmcVerificationMemcleanupTransformed}{MemoryCleanupTransformedMemsafetyMemcleanup}{Error}{OutOfMemory}{Walltime}{}{671.3086551200031}%
\StoreBenchExecResult{ExperimentCbmcVerificationMemcleanupTransformed}{MemoryCleanupTransformedMemsafetyMemcleanup}{Error}{OutOfMemory}{Walltime}{Avg}{671.3086551200031}%
\StoreBenchExecResult{ExperimentCbmcVerificationMemcleanupTransformed}{MemoryCleanupTransformedMemsafetyMemcleanup}{Error}{OutOfMemory}{Walltime}{Median}{671.3086551200031}%
\StoreBenchExecResult{ExperimentCbmcVerificationMemcleanupTransformed}{MemoryCleanupTransformedMemsafetyMemcleanup}{Error}{OutOfMemory}{Walltime}{Min}{671.3086551200031}%
\StoreBenchExecResult{ExperimentCbmcVerificationMemcleanupTransformed}{MemoryCleanupTransformedMemsafetyMemcleanup}{Error}{OutOfMemory}{Walltime}{Max}{671.3086551200031}%
\StoreBenchExecResult{ExperimentCbmcVerificationMemcleanupTransformed}{MemoryCleanupTransformedMemsafetyMemcleanup}{Error}{OutOfMemory}{Walltime}{Stdev}{0E-14}%
\StoreBenchExecResult{ExperimentCbmcVerificationMemcleanupTransformed}{MemoryCleanupTransformedMemsafetyMemcleanup}{Wrong}{}{Count}{}{2}%
\StoreBenchExecResult{ExperimentCbmcVerificationMemcleanupTransformed}{MemoryCleanupTransformedMemsafetyMemcleanup}{Wrong}{}{Cputime}{}{1.553397312}%
\StoreBenchExecResult{ExperimentCbmcVerificationMemcleanupTransformed}{MemoryCleanupTransformedMemsafetyMemcleanup}{Wrong}{}{Cputime}{Avg}{0.776698656}%
\StoreBenchExecResult{ExperimentCbmcVerificationMemcleanupTransformed}{MemoryCleanupTransformedMemsafetyMemcleanup}{Wrong}{}{Cputime}{Median}{0.776698656}%
\StoreBenchExecResult{ExperimentCbmcVerificationMemcleanupTransformed}{MemoryCleanupTransformedMemsafetyMemcleanup}{Wrong}{}{Cputime}{Min}{0.340049299}%
\StoreBenchExecResult{ExperimentCbmcVerificationMemcleanupTransformed}{MemoryCleanupTransformedMemsafetyMemcleanup}{Wrong}{}{Cputime}{Max}{1.213348013}%
\StoreBenchExecResult{ExperimentCbmcVerificationMemcleanupTransformed}{MemoryCleanupTransformedMemsafetyMemcleanup}{Wrong}{}{Cputime}{Stdev}{0.43664935700000}%
\StoreBenchExecResult{ExperimentCbmcVerificationMemcleanupTransformed}{MemoryCleanupTransformedMemsafetyMemcleanup}{Wrong}{}{Walltime}{}{1.5235476520028896}%
\StoreBenchExecResult{ExperimentCbmcVerificationMemcleanupTransformed}{MemoryCleanupTransformedMemsafetyMemcleanup}{Wrong}{}{Walltime}{Avg}{0.7617738260014448}%
\StoreBenchExecResult{ExperimentCbmcVerificationMemcleanupTransformed}{MemoryCleanupTransformedMemsafetyMemcleanup}{Wrong}{}{Walltime}{Median}{0.7617738260014448}%
\StoreBenchExecResult{ExperimentCbmcVerificationMemcleanupTransformed}{MemoryCleanupTransformedMemsafetyMemcleanup}{Wrong}{}{Walltime}{Min}{0.3202333380031632}%
\StoreBenchExecResult{ExperimentCbmcVerificationMemcleanupTransformed}{MemoryCleanupTransformedMemsafetyMemcleanup}{Wrong}{}{Walltime}{Max}{1.2033143139997264}%
\StoreBenchExecResult{ExperimentCbmcVerificationMemcleanupTransformed}{MemoryCleanupTransformedMemsafetyMemcleanup}{Wrong}{}{Walltime}{Stdev}{0.4415404879982816000000000000}%
\StoreBenchExecResult{ExperimentCbmcVerificationMemcleanupTransformed}{MemoryCleanupTransformedMemsafetyMemcleanup}{Wrong}{False}{Count}{}{1}%
\StoreBenchExecResult{ExperimentCbmcVerificationMemcleanupTransformed}{MemoryCleanupTransformedMemsafetyMemcleanup}{Wrong}{False}{Cputime}{}{0.340049299}%
\StoreBenchExecResult{ExperimentCbmcVerificationMemcleanupTransformed}{MemoryCleanupTransformedMemsafetyMemcleanup}{Wrong}{False}{Cputime}{Avg}{0.340049299}%
\StoreBenchExecResult{ExperimentCbmcVerificationMemcleanupTransformed}{MemoryCleanupTransformedMemsafetyMemcleanup}{Wrong}{False}{Cputime}{Median}{0.340049299}%
\StoreBenchExecResult{ExperimentCbmcVerificationMemcleanupTransformed}{MemoryCleanupTransformedMemsafetyMemcleanup}{Wrong}{False}{Cputime}{Min}{0.340049299}%
\StoreBenchExecResult{ExperimentCbmcVerificationMemcleanupTransformed}{MemoryCleanupTransformedMemsafetyMemcleanup}{Wrong}{False}{Cputime}{Max}{0.340049299}%
\StoreBenchExecResult{ExperimentCbmcVerificationMemcleanupTransformed}{MemoryCleanupTransformedMemsafetyMemcleanup}{Wrong}{False}{Cputime}{Stdev}{0E-14}%
\StoreBenchExecResult{ExperimentCbmcVerificationMemcleanupTransformed}{MemoryCleanupTransformedMemsafetyMemcleanup}{Wrong}{False}{Walltime}{}{0.3202333380031632}%
\StoreBenchExecResult{ExperimentCbmcVerificationMemcleanupTransformed}{MemoryCleanupTransformedMemsafetyMemcleanup}{Wrong}{False}{Walltime}{Avg}{0.3202333380031632}%
\StoreBenchExecResult{ExperimentCbmcVerificationMemcleanupTransformed}{MemoryCleanupTransformedMemsafetyMemcleanup}{Wrong}{False}{Walltime}{Median}{0.3202333380031632}%
\StoreBenchExecResult{ExperimentCbmcVerificationMemcleanupTransformed}{MemoryCleanupTransformedMemsafetyMemcleanup}{Wrong}{False}{Walltime}{Min}{0.3202333380031632}%
\StoreBenchExecResult{ExperimentCbmcVerificationMemcleanupTransformed}{MemoryCleanupTransformedMemsafetyMemcleanup}{Wrong}{False}{Walltime}{Max}{0.3202333380031632}%
\StoreBenchExecResult{ExperimentCbmcVerificationMemcleanupTransformed}{MemoryCleanupTransformedMemsafetyMemcleanup}{Wrong}{False}{Walltime}{Stdev}{0E-16}%
\StoreBenchExecResult{ExperimentCbmcVerificationMemcleanupTransformed}{MemoryCleanupTransformedMemsafetyMemcleanup}{Wrong}{True}{Count}{}{1}%
\StoreBenchExecResult{ExperimentCbmcVerificationMemcleanupTransformed}{MemoryCleanupTransformedMemsafetyMemcleanup}{Wrong}{True}{Cputime}{}{1.213348013}%
\StoreBenchExecResult{ExperimentCbmcVerificationMemcleanupTransformed}{MemoryCleanupTransformedMemsafetyMemcleanup}{Wrong}{True}{Cputime}{Avg}{1.213348013}%
\StoreBenchExecResult{ExperimentCbmcVerificationMemcleanupTransformed}{MemoryCleanupTransformedMemsafetyMemcleanup}{Wrong}{True}{Cputime}{Median}{1.213348013}%
\StoreBenchExecResult{ExperimentCbmcVerificationMemcleanupTransformed}{MemoryCleanupTransformedMemsafetyMemcleanup}{Wrong}{True}{Cputime}{Min}{1.213348013}%
\StoreBenchExecResult{ExperimentCbmcVerificationMemcleanupTransformed}{MemoryCleanupTransformedMemsafetyMemcleanup}{Wrong}{True}{Cputime}{Max}{1.213348013}%
\StoreBenchExecResult{ExperimentCbmcVerificationMemcleanupTransformed}{MemoryCleanupTransformedMemsafetyMemcleanup}{Wrong}{True}{Cputime}{Stdev}{0E-14}%
\StoreBenchExecResult{ExperimentCbmcVerificationMemcleanupTransformed}{MemoryCleanupTransformedMemsafetyMemcleanup}{Wrong}{True}{Walltime}{}{1.2033143139997264}%
\StoreBenchExecResult{ExperimentCbmcVerificationMemcleanupTransformed}{MemoryCleanupTransformedMemsafetyMemcleanup}{Wrong}{True}{Walltime}{Avg}{1.2033143139997264}%
\StoreBenchExecResult{ExperimentCbmcVerificationMemcleanupTransformed}{MemoryCleanupTransformedMemsafetyMemcleanup}{Wrong}{True}{Walltime}{Median}{1.2033143139997264}%
\StoreBenchExecResult{ExperimentCbmcVerificationMemcleanupTransformed}{MemoryCleanupTransformedMemsafetyMemcleanup}{Wrong}{True}{Walltime}{Min}{1.2033143139997264}%
\StoreBenchExecResult{ExperimentCbmcVerificationMemcleanupTransformed}{MemoryCleanupTransformedMemsafetyMemcleanup}{Wrong}{True}{Walltime}{Max}{1.2033143139997264}%
\StoreBenchExecResult{ExperimentCbmcVerificationMemcleanupTransformed}{MemoryCleanupTransformedMemsafetyMemcleanup}{Wrong}{True}{Walltime}{Stdev}{0E-16}%
\ifdefined\CbmcCbmcTotalCount\else\edef\CbmcCbmcTotalCount{0}\fi
\ifdefined\CbmcCbmcCorrectCount\else\edef\CbmcCbmcCorrectCount{0}\fi
\ifdefined\CbmcCbmcCorrectTrueCount\else\edef\CbmcCbmcCorrectTrueCount{0}\fi
\ifdefined\CbmcCbmcCorrectFalseCount\else\edef\CbmcCbmcCorrectFalseCount{0}\fi
\ifdefined\CbmcCbmcWrongCount\else\edef\CbmcCbmcWrongCount{0}\fi
\ifdefined\CbmcCbmcWrongTrueCount\else\edef\CbmcCbmcWrongTrueCount{0}\fi
\ifdefined\CbmcCbmcWrongFalseCount\else\edef\CbmcCbmcWrongFalseCount{0}\fi
\ifdefined\CbmcCbmcErrorTimeoutCount\else\edef\CbmcCbmcErrorTimeoutCount{0}\fi
\ifdefined\CbmcCbmcErrorOutOfMemoryCount\else\edef\CbmcCbmcErrorOutOfMemoryCount{0}\fi
\ifdefined\CbmcCbmcCorrectCputime\else\edef\CbmcCbmcCorrectCputime{0}\fi
\ifdefined\CbmcCbmcCorrectCputimeAvg\else\edef\CbmcCbmcCorrectCputimeAvg{None}\fi
\ifdefined\CbmcCbmcCorrectWalltime\else\edef\CbmcCbmcCorrectWalltime{0}\fi
\ifdefined\CbmcCbmcCorrectWalltimeAvg\else\edef\CbmcCbmcCorrectWalltimeAvg{None}\fi
\edef\CbmcCbmcErrorOtherInconclusiveCount{\the\numexpr \CbmcCbmcTotalCount - \CbmcCbmcCorrectCount - \CbmcCbmcWrongTrueCount - \CbmcCbmcWrongFalseCount - \CbmcCbmcErrorTimeoutCount - \CbmcCbmcErrorOutOfMemoryCount \relax}
\providecommand\StoreBenchExecResult[7]{\expandafter\newcommand\csname#1#2#3#4#5#6\endcsname{#7}}%
\StoreBenchExecResult{ExperimentCpacheckerVerificationMemcleanup}{Memcleanup}{Total}{}{Count}{}{41}%
\StoreBenchExecResult{ExperimentCpacheckerVerificationMemcleanup}{Memcleanup}{Total}{}{Cputime}{}{409.161504885}%
\StoreBenchExecResult{ExperimentCpacheckerVerificationMemcleanup}{Memcleanup}{Total}{}{Cputime}{Avg}{9.979548899634146341463414634}%
\StoreBenchExecResult{ExperimentCpacheckerVerificationMemcleanup}{Memcleanup}{Total}{}{Cputime}{Median}{7.281212641}%
\StoreBenchExecResult{ExperimentCpacheckerVerificationMemcleanup}{Memcleanup}{Total}{}{Cputime}{Min}{5.973124643}%
\StoreBenchExecResult{ExperimentCpacheckerVerificationMemcleanup}{Memcleanup}{Total}{}{Cputime}{Max}{47.825590316}%
\StoreBenchExecResult{ExperimentCpacheckerVerificationMemcleanup}{Memcleanup}{Total}{}{Cputime}{Stdev}{8.852733130018979647543450477}%
\StoreBenchExecResult{ExperimentCpacheckerVerificationMemcleanup}{Memcleanup}{Total}{}{Walltime}{}{217.0434290800258157}%
\StoreBenchExecResult{ExperimentCpacheckerVerificationMemcleanup}{Memcleanup}{Total}{}{Walltime}{Avg}{5.293742172683556480487804878}%
\StoreBenchExecResult{ExperimentCpacheckerVerificationMemcleanup}{Memcleanup}{Total}{}{Walltime}{Median}{3.80506608900032}%
\StoreBenchExecResult{ExperimentCpacheckerVerificationMemcleanup}{Memcleanup}{Total}{}{Walltime}{Min}{3.1433043100041687}%
\StoreBenchExecResult{ExperimentCpacheckerVerificationMemcleanup}{Memcleanup}{Total}{}{Walltime}{Max}{27.314573146999464}%
\StoreBenchExecResult{ExperimentCpacheckerVerificationMemcleanup}{Memcleanup}{Total}{}{Walltime}{Stdev}{5.013284709191011037734533626}%
\StoreBenchExecResult{ExperimentCpacheckerVerificationMemcleanup}{Memcleanup}{Correct}{}{Count}{}{35}%
\StoreBenchExecResult{ExperimentCpacheckerVerificationMemcleanup}{Memcleanup}{Correct}{}{Cputime}{}{289.337066201}%
\StoreBenchExecResult{ExperimentCpacheckerVerificationMemcleanup}{Memcleanup}{Correct}{}{Cputime}{Avg}{8.266773320028571428571428571}%
\StoreBenchExecResult{ExperimentCpacheckerVerificationMemcleanup}{Memcleanup}{Correct}{}{Cputime}{Median}{7.281212641}%
\StoreBenchExecResult{ExperimentCpacheckerVerificationMemcleanup}{Memcleanup}{Correct}{}{Cputime}{Min}{5.973124643}%
\StoreBenchExecResult{ExperimentCpacheckerVerificationMemcleanup}{Memcleanup}{Correct}{}{Cputime}{Max}{24.983247434}%
\StoreBenchExecResult{ExperimentCpacheckerVerificationMemcleanup}{Memcleanup}{Correct}{}{Cputime}{Stdev}{3.509796192999447252569009855}%
\StoreBenchExecResult{ExperimentCpacheckerVerificationMemcleanup}{Memcleanup}{Correct}{}{Walltime}{}{150.4588923320188763}%
\StoreBenchExecResult{ExperimentCpacheckerVerificationMemcleanup}{Memcleanup}{Correct}{}{Walltime}{Avg}{4.298825495200539322857142857}%
\StoreBenchExecResult{ExperimentCpacheckerVerificationMemcleanup}{Memcleanup}{Correct}{}{Walltime}{Median}{3.80506608900032}%
\StoreBenchExecResult{ExperimentCpacheckerVerificationMemcleanup}{Memcleanup}{Correct}{}{Walltime}{Min}{3.1433043100041687}%
\StoreBenchExecResult{ExperimentCpacheckerVerificationMemcleanup}{Memcleanup}{Correct}{}{Walltime}{Max}{12.701549876001081}%
\StoreBenchExecResult{ExperimentCpacheckerVerificationMemcleanup}{Memcleanup}{Correct}{}{Walltime}{Stdev}{1.764297909874842187255234171}%
\StoreBenchExecResult{ExperimentCpacheckerVerificationMemcleanup}{Memcleanup}{Correct}{False}{Count}{}{34}%
\StoreBenchExecResult{ExperimentCpacheckerVerificationMemcleanup}{Memcleanup}{Correct}{False}{Cputime}{}{282.832846763}%
\StoreBenchExecResult{ExperimentCpacheckerVerificationMemcleanup}{Memcleanup}{Correct}{False}{Cputime}{Avg}{8.318613140088235294117647059}%
\StoreBenchExecResult{ExperimentCpacheckerVerificationMemcleanup}{Memcleanup}{Correct}{False}{Cputime}{Median}{7.2868070335}%
\StoreBenchExecResult{ExperimentCpacheckerVerificationMemcleanup}{Memcleanup}{Correct}{False}{Cputime}{Min}{5.973124643}%
\StoreBenchExecResult{ExperimentCpacheckerVerificationMemcleanup}{Memcleanup}{Correct}{False}{Cputime}{Max}{24.983247434}%
\StoreBenchExecResult{ExperimentCpacheckerVerificationMemcleanup}{Memcleanup}{Correct}{False}{Cputime}{Stdev}{3.547805698848671079188909195}%
\StoreBenchExecResult{ExperimentCpacheckerVerificationMemcleanup}{Memcleanup}{Correct}{False}{Walltime}{}{147.0469562110156397}%
\StoreBenchExecResult{ExperimentCpacheckerVerificationMemcleanup}{Memcleanup}{Correct}{False}{Walltime}{Avg}{4.324910476794577638235294118}%
\StoreBenchExecResult{ExperimentCpacheckerVerificationMemcleanup}{Memcleanup}{Correct}{False}{Walltime}{Median}{3.8194001224983367}%
\StoreBenchExecResult{ExperimentCpacheckerVerificationMemcleanup}{Memcleanup}{Correct}{False}{Walltime}{Min}{3.1433043100041687}%
\StoreBenchExecResult{ExperimentCpacheckerVerificationMemcleanup}{Memcleanup}{Correct}{False}{Walltime}{Max}{12.701549876001081}%
\StoreBenchExecResult{ExperimentCpacheckerVerificationMemcleanup}{Memcleanup}{Correct}{False}{Walltime}{Stdev}{1.783391034308284039202015931}%

\StoreBenchExecResult{ExperimentCpacheckerVerificationMemcleanup}{Memcleanup}{Correct}{True}{Count}{}{1}%
\StoreBenchExecResult{ExperimentCpacheckerVerificationMemcleanup}{Memcleanup}{Correct}{True}{Cputime}{}{6.504219438}%
\StoreBenchExecResult{ExperimentCpacheckerVerificationMemcleanup}{Memcleanup}{Correct}{True}{Cputime}{Avg}{6.504219438}%
\StoreBenchExecResult{ExperimentCpacheckerVerificationMemcleanup}{Memcleanup}{Correct}{True}{Cputime}{Median}{6.504219438}%
\StoreBenchExecResult{ExperimentCpacheckerVerificationMemcleanup}{Memcleanup}{Correct}{True}{Cputime}{Min}{6.504219438}%
\StoreBenchExecResult{ExperimentCpacheckerVerificationMemcleanup}{Memcleanup}{Correct}{True}{Cputime}{Max}{6.504219438}%
\StoreBenchExecResult{ExperimentCpacheckerVerificationMemcleanup}{Memcleanup}{Correct}{True}{Cputime}{Stdev}{0E-14}%
\StoreBenchExecResult{ExperimentCpacheckerVerificationMemcleanup}{Memcleanup}{Correct}{True}{Walltime}{}{3.4119361210032366}%
\StoreBenchExecResult{ExperimentCpacheckerVerificationMemcleanup}{Memcleanup}{Correct}{True}{Walltime}{Avg}{3.4119361210032366}%
\StoreBenchExecResult{ExperimentCpacheckerVerificationMemcleanup}{Memcleanup}{Correct}{True}{Walltime}{Median}{3.4119361210032366}%
\StoreBenchExecResult{ExperimentCpacheckerVerificationMemcleanup}{Memcleanup}{Correct}{True}{Walltime}{Min}{3.4119361210032366}%
\StoreBenchExecResult{ExperimentCpacheckerVerificationMemcleanup}{Memcleanup}{Correct}{True}{Walltime}{Max}{3.4119361210032366}%
\StoreBenchExecResult{ExperimentCpacheckerVerificationMemcleanup}{Memcleanup}{Correct}{True}{Walltime}{Stdev}{0E-16}%

\StoreBenchExecResult{ExperimentCpacheckerVerificationMemcleanup}{Memcleanup}{Error}{}{Count}{}{6}%
\StoreBenchExecResult{ExperimentCpacheckerVerificationMemcleanup}{Memcleanup}{Error}{}{Cputime}{}{119.824438684}%
\StoreBenchExecResult{ExperimentCpacheckerVerificationMemcleanup}{Memcleanup}{Error}{}{Cputime}{Avg}{19.97073978066666666666666667}%
\StoreBenchExecResult{ExperimentCpacheckerVerificationMemcleanup}{Memcleanup}{Error}{}{Cputime}{Median}{7.1366050855}%
\StoreBenchExecResult{ExperimentCpacheckerVerificationMemcleanup}{Memcleanup}{Error}{}{Cputime}{Min}{6.16714631}%
\StoreBenchExecResult{ExperimentCpacheckerVerificationMemcleanup}{Memcleanup}{Error}{}{Cputime}{Max}{47.825590316}%
\StoreBenchExecResult{ExperimentCpacheckerVerificationMemcleanup}{Memcleanup}{Error}{}{Cputime}{Stdev}{18.62092655594042935228161856}%
\StoreBenchExecResult{ExperimentCpacheckerVerificationMemcleanup}{Memcleanup}{Error}{}{Walltime}{}{66.5845367480069394}%
\StoreBenchExecResult{ExperimentCpacheckerVerificationMemcleanup}{Memcleanup}{Error}{}{Walltime}{Avg}{11.0974227913344899}%
\StoreBenchExecResult{ExperimentCpacheckerVerificationMemcleanup}{Memcleanup}{Error}{}{Walltime}{Median}{3.7313695925040520}%
\StoreBenchExecResult{ExperimentCpacheckerVerificationMemcleanup}{Memcleanup}{Error}{}{Walltime}{Min}{3.2194764029991347}%
\StoreBenchExecResult{ExperimentCpacheckerVerificationMemcleanup}{Memcleanup}{Error}{}{Walltime}{Max}{27.314573146999464}%
\StoreBenchExecResult{ExperimentCpacheckerVerificationMemcleanup}{Memcleanup}{Error}{}{Walltime}{Stdev}{10.68305896739183906717535815}%
\StoreBenchExecResult{ExperimentCpacheckerVerificationMemcleanup}{Memcleanup}{Error}{Error}{Count}{}{6}%
\StoreBenchExecResult{ExperimentCpacheckerVerificationMemcleanup}{Memcleanup}{Error}{Error}{Cputime}{}{119.824438684}%
\StoreBenchExecResult{ExperimentCpacheckerVerificationMemcleanup}{Memcleanup}{Error}{Error}{Cputime}{Avg}{19.97073978066666666666666667}%
\StoreBenchExecResult{ExperimentCpacheckerVerificationMemcleanup}{Memcleanup}{Error}{Error}{Cputime}{Median}{7.1366050855}%
\StoreBenchExecResult{ExperimentCpacheckerVerificationMemcleanup}{Memcleanup}{Error}{Error}{Cputime}{Min}{6.16714631}%
\StoreBenchExecResult{ExperimentCpacheckerVerificationMemcleanup}{Memcleanup}{Error}{Error}{Cputime}{Max}{47.825590316}%
\StoreBenchExecResult{ExperimentCpacheckerVerificationMemcleanup}{Memcleanup}{Error}{Error}{Cputime}{Stdev}{18.62092655594042935228161856}%
\StoreBenchExecResult{ExperimentCpacheckerVerificationMemcleanup}{Memcleanup}{Error}{Error}{Walltime}{}{66.5845367480069394}%
\StoreBenchExecResult{ExperimentCpacheckerVerificationMemcleanup}{Memcleanup}{Error}{Error}{Walltime}{Avg}{11.0974227913344899}%
\StoreBenchExecResult{ExperimentCpacheckerVerificationMemcleanup}{Memcleanup}{Error}{Error}{Walltime}{Median}{3.7313695925040520}%
\StoreBenchExecResult{ExperimentCpacheckerVerificationMemcleanup}{Memcleanup}{Error}{Error}{Walltime}{Min}{3.2194764029991347}%
\StoreBenchExecResult{ExperimentCpacheckerVerificationMemcleanup}{Memcleanup}{Error}{Error}{Walltime}{Max}{27.314573146999464}%
\StoreBenchExecResult{ExperimentCpacheckerVerificationMemcleanup}{Memcleanup}{Error}{Error}{Walltime}{Stdev}{10.68305896739183906717535815}%
\ifdefined\CpacheckerCpacheckerTotalCount\else\edef\CpacheckerCpacheckerTotalCount{0}\fi
\ifdefined\CpacheckerCpacheckerCorrectCount\else\edef\CpacheckerCpacheckerCorrectCount{0}\fi
\ifdefined\CpacheckerCpacheckerCorrectTrueCount\else\edef\CpacheckerCpacheckerCorrectTrueCount{0}\fi
\ifdefined\CpacheckerCpacheckerCorrectFalseCount\else\edef\CpacheckerCpacheckerCorrectFalseCount{0}\fi
\ifdefined\CpacheckerCpacheckerWrongCount\else\edef\CpacheckerCpacheckerWrongCount{0}\fi
\ifdefined\CpacheckerCpacheckerWrongTrueCount\else\edef\CpacheckerCpacheckerWrongTrueCount{0}\fi
\ifdefined\CpacheckerCpacheckerWrongFalseCount\else\edef\CpacheckerCpacheckerWrongFalseCount{0}\fi
\ifdefined\CpacheckerCpacheckerErrorTimeoutCount\else\edef\CpacheckerCpacheckerErrorTimeoutCount{0}\fi
\ifdefined\CpacheckerCpacheckerErrorOutOfMemoryCount\else\edef\CpacheckerCpacheckerErrorOutOfMemoryCount{0}\fi
\ifdefined\CpacheckerCpacheckerCorrectCputime\else\edef\CpacheckerCpacheckerCorrectCputime{0}\fi
\ifdefined\CpacheckerCpacheckerCorrectCputimeAvg\else\edef\CpacheckerCpacheckerCorrectCputimeAvg{None}\fi
\ifdefined\CpacheckerCpacheckerCorrectWalltime\else\edef\CpacheckerCpacheckerCorrectWalltime{0}\fi
\ifdefined\CpacheckerCpacheckerCorrectWalltimeAvg\else\edef\CpacheckerCpacheckerCorrectWalltimeAvg{None}\fi
\edef\CpacheckerCpacheckerErrorOtherInconclusiveCount{\the\numexpr \CpacheckerCpacheckerTotalCount - \CpacheckerCpacheckerCorrectCount - \CpacheckerCpacheckerWrongTrueCount - \CpacheckerCpacheckerWrongFalseCount - \CpacheckerCpacheckerErrorTimeoutCount - \CpacheckerCpacheckerErrorOutOfMemoryCount \relax}
\providecommand\StoreBenchExecResult[7]{\expandafter\newcommand\csname#1#2#3#4#5#6\endcsname{#7}}%
\StoreBenchExecResult{ExperimentUautomizerVerificationMemcleanup}{Memcleanup}{Total}{}{Count}{}{41}%
\StoreBenchExecResult{ExperimentUautomizerVerificationMemcleanup}{Memcleanup}{Total}{}{Cputime}{}{9577.856732535}%
\StoreBenchExecResult{ExperimentUautomizerVerificationMemcleanup}{Memcleanup}{Total}{}{Cputime}{Avg}{233.6062617691463414634146341}%
\StoreBenchExecResult{ExperimentUautomizerVerificationMemcleanup}{Memcleanup}{Total}{}{Cputime}{Median}{27.563841008}%
\StoreBenchExecResult{ExperimentUautomizerVerificationMemcleanup}{Memcleanup}{Total}{}{Cputime}{Min}{17.307014686}%
\StoreBenchExecResult{ExperimentUautomizerVerificationMemcleanup}{Memcleanup}{Total}{}{Cputime}{Max}{961.206105615}%
\StoreBenchExecResult{ExperimentUautomizerVerificationMemcleanup}{Memcleanup}{Total}{}{Cputime}{Stdev}{370.1787107614656440335799719}%
\StoreBenchExecResult{ExperimentUautomizerVerificationMemcleanup}{Memcleanup}{Total}{}{Walltime}{}{8462.395654434010827}%
\StoreBenchExecResult{ExperimentUautomizerVerificationMemcleanup}{Memcleanup}{Total}{}{Walltime}{Avg}{206.3998940105856299268292683}%
\StoreBenchExecResult{ExperimentUautomizerVerificationMemcleanup}{Memcleanup}{Total}{}{Walltime}{Median}{16.1671023240051}%
\StoreBenchExecResult{ExperimentUautomizerVerificationMemcleanup}{Memcleanup}{Total}{}{Walltime}{Min}{10.317636909996509}%
\StoreBenchExecResult{ExperimentUautomizerVerificationMemcleanup}{Memcleanup}{Total}{}{Walltime}{Max}{925.1696505159998}%
\StoreBenchExecResult{ExperimentUautomizerVerificationMemcleanup}{Memcleanup}{Total}{}{Walltime}{Stdev}{343.9480686454906217930245130}%
\StoreBenchExecResult{ExperimentUautomizerVerificationMemcleanup}{Memcleanup}{Correct}{}{Count}{}{27}%
\StoreBenchExecResult{ExperimentUautomizerVerificationMemcleanup}{Memcleanup}{Correct}{}{Cputime}{}{1689.344415917}%
\StoreBenchExecResult{ExperimentUautomizerVerificationMemcleanup}{Memcleanup}{Correct}{}{Cputime}{Avg}{62.56831170062962962962962963}%
\StoreBenchExecResult{ExperimentUautomizerVerificationMemcleanup}{Memcleanup}{Correct}{}{Cputime}{Median}{23.560285502}%
\StoreBenchExecResult{ExperimentUautomizerVerificationMemcleanup}{Memcleanup}{Correct}{}{Cputime}{Min}{17.307014686}%
\StoreBenchExecResult{ExperimentUautomizerVerificationMemcleanup}{Memcleanup}{Correct}{}{Cputime}{Max}{618.147974944}%
\StoreBenchExecResult{ExperimentUautomizerVerificationMemcleanup}{Memcleanup}{Correct}{}{Cputime}{Stdev}{115.2120987275410484774641480}%
\StoreBenchExecResult{ExperimentUautomizerVerificationMemcleanup}{Memcleanup}{Correct}{}{Walltime}{}{1279.194434480996256}%
\StoreBenchExecResult{ExperimentUautomizerVerificationMemcleanup}{Memcleanup}{Correct}{}{Walltime}{Avg}{47.37757164744430577777777778}%
\StoreBenchExecResult{ExperimentUautomizerVerificationMemcleanup}{Memcleanup}{Correct}{}{Walltime}{Median}{13.630345240002498}%
\StoreBenchExecResult{ExperimentUautomizerVerificationMemcleanup}{Memcleanup}{Correct}{}{Walltime}{Min}{10.317636909996509}%
\StoreBenchExecResult{ExperimentUautomizerVerificationMemcleanup}{Memcleanup}{Correct}{}{Walltime}{Max}{561.6801888730042}%
\StoreBenchExecResult{ExperimentUautomizerVerificationMemcleanup}{Memcleanup}{Correct}{}{Walltime}{Stdev}{104.9126716804243534111412801}%
\StoreBenchExecResult{ExperimentUautomizerVerificationMemcleanup}{Memcleanup}{Correct}{False}{Count}{}{27}%
\StoreBenchExecResult{ExperimentUautomizerVerificationMemcleanup}{Memcleanup}{Correct}{False}{Cputime}{}{1689.344415917}%
\StoreBenchExecResult{ExperimentUautomizerVerificationMemcleanup}{Memcleanup}{Correct}{False}{Cputime}{Avg}{62.56831170062962962962962963}%
\StoreBenchExecResult{ExperimentUautomizerVerificationMemcleanup}{Memcleanup}{Correct}{False}{Cputime}{Median}{23.560285502}%
\StoreBenchExecResult{ExperimentUautomizerVerificationMemcleanup}{Memcleanup}{Correct}{False}{Cputime}{Min}{17.307014686}%
\StoreBenchExecResult{ExperimentUautomizerVerificationMemcleanup}{Memcleanup}{Correct}{False}{Cputime}{Max}{618.147974944}%
\StoreBenchExecResult{ExperimentUautomizerVerificationMemcleanup}{Memcleanup}{Correct}{False}{Cputime}{Stdev}{115.2120987275410484774641480}%
\StoreBenchExecResult{ExperimentUautomizerVerificationMemcleanup}{Memcleanup}{Correct}{False}{Walltime}{}{1279.194434480996256}%
\StoreBenchExecResult{ExperimentUautomizerVerificationMemcleanup}{Memcleanup}{Correct}{False}{Walltime}{Avg}{47.37757164744430577777777778}%
\StoreBenchExecResult{ExperimentUautomizerVerificationMemcleanup}{Memcleanup}{Correct}{False}{Walltime}{Median}{13.630345240002498}%
\StoreBenchExecResult{ExperimentUautomizerVerificationMemcleanup}{Memcleanup}{Correct}{False}{Walltime}{Min}{10.317636909996509}%
\StoreBenchExecResult{ExperimentUautomizerVerificationMemcleanup}{Memcleanup}{Correct}{False}{Walltime}{Max}{561.6801888730042}%
\StoreBenchExecResult{ExperimentUautomizerVerificationMemcleanup}{Memcleanup}{Correct}{False}{Walltime}{Stdev}{104.9126716804243534111412801}%

\StoreBenchExecResult{ExperimentUautomizerVerificationMemcleanup}{Memcleanup}{Error}{}{Count}{}{8}%
\StoreBenchExecResult{ExperimentUautomizerVerificationMemcleanup}{Memcleanup}{Error}{}{Cputime}{}{7685.965713122}%
\StoreBenchExecResult{ExperimentUautomizerVerificationMemcleanup}{Memcleanup}{Error}{}{Cputime}{Avg}{960.74571414025}%
\StoreBenchExecResult{ExperimentUautomizerVerificationMemcleanup}{Memcleanup}{Error}{}{Cputime}{Median}{960.7614681185}%
\StoreBenchExecResult{ExperimentUautomizerVerificationMemcleanup}{Memcleanup}{Error}{}{Cputime}{Min}{960.302154129}%
\StoreBenchExecResult{ExperimentUautomizerVerificationMemcleanup}{Memcleanup}{Error}{}{Cputime}{Max}{961.206105615}%
\StoreBenchExecResult{ExperimentUautomizerVerificationMemcleanup}{Memcleanup}{Error}{}{Cputime}{Stdev}{0.2964868584415146931108912351}%
\StoreBenchExecResult{ExperimentUautomizerVerificationMemcleanup}{Memcleanup}{Error}{}{Walltime}{}{7060.0005030430039}%
\StoreBenchExecResult{ExperimentUautomizerVerificationMemcleanup}{Memcleanup}{Error}{}{Walltime}{Avg}{882.5000628803754875}%
\StoreBenchExecResult{ExperimentUautomizerVerificationMemcleanup}{Memcleanup}{Error}{}{Walltime}{Median}{877.56171999350045}%
\StoreBenchExecResult{ExperimentUautomizerVerificationMemcleanup}{Memcleanup}{Error}{}{Walltime}{Min}{842.4341593960053}%
\StoreBenchExecResult{ExperimentUautomizerVerificationMemcleanup}{Memcleanup}{Error}{}{Walltime}{Max}{925.1696505159998}%
\StoreBenchExecResult{ExperimentUautomizerVerificationMemcleanup}{Memcleanup}{Error}{}{Walltime}{Stdev}{27.26481031715492350125776760}%
\StoreBenchExecResult{ExperimentUautomizerVerificationMemcleanup}{Memcleanup}{Error}{Timeout}{Count}{}{8}%
\StoreBenchExecResult{ExperimentUautomizerVerificationMemcleanup}{Memcleanup}{Error}{Timeout}{Cputime}{}{7685.965713122}%
\StoreBenchExecResult{ExperimentUautomizerVerificationMemcleanup}{Memcleanup}{Error}{Timeout}{Cputime}{Avg}{960.74571414025}%
\StoreBenchExecResult{ExperimentUautomizerVerificationMemcleanup}{Memcleanup}{Error}{Timeout}{Cputime}{Median}{960.7614681185}%
\StoreBenchExecResult{ExperimentUautomizerVerificationMemcleanup}{Memcleanup}{Error}{Timeout}{Cputime}{Min}{960.302154129}%
\StoreBenchExecResult{ExperimentUautomizerVerificationMemcleanup}{Memcleanup}{Error}{Timeout}{Cputime}{Max}{961.206105615}%
\StoreBenchExecResult{ExperimentUautomizerVerificationMemcleanup}{Memcleanup}{Error}{Timeout}{Cputime}{Stdev}{0.2964868584415146931108912351}%
\StoreBenchExecResult{ExperimentUautomizerVerificationMemcleanup}{Memcleanup}{Error}{Timeout}{Walltime}{}{7060.0005030430039}%
\StoreBenchExecResult{ExperimentUautomizerVerificationMemcleanup}{Memcleanup}{Error}{Timeout}{Walltime}{Avg}{882.5000628803754875}%
\StoreBenchExecResult{ExperimentUautomizerVerificationMemcleanup}{Memcleanup}{Error}{Timeout}{Walltime}{Median}{877.56171999350045}%
\StoreBenchExecResult{ExperimentUautomizerVerificationMemcleanup}{Memcleanup}{Error}{Timeout}{Walltime}{Min}{842.4341593960053}%
\StoreBenchExecResult{ExperimentUautomizerVerificationMemcleanup}{Memcleanup}{Error}{Timeout}{Walltime}{Max}{925.1696505159998}%
\StoreBenchExecResult{ExperimentUautomizerVerificationMemcleanup}{Memcleanup}{Error}{Timeout}{Walltime}{Stdev}{27.26481031715492350125776760}%
\StoreBenchExecResult{ExperimentUautomizerVerificationMemcleanup}{Memcleanup}{Unknown}{}{Count}{}{6}%
\StoreBenchExecResult{ExperimentUautomizerVerificationMemcleanup}{Memcleanup}{Unknown}{}{Cputime}{}{202.546603496}%
\StoreBenchExecResult{ExperimentUautomizerVerificationMemcleanup}{Memcleanup}{Unknown}{}{Cputime}{Avg}{33.75776724933333333333333333}%
\StoreBenchExecResult{ExperimentUautomizerVerificationMemcleanup}{Memcleanup}{Unknown}{}{Cputime}{Median}{27.2142200165}%
\StoreBenchExecResult{ExperimentUautomizerVerificationMemcleanup}{Memcleanup}{Unknown}{}{Cputime}{Min}{26.406985067}%
\StoreBenchExecResult{ExperimentUautomizerVerificationMemcleanup}{Memcleanup}{Unknown}{}{Cputime}{Max}{48.882752042}%
\StoreBenchExecResult{ExperimentUautomizerVerificationMemcleanup}{Memcleanup}{Unknown}{}{Cputime}{Stdev}{9.841096178583897570064710487}%
\StoreBenchExecResult{ExperimentUautomizerVerificationMemcleanup}{Memcleanup}{Unknown}{}{Walltime}{}{123.200716910010671}%
\StoreBenchExecResult{ExperimentUautomizerVerificationMemcleanup}{Memcleanup}{Unknown}{}{Walltime}{Avg}{20.53345281833511183333333333}%
\StoreBenchExecResult{ExperimentUautomizerVerificationMemcleanup}{Memcleanup}{Unknown}{}{Walltime}{Median}{16.1548137400022825}%
\StoreBenchExecResult{ExperimentUautomizerVerificationMemcleanup}{Memcleanup}{Unknown}{}{Walltime}{Min}{16.055210926002474}%
\StoreBenchExecResult{ExperimentUautomizerVerificationMemcleanup}{Memcleanup}{Unknown}{}{Walltime}{Max}{30.089550328004407}%
\StoreBenchExecResult{ExperimentUautomizerVerificationMemcleanup}{Memcleanup}{Unknown}{}{Walltime}{Stdev}{6.272077479823819376569907876}%
\StoreBenchExecResult{ExperimentUautomizerVerificationMemcleanup}{Memcleanup}{Unknown}{Unknown}{Count}{}{6}%
\StoreBenchExecResult{ExperimentUautomizerVerificationMemcleanup}{Memcleanup}{Unknown}{Unknown}{Cputime}{}{202.546603496}%
\StoreBenchExecResult{ExperimentUautomizerVerificationMemcleanup}{Memcleanup}{Unknown}{Unknown}{Cputime}{Avg}{33.75776724933333333333333333}%
\StoreBenchExecResult{ExperimentUautomizerVerificationMemcleanup}{Memcleanup}{Unknown}{Unknown}{Cputime}{Median}{27.2142200165}%
\StoreBenchExecResult{ExperimentUautomizerVerificationMemcleanup}{Memcleanup}{Unknown}{Unknown}{Cputime}{Min}{26.406985067}%
\StoreBenchExecResult{ExperimentUautomizerVerificationMemcleanup}{Memcleanup}{Unknown}{Unknown}{Cputime}{Max}{48.882752042}%
\StoreBenchExecResult{ExperimentUautomizerVerificationMemcleanup}{Memcleanup}{Unknown}{Unknown}{Cputime}{Stdev}{9.841096178583897570064710487}%
\StoreBenchExecResult{ExperimentUautomizerVerificationMemcleanup}{Memcleanup}{Unknown}{Unknown}{Walltime}{}{123.200716910010671}%
\StoreBenchExecResult{ExperimentUautomizerVerificationMemcleanup}{Memcleanup}{Unknown}{Unknown}{Walltime}{Avg}{20.53345281833511183333333333}%
\StoreBenchExecResult{ExperimentUautomizerVerificationMemcleanup}{Memcleanup}{Unknown}{Unknown}{Walltime}{Median}{16.1548137400022825}%
\StoreBenchExecResult{ExperimentUautomizerVerificationMemcleanup}{Memcleanup}{Unknown}{Unknown}{Walltime}{Min}{16.055210926002474}%
\StoreBenchExecResult{ExperimentUautomizerVerificationMemcleanup}{Memcleanup}{Unknown}{Unknown}{Walltime}{Max}{30.089550328004407}%
\StoreBenchExecResult{ExperimentUautomizerVerificationMemcleanup}{Memcleanup}{Unknown}{Unknown}{Walltime}{Stdev}{6.272077479823819376569907876}%
\ifdefined\UltimateUltimateTotalCount\else\edef\UltimateUltimateTotalCount{0}\fi
\ifdefined\UltimateUltimateCorrectCount\else\edef\UltimateUltimateCorrectCount{0}\fi
\ifdefined\UltimateUltimateCorrectTrueCount\else\edef\UltimateUltimateCorrectTrueCount{0}\fi
\ifdefined\UltimateUltimateCorrectFalseCount\else\edef\UltimateUltimateCorrectFalseCount{0}\fi
\ifdefined\UltimateUltimateWrongCount\else\edef\UltimateUltimateWrongCount{0}\fi
\ifdefined\UltimateUltimateWrongTrueCount\else\edef\UltimateUltimateWrongTrueCount{0}\fi
\ifdefined\UltimateUltimateWrongFalseCount\else\edef\UltimateUltimateWrongFalseCount{0}\fi
\ifdefined\UltimateUltimateErrorTimeoutCount\else\edef\UltimateUltimateErrorTimeoutCount{0}\fi
\ifdefined\UltimateUltimateErrorOutOfMemoryCount\else\edef\UltimateUltimateErrorOutOfMemoryCount{0}\fi
\ifdefined\UltimateUltimateCorrectCputime\else\edef\UltimateUltimateCorrectCputime{0}\fi
\ifdefined\UltimateUltimateCorrectCputimeAvg\else\edef\UltimateUltimateCorrectCputimeAvg{None}\fi
\ifdefined\UltimateUltimateCorrectWalltime\else\edef\UltimateUltimateCorrectWalltime{0}\fi
\ifdefined\UltimateUltimateCorrectWalltimeAvg\else\edef\UltimateUltimateCorrectWalltimeAvg{None}\fi
\edef\UltimateUltimateErrorOtherInconclusiveCount{\the\numexpr \UltimateUltimateTotalCount - \UltimateUltimateCorrectCount - \UltimateUltimateWrongTrueCount - \UltimateUltimateWrongFalseCount - \UltimateUltimateErrorTimeoutCount - \UltimateUltimateErrorOutOfMemoryCount \relax}
\providecommand\StoreBenchExecResult[7]{\expandafter\newcommand\csname#1#2#3#4#5#6\endcsname{#7}}%
\StoreBenchExecResult{ExperimentCbmcVerificationMemcleanup}{Memcleanup}{Total}{}{Count}{}{41}%
\StoreBenchExecResult{ExperimentCbmcVerificationMemcleanup}{Memcleanup}{Total}{}{Cputime}{}{12.620902306}%
\StoreBenchExecResult{ExperimentCbmcVerificationMemcleanup}{Memcleanup}{Total}{}{Cputime}{Avg}{0.3078268855121951219512195122}%
\StoreBenchExecResult{ExperimentCbmcVerificationMemcleanup}{Memcleanup}{Total}{}{Cputime}{Median}{0.173088739}%
\StoreBenchExecResult{ExperimentCbmcVerificationMemcleanup}{Memcleanup}{Total}{}{Cputime}{Min}{0.121981808}%
\StoreBenchExecResult{ExperimentCbmcVerificationMemcleanup}{Memcleanup}{Total}{}{Cputime}{Max}{3.350826917}%
\StoreBenchExecResult{ExperimentCbmcVerificationMemcleanup}{Memcleanup}{Total}{}{Cputime}{Stdev}{0.4973103929031568822147468998}%
\StoreBenchExecResult{ExperimentCbmcVerificationMemcleanup}{Memcleanup}{Total}{}{Walltime}{}{12.23611161599546903}%
\StoreBenchExecResult{ExperimentCbmcVerificationMemcleanup}{Memcleanup}{Total}{}{Walltime}{Avg}{0.2984417467315968056097560976}%
\StoreBenchExecResult{ExperimentCbmcVerificationMemcleanup}{Memcleanup}{Total}{}{Walltime}{Median}{0.16520884000055958}%
\StoreBenchExecResult{ExperimentCbmcVerificationMemcleanup}{Memcleanup}{Total}{}{Walltime}{Min}{0.11253105399373453}%
\StoreBenchExecResult{ExperimentCbmcVerificationMemcleanup}{Memcleanup}{Total}{}{Walltime}{Max}{3.3376688190037385}%
\StoreBenchExecResult{ExperimentCbmcVerificationMemcleanup}{Memcleanup}{Total}{}{Walltime}{Stdev}{0.4965148556282149371421656743}%
\StoreBenchExecResult{ExperimentCbmcVerificationMemcleanup}{Memcleanup}{Correct}{}{Count}{}{40}%
\StoreBenchExecResult{ExperimentCbmcVerificationMemcleanup}{Memcleanup}{Correct}{}{Cputime}{}{12.475608989}%
\StoreBenchExecResult{ExperimentCbmcVerificationMemcleanup}{Memcleanup}{Correct}{}{Cputime}{Avg}{0.311890224725}%
\StoreBenchExecResult{ExperimentCbmcVerificationMemcleanup}{Memcleanup}{Correct}{}{Cputime}{Median}{0.173324875}%
\StoreBenchExecResult{ExperimentCbmcVerificationMemcleanup}{Memcleanup}{Correct}{}{Cputime}{Min}{0.121981808}%
\StoreBenchExecResult{ExperimentCbmcVerificationMemcleanup}{Memcleanup}{Correct}{}{Cputime}{Max}{3.350826917}%
\StoreBenchExecResult{ExperimentCbmcVerificationMemcleanup}{Memcleanup}{Correct}{}{Cputime}{Stdev}{0.5028156996493233498142568808}%
\StoreBenchExecResult{ExperimentCbmcVerificationMemcleanup}{Memcleanup}{Correct}{}{Walltime}{}{12.10009647499828133}%
\StoreBenchExecResult{ExperimentCbmcVerificationMemcleanup}{Memcleanup}{Correct}{}{Walltime}{Avg}{0.30250241187495703325}%
\StoreBenchExecResult{ExperimentCbmcVerificationMemcleanup}{Memcleanup}{Correct}{}{Walltime}{Median}{0.16527458999917144}%
\StoreBenchExecResult{ExperimentCbmcVerificationMemcleanup}{Memcleanup}{Correct}{}{Walltime}{Min}{0.11253105399373453}%
\StoreBenchExecResult{ExperimentCbmcVerificationMemcleanup}{Memcleanup}{Correct}{}{Walltime}{Max}{3.3376688190037385}%
\StoreBenchExecResult{ExperimentCbmcVerificationMemcleanup}{Memcleanup}{Correct}{}{Walltime}{Stdev}{0.5020100873969847530964390012}%
\StoreBenchExecResult{ExperimentCbmcVerificationMemcleanup}{Memcleanup}{Correct}{False}{Count}{}{39}%
\StoreBenchExecResult{ExperimentCbmcVerificationMemcleanup}{Memcleanup}{Correct}{False}{Cputime}{}{12.107306284}%
\StoreBenchExecResult{ExperimentCbmcVerificationMemcleanup}{Memcleanup}{Correct}{False}{Cputime}{Avg}{0.3104437508717948717948717949}%
\StoreBenchExecResult{ExperimentCbmcVerificationMemcleanup}{Memcleanup}{Correct}{False}{Cputime}{Median}{0.173088739}%
\StoreBenchExecResult{ExperimentCbmcVerificationMemcleanup}{Memcleanup}{Correct}{False}{Cputime}{Min}{0.121981808}%
\StoreBenchExecResult{ExperimentCbmcVerificationMemcleanup}{Memcleanup}{Correct}{False}{Cputime}{Max}{3.350826917}%
\StoreBenchExecResult{ExperimentCbmcVerificationMemcleanup}{Memcleanup}{Correct}{False}{Cputime}{Stdev}{0.5091390708568240134883348942}%
\StoreBenchExecResult{ExperimentCbmcVerificationMemcleanup}{Memcleanup}{Correct}{False}{Walltime}{}{11.74292843199509663}%
\StoreBenchExecResult{ExperimentCbmcVerificationMemcleanup}{Memcleanup}{Correct}{False}{Walltime}{Avg}{0.3011007290255152982051282051}%
\StoreBenchExecResult{ExperimentCbmcVerificationMemcleanup}{Memcleanup}{Correct}{False}{Walltime}{Median}{0.16520884000055958}%
\StoreBenchExecResult{ExperimentCbmcVerificationMemcleanup}{Memcleanup}{Correct}{False}{Walltime}{Min}{0.11253105399373453}%
\StoreBenchExecResult{ExperimentCbmcVerificationMemcleanup}{Memcleanup}{Correct}{False}{Walltime}{Max}{3.3376688190037385}%
\StoreBenchExecResult{ExperimentCbmcVerificationMemcleanup}{Memcleanup}{Correct}{False}{Walltime}{Stdev}{0.5083280830034587073970380400}%
\StoreBenchExecResult{ExperimentCbmcVerificationMemcleanup}{Memcleanup}{Correct}{True}{Count}{}{1}%
\StoreBenchExecResult{ExperimentCbmcVerificationMemcleanup}{Memcleanup}{Correct}{True}{Cputime}{}{0.368302705}%
\StoreBenchExecResult{ExperimentCbmcVerificationMemcleanup}{Memcleanup}{Correct}{True}{Cputime}{Avg}{0.368302705}%
\StoreBenchExecResult{ExperimentCbmcVerificationMemcleanup}{Memcleanup}{Correct}{True}{Cputime}{Median}{0.368302705}%
\StoreBenchExecResult{ExperimentCbmcVerificationMemcleanup}{Memcleanup}{Correct}{True}{Cputime}{Min}{0.368302705}%
\StoreBenchExecResult{ExperimentCbmcVerificationMemcleanup}{Memcleanup}{Correct}{True}{Cputime}{Max}{0.368302705}%
\StoreBenchExecResult{ExperimentCbmcVerificationMemcleanup}{Memcleanup}{Correct}{True}{Cputime}{Stdev}{0E-14}%
\StoreBenchExecResult{ExperimentCbmcVerificationMemcleanup}{Memcleanup}{Correct}{True}{Walltime}{}{0.3571680430031847}%
\StoreBenchExecResult{ExperimentCbmcVerificationMemcleanup}{Memcleanup}{Correct}{True}{Walltime}{Avg}{0.3571680430031847}%
\StoreBenchExecResult{ExperimentCbmcVerificationMemcleanup}{Memcleanup}{Correct}{True}{Walltime}{Median}{0.3571680430031847}%
\StoreBenchExecResult{ExperimentCbmcVerificationMemcleanup}{Memcleanup}{Correct}{True}{Walltime}{Min}{0.3571680430031847}%
\StoreBenchExecResult{ExperimentCbmcVerificationMemcleanup}{Memcleanup}{Correct}{True}{Walltime}{Max}{0.3571680430031847}%
\StoreBenchExecResult{ExperimentCbmcVerificationMemcleanup}{Memcleanup}{Correct}{True}{Walltime}{Stdev}{0E-16}%

\StoreBenchExecResult{ExperimentCbmcVerificationMemcleanup}{Memcleanup}{Wrong}{}{Count}{}{1}%
\StoreBenchExecResult{ExperimentCbmcVerificationMemcleanup}{Memcleanup}{Wrong}{}{Cputime}{}{0.145293317}%
\StoreBenchExecResult{ExperimentCbmcVerificationMemcleanup}{Memcleanup}{Wrong}{}{Cputime}{Avg}{0.145293317}%
\StoreBenchExecResult{ExperimentCbmcVerificationMemcleanup}{Memcleanup}{Wrong}{}{Cputime}{Median}{0.145293317}%
\StoreBenchExecResult{ExperimentCbmcVerificationMemcleanup}{Memcleanup}{Wrong}{}{Cputime}{Min}{0.145293317}%
\StoreBenchExecResult{ExperimentCbmcVerificationMemcleanup}{Memcleanup}{Wrong}{}{Cputime}{Max}{0.145293317}%
\StoreBenchExecResult{ExperimentCbmcVerificationMemcleanup}{Memcleanup}{Wrong}{}{Cputime}{Stdev}{0E-14}%
\StoreBenchExecResult{ExperimentCbmcVerificationMemcleanup}{Memcleanup}{Wrong}{}{Walltime}{}{0.1360151409971877}%
\StoreBenchExecResult{ExperimentCbmcVerificationMemcleanup}{Memcleanup}{Wrong}{}{Walltime}{Avg}{0.1360151409971877}%
\StoreBenchExecResult{ExperimentCbmcVerificationMemcleanup}{Memcleanup}{Wrong}{}{Walltime}{Median}{0.1360151409971877}%
\StoreBenchExecResult{ExperimentCbmcVerificationMemcleanup}{Memcleanup}{Wrong}{}{Walltime}{Min}{0.1360151409971877}%
\StoreBenchExecResult{ExperimentCbmcVerificationMemcleanup}{Memcleanup}{Wrong}{}{Walltime}{Max}{0.1360151409971877}%
\StoreBenchExecResult{ExperimentCbmcVerificationMemcleanup}{Memcleanup}{Wrong}{}{Walltime}{Stdev}{0E-16}%
\StoreBenchExecResult{ExperimentCbmcVerificationMemcleanup}{Memcleanup}{Wrong}{False}{Count}{}{1}%
\StoreBenchExecResult{ExperimentCbmcVerificationMemcleanup}{Memcleanup}{Wrong}{False}{Cputime}{}{0.145293317}%
\StoreBenchExecResult{ExperimentCbmcVerificationMemcleanup}{Memcleanup}{Wrong}{False}{Cputime}{Avg}{0.145293317}%
\StoreBenchExecResult{ExperimentCbmcVerificationMemcleanup}{Memcleanup}{Wrong}{False}{Cputime}{Median}{0.145293317}%
\StoreBenchExecResult{ExperimentCbmcVerificationMemcleanup}{Memcleanup}{Wrong}{False}{Cputime}{Min}{0.145293317}%
\StoreBenchExecResult{ExperimentCbmcVerificationMemcleanup}{Memcleanup}{Wrong}{False}{Cputime}{Max}{0.145293317}%
\StoreBenchExecResult{ExperimentCbmcVerificationMemcleanup}{Memcleanup}{Wrong}{False}{Cputime}{Stdev}{0E-14}%
\StoreBenchExecResult{ExperimentCbmcVerificationMemcleanup}{Memcleanup}{Wrong}{False}{Walltime}{}{0.1360151409971877}%
\StoreBenchExecResult{ExperimentCbmcVerificationMemcleanup}{Memcleanup}{Wrong}{False}{Walltime}{Avg}{0.1360151409971877}%
\StoreBenchExecResult{ExperimentCbmcVerificationMemcleanup}{Memcleanup}{Wrong}{False}{Walltime}{Median}{0.1360151409971877}%
\StoreBenchExecResult{ExperimentCbmcVerificationMemcleanup}{Memcleanup}{Wrong}{False}{Walltime}{Min}{0.1360151409971877}%
\StoreBenchExecResult{ExperimentCbmcVerificationMemcleanup}{Memcleanup}{Wrong}{False}{Walltime}{Max}{0.1360151409971877}%
\StoreBenchExecResult{ExperimentCbmcVerificationMemcleanup}{Memcleanup}{Wrong}{False}{Walltime}{Stdev}{0E-16}%
\ifdefined\CbmcCbmcTotalCount\else\edef\CbmcCbmcTotalCount{0}\fi
\ifdefined\CbmcCbmcCorrectCount\else\edef\CbmcCbmcCorrectCount{0}\fi
\ifdefined\CbmcCbmcCorrectTrueCount\else\edef\CbmcCbmcCorrectTrueCount{0}\fi
\ifdefined\CbmcCbmcCorrectFalseCount\else\edef\CbmcCbmcCorrectFalseCount{0}\fi
\ifdefined\CbmcCbmcWrongCount\else\edef\CbmcCbmcWrongCount{0}\fi
\ifdefined\CbmcCbmcWrongTrueCount\else\edef\CbmcCbmcWrongTrueCount{0}\fi
\ifdefined\CbmcCbmcWrongFalseCount\else\edef\CbmcCbmcWrongFalseCount{0}\fi
\ifdefined\CbmcCbmcErrorTimeoutCount\else\edef\CbmcCbmcErrorTimeoutCount{0}\fi
\ifdefined\CbmcCbmcErrorOutOfMemoryCount\else\edef\CbmcCbmcErrorOutOfMemoryCount{0}\fi
\ifdefined\CbmcCbmcCorrectCputime\else\edef\CbmcCbmcCorrectCputime{0}\fi
\ifdefined\CbmcCbmcCorrectCputimeAvg\else\edef\CbmcCbmcCorrectCputimeAvg{None}\fi
\ifdefined\CbmcCbmcCorrectWalltime\else\edef\CbmcCbmcCorrectWalltime{0}\fi
\ifdefined\CbmcCbmcCorrectWalltimeAvg\else\edef\CbmcCbmcCorrectWalltimeAvg{None}\fi
\edef\CbmcCbmcErrorOtherInconclusiveCount{\the\numexpr \CbmcCbmcTotalCount - \CbmcCbmcCorrectCount - \CbmcCbmcWrongTrueCount - \CbmcCbmcWrongFalseCount - \CbmcCbmcErrorTimeoutCount - \CbmcCbmcErrorOutOfMemoryCount \relax}

%% file: figures/results/tex/manual-commands.tex
\newcommand{\ntasksTermination}{386}
\newcommand{\ntasksTerminationSafe}{309}
\newcommand{\ntasksTerminationUnsafe}{77}
\newcommand{\ntasksNoOverflow}{890}
\newcommand{\ntasksNoOverflowSafe}{615}
\newcommand{\ntasksNoOverflowUnsafe}{275}
\newcommand{\ntasksMemCleanup}{41}
\newcommand{\ntasksMemCleanupSafe}{2}
\newcommand{\ntasksMemCleanupUnsafe}{39}

%% file: sections/abstract.tex
Software verification is a complex problem,
and verification tools need significant tuning to achieve high 
performance.
Due to this, many verifiers choose to specialize 
on reachability properties, or invest the time to implement
known transformations from the given specification to reachability
on their internal representations.
To improve this situation, we provide
transformations as stand-alone components, modifying
the input program instead of the internal representation, enabling
their usage as a preprocessing step by other verifiers.
This way, we separate two concerns:
improving the performance of reachability analyses and
implementing efficient transformations of arbitrary specifications
to reachability.
We implement the transformations in a framework 
that is based on \emph{instrumentation automata},
inspired by the BLAST query language. 
In our initial study, we support three important concrete specifications for C~programs:
\emph{termination}, \emph{no-overflow}, and \emph{memory cleanup}.
Moreover, we discuss the broader expressiveness of our framework and
show how general liveness properties can be transformed to reachability.
We demonstrate the effectiveness and efficiency of our transformations by comparing
verifiers that support the specifications natively
with verifiers for reachability applied on the transformed programs.
The results are very promising:
Our transformations can extend existing verifiers to be effective
on specifications that they do not support natively,
and that the efficiency is often similar to verifiers that natively support the
considered specifications.

%% file: sections/introduction.tex
\section{Introduction}

Software verification is the problem to decide,
for a given program~$P$ and specification~$\varphi$,
whether the program~$P$ satisfies its specification~$\varphi$, in short: $P\models\varphi$.
Due to the complexity of the problem, verification tools need significant tuning to achieve high performance. Because of this, many verifiers choose to specialize on reachability 
properties or invest time to implement known transformations from given specifications to reachability on 
their internal representations.
Considering an overview of all tools for software verification that participated
in the competition on software verification (SV-COMP)~\cite{SVCOMP24},
we see that many verification tools (more than 30) support a basic reachability property
and have focused on tuning their algorithms towards best performance on such tasks.
A lot fewer tools also support other specifications, such as no-overflows or termination,
since adding support for these properties is a time-consuming process.

To address this issue, a software verifier
can be constructed in two ways:
(a)~implement a verification algorithm for~$P\models\varphi$ or
(b)~transform the problem in order to solve it using an existing algorithm.
The transformation-based approach consists of two steps
and assumes that an existing verifier~$v$ supports a specification~$\varphi'$.
In the first step, the problem~$P\models\varphi$ is transformed to a problem~$P'\models\varphi'$,
such that~$P\models\varphi$ holds if and only if~$P'\models\varphi'$ holds.
In the second step, the problem~$P'\models\varphi'$ is solved by the existing verifier~$v$.
The transformation is also called a \emph{reduction}
from~$P\models\varphi$ to~$P'\models\varphi'$.
Several verification tools choose to use transformations~\cite{CBMC,Kratos2,CPACHECKER-3.0-tutorial}
since transformations allow for a separation of concerns:
(1)~implement a rich set of specifications and
(2)~tune the performance of specialized algorithms 
for one particular specification.
For example, given a verifier that supports reachability,
the verifier can be extended to support other specifications, like termination
and no-overflow, for which a transformation to reachability is available.

Currently, every verification tool desiring to benefit from transformations needs to
support further properties requires a re-implementation of the transformation.
The goal of this paper is to show that it is possible
(a)~to construct verifiers in a \emph{modular} way from independent components,
that is, compose an `off-the-shelf' transformation with an `off-the-shelf' verifier,
such that a transformation can be used with arbitrary verifiers for C~programs
to support more kinds of specifications,
(b)~that transformation-based approaches can be even more efficient than native support for specifications, and 
(c)~that the standalone transformations do not necessarily lead to a performance
decrease in comparison to tool-specific implementations for checking
the input specification (with an internal transformation step).

To achieve this, we envision a transformation framework for software verification,
in which developers write specifications at a high-level as \emph{instrumentation automata (IA)}, which contain instructions to monitor the program execution.
They are inspired by the BLAST query language~\cite{BLAST-query} and
\tool{SLIC}~\cite{SLIC}, which is a
specification language for \tool{SLAM}~\cite{SLAM}.
Both specification languages are based on monitor automata
and have shown to be useful in practice,
since the convenient and succinct notation of monitor automata
is often easier to understand than LTL formulas.
Instrumentation automata
observe the state of the program in the same manner as monitor automata:
they raise an error when appropriate.
Monitor automata can be implemented 
either by instrumentation of the monitor into the program code~\cite{BLAST-query,SLIC}
or by an implicit on-the-fly product construction in which the monitor
is a separate analysis component~\cite{SeryBLAST,CPACHECKER}.
Instrumentation automata are very similar to monitor automata,
but they are used to instrument the monitor into the code.

To showcase our approach, we have implemented 
three transformations that reduce
verification problems, for which the specification is to check
\emph{termination}, \emph{memory cleanup}, or (arithmetic) \emph{no-overflows},
to verification problems for which the specification is to check reachability.
Those three kinds of specifications are important and often used,
because we want programs to not hang but progress with useful computation (termination),
we want the programs to free all allocated memory such that
we can use the programs as components as part of other programs (memory cleanup),
and we want the software to not run into undefined behavior, like signed-integer overflow,
and always have values within their types (no-overflows).

In summary, the advantage of our approach is to support a clear separation of concerns,
because the concern of optimizing a verification algorithm (for reachability) is now
completely independent from checking whether a (non-reachability) specification is satisfied.
This opens new opportunities, for example, using testing tools to check for violations of
any specification that can be instrumented into the program.
It is now possible to construct a fuzzing-based non-termination checker without spending any development
effort on the fuzzing tool.
All it takes is to develop the instrumentation automaton for our transformation framework.

\inlineheadingbf{Contributions} We make the following contributions:
\begin{itemize}
   \item We propose a verifier-independent, modular transformation framework for instrumentation of C programs,
      such that verifiers for reachability can be used also for other specifications as well
      (inspired by existing approaches~\cite{BLAST-query,LivenessAsSafety}).
   \item We offer an implementation of our transformation framework in tool \transver~\footnote{\url{https://gitlab.com/sosy-lab/software/transver}} that can be used
      to extend existing verifiers ---in a black-box manner--- to support more specifications
      without changing the code of the existing verifier.
   \item We conduct an experimental evaluation on a large benchmark set of C programs, which shows that
      we can effectively extend existing verifiers to specifications that they
      did not support before (\rqapplicability{}),
      verifiers combined with transformations are performing really well
      and often outperform tools specialized in verifying
      the original property (\rqeffectiveness{}, \rqefficiency{}), and
      it does not pay off to implement transformations inside a verifier,
      that is, verifiers with internal transformations are not necessarily more efficient than
      a composition using our transformation framework and an `off-the-shelf' verifier (\rqmodularity).
\end{itemize}

%% file: sections/related-work.tex
\section{Related Work}

Program transformations have a wide variety
of applications~\cite{ProgramTransformationSystems, ProgramTransformationSystemsSurvey}.
We use the following classification of transformers in 
formal methods~\cite{TransformationGame}: transformers that
(a) simplify a model (in the same language) so that
    it is easier to verify, named \emph{Reducers},
(b) convert a verification task to an equisatisfiable
    task with a different specification, named \emph{Specification Transformers}, and 
(c) expand a model (in the same language) to
    record information for further analysis, called \emph{Instrumentors}.
Our approach is a specification transformer that uses an instrumentor to expand the program.

There are \emph{reducers} that remove
complicated language constructs, for example, by
sequentializing concurrent programs~\cite{CSEQ2013,LAZYCSEQ-ASE},
by using shadow memory~\cite{ShadowMemory2015,ShadowMemory2017},
by reducing the program to a simplified 
syntax~\cite{CIL,CBMC}, or by merging multiple loops 
into one single loop~\cite{DragonBook,KroeningATVA}.
There are also reducers that focus on improving the performance of
the verification process, replacing program constructs
(for example, loops) by constructs that are easier to
verify~\cite{LoopAbstractionCEGAR,CEGAR-PT,
AbstractAccelerationGeneralLinearLoops,Silverman19,CalculusModularLoopAcceleration,
AcceleratingInvariantGeneration}. 
Sometimes they use information from run-time verification
to ease the static analysis~\cite{Bodden2012}.

\emph{Specifications transformers} convert a problem $P\models\varphi$
to a problem $P'\models\varphi'$.
This makes it possible to use algorithms for the verification of $\varphi'$
to also verify $\varphi$~\cite{BLAST-memsafety, EndWatch, LivenessAsSafety, Symbiotic5ms, MetAcsl2019, MetAcsl2021, RPP2017, RPP2018,Kratos2}.
Specifications transformers are also used for testing,
in order to transform a program and a coverage specification to another
program and coverage specification, such that existing tools for
test generation or test-suite analysis can be
used~\cite{TestabilityTransformation,TestabilityTransformation-keynote}.
Our work focuses on this kind of program transformation.
We improve over existing works by two aspects: modularity and generality.
Outputting a modified C program makes the application of any
C~verifier that supports reachability effortless. Moreover,
we demonstrate that our framework supports transformations of multiple 
properties.

\emph{Instrumentors} are used to add code to the program that is used to collect information for further analysis.
A recently proposed instrumentation approach~\cite{InstrumnetationRummer}
can speed up the verification 
of programs containing operations over arrays using ghost variables and rewriting rules. 
The two main differences between our and the mentioned approach are
(a) that we transform the task from other specifications to reachability specifications,
    and the above-mentioned approach transforms programs with extended quantifiers in assertions
    to programs with simpler assertions
and (b) the ordering of instrumentation. 
The approach~\cite{InstrumnetationRummer} provides a set of rewrite rules
that are applied based on the syntax of the program. 
Under some conditions, the instrumentations can only introduce false errors
but will not miss the errors that were in the original program (soundness). 
Therefore, the approach tries multiple orderings of the application of the rewriting rules.
Then, if a counterexample is found, they follow the counterexample-guided abstraction refinement (CEGAR)
approach to try different orders of the rewriting rules.
Our instrumentation framework monitors the syntax of the program and also semantic structures. 
For example, the termination transformation would not be correct if we cannot monitor loops implied by recursive calls of functions or goto jumps. 
Moreover, the instrumentation automata provide an implicit ordering of the instrumentation. 
Hence, we construct only one instrumented program and do not need to refine it based on counterexamples. 

Other approaches use instrumentors to express the verification goal as part of the 
program to be verified. This has been studied
for a variety of applications. One such
example is in the context of
verification witnesses~\cite{WitnessesJournal,VerificationWitnesses-2.0}
in the case of MetaVal~\cite{MetaVal}, which creates a product of
the witness and the program.
Another example is proof-carrying code~\cite{PCC},
where the proof is embedded into the program.
Furthermore, adding the specification
into the program allows for its
run-time monitoring~\cite{EACSL,CCured02,CCured03}.

%% file: sections/background.tex
\section{Background}

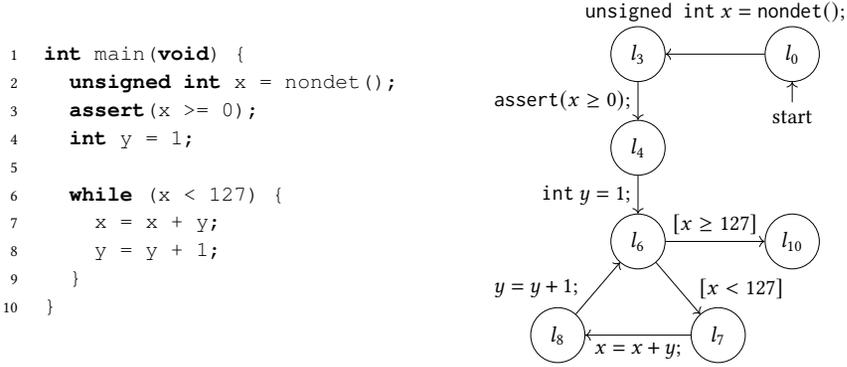
\begin{figure}[t]
    \begin{minipage}{0.45\textwidth}
\begin{lstlisting}[style=c]
int main(void) {
  unsigned int x = nondet();
  assert(x >= 0);
  int y = 1;

  while (x < 127) {
    x = x + y;
    y = y + 1;
  }
}
\end{lstlisting}
    \end{minipage}
    \begin{minipage}{0.45\textwidth}
        \resizebox{0.8\textwidth}{!}{
            \input{figures/examples/example-CFA.tex}    
        }
    \end{minipage}
    \caption{An example program (left) with a corresponding 
    CFA (right)}
    \label{fig:background:running-example}
\end{figure}

\inlineheadingbf{Control-Flow Automata (CFA)}
We model the control-flow of C programs as
\emph{control-flow automaton}~\cite{HBMC-dataflow}. 
A CFA is a tuple $(L, l_0, G)$, where $L$ is a finite set of locations (or nodes),
$l_0\in L$ is the initial location and $G\subseteq L\times Opt\times L$
is a set of edges between locations, which represent the
operations in the program.
We consider the following special functions; \texttt{nondet()} which
returns a nondeterministic value
and \texttt{assert($\pi$)} describes a safety condition $\pi$
on the variables that should always hold at its location.
In the case of memory-cleanup transformation, we also allow
calls to the standard memory-allocation functions
\texttt{malloc}, \texttt{calloc}, \texttt{realloc} and \texttt{free}.
\Cref{fig:background:running-example} shows an
example of a program and a corresponding CFA.
In practice, we also store the information about type
of variables and expressions used on the edges of CFA.
In this paper, a \emph{state} of a program is a 
mapping of the variables to values.

We assume that the CFA is constructed from a C program
such that every edge contains at most a single operation as
is done by \cpachecker~\cite{CPACHECKER}.

\inlineheadingbf{Specifications}
One of the most prominent specification languages for behavioral properties
is linear-time temporal logic (LTL)~\cite{HBMC-TemporalLogic}.
One standard way to implement a model checker for LTL is
to transform the LTL formula into a Büchi automaton
and check the product of its complement and the model of the implementation for emptiness.
Several transformation tools that transform specifications from LTL to Büchi automata
are available (e.g., Spot~\cite{SPOT} and LTL2BA~\cite{LTL2BA}).
There are also model checkers that offer to write down specifications directly as
monitor automata, such as \blast~\cite{BLAST} with the BLAST query language~\cite{BLAST-query}
or \tool{SLAM} with its specification language \tool{SLIC}.

\begin{table}[t]
    \setlength{\tabcolsep}{3.5pt}
    \setlength{\multilen}{9cm}
    \caption{Specifications considered in this work}
    \label{tab:specifications}
    \centering
\begin{tabular}{ll}
    \toprule
    Specification & Explanation \\
    \midrule
    \emph{reachability} & \multi{No execution of the program violates condition $\pi$ at the location with operation \ttbox{assert($\pi$)}.} \\
    \emph{no-overflow} & \multi{No execution of the program produces a signed integer result of an operation whose value is outside the range of the C~type \ttbox{int}.} \\
    \emph{termination} & \multi{Every execution of the program eventually terminates.} \\
    \emph{memory cleanup} & \multi{No execution of the program allocates a pointer and then finishes without freeing it, or frees the same pointer twice.} \\
    \bottomrule
    \emph{explicit liveness} & \multi{Every infinite execution of the program satisfies $\pi$ at the location with \ttbox{assert\_live($\pi$)} infinitely often.} \\
    \bottomrule
\end{tabular}
\end{table}

In the International Competition on Software Verification (SVCOMP)~\cite{SVCOMP24}
tools compete in verifying multiple practical LTL properties\footnote{\url{https://sv-comp.sosy-lab.org/2024/rules.php}}.
\Cref{tab:specifications} lists the properties from \svcomp, for which we demonstrate transformations
within our framework in~\cref{sec:showcase}.
Coming back to LTL, there are two important subclasses of LTL formulas: \emph{safety} and \emph{liveness}. 
Any LTL formula can be decomposed into a conjunction of subformulas of these types~\cite{PETRICMARETIC2014408}. 
Furthermore it has been shown even liveness properties, an important subclass of 
LTL properties, can be reduced to safety~\cite{LivenessAsSafety}.
Therefore, if a verifier supports general safety specifications, in principle it can verify any LTL formula.

\emph{Safety} properties hold for every reachable state in every execution.
They are violated if there is a reachable state that violates this property.
In SV-COMP the most general version of this property is reachability.
It is a very modular property, since one can explicitly
encode multiple properties as assertions.
Therefore, it is very natural to express other specifications as reachability.
Further reasons to reduce specifications to this property is the high interest in reachability 
by the software verification~\cite{SVCOMP24}, hardware
verification~\cite{McMillanCraig,IC3}, and automated testing community~\cite{AFLfast, FUSEBMC-long,TESTCOMP24}.

\emph{Liveness} properties describe desirable properties on states that should be satisfied by every infinite execution infinitely often. 
A liveness property is violated if there exists an infinite execution that, from some point, never satisfies the specified property on the states.
The main example of this property in \svcomp is termination.
However, \svcomp does not introduce a general liveness property similar to reachability for safety. 
Therefore, we define \emph{explicit liveness} that expresses the conditions on the states explicitly with special labels similar to \texttt{assert} for reachability.
While liveness properties may appear more complex than safety properties,
they can be reduced to them for any finite system~\cite{LivenessAsSafety}.
We show, how to perform such transformation in our framework in \cref{sec:liveness}.
Therefore it is feasible for verifiers to concentrate on safety
properties and use the transformation to verify liveness properties.

%% file: figures/examples/example-CFA.tex
    \begin{tikzpicture}[]
        \node[state] (l2) at (2.5, 3) {$l_0$};
        \node[] (l1) at (2.5, 2) {start};
        \node[state] (l3) at (0, 3) {$l_3$};
        \node[state] (l4) at (0, 1.5) {$l_4$};
        \node[state] (l7) at (0, 0) {$l_6$};
        \node[state] (l8) at (1.3, -1.5) {$l_7$};
        \node[state] (l9) at (-1.3, -1.5) {$l_8$};
        \node[state] (l10) at (2.5, 0) {$l_{10}$};
        \path[->,auto]
        (l1) edge node[above]{} (l2)
        (l2) edge node[above=0.4]{$\texttt{unsigned int}~x = \texttt{nondet}();$} (l3)
        (l3) edge node[left]{$\texttt{assert}(x \geq 0);$} (l4)
        (l4) edge node[left]{$\texttt{int}~y = 1;$} (l7)
        (l7) edge node[right=0.2]{$[x < 127]$} (l8)
        (l7) edge node[above]{$[x \geq 127]$} (l10)
        (l8) edge node[below]{$x = x + y;$} (l9)
        (l9) edge node[left=0.2]{$y = y + 1;$} (l7);
    \end{tikzpicture}

%% file: sections/transformation.tex
\section{Transformation Framework}\label{sec:framework}

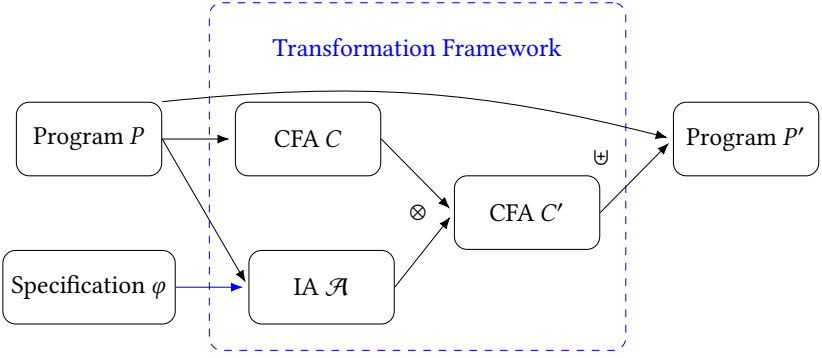
\begin{figure}[t]
    \centering
    \resizebox{0.8\textwidth}{!}{
        \input{figures/examples/framework.tex}
    }
    \caption{Workflow of the program transformation}
    \label{fig:workflow}
\end{figure}

\Cref{fig:workflow} presents the workflow of our transformation
framework. It takes as an input a program $P$ and
a specification $\varphi$. 
The user first needs to formalize the property in the
context of instrumentation automata (IA) and implement it inside
\cpachecker, marked by the blue arrow. 
We show how to do the formalization 
for three properties in \cref{sec:showcase}.
The framework
then transforms the program $P$ into an equivalent CFA $C$
and combines it with 
the IA $\mathcal{A}$ to obtain a new CFA $C'$ using the sequentialization 
operator $\otimes$.
Finally, the resulting CFA $C'$ is used to instrument the original
program $P$ with the instrumentation operator $\uplus$ preserving the 
structure of the original program as much as possible.

\inlineheadingbf{Instrumentation Automata}
An instrumentation automaton specifies how an input program
needs to be instrumented in order to make the original property
explicit using assertions. It is inspired by the observer automata from 
the BLAST query language~\cite{BLAST-query}. 
The observer automata are expressive enough for all temporal safety and liveness properties, and their
syntax resembles C programs, which makes them convenient to use for
software engineers. They allow syntax pattern matching
of the operations used in a given C program followed up by the corresponding
event actions or assertions. 
To be able to transform more complex properties,
we allow the pattern matching over the structure of the CFA, which considers not only the syntax of the program but also its semantics. 
It enables matching on the semantic loops (implied by recursive calls or goto jumps).
Therefore, we extend our modeling formalism to queries over
locations in the CFA.

An \emph{instrumentation automaton} is a tuple $(Q, q_0, \mathit{Var}, \delta, \alpha)$,
where $Q$ is a set of states,
$q_0$ is an initial state,
$\mathit{Var}$ is a set of variables of the automaton,
$\delta\subseteq Q\times\mathit{PATTERNS}
\times\mathit{Opt}\times\{A,B\}\times Q$ is a transition relation,
and $\alpha$ is a state annotation.
Transition $(q, \rho, op, X, p)$ from state $q$ to state $p$ 
specifies:
\begin{itemize}
    \item Pattern $\rho$ is a C expression used to match an 
        operation on a CFA edge while allowing terms like
        $\$x_0$, to match variables from the CFA and then read their value in $op$.
        Similarly, as for regular expressions, we define a special pattern 
        $.*$ that matches any number of any symbols.
        For example, $\rho=.*\$x_0 + \$x_1.*;$ matches an expression that contains
        plus and it passes its operands to $op$.
        Further, special patterns \texttt{cond} and \texttt{!cond} match
        a condition and its negation in a location with branching.
    \item Operation $op$ can read and write to variables from $\mathit{Var}$,
        but it can only read values from CFA variables matched by $\rho$. In particular,
        it can introduce assertions.
    \item Symbol $X\in\{A,B\}$ specifies whether the operation 
    should be placed before (B) or after (A) the matched edge in the CFA.
\end{itemize}
The state annotation $\alpha: Q\rightarrow \{\texttt{true}, \texttt{loop\_head}, \texttt{init}, \texttt{end}\}$ 
assigns a predicate to every state. 
Currently, four predicates are used: 
\texttt{true} holds for any location, 
\texttt{loop\_head} is true for locations at the beginning of a loop,
\texttt{init} is true for any initial location, and
\texttt{end} is true for any final location and corresponds to
a final \texttt{at\_exit} call.

\begin{algorithm}[t]
    \caption{Sequentialization operator $\otimes$}
    \label{alg:sequentialization}
    \input{figures/examples/sequentialization-new.tex}
\end{algorithm}

\inlineheadingbf{Sequentialization Operator}\label{ssec:sequentialization-operator}
The sequentialization operator $\otimes$ is responsible for taking the operations
from a given IA and placing them in the indicated places in an input CFA. 
The operator implicitly 
traverses both the CFA and IA in parallel. 
During the traversal, it
checks for states that are matched by the locations in the annotation $\alpha$ and 
inserts the operations from the transitions in the IA that are matched with the
transitions in the CFA.
Whenever there is a match of the transitions, it progresses only in the IA by
including all the successors of the matched transitions of a state into a waitlist. 
If there is no match for an edge in the CFA, the algorithm traverses only the CFA.

The sequentialization operator is realized by the 
procedure in \cref{alg:sequentialization}. 
The algorithm starts with the initialization step, 
where it traverses the CFA and collects additional 
information about the program that can be then used to 
construct specific instrumentation automata for a 
given program. For example, a transformation for termination 
initializes one automaton (\cref{fig:showcase-termination}) per loop in
the CFA, 
collects all the variables used in the respective loop, 
and initializes one ghost variable for each. 
Another example is the liveness automaton (\cref{fig:liveness}), which 
initializes one variable $l_i'$ per each 
\texttt{assert_live}. Moreover, 
\texttt{initialize\_automata} also returns 
a location from the CFA for each automaton, 
labeled as the initial location for the automaton. 
This optimization is used to skip parts of the CFA, 
for example it is used for the 
automatan for termination to only monitor its own loop.

The while
loop in \cref{line:comoposition-loop} traverses both 
the input CFA and the IA in parallel. 
States and all the other components from automaton 
$\mathcal{A}_i$, $0 \leq i \leq k$ are marked with 
$i$ in the superscript.
In one iteration,
it processes one pair of a location and a state $(l,q^i)$.

It processes each pair $(l,q^i)$ by first computing the new pairs locations and states that have not been visited before, 
as seen in \cref{line:new-pairs}. 
For this it uses the function \texttt{succ} shown in~\cref{eq:succ}.

\begin{equation}\label{eq:succ-IA}
    \texttt{succ\_IA}((l, q^i))=\{(l, p^i)\mid \exists (l, op^{C}, l')\in G, \exists (q^i, \rho, op^{\mathcal{A}_i}, X, p^i)\in\delta^i . \texttt{match}(op^C,\rho)\}  
\end{equation}

\begin{equation}\label{eq:succ}
    \texttt{succ}((l, q^i)) =
    \begin{cases}
      \{(l', q^i)\mid \exists (l, op, l')\in G & if~\neg\alpha^i(q^i)(l)\lor\texttt{succ\_IA}(l, q^i)=\{\} \\
      \texttt{succ\_IA}(l, q^i) & otherwise\\
    \end{cases}
\end{equation}

To compute the successors using~\cref{eq:succ} there are two cases. 
In the first case, the annotation of the state does not hold 
for the location, or none of the outgoing edges from $q^i$ matches an outgoing edge from $l$. 
Therefore, the algorithm progresses only in the CFA and creates new pairs of $q^i$ with all the successors of $l$. 
In the second case, the annotation holds $\alpha^i(q^i)(l)$, and there are transitions from $q^i$ with pattern that matches some edges from $l$. 
The case results in progressing only with the IA and pairing all the successors of the transitions that matched some edge with $l$, which is
expressed through the function \texttt{succ\_IA} in \cref{eq:succ-IA}.

The condition in \cref{line:new-edges} is responsible for adding new edges into the resulting CFA. 
Similarly, as in the first case of successors function \texttt{succ}, 
if no outgoing edge is matched or the state annotation does not hold for the location, 
the algorithm adds all the edges from the input CFA C. 
In the else case, the function \texttt{NE} (new edges) presented 
in \cref{eq:new-edges} computes all the edges that instrument the original program.

Inside the function in \cref{eq:new-edges} the predicate 
$MV$ (\emph{Matched Variables}) presented in \cref{eq:mv} 
is true if it the two operations match through the correct substitution 
of the placeholder variable.
This is done by seeing if an operation $op^C$ from a CFA edge and an operation $op^{\mathcal{A}_i}$ match. 
Where $op^{\mathcal{A}_i}$ is constructed by replacing wildcard placeholders 
in the operation $\overline{op}^{\mathcal{A}_i}$ taken from a transition in the IA with the matched variables. 
For example, let us assume CFA edge $(l, y = z + 42, l')$ and an IA transition $(q, .*\$x1+\$x2.*, \texttt{assert}(x1 > x2), B, q')$, then 
$MV(y = z + 42, \texttt{assert}(z > 42), B)$ evaluates to true. Function $\texttt{NE}$ uses the predicate $MV$ to get the matched operations with replaced variables based on the pattern $\rho$. 
Further, the algorithm places the new operation after ($A$) or before ($B$) the edge in the original CFA C. 
It adds a new location $l^{new}$ and either orders the operations $op^{\mathcal{A}_i}, op^C$ ($A$) or $op^C, op^{\mathcal{A}_i}$ ($B$). 
Lastly, the algorithm adds the new locations into $C'$ in line \cref{line:new-states}.

\begin{align}\label{eq:mv}
    \begin{split}
        MV(op^C, op^{\mathcal{A}_i},X)=\exists(q^i, \rho, \overline{op}^{\mathcal{A}_i}, X', p^i)\in\delta^i\,.\,
        & \texttt{match}(op^{C},\rho) \land \\
        & X = X' \land \\
        & op^{\mathcal{A}_i} = \texttt{vars}(op^C,\overline{op}^{\mathcal{A}_i},\rho)
    \end{split} 
\end{align}

\begin{equation}\label{eq:new-edges}
    \texttt{NE}((l, q^i)) = 
    \bigcup_{\substack{(x,y,X)\in\\ \{(\mathcal{A}_i,C,B),\\(C,\mathcal{A}_i,A)\}}}
    \{(l, op^x, l^{new}),(l^{new}, op^y, l')\mid (l, op^{C}, l')\in G~\land~MV(op^C, op^{\mathcal{A}_i},X)\}
\end{equation}

\inlineheadingbf{Instrumentation Operator}
The instrumentation operator $\uplus$ takes as an input
the output CFA from the sequentialization operator and the
original program. It iterates through every statement of the program
and checks if there are some newly inserted operations in the 
corresponding CFA node. It puts them before or after the 
operation in the original program based on their order in the input CFA.

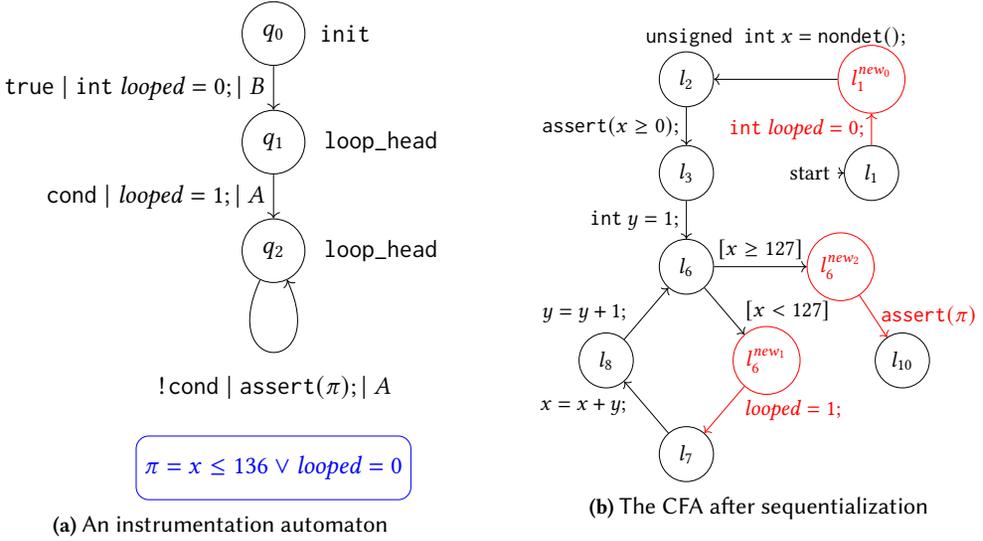
\begin{figure}[t]
    \label{fig:transformation:running-example}
    \centering
    \subfloat[An instrumentation automaton]{%
        \begin{minipage}{0.485\textwidth}
            \centering
            \resizebox{0.9\textwidth}{!}{
                \input{figures/examples/example-EOA.tex}%
            }
            \label{subfig:transformation:running-example:example-EOA}
        \end{minipage}%
    }\hfill
    \subfloat[The CFA after sequentialization]{%
        \begin{minipage}{0.485\textwidth}
            \centering
            \resizebox{0.9\textwidth}{!}{
                \input{figures/examples/example-Sequentialization.tex}%
            }
            \label{subfig:transformation:running-example:cfa}
        \end{minipage}%
    }
    \vspace{-0.2cm}
    \caption{An example IA (left) and 
    the corresponding CFA after sequentialization with the CFA from
    \cref{fig:background:running-example} (right)}
\end{figure}

\inlineheadingbf{Example}
Consider the program from \cref{fig:background:running-example} and the
property \emph{"If the execution of a program does at least one
iteration of the loop, the value of x will always be at most 136"}. The
property can be formalized  as an 
instrumentation automaton~\cref{subfig:transformation:running-example:example-EOA}. 
Applying the 
sequentialization operator 
to the IA \cref{subfig:transformation:running-example:example-EOA} 
and the CFA from \cref{fig:background:running-example} 
produces the CFA \cref{subfig:transformation:running-example:cfa}.
The function \texttt{initeliaze\_automata} returns the same IA as 
there is nothing to be initialized in this example. It then
initialize \textit{waitlist} with the only pair $(l_1, q_0)$.
The most important explored pairs are $(l_1, q_0),~(l_6,q_1)$ and $(l_6,q_2)$ as
they are the only pairs for which the annotation $\alpha$ holds.
For example, let us focus on the edge $(l_6, [x < 127], l_7)$ and the
transition from $q_1$ to $q_2$. $MV([x < 127], \mathit{looped}=1;A)$ is satisfied
because the edge matches pattern \texttt{cond}, therefore 
\texttt{new\_edges}$((l_6,q_1)) = \{(l_6, [x < 127], l^{new}_6),(l^{new}_6, \textit{looped}=1;, l_7)\}$
The output program after $\uplus$ applied on the
CFA in \cref{subfig:transformation:running-example:cfa} and the 
original program from \cref{fig:background:running-example}
results in \cref{fig:transformation:instrumented}.

\begin{figure}[t]
    \lstset{commentstyle=\color{commentgreen}}
    \begin{minipage}{0.8\textwidth}
\begin{lstlisting}[style=c]
int main(void) {
  int looped = 0; // added
  unsigned int x = nondet();
  x = x * x;
  int y = 1;
  assert(x >= 0);

  while (x < 127) {
    looped = 1; // added
    x = x + y;
    y = y + 1;
  }
  assert(x <= 136 || looped == 0); // added
}
\end{lstlisting}
    \end{minipage}
    \vspace{-0.4cm}
    \caption{An example of an instrumented program}
    \label{fig:transformation:instrumented}
    \vspace{0.1cm}
\end{figure}

%% file: figures/examples/framework.tex
\begin{tikzpicture}[
    node distance=1cm,
    box/.style={draw, rectangle, minimum width=2cm, minimum height=1cm, rounded corners=5pt},
    arrow/.style={-Latex, shorten >=2pt},
    background/.style={draw=blue, rounded corners=5pt, inner sep=10pt, fit=#1, dashed}
]

\node[box] (program) {Program $P$};
\node[box, below=of program] (specification) {Specification $\varphi$};

\node[box, right=of program] (cfa) {CFA $C$};
\node[box, right=of specification] (eoa) {IA $\mathcal{A}$};
\node[box, right=of cfa, yshift=-1cm] (cfap) {CFA $C'$};
\node[box, right=of cfap, yshift=1cm] (programp) {Program $P'$};
\node[left=of cfap, xshift=.75cm] (Sequentialization) {$\otimes$};
\node[left=of programp, xshift=.25cm, yshift=-.25cm] (instrumentation) {$\uplus$};
\node[color=blue,above=of cfa, xshift=1.50cm, yshift=-.50cm] (framework) {Transformation Framework};

\begin{scope}[on background layer]
    \node[background={(cfa)(cfap)(eoa)(framework)}] (back) {};
\end{scope}


\draw[arrow] (program.east) -- (cfa.west);
\draw[arrow] (program.east) -- (eoa.west);
\draw[arrow,color=blue] (specification.east) -- (eoa.west);

\draw[arrow] (cfa.east) -- (cfap.west);
\draw[arrow] (eoa.east) -- (cfap.west);
\draw[arrow] (program.north east) to[bend left=10] (programp.west);

\draw[arrow] (cfap.east) -- (programp.west);

\end{tikzpicture}

%% file: figures/examples/sequentialization-new.tex
\begin{algorithmic}[1]
    \REQUIRE
    a CFA $C = (L, l_0, G)$, an IA $\mathcal{A} = (Q, \mathit{Var}, \delta, q_0, \alpha)$
    \ENSURE
    CFA $C' = (L', l_0', G')$ which can be used for the instrumentation of the original program

    \STATE $(\mathcal{A}_0, l_{init}^0),(\mathcal{A}_1, l_{init}^1),\dots,(\mathcal{A}_k, l_{init}^k) \gets \texttt{initialize\_automata}(C, \mathcal{A});$ \label{line:initialization}
    \STATE $L', G' \gets \{l_0'\}, \{\};$
    \STATE $\mathit{waitlist}, \mathit{finished} \gets \{(l_{init}^0, q_0^0), (l_{init}^1, q_0^1),\dots,(l_{init}^k, q_0^k)\}, \{\};$

    \WHILE{$\mathit{waitlist} \neq \emptyset$} \label{line:comoposition-loop}
      \STATE $(l,q^i) \gets \mathit{waitlist}.\mathit{pop}();$
      \IF {($(l,q^i) \in \mathit{finished}$)}
          \STATE $\texttt{continue};$
      \ENDIF
      \STATE $\mathit{finished}.add((l,q^i));$
      \STATE $\mathit{waitlist} \gets \mathit{waitlist}\cup \texttt{succ}((l, q^i));$ \label{line:new-pairs}
      \IF {$\neg\alpha^i(q^i)(l)\lor\texttt{succ\_IA}(l, q^i)=\{\}$} \label{line:new-edges}
        \STATE $G'\gets G' \cup \{(l, op^C, l')\in G\};$ 
      \ELSE
        \STATE $G'\gets G'\cup \texttt{NE}((l, q^i));$ 
      \ENDIF
      \STATE $L'\gets L'\cup \{l^{new}\mid l^{new}\notin L' \land \exists l',op \big((l',op,l^{new})\in G' \lor (l^{new},op,l')\in G'\big)\};$ \label{line:new-states}
    \ENDWHILE
    \RETURN $(L', l_0', G');$
\end{algorithmic}

%% file: figures/examples/example-EOA.tex
    \begin{tikzpicture}[]
        \node[state,rectangle,color=blue, rounded corners=5pt] (legend) at (0, -3.5) {$\pi = x\leq 136 \lor \mathit{looped}=0$};
        \node[state] (l1) at (0, 2.5) {$q_0$};
        \node[] (a1) at (1, 2.5) {$\texttt{init}$};
        \node[state] (l2) at (0, 1) {$q_1$};
        \node[] (a1) at (1.5, 1) {$\texttt{loop\_head}$};
        \node[state] (l3) at (0, -0.5) {$q_2$};
        \node[] (a1) at (1.5, -0.5) {$\texttt{loop\_head}$};
        \path[->,auto]
        (l1) edge node[left]{$\texttt{true}\mid \texttt{int}~\mathit{looped}=0;
                             \mid B$} (l2)
        (l2) edge node[left]{$\texttt{cond}\mid \mathit{looped}=1;
                             \mid A$} (l3)
        (l3) edge[out=245,in=295,looseness=10] node[below=0.2]{$\texttt{!cond}
                              \mid \texttt{assert}(\pi); \mid A$} (l3);
    \end{tikzpicture}

%% file: figures/examples/example-Sequentialization.tex
    \begin{tikzpicture}[]
        \node[state] (l0) at (3, 1.5) {$l_1$};
        \node[] (l) at (2, 1.5) {start};
        \node[state,color=red] (l1) at (3, 3) {$l_1^{\mathit{new}_0}$};
        \node[state] (l2) at (0, 3) {$l_2$};
        \node[state] (l3) at (0, 1.5) {$l_3$};
        \node[state] (l4) at (0, 0) {$l_6$};
        \node[state,color=red] (l6) at (1.3, -1.5) {$l_6^{\mathit{new}_1}$};
        \node[state] (l7) at (0, -3) {$l_7$};
        \node[state] (l8) at (-1.3, -1.5) {$l_8$};
        \node[state,color=red] (l9) at (2.5, 0) {$l_{6}^{\mathit{new}_2}$};
        \node[state] (l10) at (3.5, -1.5) {$l_{10}$};
        \path[->,auto]
        (l) edge node[left]{} (l0)
        (l0) edge[color=red] node[left,color=red]{$\texttt{int}~\mathit{looped} = 0;$} (l1)
        (l1) edge node[above=0.4]{$\texttt{unsigned int}~x = \texttt{nondet}();$} (l2)        
        (l2) edge node[left]{$\texttt{assert}(x \geq 0);$} (l3)
        (l3) edge node[left]{$\texttt{int}~y = 1;$} (l4)
        (l4) edge node[right=0.2]{$[x < 127]$} (l6)
        (l6) edge[color=red] node[right=0.2,color=red]{$\mathit{looped} = 1;$} (l7)
        (l7) edge node[left=0.2]{$x = x + y;$} (l8)
        (l8) edge node[left=0.2]{$y = y + 1;$} (l4)
        (l4) edge node[above]{$[x \geq 127]$} (l9)
        (l9) edge[color=red] node[right,color=red]{$\texttt{assert}(\pi)$} (l10);
    \end{tikzpicture}

%% file: sections/showcase.tex
\section{Specifications as Instrumentation Automata}\label{sec:showcase}
To study the performance of reachability analyzers when applied to
different specifications, we focus our experiments to 
the transformation of
three specifications from \svcomp~\cite{SVCOMP24}.
This section shows how to formalize them as IA.

\inlineheadingbf{No-Overflow}
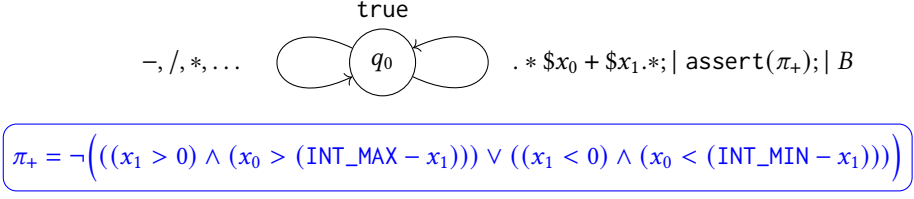
\begin{figure}[t]
    \centering
    \input{figures/examples/no-overflow.tex}
    \caption{An excerpt of the IA for the \emph{no-overflow} property}
    \label{fig:showcase-no-overflow}
\end{figure}
According to \svcomp, the specification \emph{no-overflow} is violated by
a given program if there exists an execution of the program that
executes an operation with a signed-integer result,
but the resulting value does not fit within the range of the signed integer. 
To simplify the instrumentation, let us assume that a program consists only of
signed integer variables and, at most, one arithmetic operation per line.
In our implementation, we enforce these assumptions with \cpachecker~\cite{CPACHECKER} 
by
tracking the types of the variables and the expressions on the edges of the CFA, and instrumenting only
the operations with signed-integer results.
We also decompose complex arithmetic expressions
into multiple CFA edges, each containing one operation, using intermediate
results.
\Cref{fig:showcase-no-overflow} shows the IA
corresponding to the no-overflow specification.
We display only the transition for addition
since the transitions for the other operations are analogous.
The automaton matches every operation and adds the corresponding
condition as an \texttt{assert} before the operation. For example,
before doing an addition, the result of $x_0 + x_1$ should not be larger than
\texttt{INT\_MAX} and if $x_1$ is negative, then $x_0 + x_1$
should not be smaller than \texttt{INT\_MIN}.
The operations
together with the necessary conditions to prevent the overflows
can be found listed online~\cite{overflow-conditions}.
Assuming that we handle every arithmetic operation with the resulting
type of signed integer, this transformation is sound and complete.

\begin{figure}[t]
    \centering
    \input{figures/examples/termination.tex}
    \caption{An example of an IA for the \emph{termination} property}
    \label{fig:showcase-termination}
\end{figure}
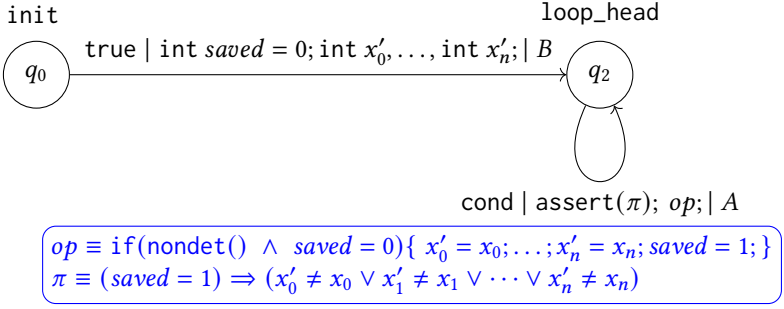

\inlineheadingbf{Termination}
The representation of the \emph{termination} property as an IA
is shown in \cref{fig:showcase-termination} and based
upon the work of Schuppan and Biere~\cite{LivenessAsSafety}, who
propose a transformation of liveness properties to safety properties
for finite systems.
To find an infinite execution, the instrumentation
monitors the visited states of the execution.
If a state is encountered twice inside a loop, a non-terminating
execution has been found.
As a preprocessing step in \cref{line:initialization}, our implementation 
traverses the whole CFA and initializes an automaton for each loop.
For every loop, it collects all the used variables and initializes
their shadow copies $x_0',\dots,x_n'$. In practice, we use the 
information about the types of variables stored in the edges of CFA 
and can monitor even variables with more complex types like pointers.
The loop-head of the loop is the initial location for every such automaton.
Each time the loop-head is visited, the instrumented program can 
make a non-deterministic choice to save the state if no state
has been saved before. This is performed by the operation $op$.
If the state was saved previously, an assertion 
ensures that the current state is different. 
The violation of the assertion means that the execution
encountered the same state twice, and hence, it can make the same 
non-deterministic decisions to repeat the loop infinitely often.
Notice that the transformation is complete but not sound if we consider 
dynamic structures like linked lists because such programs can 
have an infinite execution without visiting the same state twice.
For programs with 
variables of the types with finite ranges, this transformation is sound and complete.

\begin{figure}[t]
    \centering
    \input{figures/examples/mem-cleanup.tex}
    \caption{An example of an IA for the \emph{memory cleanup} property}
    \label{fig:showcase-mem-cleanup}  
\end{figure}
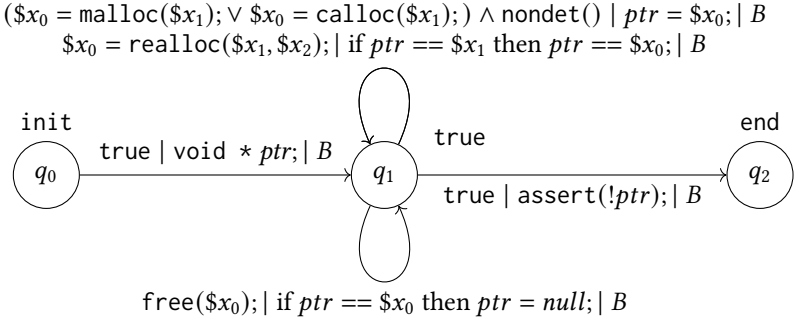

\inlineheadingbf{Memory Cleanup}
\Cref{fig:showcase-mem-cleanup} shows an IA for this property. 
It nondeterministically decides to track the pointer being allocated
and checks if the tracked pointer has been deallocated when the program terminates.
The transition from $q_0$ to $q_1$ initializes the tracking of the pointer. 
The result of every memory allocating function i.e. 
\texttt{malloc}, \texttt{calloc} is nondeterministically assigned to the tracking pointer \texttt{ptr}. For \texttt{realloc}, we first see if the
pointer being reallocated is the same as the one being tracked. If it is, we update the tracking pointer.
These checks are handled by the upwards facing self-loop.
When freeing memory by using $\texttt{free}$, we have to see if the 
pointer is the one currently being tracked. If it is, we set it to null i.e. we are currently not tracking any pointer.
Finally, when exiting the program, demonstrated through the node $q_2$, we see if the tracking pointer is null i.e. all allocated memory has been deallocated. This is handled by the edge between the nodes $q_1$ and $q_2$.
Since we handle all the relevant standard functions for memory
allocation, this transformation is sound and complete.

%% file: figures/examples/no-overflow.tex
\begin{tikzpicture}[]
    \node[state,rectangle,color=blue, rounded corners=5pt] (legend) at (4, 0.75) {$\pi_+ = \neg\Big(((x_1>0)\land(x_0>(\texttt{INT\_MAX}-x_1)))
    \lor ((x_1<0)\land(x_0<(\texttt{INT\_MIN}-x_1)))\Big)$};
    \node[state] (q1) at (3, 2) {$q_0$};
    \node[] (b1) at (3, 2.7) {$\texttt{true}$};
    \path[->,auto]
    (q1) edge[out=335,in=385,looseness=10] node[right=0.2]{$.*\$x_0 + \$x_1.*;
                            \mid \texttt{assert}(\pi_+); \mid B$} (q1)
    (q1) edge[out=155,in=205,looseness=10] node[left=0.3]{$-,/,*,\dots$} (q1);
\end{tikzpicture}

%% file: figures/examples/termination.tex
\begin{tikzpicture}[]
    \node[state] (q1) at (3.5, 0) {$q_2$};
    \node[state,rectangle,color=blue, rounded corners=5pt, text width=9.6cm] (legend) at (1,-2.5) 
    {$op \equiv \texttt{if}(\texttt{nondet}()~\land~\mathit{saved}=0) \{~x_0'=x_0;\dots;x_n'=x_n;\mathit{saved}=1;\}\newline\pi \equiv(\mathit{saved} = 1) \implies (x_0' \neq x_0 \lor x_1' \neq x_1\lor\dots\lor x_n'\neq x_n)$};
    \node[] (a1) at (3.5, 0.8) {$\texttt{loop\_head}$};
    \node[state] (q2) at (-4, 0) {$q_0$};
    \node[] (a2) at (-4.05, 0.8) {$\texttt{init}$};
    \path[->,auto]
    (q1) edge[out=245,in=295,looseness=10] node[below]{$\texttt{cond} \mid \texttt{assert}(\pi);~op; \mid A$} (q1)
    (q2) edge node[above]{$\texttt{true}\mid \texttt{int}~saved=0; \texttt{int}~x_0',\dots,\texttt{int}~x_n';
                        \mid B$} (q1);
\end{tikzpicture}

%% file: figures/examples/mem-cleanup.tex
\begin{tikzpicture}[]
    \node[state] (l1) at (0, -2) {$q_0$};
    \node[] (a1) at (0, -1.3) {$\texttt{init}$};
    \node[state] (l2) at (4.5, -2) {$q_1$};
    \node[] (a1) at (5.5, -1.5) {$\texttt{true}$};
    \node[state] (l3) at (9.5, -2) {$q_2$};
    \node[] (a1) at (9.5, -1.3) {$\texttt{end}$};
    \path[->,auto]
    (l1) edge node[above]{$\texttt{true}\mid \texttt{void *}~\mathit{ptr};
                            \mid B$} (l2)
    (l2) edge[out=65,in=115,looseness=10] node [above] {$\$x_0=\texttt{realloc}(\$x_1, \$x_2);\mid 
    \text{if } ptr == \$x_1 \text{ then } ptr == \$x_0;  \mid B$} (l2)
    (l2) edge[out=65,in=115,looseness=10] node [above=0.4] {$(\$x_0=\texttt{malloc}(\$x_1); \lor ~\$x_0=\texttt{calloc}(\$x_1);) \land \texttt{nondet}()\mid 
    ptr = \$x_0; \mid B$} (l2)
    (l2) edge[out=245,in=295,looseness=10] node[below]{$\texttt{free}(\$x_0); \mid \text{if } ptr == \$x_0 \text{ then } ptr = null; \mid B$} (l2)
    (l2) edge[above] node[below] {$\texttt{true} \mid \texttt{assert}(!ptr); \mid B$} (l3);
\end{tikzpicture}

%% file: sections/expressiveness.tex
\section{Expressiveness Beyond the Presented Transformations}
\label{sec:liveness}

So far, we have shown the transformation of 
three practical specifications to reachability using 
instrumentation automata. However, the question 
arises of whether the framework is general enough to 
unite the transformations of a larger class of 
specifications into reachability. 

In the verification community, linear-time temporal logic (LTL)~\cite{HBMC-TemporalLogic} 
is one of the most prominent languages for expressing specifications. 
There are two important subclasses of LTL formulas: (a) safety 
formulas express that a desired property always holds, 
whereas (b) liveness formulas express that the property 
will eventually hold in every execution.
Expressing properties from these classes is relevant 
because as Maretić, Dashti, and Basin~\cite{PETRICMARETIC2014408} have shown, any 
LTL formula is decomposable into a conjunction of safety and liveness formulas. 
Therefore, supporting formulas from these classes leads to supporting full LTL.
In practice, any safety LTL formula can be encoded as a 
reachability problem. 
To further reason about liveness formulas, 
we define an analogy to reachability in \cref{tab:specifications} - explicit liveness, which is a concept which has successfully been used by tools like \tool{K2}~\cite{Kratos2}. 
Termination is an example of liveness specification and 
it can be expressed as explicit liveness by just adding 
$\texttt{assert_{live}}$ with the negation of the loop 
condition after every loop in the program. If the program is non-terminating,
then there is at least one execution that stays infinitely inside the loop and never 
satisfies the $\texttt{assert_{live}}$ condition. If it is terminating, it eventually
ends up in the artificial state that satisfies $\texttt{assert_{live}}$ and loops there infinitely.

\begin{figure}[t]
    \centering
    \input{figures/examples/liveness.tex}
    \caption{An example of an IA for the \emph{explicit liveness} property}
    \label{fig:liveness}
\end{figure}
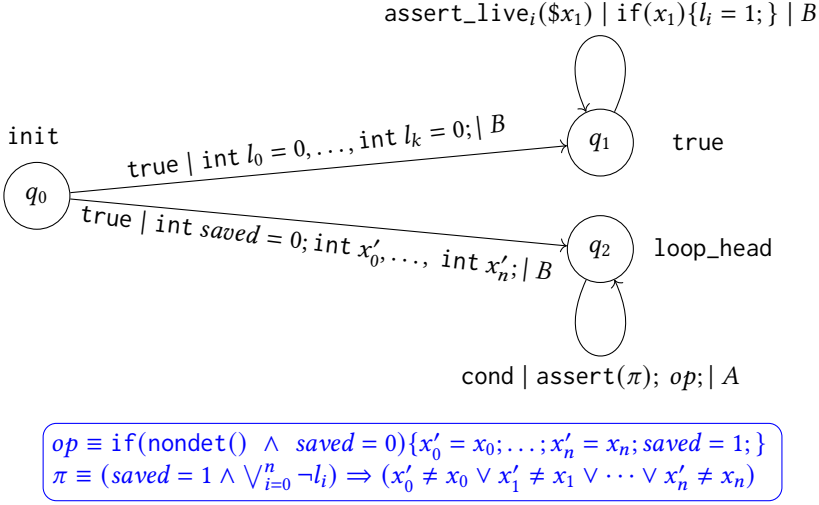

\inlineheadingbf{Explicit Liveness Transformation}
\Cref{fig:liveness} shows an instrumentation automaton for transforming explicit liveness to reachability.
It utilizes the transformation proposed by Biere and Schuppan~\cite{LivenessAsSafety}. 
Similar to the termination automaton~\cref{fig:showcase-termination}, 
it monitors the program for a possible infinite execution.
However, in this case, the infinite execution must not satisfy at least one of the \texttt{assert\_live} conditions. 
Such execution would violate the liveness property because there is an infinite execution 
of the program that never satisfies the property.
The automaton introduces ghost variables for every variable in the program 
which are used to track an infinite execution of the program. 
Explicit liveness is violated if there exists an infinite execution that never 
satisfies at least one of the $\texttt{assert_{live}}(\pi)$ conditions. 
Therefore, the automaton introduces a flag $l_i$ for every liveness assertion, and the flag is set, 
whenever the condition $\pi$ is satisfied.
This is handled by the self-loop at $q_1$. 
If there exists an execution that visits the same state 
twice and does not satisfy all of the assertions, the property is violated. 
This is captured by the $\texttt{assert}$ on the self-loop in $q_2$.
Since both transitions from $q_0$ match an initial edge in the input CFA, they can
look for edges with \texttt{assert\_live} to introduce the flags and for the loops in parallel.

%% file: figures/examples/liveness.tex
\begin{tikzpicture}[]
    \node[state] (q2) at (3.5, 0.3) {$q_2$};
    \node[state] (q1) at (3.5, 1.7) {$q_1$};
    \node[state,rectangle,color=blue, rounded corners=5pt, text width=9.6cm] (legend) at (1,-2.5) 
    {$op \equiv \texttt{if}(\texttt{nondet}()~\land~\mathit{saved}=0) \{x_0'=x_0;\dots;x_n'=x_n;\mathit{saved}=1;\}\newline\pi \equiv(\mathit{saved} = 1\land \bigvee_{i=0}^n \neg l_i) \implies (x_0' \neq x_0 \lor x_1' \neq x_1\lor\dots\lor x_n'\neq x_n)$};
    \node[] (a1) at (5, 0.3) {$\texttt{loop\_head}$};
    \node[] (a3) at (4.8, 1.7) {$\texttt{true}$};
    \node[state] (q0) at (-4, 1) {$q_0$};
    \node[] (a2) at (-4.05, 1.8) {$\texttt{init}$};
    \path[->,auto]
    (q2) edge[out=245,in=295,looseness=10] node[below]{$\texttt{cond} \mid \texttt{assert}(\pi);~op; \mid A$} (q2)
    (q1) edge[out=65,in=115,looseness=10] node[above]{$\texttt{assert\_live}_i(\$x_1) \mid \texttt{if} (x_1) \{l_i = 1;\} \mid B$} (q1)
    (q0) edge node[below,rotate=-7]{$\texttt{true}\mid \texttt{int}~saved=0; \texttt{int}~x_0',\dots,~\texttt{int}~x_n';
                        \mid B$} (q2)
    (q0) edge node[above,rotate=7]{$\texttt{true}\mid \texttt{int}~l_0=0,\dots,\texttt{int}~l_k=0;
                        \mid B$} (q1);
\end{tikzpicture}

%% file: sections/evaluation.tex
\section{Evaluation}
\label{sec:evaluation}

The evaluation of our approach addresses
the following research questions.

\begin{itemize}
    \item[] \textbf{\rqapplicability{} (Modularity)}: 
        To what extent do the transformations make
        verifiers, natively supporting only reachability,
        competitive in the verification of unsupported 
        properties?
    \item[] \textbf{\rqeffectiveness{} (Effectiveness)}: 
        To what extent are state-of-the-art reachability verifiers
        using transformation \textit{effective} compared to 
        state-of-the-art
        verifiers for the original specification?
    \item[] \textbf{\rqefficiency{} (Efficiency)}: 
        To what extent are state-of-the-art reachability verifiers
        using transformation \textit{efficient} 
        compared to state-of-the-art
        verifiers for the original specification?
    \item[] \textbf{\rqmodularity{} (No Degradation)}: 
        Is there a difference in the performance
        of a verifier when the program transformation
        is done on the input program level instead of
        during the analysis?
\end{itemize}

The proposed research questions aim to divide the evaluation into three parts to provide the answers for our initial motivation.
First, \rqapplicability{} aims to show the modularity of our approach 
i.e. can verifiers supporting only reachability be adapted to verify other properties without additional engineering effort?
Second, \rqeffectiveness{} and \rqefficiency{} aim to evaluate whether the fine-tuned reachability analyses
of the best-performing tools can be as effective and efficient as the state-of-the-art verifiers of the original properties. 
This supports our motivation for the clear separation of concerns. 
It suffices to improve the reachability analysis to perform well for multiple specifications. 
Last, \rqmodularity{} evaluates, whether encoding the transformation directly into a C program loses performance 
when compared to the encoding of the transformation inside the verification algorithm.

While the transformations for no-overflow and termination
significantly simplify the implementation of the
specification in the tool. For memory cleanup
the verifiers need to handle memory allocation and deallocation
correctly even with the transformation. Therefore, we only
answer \rqeffectiveness{} and \rqefficiency{} for memcleanup.
Since the tools used as reachability analyzers 
in \rqapplicability{} do not support any of the tasks even in their
original form. We also exclude \rqmodularity{} since there is
no internal transformation of memcleanup inside \cpachecker.

\inlineheadingbf{Benchmark Dataset} To answer the 
proposed research questions, we use a subset with \ntasksNoOverflow~tasks
for no-overflow, a subset with \ntasksTermination~tasks for termination,
and the full set with \ntasksMemCleanup~tasks for memory-cleanup of SV-Benchmarks at its SV-COMP24 version~\cite{SVCOMP24-SVBENCHMARKS-artifact},
the largest dataset of C programs with their 
verification verdicts for multiple properties.
The chosen subset for termination and no-overflow 
removes programs containing structures, and arrays,
since our current implementation does not support them.

\inlineheadingbf{Tools Used}\label{ssec:tools} \Cref{tab:tools} lists all the
sound\footnote{We removed \proton~\cite{PROTON-SVCOMP24} from the comparison, even though
    it performed best in the termination track of \svcomp 2024,
    because it performs known unsound 
    guessing if a non-termination argument is not found.}
and open-source\footnote{\veriabsl~\cite{VERIABSL-SVCOMP23} 
    and \veriabs~\cite{VERIABS} 
    performed very well in the reachability 
    category but are not open-source.}
tools with their
supported properties that we used in our experiments.
All those tools participated in \svcomp 2024~\cite{SVCOMP24} and scored among the best in 
their respective category.
We used the version submitted to \svcomp 2024 in our comparison.
We denote our verifier compositions with transformation to reachability
by adding \emph{-R} to the verifier's name,
for example, \cpv-R means that we applied the transformation and then the reachability analysis of \cpv.

\begin{table}[t]
    \caption{Tools used in the experiments}
    \label{tab:tools}
    \centering
\begin{tabular}{l@{\hspace{3mm}}c@{\hspace{3mm}}c@{\hspace{3mm}}c@{\hspace{3mm}}c}
    \toprule
    Tool & reachability & no-overflow & memory cleanup & termination \\
    \midrule
    \cpachecker~\cite{CPACHECKER-SVCOMP24} & \cmark & \cmark & \cmark & \cmark \\
    \uautomizer~\cite{ULTIMATEAUTOMIZER-SVCOMP24} & \cmark & \cmark & \cmark & \cmark \\
    \utaipan~\cite{UTAIPAN-SVCOMP23} & \cmark & \cmark & \xmark & \xmark \\
    \twols~\cite{2LS-SVCOMP23} & \cmark & \xmark & \xmark & \cmark \\
    \predatorhp~\cite{PREDATORHP-SVCOMP20} & \cmark & \xmark & \cmark & \xmark \\
    \symbiotic~\cite{SYMBIOTIC-SVCOMP24} & \cmark & \cmark & \cmark & \cmark \\
    \thetatool~\cite{THETA-SVCOMP24} & \cmark & \xmark & \xmark & \xmark \\
    \emergenthetatool~\cite{EMERGENTHETA-SVCOMP24} & \cmark & \xmark & \xmark & \xmark \\
    \cpv~\cite{CPV-SVCOMP24} & \cmark & \xmark & \xmark & \xmark \\
    \bottomrule
\end{tabular}
\end{table}

\inlineheadingbf{Benchmark Environment}\label{ssec:benchmark-environment}
For conducting our evaluation, we use~\benchexec to ensure
reliable benchmarking~\cite{Benchmarking-STTT}.
All benchmarks are performed on machines with
an Intel Xeon E5-1230 CPU (4 physical cores with 2 processing units each),
\SI{33}{GB} of RAM, and running
Ubuntu 22.04 as operating system.
Each verification task is executed with resource limits 
similar to the ones used in SV-COMP,
i.e., 900\,s of CPU time,
15 GB of memory,
and 1 physical core 
(2 processing units)\footnote{SV-COMP 2024 used 2 physical cores (4 processing units).}.
\subsection{\rqapplicability{}: Modularity}

\begin{figure}[t]
    \centering
    \subfloat[No-overflow]{\scalebox{.83}{\input{\plotpath/quantile-cputime-overflow}}\label{fig:evaluation-quantile-overflow}}
    \hfill
    \subfloat[Termination]{\scalebox{.83}{\input{\plotpath/quantile-cputime-termination}}\label{fig:evaluation-quantile-termination}}
    \caption{Quantile plots for (a)~no-overflow and (b)~termination tasks showing verifiers
            not supporting the property natively on the transformed tasks}
    \label{fig:evaluation:quantile-cputime-app}
\end{figure}
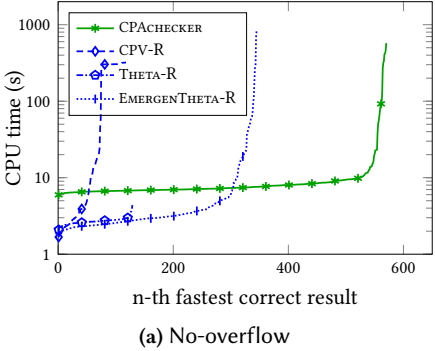
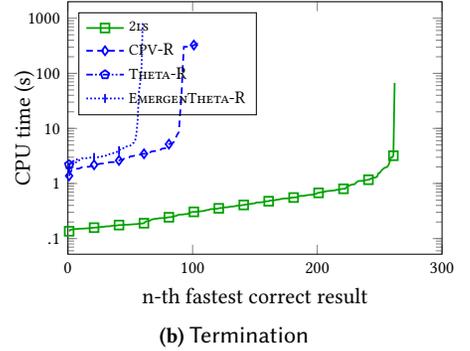

Since our approach produces transformed C programs, it directly allows
any verifier for C programs that supports reachability, to also analyze other specifications.
We consider three verifiers from \svcomp 2024: \cpv,~\thetatool~and \emergenthetatool,
which support only reachability.
\Cref{fig:evaluation-quantile-overflow,fig:evaluation-quantile-termination} 
compare these tools against the third-best
tool in the no-overflow category and
the termination category of \svcomp 2024, respectively.
We choose the third-best tool, since they are still competitive
i.e. are unlikely to contain bugs but are also comparable
to the ranking of the tools supporting reachability in that
category.

For no-overflow, \emergenthetatool-R was 226~tasks behind \cpachecker,
and for termination, \cpv-R was only 155~tasks behind \twols.
The tools supporting only reachability could solve roughly half
of the tasks that the third best-performing tool could solve in the 
respective category. 
Notable is that \cpv~was 5th, \thetatool~was 18th, and
\emergenthetatool~was 20th for reachability
in SV-COMP24~\cite{SVCOMP24}. This means that
even though they did not perform as well as the best 
performing verifiers in the reachability category, they can
still be relatively successful in the verification of other specifications.

\llbox{Our transformation framework makes it 
    possible for reachability verifiers to successfully support 
    properties which they do not natively support and to be competitive
    in verifying them.}

\subsection{\rqeffectiveness{}: Effectiveness}

In order to \rqeffectiveness{}, we compare the performance of the tools on the transformed and the original tasks.

\inlineheadingbf{No-Overflow}
As \cref{tab:statuses-overflow} shows,
\utaipan and \uautomizer were able to provide 692 $(78\,\%)$ and 676 $(76\,\%)$ correct results, respectively.
We observe only a slight degradation in the numbers of the solved tasks
if \uautomizer is applied as a reachability analyzer - 654 $(73\,\%)$.
Thanks to the modularity of the approach, \cpachecker as a reachability analyzer was able to find $88$ more
proofs and $15$ more alarms than its integrated no-overflow analysis.
Moreover, it solved only $2$ tasks less than the second-best tool \uautomizer.
We inspected all incorrect results and
concluded that they were not caused by our transformation, since the other reachability analyzers were able
to solve them correctly.
It is expected that reachability and no-overflow algorithms are conceptually similar as they are both
safety properties. However, the transformation allows us to use other
algorithms like K-Induction~\cite{K-Induction} for reachability
which leads to the increase in the performance for \cpachecker.
In this particular case the difference is likely due to the overflow
analysis of \cpachecker being based on predicate analysis and the
reachability analysis being a portfolio approach including
K-Induction, predicate analysis and value analysis.

\begin{table}[t]
    \centering
    \caption{Summary of the results for transformed \ntasksNoOverflow~no-overflow tasks}
    \label{tab:statuses-overflow}
    \input{figures/tables/results-overflow-rq1.tex}
\end{table}

\inlineheadingbf{Termination} 
\Cref{tab:statuses-termination} shows the results for termination.
\uautomizer as a reachability analyzer
performs the best among all the tools.
It was able to solve 327 $(85\,\%)$ of the tasks, and provide $13$ more proofs and $9$ more alarms than
\uautomizer as the termination analyzer.
In contrast to no-overflow, where the performance gain could be
attributed to the used algorithms, 
the difference in the performance for termination
is more likely due to the conceptual differences between the verification
approaches, because the algorithms developed to analyze termination
are usually very different from the algorithms for reachability.

Notably, our evaluation dataset contained $403$ tasks in the beginning. 
However, thanks to the transformation framework, 
we have found that $17$ of these
tasks contained undefined behavior in the 
form of signed-integer overflows. Since the termination
behavior is not defined in this case, we had to remove them from the comparison.
A merge request\footnote{\url{https://gitlab.com/sosy-lab/benchmarking/sv-benchmarks/-/merge_requests/1543}}
with the detailed description was created to remove
the currently wrong expected verdict for termination.

\begin{table}[t]
    \centering
    \caption{Summary of the results for transformed \ntasksTermination~termination tasks}
    \label{tab:statuses-termination}
    \input{figures/tables/results-termination-rq1.tex}
\end{table}

\inlineheadingbf{Memory-Cleanup}
\Cref{tab:statuses-memcleanup} shows the results of the comparison
of the tools on the transformed memory cleanup tasks.
The results show no large discrepancy between the different tools.
Since the dataset consists of only $41$ tasks it is difficult
to draw a general conclusion from the results. However, it is notable 
that \cpachecker-R could outperform \uautomizer in this case and that
it is close to the performance of \predatorhp which is a verifier
specializing in memory analysis.
The wrong verdict produced by \cpachecker-R is due to
its incomplete handling of function pointers.

\begin{table}[t]
    \centering
    \caption{Summary of the results for transformed \ntasksMemCleanup~Memory Cleanup tasks}
    \label{tab:statuses-memcleanup}
    \input{figures/tables/results-memcleanup-rq1.tex}
\end{table}

\llbox{For the no-overflow tasks,
    the performance of reachability analyzers is comparable to the verifiers that natively support it.
    For termination, the performance is better in some cases.
    For memory cleanup, some tools perform better than others in both directions.
}

\subsection{\rqefficiency{}: Efficiency}

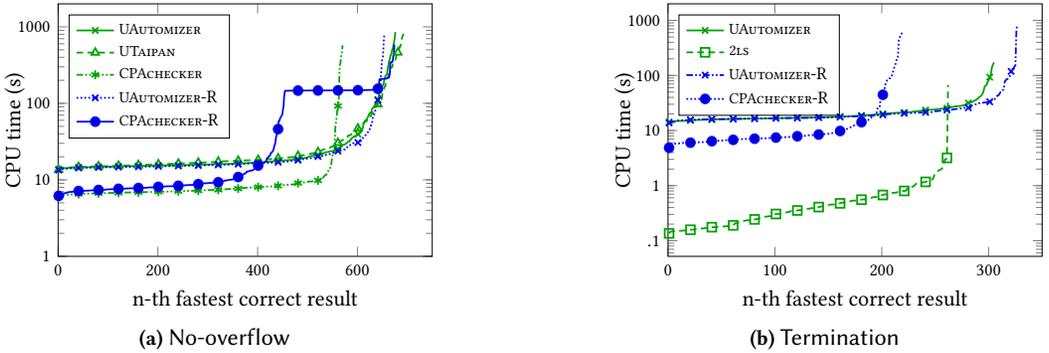
\begin{figure}[t]
    \centering
    \subfloat[No-overflow]{\scalebox{.83}{\input{\plotpath/quantile-cputime-efficiency-overflow}}\label{fig:evaluation-quantile-efficiency-overflows}}
    \hfill
    \subfloat[Termination]{\scalebox{.83}{\input{\plotpath/quantile-cputime-efficiency-termination}}\label{fig:evaluation-quantile-efficiency-termination}}
    \vspace{-2mm}
    \caption{Quantile plots for (a)~no-overflow and (b)~termination tasks comparing
            verifiers on the original and transformed tasks}
    \label{fig:evaluation:quantile-cputime}
\end{figure}

\inlineheadingbf{No-Overflow}
\Cref{fig:evaluation-quantile-efficiency-overflows} shows that there are no significant
differences in the efficiency of the approaches for the \ultimate tools.
The difference in efficiency between \cpachecker and \mbox{\cpachecker-R}
is due to the fact that the former uses only predicate analysis,
while, thanks to the flexibility gained by the transformation,
we can now also use \cpachecker-R with its sequential portfolio of reachability analyses;
the switch from one analysis to the next is clearly visible in the quantile plot
as a bend of the graph at around 100\,s.

\inlineheadingbf{Termination}
\Cref{fig:evaluation-quantile-efficiency-termination} shows
a comparison for \uautomizer and \twols, which were 2nd and 3rd in \svcomp 2024.
\twols consumes the least amount of CPU time, which is expected
since the other tools are Java-based and have a long startup
time due to the JVM.
However \twols cannot solve as many tasks as \uautomizer and \uautomizer-R.
For \uautomizer, there is a significant improvement in effectiveness
when using transformation, but the efficiency does not change much.

\begin{figure}[t]
    \centering
    \scalebox{.83}{\input{\plotpath/quantile-cputime-memcleanup}}
    \vspace{-3mm}
    \caption{Quantile plot for all correct results on memcleanup tasks}
    \label{fig:evaluation:quantile-cputime-memcleanup}
    \vspace{-2mm}
\end{figure}
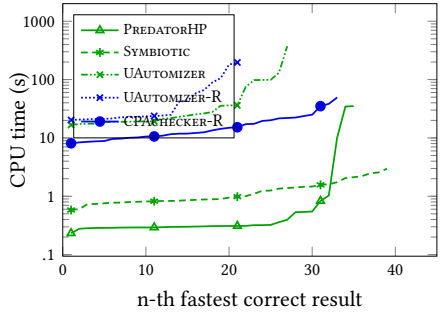

\inlineheadingbf{Memory-Cleanup}
\Cref{fig:evaluation:quantile-cputime-memcleanup} shows 
the quantile plot
for the correct results of the memory cleanup tasks.
The plot shows that there is no large discrepancy between tools,
accounting for around 10 seconds of JVM startup time for
\uautomizer and \cpachecker,
at the beginning but the discrepancy increases for 
tasks requiring a longer run-time. 
This is to be expected since the transformation
introduces a nondeterministic choice for each allocation 
making task more difficult.
Due to the small size of the dataset, it is difficult to draw
a general conclusion from the results.

\llbox{Verifying the transformed programs with reachability
 is as efficient as verifying the original programs against the original specification in many cases.
}

\subsection{\rqmodularity{}: No Degradation}

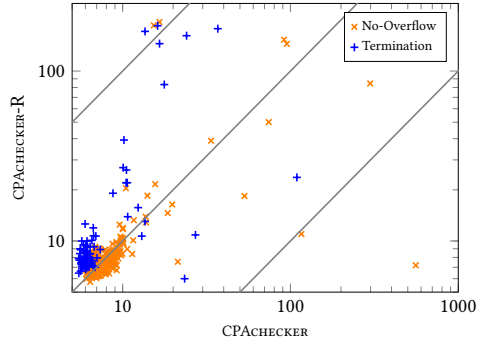
\begin{figure}[t]
    \vspace{3mm}
    \centering
    \scalebox{.75}{\input{\plotpath/scatter-cputime-nooverflow}}\label{fig:evaluation-scatter-nooverflow}
    \vspace{-2mm}
    \caption{Comparison of the CPU time of \cpachecker when transforming the properties 
        internally and when doing the transformation in the input program for no-overflow and termination tasks}
    \label{fig:evaluation:scatter-cputime}
\end{figure}

Some of the verifiers internally transform various properties to reachability.
They usually do it by instrumenting their intermediate representation of an
input program or by reflecting the assertion checks in their analyzing algorithm.
There are two algorithms in \cpachecker doing the latter. We compare how costly 
it is when we encode these checks as assertions in the input program instead
of doing the checks during the analysis.
\Cref{fig:evaluation:scatter-cputime} shows the CPU time in seconds for 
correctly solved tasks by both approaches. Notably, there was an increase in the amount of solved tasks for both properties when using the reachability analysis.

For termination~(\color{blue}$\bm{+}$\color{black}),
both approaches used bounded model checking~\cite{BMCJournal} as the reachability algorithm. 
It usually solves the task very quickly or does not solve it at all.
We can see that for most of the tasks, there is an overhead between
1-20 seconds if we represent the specification in the input program which is not
too much in relation to the 900 seconds time limit. There were a few outliers, where
the reachability analysis took approximately 100 seconds more.

For no-overflow~(\color{orange}$\times$\color{black}),
both approaches used predicate analysis~\cite{AlgorithmComparison-JAR} 
as the reachability algorithm.
The sample of commonly solved tasks is much larger than for termination.
For most of the tasks, there is no clear overhead in either direction.
There are a few outliers in both directions with significant overheads.

\llbox{We do not observe any clear degradation of efficiency when 
    transforming
    no-overflow to reachability in the input program. 
    For termination, in most cases, the overhead is relatively small. 
    In sum, the transformation inside the verifier does not 
    provide significant boost in the performance.
}

%% file: figures/results/tex/quantile-cputime-overflow.tex
\begin{tikzpicture}
\begin{semilogyaxis}[
    xlabel=n-th fastest correct result,
    ylabel=CPU time (\second),
    quantile plot,
    mark repeat=40,
    /pgf/number format/1000 sep={},
    legend columns=1,
    xmax=650,
    ]
    \addgraphtable{color=green, mark=asterisk}{}{\cpachecker}{../../../csv/cpachecker.overflow.quantile-cputime.csv}
    \addgraphtable{color=blue, mark=diamond}{}{\cpv-R}{../../../csv/cpv.o2r-reachability.quantile-cputime.csv}
    \addgraphtable{color=blue, mark=pentagon}{}{\thetatool-R}{../../../csv/theta.o2r-reachability.quantile-cputime.csv}
    \addgraphtable{color=blue, mark=|}{}{\emergenthetatool-R}{../../../csv/emergentheta.o2r-reachability.quantile-cputime.csv}
\end{semilogyaxis}
\end{tikzpicture}

%% file: figures/results/tex/quantile-cputime-termination.tex
\begin{tikzpicture}
\begin{semilogyaxis}[
    xlabel=n-th fastest correct result,
    ylabel=CPU time (\second),
    quantile plot,
    mark repeat=20,
    /pgf/number format/1000 sep={},
    legend columns=1,
    ymin=0,
    xmax=300,
    ]
    \addgraphtable{color=green, mark=square}{}{\twols}{../../../csv/2ls.termination.quantile-cputime.csv}
    \addgraphtable{color=blue, mark=diamond}{}{\cpv-R}{../../../csv/cpv.t2r-reachability.quantile-cputime.csv}
    \addgraphtable{color=blue, mark=pentagon}{}{\thetatool-R}{../../../csv/theta.t2r-reachability.quantile-cputime.csv}
    \addgraphtable{color=blue, mark=|}{}{\emergenthetatool-R}{../../../csv/emergentheta.t2r-reachability.quantile-cputime.csv}
\end{semilogyaxis}
\end{tikzpicture}

%% file: figures/tables/results-overflow-rq1.tex
\newcommand\precnum[1]{\tablenum[table-format=4]{#1}}
\begin{tabular}{l@{\hspace{-2mm}}c@{\hspace{1mm}}|@{\hspace{2mm}}c@{\hspace{2mm}}c@{\hspace{2mm}}c@{\hspace{2mm}}|@{\hspace{2mm}}c@{\hspace{2mm}}c@{\hspace{2mm}}}
    \toprule
    Tools          & \textcolor{gray}{\smaller (\#Tasks)} & \uautomizer & \utaipan & \cpachecker & \uautomizer-R & \cpachecker-R \\
    \midrule
    Correct        & \textcolor{gray}{\smaller\precnum{\ntasksNoOverflow}}
                        & \precnum{\UltimateBenchmarkDefinitionORNooverflowReachsafetyLoopsCorrectCount}
                        & \textbf{\precnum{\TaipanBenchmarkDefinitionORNooverflowReachsafetyLoopsCorrectCount}}
                        & \precnum{\CpacheckerBenchmarkDefinitionORNooverflowReachsafetyLoopsCorrectCount}
                        & \precnum{\UltimateBenchmarkDefinitionORReachabilityReachsafetyLoopsCorrectCount}
                        & \precnum{\CpacheckerBenchmarkDefinitionORReachabilityReachsafetyLoopsCorrectCount}
    \\
    \quad Proofs          & \textcolor{gray}{\smaller\precnum{\ntasksNoOverflowSafe}}
                        & \precnum{\UltimateBenchmarkDefinitionORNooverflowReachsafetyLoopsCorrectTrueCount}
                        & \textbf{\precnum{\TaipanBenchmarkDefinitionORNooverflowReachsafetyLoopsCorrectTrueCount}}
                        & \precnum{\CpacheckerBenchmarkDefinitionORNooverflowReachsafetyLoopsCorrectTrueCount}
                        & \precnum{\UltimateBenchmarkDefinitionORReachabilityReachsafetyLoopsCorrectTrueCount}
                        & \textbf{\precnum{\CpacheckerBenchmarkDefinitionORReachabilityReachsafetyLoopsCorrectTrueCount}}
    \\
    \quad Alarms          & \textcolor{gray}{\smaller\precnum{\ntasksNoOverflowUnsafe}}
                        & \precnum{\UltimateBenchmarkDefinitionORNooverflowReachsafetyLoopsCorrectFalseCount}
                        & \textbf{\precnum{\TaipanBenchmarkDefinitionORNooverflowReachsafetyLoopsCorrectFalseCount}}
                        & \precnum{\CpacheckerBenchmarkDefinitionORNooverflowReachsafetyLoopsCorrectFalseCount}
                        & \precnum{\UltimateBenchmarkDefinitionORReachabilityReachsafetyLoopsCorrectFalseCount}
                        & \precnum{\CpacheckerBenchmarkDefinitionORReachabilityReachsafetyLoopsCorrectFalseCount}
    \\
    Incorrect      &
                        & \precnum{\UltimateUltimateWrongCount}
                        & \precnum{\TaipanTaipanWrongCount}
                        & \precnum{\CpacheckerCpacheckerWrongCount}
                        & \precnum{\UltimateBenchmarkDefinitionORReachabilityReachsafetyLoopsWrongCount}
                        & \precnum{\CpacheckerBenchmarkDefinitionORReachabilityReachsafetyLoopsWrongCount}
    \\
    \quad Proofs          &
                        & \precnum{\UltimateUltimateWrongTrueCount}
                        & \precnum{\TaipanTaipanWrongTrueCount}
                        & \precnum{\CpacheckerCpacheckerWrongTrueCount}
                        & \precnum{\UltimateBenchmarkDefinitionORReachabilityReachsafetyLoopsWrongTrueCount}
                        & \precnum{\CpacheckerBenchmarkDefinitionORReachabilityReachsafetyLoopsWrongTrueCount}
    \\
    \quad Alarms          &
                        & \precnum{\UltimateUltimateWrongFalseCount}
                        & \precnum{\TaipanTaipanWrongFalseCount}
                        & \precnum{\CpacheckerCpacheckerWrongFalseCount}
                        & \precnum{\UltimateUltimateWrongTrueCount}
                        & \precnum{\CpacheckerBenchmarkDefinitionORReachabilityReachsafetyLoopsWrongFalseCount}
    \\
    \bottomrule
\end{tabular}

%% file: figures/tables/results-termination-rq1.tex
\newcommand\precnum[1]{\tablenum[table-format=4]{#1}}
\begin{tabular}{l@{\hspace{-2mm}}c@{\hspace{1mm}}|@{\hspace{2mm}}c@{\hspace{2mm}}c@{\hspace{2mm}}|@{\hspace{2mm}}c@{\hspace{2mm}}c@{\hspace{2mm}}}
    \toprule
    Results          & \textcolor{gray}{\smaller (\#tasks)} & \uautomizer & \twols & \uautomizer-R & \cpachecker-R \\
    \midrule
    Correct        & \textcolor{gray}{\smaller\precnum{\ntasksTermination}}
                        & \precnum{\UltimateBenchmarkDefinitionterminationTRTerminationReachsafetyLoopsCorrectCount}
                        & \precnum{\LsBenchmarkDefinitionterminationTRTerminationReachsafetyLoopsCorrectCount}
                        & \textbf{\precnum{\UltimateBenchmarkDefinitionTRReachabilityReachsafetyLoopsCorrectCount}}
                        & \precnum{\CpacheckerBenchmarkDefinitionTRReachabilityReachsafetyLoopsCorrectCount}
    \\
    \quad Proofs          & \textcolor{gray}{\smaller\precnum{\ntasksTerminationSafe}}
                        & \precnum{\UltimateBenchmarkDefinitionterminationTRTerminationReachsafetyLoopsCorrectTrueCount}
                        & \precnum{\LsBenchmarkDefinitionterminationTRTerminationReachsafetyLoopsCorrectTrueCount}
                        & \textbf{\precnum{\UltimateBenchmarkDefinitionTRReachabilityReachsafetyLoopsCorrectTrueCount}}
                        & \precnum{\CpacheckerBenchmarkDefinitionTRReachabilityReachsafetyLoopsCorrectTrueCount}
    \\
    \quad Alarms          & \textcolor{gray}{\smaller\precnum{\ntasksTerminationUnsafe}}
                        & \precnum{\UltimateBenchmarkDefinitionterminationTRTerminationReachsafetyLoopsCorrectFalseCount}
                        & \textbf{\precnum{\LsBenchmarkDefinitionterminationTRTerminationReachsafetyLoopsCorrectFalseCount}}
                        & \precnum{\UltimateBenchmarkDefinitionTRReachabilityReachsafetyLoopsCorrectFalseCount}
                        & \precnum{\CpacheckerBenchmarkDefinitionTRReachabilityReachsafetyLoopsCorrectFalseCount}
    \\
    \bottomrule
\end{tabular}

%% file: figures/tables/results-memcleanup-rq1.tex
\newcommand\precnum[1]{\tablenum[table-format=4]{#1}}
\begin{tabular}{l@{\hspace{-3mm}}c@{\hspace{2mm}}|@{\hspace{2mm}}c@{\hspace{2mm}}c@{\hspace{2mm}}c@{\hspace{2mm}}|@{\hspace{2mm}}c@{\hspace{2mm}}c@{\hspace{2mm}}}
    \toprule
    Tools          & \textcolor{gray}{\smaller (\#tasks)} & \predatorhp & \symbiotic & \uautomizer & \uautomizer-R & \cpachecker-R \\
    \midrule
    Correct results       & \textcolor{gray}{\smaller\precnum{\ntasksMemCleanup}}
                        & \precnum{35}
                        & \textbf{\precnum{39}}
                        & \precnum{27}
                        & \precnum{24}
                        & \precnum{33}
    \\
    \quad proofs          & \textcolor{gray}{\smaller\precnum{\ntasksMemCleanupSafe}}
                        & \precnum{1}
                        & \textbf{\precnum{2}}
                        & \precnum{0}
                        & \precnum{0}
                        & \precnum{0}
    \\
    \quad alarms          & \textcolor{gray}{\smaller\precnum{\ntasksMemCleanupUnsafe}}
                        & \precnum{34}
                        & \textbf{\precnum{37}}
                        & \precnum{27}
                        & \precnum{24}
                        & \precnum{33}
    \\
    Incorrect results     &
                        & \precnum{0}
                        & \precnum{0}
                        & \precnum{0}
                        & \precnum{0}
                        & \precnum{1}
    \\
    \quad proofs          &
                        & \precnum{0}
                        & \precnum{0}
                        & \precnum{0}
                        & \precnum{0}
                        & \precnum{0}
    \\
    \quad alarms          &
                        & \precnum{0}
                        & \precnum{0}
                        & \precnum{0}
                        & \precnum{0}
                        & \precnum{1}
    \\
    \bottomrule
\end{tabular}

%% file: figures/results/tex/quantile-cputime-efficiency-overflow.tex
\begin{tikzpicture}
\begin{semilogyaxis}[
    xlabel=n-th fastest correct result,
    ylabel=CPU time (\second),
    quantile plot,
    mark repeat=40,
    /pgf/number format/1000 sep={},
    legend columns=1,
    xmax=750,
    ]
    \addgraphtable{color=green, mark=x}{}{\uautomizer}{../../../csv/ultimate.overflow.quantile-cputime.csv}
    \addgraphtable{color=green, mark=triangle}{}{\utaipan}{../../../csv/taipan.overflow.quantile-cputime.csv}
    \addgraphtable{color=green, mark=asterisk}{}{\cpachecker}{../../../csv/cpachecker.overflow.quantile-cputime.csv}
    \addgraphtable{color=blue, mark=x}{x index=0,y index=3}{\uautomizer-R}{../../../csv/ultimate.o2r-reachability.quantile-cputime.csv}
    \addgraphtable{color=blue, mark=*}{x index=0,y index=3}{\cpachecker-R}{../../../csv/cpachecker.o2r-reachability.quantile-cputime.csv}
\end{semilogyaxis}
\end{tikzpicture}

%% file: figures/results/tex/quantile-cputime-efficiency-termination.tex
\begin{tikzpicture}
\begin{semilogyaxis}[
    xlabel=n-th fastest correct result,
    ylabel=CPU time (\second),
    quantile plot,
    mark repeat=20,
    /pgf/number format/1000 sep={},
    legend columns=1,
    ymin=0,
    xmax=350,
    ]
    \addgraphtable{color=green, mark=x}{}{\uautomizer}{../../../csv/ultimate.termination.quantile-cputime.csv}
    \addgraphtable{color=green, mark=square}{}{\twols}{../../../csv/2ls.termination.quantile-cputime.csv}
    \addgraphtable{color=blue, mark=x}{x index=0,y index=4}{\uautomizer-R}{../../../csv/ultimate.reachability.quantile-cputime.csv}
    \addgraphtable{color=blue, mark=*}{x index=0,y index=4}{\cpachecker-R}{../../../csv/predAnalysis-linear.reachability.quantile-cputime.csv}
\end{semilogyaxis}
\end{tikzpicture}

%% file: figures/results/tex/quantile-cputime-memcleanup.tex
\begin{tikzpicture}
\begin{semilogyaxis}[
    xlabel=n-th fastest correct result,
    ylabel=CPU time (\second),
    quantile plot,
    mark repeat=10,
    /pgf/number format/1000 sep={},
    legend columns=1,
    xmax=45,
    ymin = 0,
    ]
    \addgraphtable{color=green, mark=triangle}{x index=0,y index=3}{\predatorhp}{../../../csv/predatorhp.memcleanup.quantile-cputime.csv}
    \addgraphtable{color=green, mark=asterisk}{x index=0,y index=3}{\symbiotic}{../../../csv/symbiotic.memcleanup.quantile-cputime.csv}
    \addgraphtable{color=green, mark=x}{x index=0,y index=3}{\uautomizer}{../../../csv/uautomizer.memcleanup.quantile-cputime.csv}
    \addgraphtable{color=blue, mark=x}{x index=0,y index=3}{\uautomizer-R}{../../../csv/uautomizer.m2r-reachability.quantile-cputime.csv}
    \addgraphtable{color=blue, mark=*}{x index=0,y index=3}{\cpachecker-R}{../../../csv/cpachecker.m2r-reachability.quantile-cputime.csv}
\end{semilogyaxis}
\end{tikzpicture}

%% file: figures/results/tex/scatter-cputime-nooverflow.tex
\begin{tikzpicture}
\begin{loglogaxis}[
    xlabel=\cpachecker,
    ylabel=\cpachecker-R,
    xmin=5,
    xmax=1000,
    ymin=5,
    ymax=250,
    domain=1:1901,
    clip mode=individual,
    axis equal image,
    legend pos=north east,
    legend style={font=\scriptsize},
    legend cell align={left},
    /pgf/number format/1000 sep={},
    ]
    \addplot+[orange, mark=x, only marks]
        table[
            header=false,
            x index=6, 
            y index=10  
        ] {csv/nooverflow.scatter-cputime-final.table.csv};
    \addlegendentry{No-Overflow}
    \addplot+[blue, mark=+, only marks]
        table[
            header=false,
            x index=6, 
            y index=10  
        ] {csv/termination.scatter-cputime-final.table.csv};
    \addlegendentry{Termination}
    \addplot[gray] {x};
    \addplot[gray] {10*x};
    \addplot[gray] {x/10};
\end{loglogaxis}
\end{tikzpicture}

%% file: sections/threats-to-validity.tex
\subsection{Threats to Validity}

\inlineheadingbf{Internal Validity}
We used the benchmarking framework \benchexec~\cite{Benchmarking-STTT} to run the experiments,
which uses the most modern Linux features for reliable benchmarking.
This tool also makes sure to never run two different executions on the
same physical core, to avoid interference of shared computing resources.
On the other hand, the implementation of the transformations may contain bugs,
which could lead to incorrect results. However, we checked all the incorrect
results in our experiments and none of them were caused by the transformation.

\inlineheadingbf{External Validity}
The conclusions about the benefit of the transformations might
not hold for other programs and other verifiers.
However, we evaluated on state-of-the-art verifiers and a large benchmark set,
which reduces this risk.
Furthermore, we considered only three popular specifications, and
it could be that the benefits described are different for other specifications.

%% file: sections/conclusion.tex
\section{Conclusion}

Developing a tool for software verification is challenging and requires 
a large engineering effort.
The effort is even larger for supporting various specifications.
Verification tools sometimes use internal transformations
to mitigate the development time.
However, these transformations are usually not modular and have to be done separately 
for every verifier, and for each specification.
Our contribution offers a new modular
framework that separates the concern of a reachability algorithm
from supporting other specifications.
We demonstrated how the framework works by implementing
the transformations for three interesting specifications.

We showed that the construction of new verifiers by the transformation
followed by reachability analysis is usually also efficient
and effective, and can compete with (and sometimes outperform)
state-of-the-art verifiers for no-overflow 
and termination analysis.
Furthermore, our approach enabled
tools like \cpv or \thetatool to be competitive in the verification
of properties that they do not natively support so far.
Lastly, we did not observe
a significant degradation in the performance when we transformed the input program
instead of the intermediate representation or changed the analyzing algorithm.

\inlineheadingbf{Future Work} 
Currently, only three transformations have been implemented in the framework.
In the future, we plan to extend the set of provided instrumentation automata
to contain more specifications, and the framework to support more kinds of specifications,
for example, memory safety and the sequentialization of concurrency.

In the future, we plan to implement an interface for user-defined instrumentation automata.
This would make
it easier to add new specifications and transformations to the framework
and compare the performance of different transformations for
the same property.

%% file: main.bbl

\providecommand{\serysort}{}\providecommand{\svejdasort}{}
\begin{thebibliography}{73}


\ifx \showCODEN    \undefined \def \showCODEN     #1{\unskip}     \fi
\ifx \showDOI      \undefined \def \showDOI       #1{#1}\fi
\ifx \showISBNx    \undefined \def \showISBNx     #1{\unskip}     \fi
\ifx \showISBNxiii \undefined \def \showISBNxiii  #1{\unskip}     \fi
\ifx \showISSN     \undefined \def \showISSN      #1{\unskip}     \fi
\ifx \showLCCN     \undefined \def \showLCCN      #1{\unskip}     \fi
\ifx \shownote     \undefined \def \shownote      #1{#1}          \fi
\ifx \showarticletitle \undefined \def \showarticletitle #1{#1}   \fi
\ifx \showURL      \undefined \def \showURL       {\relax}        \fi
\providecommand\bibfield[2]{#2}
\providecommand\bibinfo[2]{#2}
\providecommand\natexlab[1]{#1}
\providecommand\showeprint[2][]{arXiv:#2}

\bibitem[ove({[n.\,d.]})]%
        {overflow-conditions}
 \bibinfo{year}{[n.\,d.]}\natexlab{}.
\newblock \bibinfo{title}{INT32-C. Ensure that operations on signed integers do not result in overflow}.
\newblock \bibinfo{howpublished}{\url{https://wiki.sei.cmu.edu/confluence/display/c/INT32-C.+Ensure+that+operations+on+signed+integers+do+not+result+in+overflow}}.
\newblock
\newblock
\shownote{[Accessed 28-08-2024]}.


\bibitem[Afzal et~al\mbox{.}(2019)]%
        {VERIABS}
\bibfield{author}{\bibinfo{person}{M. Afzal}, \bibinfo{person}{A. Asia}, \bibinfo{person}{A. Chauhan}, \bibinfo{person}{B. Chimdyalwar}, \bibinfo{person}{P. Darke}, \bibinfo{person}{A. Datar}, \bibinfo{person}{S. Kumar}, {and} \bibinfo{person}{R. Venkatesh}.} \bibinfo{year}{2019}\natexlab{}.
\newblock \showarticletitle{\textsc{VeriAbs}: {V}erification by Abstraction and Test Generation}. In \bibinfo{booktitle}{\emph{Proc.\ ASE}}. \bibinfo{publisher}{IEEE}, \bibinfo{pages}{1138--1141}.
\newblock
\urldef\tempurl%
\url{https://doi.org/10.1109/ASE.2019.00121}
\showDOI{\tempurl}


\bibitem[Aho et~al\mbox{.}(1986)]%
        {DragonBook}
\bibfield{author}{\bibinfo{person}{A.~V. Aho}, \bibinfo{person}{R. Sethi}, {and} \bibinfo{person}{J.~D. Ullman}.} \bibinfo{year}{1986}\natexlab{}.
\newblock \bibinfo{booktitle}{\emph{Compilers: Principles, Techniques, and Tools}}.
\newblock \bibinfo{publisher}{Addison-Wesley}.
\newblock
\showISBNx{978-0-201-10088-4}
\urldef\tempurl%
\url{https://www.worldcat.org/isbn/978-0-201-10088-4}
\showURL{%
\tempurl}


\bibitem[Alglave et~al\mbox{.}(2011)]%
        {KroeningATVA}
\bibfield{author}{\bibinfo{person}{J. Alglave}, \bibinfo{person}{A.~F. Donaldson}, \bibinfo{person}{D. Kröning}, {and} \bibinfo{person}{M. Tautschnig}.} \bibinfo{year}{2011}\natexlab{}.
\newblock \showarticletitle{Making Software Verification Tools Really Work}. In \bibinfo{booktitle}{\emph{Proc.\ ATVA}} \emph{(\bibinfo{series}{LNCS~6996})}. \bibinfo{publisher}{Springer}, \bibinfo{pages}{28--42}.
\newblock
\urldef\tempurl%
\url{https://doi.org/10.1007/978-3-642-24372-1_3}
\showDOI{\tempurl}


\bibitem[Alshmrany et~al\mbox{.}(2021)]%
        {FUSEBMC-long}
\bibfield{author}{\bibinfo{person}{K.~M. Alshmrany}, \bibinfo{person}{M. Aldughaim}, \bibinfo{person}{A. Bhayat}, {and} \bibinfo{person}{L.~C. Cordeiro}.} \bibinfo{year}{2021}\natexlab{}.
\newblock \showarticletitle{\textsc{FuSeBMC}: {A}n energy-efficient test generator for finding security vulnerabilities in {C} programs}. In \bibinfo{booktitle}{\emph{Proc.\ TAP}}. \bibinfo{publisher}{Springer}, \bibinfo{pages}{85--105}.
\newblock
\urldef\tempurl%
\url{https://doi.org/10.1007/978-3-030-79379-1_6}
\showDOI{\tempurl}


\bibitem[Amilon et~al\mbox{.}(2023)]%
        {InstrumnetationRummer}
\bibfield{author}{\bibinfo{person}{Jesper Amilon}, \bibinfo{person}{Zafer Esen}, \bibinfo{person}{Dilian Gurov}, \bibinfo{person}{Christian Lidstr{\"o}m}, {and} \bibinfo{person}{Philipp R{\"u}mmer}.} \bibinfo{year}{2023}\natexlab{}.
\newblock \showarticletitle{Automatic {Program} {Instrumentation} for {Automatic} {Verification}}. In \bibinfo{booktitle}{\emph{Proc.\ CAV}}. \bibinfo{pages}{281--304}.
\newblock
\showISBNx{978-3-031-37709-9}


\bibitem[Ayaziová et~al\mbox{.}(2024)]%
        {VerificationWitnesses-2.0}
\bibfield{author}{\bibinfo{person}{P. Ayaziová}, \bibinfo{person}{D. Beyer}, \bibinfo{person}{M. Lingsch-Rosenfeld}, \bibinfo{person}{M. Spiessl}, {and} \bibinfo{person}{J. Strejček}.} \bibinfo{year}{2024}\natexlab{}.
\newblock \showarticletitle{Software Verification Witnesses~2.0}. In \bibinfo{booktitle}{\emph{Proc.\ SPIN}}. \bibinfo{publisher}{Springer}.
\newblock


\bibitem[Baier et~al\mbox{.}(2024)]%
        {CPACHECKER-3.0-tutorial}
\bibfield{author}{\bibinfo{person}{D. Baier}, \bibinfo{person}{D. Beyer}, \bibinfo{person}{P.-C. Chien}, \bibinfo{person}{M.-C. Jakobs}, \bibinfo{person}{M. Jankola}, \bibinfo{person}{M. Kettl}, \bibinfo{person}{N.-Z. Lee}, \bibinfo{person}{T. Lemberger}, \bibinfo{person}{M. Lingsch-Rosenfeld}, \bibinfo{person}{H. Wachowitz}, {and} \bibinfo{person}{P. Wendler}.} \bibinfo{year}{2024}\natexlab{}.
\newblock \showarticletitle{Software Verification with \textsc{CPAchecker} 3.0: {Tutorial} and User Guide}. In \bibinfo{booktitle}{\emph{Proc.\ FM}} \emph{(\bibinfo{series}{LNCS~14934})}. \bibinfo{publisher}{Springer}.
\newblock
\urldef\tempurl%
\url{https://doi.org/10.1007/978-3-031-71177-0_30}
\showDOI{\tempurl}


\bibitem[Baier et~al\mbox{.}(2024)]%
        {CPACHECKER-SVCOMP24}
\bibfield{author}{\bibinfo{person}{D. Baier}, \bibinfo{person}{D. Beyer}, \bibinfo{person}{P.-C. Chien}, \bibinfo{person}{M. Jankola}, \bibinfo{person}{M. Kettl}, \bibinfo{person}{N.-Z. Lee}, \bibinfo{person}{T. Lemberger}, \bibinfo{person}{M. Lingsch-Rosenfeld}, \bibinfo{person}{M. Spiessl}, \bibinfo{person}{H. Wachowitz}, {and} \bibinfo{person}{P. Wendler}.} \bibinfo{year}{2024}\natexlab{}.
\newblock \showarticletitle{\textsc{CPAchecker} 2.3 with Strategy Selection (Competition Contribution)}. In \bibinfo{booktitle}{\emph{Proc.\ TACAS~(3)}} \emph{(\bibinfo{series}{LNCS~14572})}. \bibinfo{publisher}{Springer}, \bibinfo{pages}{359--364}.
\newblock
\urldef\tempurl%
\url{https://doi.org/10.1007/978-3-031-57256-2_21}
\showDOI{\tempurl}


\bibitem[Bajczi et~al\mbox{.}(2024b)]%
        {EMERGENTHETA-SVCOMP24}
\bibfield{author}{\bibinfo{person}{L. Bajczi}, \bibinfo{person}{D. Szekeres}, \bibinfo{person}{M. Mondok}, \bibinfo{person}{Z. Ádám}, \bibinfo{person}{M. Somorjai}, \bibinfo{person}{C. Telbisz}, \bibinfo{person}{M. Dobos-Kovács}, {and} \bibinfo{person}{V. Molnár}.} \bibinfo{year}{2024}\natexlab{b}.
\newblock \showarticletitle{\textsc{EmergenTheta}: Verification Beyond Abstraction Refinement (Competition Contribution)}. In \bibinfo{booktitle}{\emph{Proc.\ TACAS~(3)}} \emph{(\bibinfo{series}{LNCS~14572})}. \bibinfo{publisher}{Springer}, \bibinfo{pages}{371--375}.
\newblock
\urldef\tempurl%
\url{https://doi.org/10.1007/978-3-031-57256-2_23}
\showDOI{\tempurl}


\bibitem[Bajczi et~al\mbox{.}(2024a)]%
        {THETA-SVCOMP24}
\bibfield{author}{\bibinfo{person}{L. Bajczi}, \bibinfo{person}{C. Telbisz}, \bibinfo{person}{M. Somorjai}, \bibinfo{person}{Z. Ádám}, \bibinfo{person}{M. Dobos-Kovács}, \bibinfo{person}{D. Szekeres}, \bibinfo{person}{M. Mondok}, {and} \bibinfo{person}{V. Molnár}.} \bibinfo{year}{2024}\natexlab{a}.
\newblock \showarticletitle{\textsc{Theta}: Abstraction Based Techniques for Verifying Concurrency (Competition Contribution)}. In \bibinfo{booktitle}{\emph{Proc.\ TACAS~(3)}} \emph{(\bibinfo{series}{LNCS~14572})}. \bibinfo{publisher}{Springer}, \bibinfo{pages}{412--417}.
\newblock
\urldef\tempurl%
\url{https://doi.org/10.1007/978-3-031-57256-2_30}
\showDOI{\tempurl}


\bibitem[Ball and Rajamani(2002a)]%
        {SLIC}
\bibfield{author}{\bibinfo{person}{T. Ball} {and} \bibinfo{person}{S.~K. Rajamani}.} \bibinfo{year}{2002}\natexlab{a}.
\newblock \bibinfo{booktitle}{\emph{{SLIC}: A Specification Language for Interface Checking (of~{C})}}.
\newblock \bibinfo{type}{{T}echnical {R}eport} MSR-TR-2001-21. \bibinfo{institution}{Microsoft Research}.
\newblock
\urldef\tempurl%
\url{https://www.microsoft.com/en-us/research/publication/slic-a-specification-language-for-interface-checking-of-c/}
\showURL{%
\tempurl}


\bibitem[Ball and Rajamani(2002b)]%
        {SLAM}
\bibfield{author}{\bibinfo{person}{T. Ball} {and} \bibinfo{person}{S.~K. Rajamani}.} \bibinfo{year}{2002}\natexlab{b}.
\newblock \showarticletitle{The \textsc{Slam} project: Debugging System Software via Static Analysis}. In \bibinfo{booktitle}{\emph{Proc.\ POPL}}. \bibinfo{publisher}{{ACM}}, \bibinfo{pages}{1--3}.
\newblock
\urldef\tempurl%
\url{https://doi.org/10.1145/503272.503274}
\showDOI{\tempurl}


\bibitem[Beyer(2024)]%
        {TESTCOMP24}
\bibfield{author}{\bibinfo{person}{D. Beyer}.} \bibinfo{year}{2024}\natexlab{}.
\newblock \showarticletitle{Automatic Testing of {C} Programs: {Test-Comp} 2024}. In \bibinfo{booktitle}{\emph{TBA}}. \bibinfo{publisher}{Springer}.
\newblock


\bibitem[Beyer(2024)]%
        {SVCOMP24}
\bibfield{author}{\bibinfo{person}{D. Beyer}.} \bibinfo{year}{2024}\natexlab{}.
\newblock \showarticletitle{State of the Art in Software Verification and Witness Validation: {SV-COMP 2024}}. In \bibinfo{booktitle}{\emph{Proc.\ TACAS~(3)}} \emph{(\bibinfo{series}{LNCS~14572})}. \bibinfo{publisher}{Springer}, \bibinfo{pages}{299--329}.
\newblock
\urldef\tempurl%
\url{https://doi.org/10.1007/978-3-031-57256-2_15}
\showDOI{\tempurl}


\bibitem[Beyer(2024)]%
        {SVCOMP24-SVBENCHMARKS-artifact}
\bibfield{author}{\bibinfo{person}{D. Beyer}.} \bibinfo{year}{2024}\natexlab{}.
\newblock \bibinfo{title}{{SV-Benchmarks}: {Benchmark} Set for Software Verification ({SV-COMP 2024})}.
\newblock \bibinfo{howpublished}{Zenodo}.
\newblock
\urldef\tempurl%
\url{https://doi.org/10.5281/zenodo.10669723}
\showDOI{\tempurl}


\bibitem[Beyer et~al\mbox{.}(2004)]%
        {BLAST-query}
\bibfield{author}{\bibinfo{person}{D. Beyer}, \bibinfo{person}{A.~J. Chlipala}, \bibinfo{person}{T.~A. Henzinger}, \bibinfo{person}{R. Jhala}, {and} \bibinfo{person}{R. Majumdar}.} \bibinfo{year}{2004}\natexlab{}.
\newblock \showarticletitle{The \textsc{Blast} Query Language for Software Verification}. In \bibinfo{booktitle}{\emph{Proc.\ SAS}} \emph{(\bibinfo{series}{LNCS~3148})}. \bibinfo{publisher}{Springer}, \bibinfo{pages}{2--18}.
\newblock
\urldef\tempurl%
\url{https://doi.org/10.1007/978-3-540-27864-1_2}
\showDOI{\tempurl}


\bibitem[Beyer et~al\mbox{.}(2022)]%
        {WitnessesJournal}
\bibfield{author}{\bibinfo{person}{D. Beyer}, \bibinfo{person}{M. Dangl}, \bibinfo{person}{D. Dietsch}, \bibinfo{person}{M. Heizmann}, \bibinfo{person}{T. Lemberger}, {and} \bibinfo{person}{M. Tautschnig}.} \bibinfo{year}{2022}\natexlab{}.
\newblock \showarticletitle{Verification Witnesses}.
\newblock \bibinfo{journal}{\emph{ACM Trans. Softw. Eng. Methodol.}} \bibinfo{volume}{31}, \bibinfo{number}{4} (\bibinfo{year}{2022}), \bibinfo{pages}{57:1--57:69}.
\newblock
\urldef\tempurl%
\url{https://doi.org/10.1145/3477579}
\showDOI{\tempurl}


\bibitem[Beyer et~al\mbox{.}(2018)]%
        {AlgorithmComparison-JAR}
\bibfield{author}{\bibinfo{person}{D. Beyer}, \bibinfo{person}{M. Dangl}, {and} \bibinfo{person}{P. Wendler}.} \bibinfo{year}{2018}\natexlab{}.
\newblock \showarticletitle{A Unifying View on {SMT}-Based Software Verification}.
\newblock \bibinfo{journal}{\emph{J. Autom. Reasoning}} \bibinfo{volume}{60}, \bibinfo{number}{3} (\bibinfo{year}{2018}), \bibinfo{pages}{299--335}.
\newblock
\showISSN{1573-0670}
\urldef\tempurl%
\url{https://doi.org/10.1007/s10817-017-9432-6}
\showDOI{\tempurl}


\bibitem[Beyer et~al\mbox{.}(2018)]%
        {HBMC-dataflow}
\bibfield{author}{\bibinfo{person}{D. Beyer}, \bibinfo{person}{S. Gulwani}, {and} \bibinfo{person}{D. Schmidt}.} \bibinfo{year}{2018}\natexlab{}.
\newblock \showarticletitle{Combining Model Checking and Data-Flow Analysis}.
\newblock In \bibinfo{booktitle}{\emph{Handbook of Model Checking}}. \bibinfo{publisher}{Springer}, \bibinfo{pages}{493--540}.
\newblock
\urldef\tempurl%
\url{https://doi.org/10.1007/978-3-319-10575-8_16}
\showDOI{\tempurl}


\bibitem[Beyer et~al\mbox{.}(2005)]%
        {BLAST-memsafety}
\bibfield{author}{\bibinfo{person}{D. Beyer}, \bibinfo{person}{T.~A. Henzinger}, \bibinfo{person}{R. Jhala}, {and} \bibinfo{person}{R. Majumdar}.} \bibinfo{year}{2005}\natexlab{}.
\newblock \showarticletitle{Checking Memory Safety with \textsc{Blast}}. In \bibinfo{booktitle}{\emph{Proc.\ FASE}} \emph{(\bibinfo{series}{LNCS~3442})}. \bibinfo{publisher}{Springer}, \bibinfo{pages}{2--18}.
\newblock
\urldef\tempurl%
\url{https://doi.org/10.1007/978-3-540-31984-9_2}
\showDOI{\tempurl}


\bibitem[Beyer et~al\mbox{.}(2007)]%
        {BLAST}
\bibfield{author}{\bibinfo{person}{D. Beyer}, \bibinfo{person}{T.~A. Henzinger}, \bibinfo{person}{R. Jhala}, {and} \bibinfo{person}{R. Majumdar}.} \bibinfo{year}{2007}\natexlab{}.
\newblock \showarticletitle{The Software Model Checker \textsc{Blast}}.
\newblock \bibinfo{journal}{\emph{Int.\ J.\ Softw.\ Tools Technol.\ Transfer}} \bibinfo{volume}{9}, \bibinfo{number}{5-6} (\bibinfo{year}{2007}), \bibinfo{pages}{505--525}.
\newblock
\urldef\tempurl%
\url{https://doi.org/10.1007/s10009-007-0044-z}
\showDOI{\tempurl}


\bibitem[Beyer and Keremoglu(2011)]%
        {CPACHECKER}
\bibfield{author}{\bibinfo{person}{D. Beyer} {and} \bibinfo{person}{M.~E. Keremoglu}.} \bibinfo{year}{2011}\natexlab{}.
\newblock \showarticletitle{\textsc{CPAchecker}: A Tool for Configurable Software Verification}. In \bibinfo{booktitle}{\emph{Proc.\ CAV}} \emph{(\bibinfo{series}{LNCS~6806})}. \bibinfo{publisher}{Springer}, \bibinfo{pages}{184--190}.
\newblock
\urldef\tempurl%
\url{https://doi.org/10.1007/978-3-642-22110-1_16}
\showDOI{\tempurl}


\bibitem[Beyer and Lee(2024)]%
        {TransformationGame}
\bibfield{author}{\bibinfo{person}{D. Beyer} {and} \bibinfo{person}{N.-Z. Lee}.} \bibinfo{year}{2024}\natexlab{}.
\newblock \showarticletitle{The Transformation Game: Joining Forces for Verification}. \bibinfo{publisher}{Springer}.
\newblock
\urldef\tempurl%
\url{https://www.sosy-lab.org/research/pub/2024-Katoen60.The_Transformation_Game_Joining_Forces_for_Verification.pdf}
\showURL{%
\tempurl}


\bibitem[Beyer et~al\mbox{.}(2022)]%
        {LoopAbstractionCEGAR}
\bibfield{author}{\bibinfo{person}{D. Beyer}, \bibinfo{person}{M. Lingsch-Rosenfeld}, {and} \bibinfo{person}{M. Spiessl}.} \bibinfo{year}{2022}\natexlab{}.
\newblock \showarticletitle{A Unifying Approach for Control-Flow-Based Loop Abstraction}. In \bibinfo{booktitle}{\emph{Proc.\ SEFM}} \emph{(\bibinfo{series}{LNCS~13550})}. \bibinfo{publisher}{Springer}, \bibinfo{pages}{3--19}.
\newblock
\urldef\tempurl%
\url{https://doi.org/10.1007/978-3-031-17108-6_1}
\showDOI{\tempurl}


\bibitem[Beyer et~al\mbox{.}(2023)]%
        {CEGAR-PT}
\bibfield{author}{\bibinfo{person}{D. Beyer}, \bibinfo{person}{M. Lingsch-Rosenfeld}, {and} \bibinfo{person}{M. Spiessl}.} \bibinfo{year}{2023}\natexlab{}.
\newblock \showarticletitle{{CEGAR-PT}: {A} Tool for Abstraction by Program Transformation}. In \bibinfo{booktitle}{\emph{Proc.\ ASE}}. \bibinfo{publisher}{IEEE}, \bibinfo{pages}{2078--2081}.
\newblock
\urldef\tempurl%
\url{https://doi.org/10.1109/ASE56229.2023.00215}
\showDOI{\tempurl}


\bibitem[Beyer et~al\mbox{.}(2019)]%
        {Benchmarking-STTT}
\bibfield{author}{\bibinfo{person}{D. Beyer}, \bibinfo{person}{S. Löwe}, {and} \bibinfo{person}{P. Wendler}.} \bibinfo{year}{2019}\natexlab{}.
\newblock \showarticletitle{Reliable Benchmarking: {R}equirements and Solutions}.
\newblock \bibinfo{journal}{\emph{Int.\ J.\ Softw.\ Tools Technol.\ Transfer}} \bibinfo{volume}{21}, \bibinfo{number}{1} (\bibinfo{year}{2019}), \bibinfo{pages}{1--29}.
\newblock
\urldef\tempurl%
\url{https://doi.org/10.1007/s10009-017-0469-y}
\showDOI{\tempurl}


\bibitem[Beyer and Spiessl(2020)]%
        {MetaVal}
\bibfield{author}{\bibinfo{person}{D. Beyer} {and} \bibinfo{person}{M. Spiessl}.} \bibinfo{year}{2020}\natexlab{}.
\newblock \showarticletitle{\textsc{MetaVal}: {W}itness Validation via Verification}. In \bibinfo{booktitle}{\emph{Proc.\ CAV}} \emph{(\bibinfo{series}{LNCS~12225})}. \bibinfo{publisher}{Springer}, \bibinfo{pages}{165--177}.
\newblock
\urldef\tempurl%
\url{https://doi.org/10.1007/978-3-030-53291-8_10}
\showDOI{\tempurl}


\bibitem[Biere et~al\mbox{.}(2003)]%
        {BMCJournal}
\bibfield{author}{\bibinfo{person}{Armin Biere}, \bibinfo{person}{Alessandro Cimatti}, \bibinfo{person}{Edmund~M. Clarke}, \bibinfo{person}{Ofer Strichman}, {and} \bibinfo{person}{Yunshan Zhu}.} \bibinfo{year}{2003}\natexlab{}.
\newblock \showarticletitle{Bounded model checking}.
\newblock \bibinfo{journal}{\emph{Advances in Computers}}  \bibinfo{volume}{58} (\bibinfo{year}{2003}), \bibinfo{pages}{117--148}.
\newblock
\urldef\tempurl%
\url{https://doi.org/10.1016/S0065-2458(03)58003-2}
\showDOI{\tempurl}


\bibitem[Blatter et~al\mbox{.}(2017)]%
        {RPP2017}
\bibfield{author}{\bibinfo{person}{Lionel Blatter}, \bibinfo{person}{Nikolai Kosmatov}, \bibinfo{person}{Pascale Le~Gall}, {and} \bibinfo{person}{Virgile Prevosto}.} \bibinfo{year}{2017}\natexlab{}.
\newblock \showarticletitle{RPP: {Automatic} Proof of Relational Properties by Self-composition}. In \bibinfo{booktitle}{\emph{Tools and Algorithms for the Construction and Analysis of Systems}}. \bibinfo{pages}{391--397}.
\newblock
\showISBNx{978-3-662-54577-5}


\bibitem[Blatter et~al\mbox{.}(2018)]%
        {RPP2018}
\bibfield{author}{\bibinfo{person}{Lionel Blatter}, \bibinfo{person}{Nikolai Kosmatov}, \bibinfo{person}{Pascale Le~Gall}, \bibinfo{person}{Virgile Prevosto}, {and} \bibinfo{person}{Guillaume Petiot}.} \bibinfo{year}{2018}\natexlab{}.
\newblock \showarticletitle{Static and Dynamic Verification of Relational Properties on Self-composed {C} Code}. In \bibinfo{booktitle}{\emph{Tests and Proofs}}. \bibinfo{pages}{44--62}.
\newblock
\showISBNx{978-3-319-92994-1}


\bibitem[Bodden and Hendren(2012)]%
        {Bodden2012}
\bibfield{author}{\bibinfo{person}{Eric Bodden} {and} \bibinfo{person}{Laurie Hendren}.} \bibinfo{year}{2012}\natexlab{}.
\newblock \showarticletitle{The {Clara} framework for hybrid typestate analysis}.
\newblock \bibinfo{journal}{\emph{International Journal on Software Tools for Technology Transfer}} \bibinfo{volume}{14}, \bibinfo{number}{3} (\bibinfo{date}{01 Jun} \bibinfo{year}{2012}), \bibinfo{pages}{307--326}.
\newblock
\showISSN{1433-2787}
\urldef\tempurl%
\url{https://doi.org/10.1007/s10009-010-0183-5}
\showDOI{\tempurl}


\bibitem[B{\"{o}}hme et~al\mbox{.}(2016)]%
        {AFLfast}
\bibfield{author}{\bibinfo{person}{M. B{\"{o}}hme}, \bibinfo{person}{V.{-}T. Pham}, {and} \bibinfo{person}{A. Roychoudhury}.} \bibinfo{year}{2016}\natexlab{}.
\newblock \showarticletitle{Coverage-Based Greybox Fuzzing as Markov Chain}. In \bibinfo{booktitle}{\emph{Proc.\ {SIGSAC}}}. \bibinfo{publisher}{{ACM}}, \bibinfo{address}{New York, NY, USA}, \bibinfo{pages}{1032--1043}.
\newblock
\urldef\tempurl%
\url{https://doi.org/10.1145/2976749.2978428}
\showDOI{\tempurl}


\bibitem[Bradley(2011)]%
        {IC3}
\bibfield{author}{\bibinfo{person}{A.~R. Bradley}.} \bibinfo{year}{2011}\natexlab{}.
\newblock \showarticletitle{{SAT}-Based model checking without unrolling}. In \bibinfo{booktitle}{\emph{Proc.\ VMCAI}} \emph{(\bibinfo{series}{LNCS~6538})}. \bibinfo{publisher}{Springer}, \bibinfo{pages}{70--87}.
\newblock
\urldef\tempurl%
\url{https://doi.org/10.1007/978-3-642-18275-4_7}
\showDOI{\tempurl}


\bibitem[Chalupa et~al\mbox{.}(2018)]%
        {Symbiotic5ms}
\bibfield{author}{\bibinfo{person}{M. Chalupa}, \bibinfo{person}{J. Strejček}, {and} \bibinfo{person}{M. Vitovsk{\'{a}}}.} \bibinfo{year}{2018}\natexlab{}.
\newblock \showarticletitle{Joint Forces for Memory Safety Checking}. In \bibinfo{booktitle}{\emph{Proc.\ SPIN}}. \bibinfo{publisher}{Springer}, \bibinfo{pages}{115--132}.
\newblock
\urldef\tempurl%
\url{https://doi.org/10.1007/978-3-319-94111-0_7}
\showDOI{\tempurl}


\bibitem[Chien and Lee(2024)]%
        {CPV-SVCOMP24}
\bibfield{author}{\bibinfo{person}{P.-C. Chien} {and} \bibinfo{person}{N.-Z. Lee}.} \bibinfo{year}{2024}\natexlab{}.
\newblock \showarticletitle{\textsc{CPV}: {A} Circuit-Based Program Verifier (Competition Contribution)}. In \bibinfo{booktitle}{\emph{Proc.\ TACAS~(3)}} \emph{(\bibinfo{series}{LNCS~14572})}. \bibinfo{publisher}{Springer}, \bibinfo{pages}{365--370}.
\newblock
\urldef\tempurl%
\url{https://doi.org/10.1007/978-3-031-57256-2_22}
\showDOI{\tempurl}


\bibitem[Clarke et~al\mbox{.}(2004)]%
        {CBMC}
\bibfield{author}{\bibinfo{person}{E.~M. Clarke}, \bibinfo{person}{D. Kröning}, {and} \bibinfo{person}{F. Lerda}.} \bibinfo{year}{2004}\natexlab{}.
\newblock \showarticletitle{A Tool for Checking {ANSI-C} Programs}. In \bibinfo{booktitle}{\emph{Proc.\ TACAS}} \emph{(\bibinfo{series}{LNCS~2988})}. \bibinfo{publisher}{Springer}, \bibinfo{pages}{168--176}.
\newblock
\urldef\tempurl%
\url{https://doi.org/10.1007/978-3-540-24730-2_15}
\showDOI{\tempurl}


\bibitem[Condit et~al\mbox{.}(2003)]%
        {CCured03}
\bibfield{author}{\bibinfo{person}{J. Condit}, \bibinfo{person}{M. Harren}, \bibinfo{person}{S. McPeak}, \bibinfo{person}{G.~C. Necula}, {and} \bibinfo{person}{W. Weimer}.} \bibinfo{year}{2003}\natexlab{}.
\newblock \showarticletitle{\textsc{CCured} in the real world}. In \bibinfo{booktitle}{\emph{Proc.\ PLDI}}. \bibinfo{publisher}{{ACM}}, \bibinfo{pages}{232--244}.
\newblock


\bibitem[Darke et~al\mbox{.}(2023)]%
        {VERIABSL-SVCOMP23}
\bibfield{author}{\bibinfo{person}{P. Darke}, \bibinfo{person}{B. Chimdyalwar}, \bibinfo{person}{S. Agrawal}, \bibinfo{person}{R. Venkatesh}, \bibinfo{person}{S. Chakraborty}, {and} \bibinfo{person}{S. Kumar}.} \bibinfo{year}{2023}\natexlab{}.
\newblock \showarticletitle{\textsc{VeriAbsL}: {Scalable} Verification by Abstraction and Strategy Prediction (Competition Contribution)}. In \bibinfo{booktitle}{\emph{Proc.\ TACAS~(2)}} \emph{(\bibinfo{series}{LNCS~13994})}. \bibinfo{publisher}{Springer}, \bibinfo{pages}{588--593}.
\newblock
\urldef\tempurl%
\url{https://doi.org/10.1007/978-3-031-30820-8_41}
\showDOI{\tempurl}


\bibitem[Dietsch et~al\mbox{.}(2023)]%
        {UTAIPAN-SVCOMP23}
\bibfield{author}{\bibinfo{person}{D. Dietsch}, \bibinfo{person}{M. Heizmann}, \bibinfo{person}{D. Klumpp}, \bibinfo{person}{F. Schüssele}, {and} \bibinfo{person}{A. Podelski}.} \bibinfo{year}{2023}\natexlab{}.
\newblock \showarticletitle{\textsc{Ultimate Taipan} 2023 (Competition Contribution)}. In \bibinfo{booktitle}{\emph{Proc.\ TACAS~(2)}} \emph{(\bibinfo{series}{LNCS~13994})}. \bibinfo{publisher}{Springer}, \bibinfo{pages}{582--587}.
\newblock
\urldef\tempurl%
\url{https://doi.org/10.1007/978-3-031-30820-8_40}
\showDOI{\tempurl}


\bibitem[Donaldson et~al\mbox{.}(2011)]%
        {K-Induction}
\bibfield{author}{\bibinfo{person}{A.~F. Donaldson}, \bibinfo{person}{L. Haller}, \bibinfo{person}{D. Kröning}, {and} \bibinfo{person}{P. R{\"{u}}mmer}.} \bibinfo{year}{2011}\natexlab{}.
\newblock \showarticletitle{Software Verification Using k-Induction}. In \bibinfo{booktitle}{\emph{Proc.\ SAS}} \emph{(\bibinfo{series}{LNCS~6887})}. \bibinfo{publisher}{Springer}, \bibinfo{pages}{351--368}.
\newblock
\urldef\tempurl%
\url{https://doi.org/10.1007/978-3-642-23702-7_26}
\showDOI{\tempurl}


\bibitem[Duret-Lutz and Poitrenaud(2004)]%
        {SPOT}
\bibfield{author}{\bibinfo{person}{Alexandre Duret-Lutz} {and} \bibinfo{person}{Denis Poitrenaud}.} \bibinfo{year}{2004}\natexlab{}.
\newblock \showarticletitle{SPOT: An Extensible Model Checking Library Using Transition-Based Generalized B\"{u}chi Automata}. In \bibinfo{booktitle}{\emph{Proc.\ MASCOTS}}. \bibinfo{publisher}{IEEE}, \bibinfo{pages}{76--83}.
\newblock
\urldef\tempurl%
\url{https://doi.org/10.1109/MASCOT.2004.1348184}
\showDOI{\tempurl}


\bibitem[Fischer et~al\mbox{.}(2013)]%
        {CSEQ2013}
\bibfield{author}{\bibinfo{person}{Bernd Fischer}, \bibinfo{person}{Omar Inverso}, {and} \bibinfo{person}{Gennaro Parlato}.} \bibinfo{year}{2013}\natexlab{}.
\newblock \showarticletitle{\textsc{CSeq}: {A} concurrency pre-processor for sequential {C} verification tools}. In \bibinfo{booktitle}{\emph{Proc.\ ASE}}. \bibinfo{publisher}{IEEE}, \bibinfo{pages}{710--713}.
\newblock
\urldef\tempurl%
\url{https://doi.org/10.1109/ASE.2013.6693139}
\showDOI{\tempurl}


\bibitem[Frohn(2020)]%
        {CalculusModularLoopAcceleration}
\bibfield{author}{\bibinfo{person}{F. Frohn}.} \bibinfo{year}{2020}\natexlab{}.
\newblock \showarticletitle{A Calculus for Modular Loop Acceleration}. In \bibinfo{booktitle}{\emph{Proc. TACAS (1)}} \emph{(\bibinfo{series}{LNCS~12078})}. \bibinfo{publisher}{Springer}, \bibinfo{pages}{58--76}.
\newblock
\urldef\tempurl%
\url{https://doi.org/10.1007/978-3-030-45190-5_4}
\showDOI{\tempurl}


\bibitem[Gastin and Oddoux(2001)]%
        {LTL2BA}
\bibfield{author}{\bibinfo{person}{Paul Gastin} {and} \bibinfo{person}{Denis Oddoux}.} \bibinfo{year}{2001}\natexlab{}.
\newblock \showarticletitle{Fast LTL to B{\"u}chi Automata Translation}. In \bibinfo{booktitle}{\emph{Proc.\ CAV}}. \bibinfo{publisher}{Springer}, \bibinfo{pages}{53--65}.
\newblock
\urldef\tempurl%
\url{https://doi.org/10.1007/3-540-44585-4_6}
\showDOI{\tempurl}


\bibitem[Griggio and Jon\'{a}\v{s}(2023)]%
        {Kratos2}
\bibfield{author}{\bibinfo{person}{A. Griggio} {and} \bibinfo{person}{M. Jon\'{a}\v{s}}.} \bibinfo{year}{2023}\natexlab{}.
\newblock \showarticletitle{\textsc{Kratos2}: An {SMT}-Based Model Checker for Imperative Programs}. In \bibinfo{booktitle}{\emph{Proc.\ CAV}}. \bibinfo{publisher}{Springer}, \bibinfo{pages}{423--436}.
\newblock
\showISBNx{978-3-031-37708-2}
\urldef\tempurl%
\url{https://doi.org/10.1007/978-3-031-37709-9_20}
\showDOI{\tempurl}


\bibitem[Harman(2018)]%
        {TestabilityTransformation-keynote}
\bibfield{author}{\bibinfo{person}{Mark Harman}.} \bibinfo{year}{2018}\natexlab{}.
\newblock \showarticletitle{We Need a Testability Transformation Semantics}. In \bibinfo{booktitle}{\emph{Proc.\ SEFM}} \emph{(\bibinfo{series}{LNCS~10886})}. \bibinfo{publisher}{Springer}, \bibinfo{pages}{3--17}.
\newblock
\urldef\tempurl%
\url{https://doi.org/10.1007/978-3-319-92970-5_1}
\showDOI{\tempurl}


\bibitem[Harman et~al\mbox{.}(2004)]%
        {TestabilityTransformation}
\bibfield{author}{\bibinfo{person}{M. Harman}, \bibinfo{person}{L. Hu}, \bibinfo{person}{R.~M. Hierons}, \bibinfo{person}{J. Wegener}, \bibinfo{person}{H. Sthamer}, \bibinfo{person}{A. Baresel}, {and} \bibinfo{person}{M. Roper}.} \bibinfo{year}{2004}\natexlab{}.
\newblock \showarticletitle{Testability Transformation}.
\newblock \bibinfo{journal}{\emph{{IEEE} Trans. Softw. Eng.}} \bibinfo{volume}{30}, \bibinfo{number}{1} (\bibinfo{year}{2004}), \bibinfo{pages}{3--16}.
\newblock
\urldef\tempurl%
\url{https://doi.org/10.1109/TSE.2004.1265732}
\showDOI{\tempurl}


\bibitem[Heizmann et~al\mbox{.}(2024)]%
        {ULTIMATEAUTOMIZER-SVCOMP24}
\bibfield{author}{\bibinfo{person}{M. Heizmann}, \bibinfo{person}{M. Bentele}, \bibinfo{person}{D. Dietsch}, \bibinfo{person}{X. Jiang}, \bibinfo{person}{D. Klumpp}, \bibinfo{person}{F. Schüssele}, {and} \bibinfo{person}{A. Podelski}.} \bibinfo{year}{2024}\natexlab{}.
\newblock \showarticletitle{Ultimate Automizer and the Abstraction of Bitwise Operations (Competition Contribution)}. In \bibinfo{booktitle}{\emph{Proc.\ TACAS~(3)}} \emph{(\bibinfo{series}{LNCS~14572})}. \bibinfo{publisher}{Springer}, \bibinfo{pages}{418--423}.
\newblock
\urldef\tempurl%
\url{https://doi.org/10.1007/978-3-031-57256-2_31}
\showDOI{\tempurl}


\bibitem[Inverso et~al\mbox{.}(2015)]%
        {LAZYCSEQ-ASE}
\bibfield{author}{\bibinfo{person}{Omar Inverso}, \bibinfo{person}{T.~L. Nguyen}, \bibinfo{person}{Bernd Fischer}, \bibinfo{person}{S. {La~Torre}}, {and} \bibinfo{person}{Gennaro Parlato}.} \bibinfo{year}{2015}\natexlab{}.
\newblock \showarticletitle{\textsc{Lazy-CSeq}: {A} Context-Bounded Model Checking Tool for Multi-threaded {C} Programs}. In \bibinfo{booktitle}{\emph{Proc.\ ASE}}. \bibinfo{publisher}{{IEEE}}, \bibinfo{pages}{807--812}.
\newblock
\urldef\tempurl%
\url{https://doi.org/10.1109/ASE.2015.108}
\showDOI{\tempurl}


\bibitem[Jakobsson et~al\mbox{.}(2015)]%
        {ShadowMemory2015}
\bibfield{author}{\bibinfo{person}{Arvid Jakobsson}, \bibinfo{person}{Nikolai Kosmatov}, {and} \bibinfo{person}{Julien Signoles}.} \bibinfo{year}{2015}\natexlab{}.
\newblock \showarticletitle{Fast as a shadow, expressive as a tree: hybrid memory monitoring for {C}}. In \bibinfo{booktitle}{\emph{Proceedings of the 30th Annual ACM Symposium on Applied Computing}} \emph{(\bibinfo{series}{SAC '15})}. \bibinfo{publisher}{ACM}, \bibinfo{pages}{1765--1772}.
\newblock
\showISBNx{9781450331968}
\urldef\tempurl%
\url{https://doi.org/10.1145/2695664.2695815}
\showDOI{\tempurl}


\bibitem[Jeannet et~al\mbox{.}(2014)]%
        {AbstractAccelerationGeneralLinearLoops}
\bibfield{author}{\bibinfo{person}{B. Jeannet}, \bibinfo{person}{P. Schrammel}, {and} \bibinfo{person}{S. Sankaranarayanan}.} \bibinfo{year}{2014}\natexlab{}.
\newblock \showarticletitle{Abstract acceleration of general linear loops}. In \bibinfo{booktitle}{\emph{Proc. POPL}}. \bibinfo{publisher}{ACM}, \bibinfo{pages}{529--540}.
\newblock
\urldef\tempurl%
\url{https://doi.org/10.1145/2535838.2535843}
\showDOI{\tempurl}


\bibitem[Jonáš et~al\mbox{.}(2024)]%
        {SYMBIOTIC-SVCOMP24}
\bibfield{author}{\bibinfo{person}{M. Jonáš}, \bibinfo{person}{K. Kumor}, \bibinfo{person}{J. Novák}, \bibinfo{person}{J. Sedláček}, \bibinfo{person}{M. Trtík}, \bibinfo{person}{L. Zaoral}, \bibinfo{person}{P. Ayaziová}, {and} \bibinfo{person}{J. Strejček}.} \bibinfo{year}{2024}\natexlab{}.
\newblock \showarticletitle{\textsc{Symbiotic 10}: {Lazy} Memory Initialization and Compact Symbolic Execution (Competition Contribution)}. In \bibinfo{booktitle}{\emph{Proc.\ TACAS~(3)}} \emph{(\bibinfo{series}{LNCS~14572})}. \bibinfo{publisher}{Springer}, \bibinfo{pages}{406--411}.
\newblock
\urldef\tempurl%
\url{https://doi.org/10.1007/978-3-031-57256-2_29}
\showDOI{\tempurl}


\bibitem[Julien(2022)]%
        {EACSL}
\bibfield{author}{\bibinfo{person}{S. Julien}.} \bibinfo{year}{2022}\natexlab{}.
\newblock \bibinfo{title}{{E-ACSL}: {Executable} {ANSI}/{ISO} {C} Specification Language}.
\newblock
\newblock
\newblock
\shownote{Available at \url{http://frama-c.com/download/e-acsl/e-acsl.pdf}}.


\bibitem[Madhukar et~al\mbox{.}(2015)]%
        {AcceleratingInvariantGeneration}
\bibfield{author}{\bibinfo{person}{K. Madhukar}, \bibinfo{person}{B. Wachter}, \bibinfo{person}{D. Kröning}, \bibinfo{person}{M. Lewis}, {and} \bibinfo{person}{M.~K. Srivas}.} \bibinfo{year}{2015}\natexlab{}.
\newblock \showarticletitle{Accelerating Invariant Generation}. In \bibinfo{booktitle}{\emph{Proc. FMCAD}}. \bibinfo{publisher}{IEEE}, \bibinfo{pages}{105--111}.
\newblock


\bibitem[Malík et~al\mbox{.}(2023)]%
        {2LS-SVCOMP23}
\bibfield{author}{\bibinfo{person}{V. Malík}, \bibinfo{person}{P. Schrammel}, \bibinfo{person}{T. Vojnar}, {and} \bibinfo{person}{F. Nečas}.} \bibinfo{year}{2023}\natexlab{}.
\newblock \showarticletitle{\textsc{2LS}: {Arrays} and Loop Unwinding (Competition Contribution)}. In \bibinfo{booktitle}{\emph{Proc.\ TACAS~(2)}} \emph{(\bibinfo{series}{LNCS~13994})}. \bibinfo{publisher}{Springer}, \bibinfo{pages}{529--534}.
\newblock
\urldef\tempurl%
\url{https://doi.org/10.1007/978-3-031-30820-8_31}
\showDOI{\tempurl}


\bibitem[McMillan(2003)]%
        {McMillanCraig}
\bibfield{author}{\bibinfo{person}{K.~L. McMillan}.} \bibinfo{year}{2003}\natexlab{}.
\newblock \showarticletitle{Interpolation and {SAT}-Based Model Checking}. In \bibinfo{booktitle}{\emph{Proc.\ CAV}} \emph{(\bibinfo{series}{LNCS~2725})}. \bibinfo{publisher}{Springer}, \bibinfo{pages}{1--13}.
\newblock
\urldef\tempurl%
\url{https://doi.org/10.1007/978-3-540-45069-6_1}
\showDOI{\tempurl}


\bibitem[Metta et~al\mbox{.}(2024)]%
        {PROTON-SVCOMP24}
\bibfield{author}{\bibinfo{person}{R. Metta}, \bibinfo{person}{H. Karmarkar}, \bibinfo{person}{K. Madhukar}, \bibinfo{person}{R. Venkatesh}, {and} \bibinfo{person}{S. Chakraborty}.} \bibinfo{year}{2024}\natexlab{}.
\newblock \showarticletitle{\textsc{Proton}: {Probes} for Non-termination and Termination (Competition Contribution)}. In \bibinfo{booktitle}{\emph{Proc.\ TACAS~(3)}} \emph{(\bibinfo{series}{LNCS~14572})}. \bibinfo{publisher}{Springer}, \bibinfo{pages}{393--398}.
\newblock
\urldef\tempurl%
\url{https://doi.org/10.1007/978-3-031-57256-2_27}
\showDOI{\tempurl}


\bibitem[Necula(1997)]%
        {PCC}
\bibfield{author}{\bibinfo{person}{G.~C. Necula}.} \bibinfo{year}{1997}\natexlab{}.
\newblock \showarticletitle{Proof-Carrying Code}. In \bibinfo{booktitle}{\emph{Proc.\ POPL}}. \bibinfo{publisher}{{ACM}}, \bibinfo{pages}{106--119}.
\newblock
\urldef\tempurl%
\url{https://doi.org/10.1145/263699.263712}
\showDOI{\tempurl}


\bibitem[Necula et~al\mbox{.}(2002)]%
        {CIL}
\bibfield{author}{\bibinfo{person}{G.~C. Necula}, \bibinfo{person}{S. McPeak}, \bibinfo{person}{S.~P. Rahul}, {and} \bibinfo{person}{W. Weimer}.} \bibinfo{year}{2002}\natexlab{}.
\newblock \showarticletitle{\textsc{Cil}: {I}ntermediate Language and Tools for Analysis and Transformation of {C} Programs}. In \bibinfo{booktitle}{\emph{Proc.\ CC}} \emph{(\bibinfo{series}{LNCS~2304})}. \bibinfo{publisher}{Springer}, \bibinfo{pages}{213--228}.
\newblock
\urldef\tempurl%
\url{https://doi.org/10.1007/3-540-45937-5_16}
\showDOI{\tempurl}


\bibitem[Necula et~al\mbox{.}(2002)]%
        {CCured02}
\bibfield{author}{\bibinfo{person}{G.~C. Necula}, \bibinfo{person}{S. McPeak}, {and} \bibinfo{person}{W. Weimer}.} \bibinfo{year}{2002}\natexlab{}.
\newblock \showarticletitle{\textsc{CCured}: Type-Safe Retrofitting of Legacy Code}. In \bibinfo{booktitle}{\emph{Proc.\ POPL}}. \bibinfo{publisher}{{ACM}}, \bibinfo{pages}{128--139}.
\newblock
\urldef\tempurl%
\url{https://doi.org/10.1145/503272.503286}
\showDOI{\tempurl}


\bibitem[Partsch and Steinbrüggen(1983)]%
        {ProgramTransformationSystems}
\bibfield{author}{\bibinfo{person}{Helmuth Partsch} {and} \bibinfo{person}{Ralf Steinbrüggen}.} \bibinfo{year}{1983}\natexlab{}.
\newblock \showarticletitle{Program Transformation Systems}.
\newblock \bibinfo{journal}{\emph{ACM Comput. Surv.}} \bibinfo{volume}{15}, \bibinfo{number}{3} (\bibinfo{year}{1983}), \bibinfo{pages}{199--236}.
\newblock
\urldef\tempurl%
\url{https://doi.org/10.1145/356914.356917}
\showDOI{\tempurl}


\bibitem[Peringer et~al\mbox{.}(2020)]%
        {PREDATORHP-SVCOMP20}
\bibfield{author}{\bibinfo{person}{P. Peringer}, \bibinfo{person}{V. Šoková}, {and} \bibinfo{person}{T. Vojnar}.} \bibinfo{year}{2020}\natexlab{}.
\newblock \showarticletitle{\textsc{PredatorHP} Revamped (Not Only) for Interval-Sized Memory Regions and Memory Reallocation (Competition Contribution)}. In \bibinfo{booktitle}{\emph{Proc.\ TACAS~(2)}} \emph{(\bibinfo{series}{LNCS~12079})}. \bibinfo{publisher}{Springer}, \bibinfo{pages}{408--412}.
\newblock
\urldef\tempurl%
\url{https://doi.org/10.1007/978-3-030-45237-7_30}
\showDOI{\tempurl}


\bibitem[{Petric Maretić} et~al\mbox{.}(2014)]%
        {PETRICMARETIC2014408}
\bibfield{author}{\bibinfo{person}{Grgur {Petric Maretić}}, \bibinfo{person}{Mohammad {Torabi Dashti}}, {and} \bibinfo{person}{David Basin}.} \bibinfo{year}{2014}\natexlab{}.
\newblock \showarticletitle{LTL is closed under topological closure}.
\newblock \bibinfo{journal}{\emph{Inform. Process. Lett.}} \bibinfo{volume}{114}, \bibinfo{number}{8} (\bibinfo{year}{2014}), \bibinfo{pages}{408--413}.
\newblock
\showISSN{0020-0190}
\urldef\tempurl%
\url{https://doi.org/10.1016/j.ipl.2014.03.001}
\showDOI{\tempurl}


\bibitem[Piterman and Pnueli(2018)]%
        {HBMC-TemporalLogic}
\bibfield{author}{\bibinfo{person}{Nir Piterman} {and} \bibinfo{person}{Amir Pnueli}.} \bibinfo{year}{2018}\natexlab{}.
\newblock \showarticletitle{Temporal Logic and Fair Discrete Systems}.
\newblock In \bibinfo{booktitle}{\emph{Handbook of Model Checking}}. \bibinfo{publisher}{Springer}, \bibinfo{pages}{27--73}.
\newblock
\urldef\tempurl%
\url{https://doi.org/10.1007/978-3-319-10575-8_2}
\showDOI{\tempurl}


\bibitem[Robles et~al\mbox{.}(2021)]%
        {MetAcsl2021}
\bibfield{author}{\bibinfo{person}{Virgile Robles}, \bibinfo{person}{Nikolai Kosmatov}, \bibinfo{person}{Virgile Prevosto}, \bibinfo{person}{Louis Rilling}, {and} \bibinfo{person}{Pascale~Le Gall}.} \bibinfo{year}{2021}\natexlab{}.
\newblock \showarticletitle{Methodology for Specification and Verification of High-Level Requirements with \textsc{MetAcsl}}. In \bibinfo{booktitle}{\emph{FormaliSE}}. \bibinfo{pages}{54--67}.
\newblock
\urldef\tempurl%
\url{https://doi.org/10.1109/FormaliSE52586.2021.00012}
\showDOI{\tempurl}


\bibitem[Robles et~al\mbox{.}(2019)]%
        {MetAcsl2019}
\bibfield{author}{\bibinfo{person}{Virgile Robles}, \bibinfo{person}{Nikolai Kosmatov}, \bibinfo{person}{Virgile Prevosto}, \bibinfo{person}{Louis Rilling}, {and} \bibinfo{person}{Pascale Le~Gall}.} \bibinfo{year}{2019}\natexlab{}.
\newblock \showarticletitle{\textsc{MetAcsl}: {Specification} and Verification of High-Level Properties}. In \bibinfo{booktitle}{\emph{Tools and Algorithms for the Construction and Analysis of Systems}}. \bibinfo{publisher}{Springer}, \bibinfo{pages}{358--364}.
\newblock
\showISBNx{978-3-030-17462-0}


\bibitem[Schuppan and Biere(2006)]%
        {LivenessAsSafety}
\bibfield{author}{\bibinfo{person}{Viktor Schuppan} {and} \bibinfo{person}{Armin Biere}.} \bibinfo{year}{2006}\natexlab{}.
\newblock \showarticletitle{Liveness Checking as Safety Checking for Infinite State Spaces}.
\newblock \bibinfo{journal}{\emph{Electr. Notes Theor. Comput. Sci.}} \bibinfo{volume}{149}, \bibinfo{number}{1} (\bibinfo{year}{2006}), \bibinfo{pages}{79--96}.
\newblock
\urldef\tempurl%
\url{https://doi.org/10.1016/j.entcs.2005.11.018}
\showDOI{\tempurl}


\bibitem[\serysort{}O. {\^{S}}erý(2009)]%
        {SeryBLAST}
\bibfield{author}{\bibinfo{person}{\serysort{}O. {\^{S}}erý}.} \bibinfo{year}{2009}\natexlab{}.
\newblock \showarticletitle{Enhanced Property Specification and Verification in \textsc{Blast}}. In \bibinfo{booktitle}{\emph{Proc.\ FASE}} \emph{(\bibinfo{series}{LNCS~5503})}. \bibinfo{publisher}{Springer}, \bibinfo{pages}{456--469}.
\newblock
\urldef\tempurl%
\url{https://doi.org/10.1007/978-3-642-00593-0_32}
\showDOI{\tempurl}


\bibitem[Silverman and Kincaid(2019)]%
        {Silverman19}
\bibfield{author}{\bibinfo{person}{J. Silverman} {and} \bibinfo{person}{Z. Kincaid}.} \bibinfo{year}{2019}\natexlab{}.
\newblock \showarticletitle{Loop Summarization with Rational Vector Addition Systems}. In \bibinfo{booktitle}{\emph{Proc.\ CAV, Part~2}} \emph{(\bibinfo{series}{LNCS~11562})}. \bibinfo{publisher}{Springer}, \bibinfo{pages}{97--115}.
\newblock
\urldef\tempurl%
\url{https://doi.org/10.1007/978-3-030-25543-5_7}
\showDOI{\tempurl}


\bibitem[Visser(2001)]%
        {ProgramTransformationSystemsSurvey}
\bibfield{author}{\bibinfo{person}{Eelco Visser}.} \bibinfo{year}{2001}\natexlab{}.
\newblock \showarticletitle{A Survey of Strategies in Program Transformation Systems}. In \bibinfo{booktitle}{\emph{Proc.\ WRS}} \emph{(\bibinfo{series}{ENTCS~57})}. \bibinfo{publisher}{Elsevier}, \bibinfo{pages}{109--143}.
\newblock
\urldef\tempurl%
\url{https://doi.org/10.1016/S1571-0661(04)00270-1}
\showDOI{\tempurl}


\bibitem[Vorobyov et~al\mbox{.}(2017)]%
        {ShadowMemory2017}
\bibfield{author}{\bibinfo{person}{Kostyantyn Vorobyov}, \bibinfo{person}{Julien Signoles}, {and} \bibinfo{person}{Nikolai Kosmatov}.} \bibinfo{year}{2017}\natexlab{}.
\newblock \showarticletitle{Shadow state encoding for efficient monitoring of block-level properties}. In \bibinfo{booktitle}{\emph{Proceedings of the 2017 ACM SIGPLAN International Symposium on Memory Management}}. \bibinfo{publisher}{ACM}, \bibinfo{pages}{47--58}.
\newblock
\showISBNx{9781450350440}
\urldef\tempurl%
\url{https://doi.org/10.1145/3092255.3092269}
\showDOI{\tempurl}


\bibitem[Zhang et~al\mbox{.}(2023)]%
        {EndWatch}
\bibfield{author}{\bibinfo{person}{Yao Zhang}, \bibinfo{person}{Xiaofei Xie}, \bibinfo{person}{Yi Li}, \bibinfo{person}{Sen Chen}, \bibinfo{person}{Cen Zhang}, {and} \bibinfo{person}{Xiaohong Li}.} \bibinfo{year}{2023}\natexlab{}.
\newblock \showarticletitle{{EndWatch}: {A} Practical Method for Detecting Non-Termination in Real-World Software}. In \bibinfo{booktitle}{\emph{Proc.\ ASE}}. \bibinfo{pages}{686--697}.
\newblock
\urldef\tempurl%
\url{https://doi.org/10.1109/ASE56229.2023.00061}
\showDOI{\tempurl}


\end{thebibliography}
